\documentclass[fleqn,usenatbib]{mnras}

\usepackage{newtxtext,newtxmath}

\usepackage[T1]{fontenc}

\DeclareRobustCommand{\VAN}[3]{#2}
\let\VANthebibliography\thebibliography
\def\thebibliography{\DeclareRobustCommand{\VAN}[3]{##3}\VANthebibliography}

\usepackage{graphicx}	

\usepackage[labelfont=bf]{caption}  
\usepackage{threeparttable} 
\usepackage{subcaption} 

\title[Physical modelling of NEA (23187) 2000 PN9]{Physical modelling of near-Earth asteroid (23187) 2000 PN9 with ground-based optical and radar observations}

\author[L. Dover]{
L. Dover$^{1}$\thanks{E-mail: lorddover@me.com},
S.~C.~Lowry$^{1}$,
A.~Ro\.zek$^{2,1}$,
B.~Rozitis$^{3}$,
S. L. Jackson$^{3}$,
T.~Zegmott$^{1}$,
Yu.~N.~Krugly$^{4,5}$,
\newauthor
I.~N.~Belskaya$^ {4,6}$,
A.~Fitzsimmons$^{7}$,
S.~F.~Green$^{3}$,
C.~Snodgrass$^{2}$,
P.~R.~Weissman$^{8}$,
M.~Brozovi\'c$^{9}$,
\newauthor
L. A. M. Benner$^{9}$,
M. W. Busch$^{10}$,
V. R. Ayvazian$^{11,12}$,
V. Chiorny$^{4}$,
R. Ya. Inasaridze$^{11,12}$,
M. Krugov$^{13}$,
\newauthor
S. Mykhailova$^{4,5}$,
I. Reva$^{13}$,
and J. Hibbert$^{14}$
\\
$^{1}$Centre for Astrophysics and Planetary Science, University of Kent, Canterbury, UK\\
$^{2}$Institute for Astronomy, University of Edinburgh, Royal Observatory, Edinburgh, UK\\
$^{3}$Planetary and Space Sciences, School of Physical Sciences, The Open University, Milton Keynes, UK\\
$^{4}$Institute of Astronomy, V. N. Karazin Kharkiv National University, Kharkiv, Ukraine\\
$^{5}$Astronomical Observatory Institute, Faculty of Physics, A. Mickiewicz University, Poznan, Poland \\
$^{6}$LESIA, Observatoire de Paris, Université PSL, CNRS, Université Paris Cité, Sorbonne Université, Meudon, France\\
$^{7}$Astrophysics Research Centre, Queens University Belfast, Belfast, UK\\
$^{8}$Planetary Sciences Institute, Tucson, Arizona, USA\\
$^{9}$Jet Propulsion Laboratory, California Institute of Technology, USA\\
$^{10}$SETI Institute, Mountain View, California, USA\\
$^{11}$E. Kharadze Georgian National Astrophysical Observatory, Abastumani, Georgia\\
$^{12}$Samtskhe-Javakheti State University, Akhaltsikhe, Georgia\\
$^{13}$Fesenkov Astrophysical Institute, Almaty, Kazakhstan\\
$^{14}$Isaac Newton Group, Apartado de correos 321, 38700, Santa Cruz de La Palma, Canary Islands, Spain\\
}

\date{Accepted XXX. Received YYY; in original form ZZZ}

\pubyear{2023}

\begin{document}
\label{firstpage}
\pagerange{\pageref{firstpage}--\pageref{lastpage}}
\maketitle

\begin{abstract}
We present a physical model and spin-state analysis of the potentially hazardous asteroid (23187) 2000 PN9. As part of a long-term campaign to make direct detections of the YORP effect, we collected optical lightcurves of the asteroid between 2006 and 2020. These observations were combined with planetary radar data to develop a detailed shape model which was used to search for YORP acceleration. We report that 2000 PN9 is a relatively large top-shaped body with a sidereal rotation period of 2.53216$\pm$0.00015 h. Although we find no evidence for rotational acceleration, YORP torques smaller than $\sim$10$^{-8}$$\,\rm rad/day^{2}$ cannot be ruled out. It is likely that 2000 PN9 is a YORP-evolved object, and may be an example of YORP equilibrium or self limitation. 
\end{abstract}




\section{Introduction}
The Yarkovsky-O’Keefe-Radzievskii-Paddack (YORP) effect is a thermal torque caused by the reflection, absorption and anisotropic re-emission of Solar radiation \citep{rubincam_radiative_2000}.  The YORP effect can change the rotation period and spin axis orientation of small bodies, and is a key mechanism in their evolution. It can trigger the formation of binary asteroids, deliver asteroids to Earth-crossing orbits \citep{bottke_yarkovsky_2006} and cause spin-axis alignment in asteroidal families \citep{vokrouhlicky_vector_2003}. 
There have been eleven direct detections of the YORP effect to date: (54509) YORP, (1862) Apollo, (1620) Geographos, (3103) Eger, (25143) Itokawa, (161989) Cacus, (101955) Bennu, (68346) 2001 KZ66, (10115) 1992 SK and (1685) Toro \citep{lowry_direct_2007, taylor_spin_2007, 2007Natur.446..420K, Durech:2008di, Durech:2012bq, lowry_internal_2014, Durech:2018gg, Nolan:2019eib,2021MNRAS.507.4914Z,2022A&A...657A...5D}. Ten of these detections have a YORP acceleration below 10$^{-7}$$\,\rm rad/day^{2}$, with the much smaller asteroid (54509) YORP having a detected YORP spin-up rate of 3.49$\times$10$^{-6}$$\,\rm rad/day^{2}$.
The YORP effect is expected to appear in both spin-up and spin-down configurations, yet all detections to date are in the spin-up case. This could be due to a physical process causing an excess of positive torques, such as such as tangential YORP \citep{golubov_tangential_2012}. The lack of spin-down detections may also be a consequence of observational bias, as objects experiencing rotational deceleration will generally have longer periods. For periods greater than $\sim$8 h, it is difficult to obtain full rotational coverage with a single lightcurve. Asteroids with shorter rotation periods can more readily be observed in a single night, making them more lucrative targets. Objects with fast rotation are more likely to be in a spin-up configuration, hence there is a bias towards detections of spin-up YORP.
In order to better understand the YORP effect and its important influence on the evolution of the Solar System, further detections (and non-detections) must be made. As the process of making a YORP detection includes the development of a detailed physical model, the pursuit of YORP detections provides wider benefits to the overall understanding of asteroid evolution.
Since April 2010, our group has been monitoring a selection of small asteroids that are strong candidates for direct detection of the YORP effect. The majority of these observations were conducted through a European Southern Observatory Large Programme with the 3.6 m New Technology Telescope (NTT) at La Silla, Chile. Accompanying observations have been made with various small and medium sized telescopes, with most imaging conducted at optical wavelengths. Asteroids that are closer to the Sun are exposed to more solar insolation, which in turn increases the strength of the YORP effect, thus all of the targets are near-Earth asteroids (NEAs). Asteroids were selected for their long-term observability, the range of achievable viewing geometries and their short rotation periods. By selecting targets with short rotation periods, we could ensure that several full rotations could be observed over the course of a single night, or several nights with lightcurve folding. 
Aside from observational constraints, the short-period (<8 h) regime is critical to understanding the fate of asteroids in a spin-up configuration. Objects must either reach a state of rotational equilibrium or accelerate beyond the spin-breakup barrier and experience a disruptive event. Probing asteroids that are close to the breakup limit thus makes it possible to link each asteroid's physical properties not only to its YORP state, but to its evolutionary track.
This study focuses on one target from our campaign, (23187) 2000 PN9 (herafter PN9), which was observed using optical and planetary radar facilities between 2001 and 2020. PN9 is an Apollo-class NEA that has been designated as a potentially hazardous asteroid. It was discovered by the Lincoln Near-Earth Asteroid Research (LINEAR) tracking programme in August 2000 \citep{2000MPEC....P...48M} with a semi-major axis of 1.85 AU and an eccentricity of 0.59.

Using optical observations in 2001 and 2006, \cite{belskaya_polarimetry_2009} determined a synodic rotation period of 2.5325$\pm$0.0004 h, a lightcurve amplitude of 0.13 mag, a polarimetrically-derived albedo of 0.24$\pm$0.06 and an absolute magnitude $H=16.2$, resulting in a diameter of 1.6$\pm$0.3 km. \cite{2006SASS...25..169B} determined that the asteroid is roughly spherical with an approximate diameter of 2 km from radar observations made in 2001. A synodic rotation period of 2.537$\pm$0.002 h was reported from optical observations in 2016 \citep{2016MPBu...43..240W}. MIT-Hawaii Near-Earth Object Spectroscopic Survey (MITHNEOS) observations have lead to PN9 being  classified as belonging to either the S/Sq, Sq or Sq/Q taxonomic types \citep{2014Icar..228..217T,2019Icar..324...41B}.

In Section (\ref{sec:observations}) of this paper, we describe the optical and radar observations that were used in this analysis. Section (\ref{sec:modelling}) describes the process of developing a physical model for PN9, presenting both a lightcurve-only model and a combined optical and radar model. We also describe an analysis of PN9's rotational phase to search for evidence of YORP acceleration. In Section (\ref{sec:analysis}) we discuss the physical parameters of PN9, the implications of our YORP non-detection, and the importance of studying more top-shaped asteroids similar to PN9.

\section{Observations of (23187) 2000 PN9}\label{sec:observations}
\subsection{Optical lightcurves}
\begin{figure*}
	\includegraphics[width=16cm,trim=0 0 0 0,clip=true]{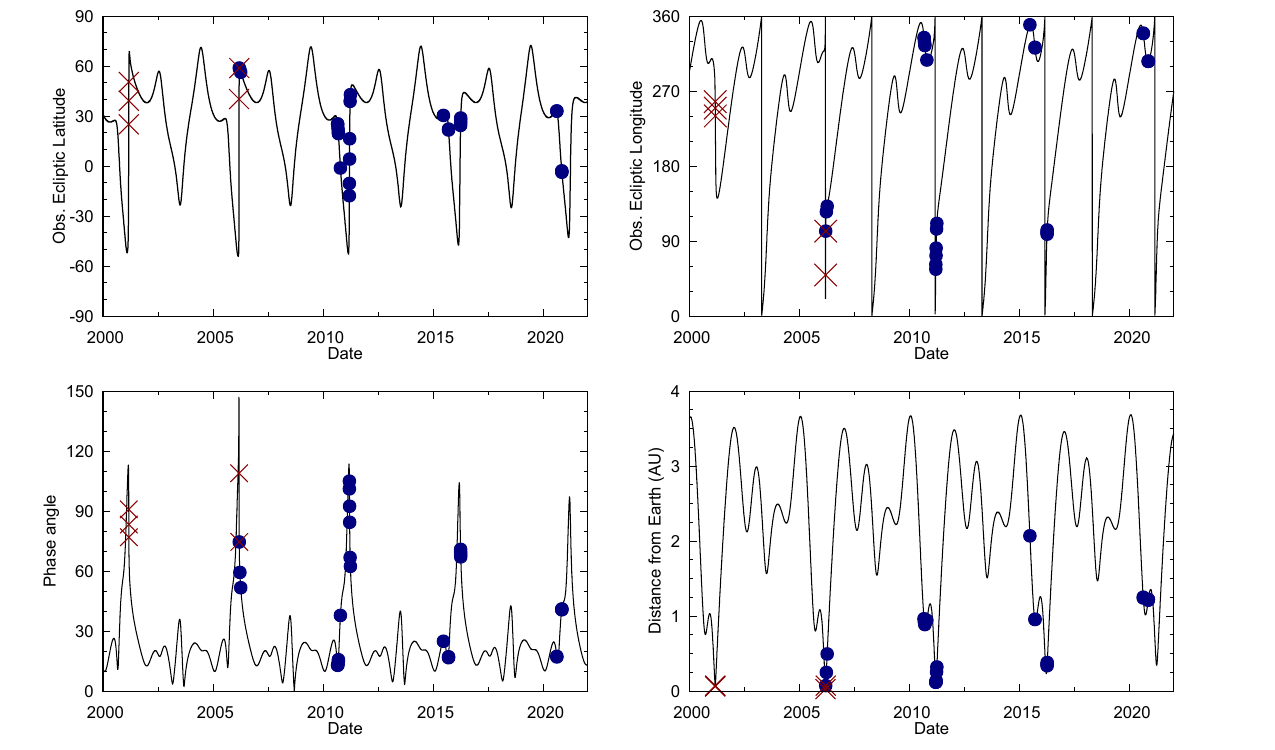}
	\caption{Observing geometries for the asteroid (23187) 2000 PN9 from the 2000 to the start of 2022. The top panels show the position of the asteroid in the ecliptic coordinate system (latitude and longitude) as observed from Earth. The bottom left panel shows the solar phase angle while the bottom right panel shows the geocentric distance to the asteroid. The marked points denote observations of the asteroid. Optical lightcurves are marked as blue circles and radar observations are represented by red crosses. 
	}
	\label{fig:viewgeom}
\end{figure*}

Our optical lightcurve dataset for PN9 spans fourteen years, from March 2006 to November 2020. Each lightcurve is summarised in Table \ref{tab:obstable}, with indications of how each lightcurve was used in this work. Some low-quality lightcurves were not used for modelling or spin-state analysis. The subset of lightcurves used in modelling was used for both the lightcurve-only (Sect. \ref{sec:LCmodel}) and combined lightcurve and radar (Sect. \ref{sec:shape}) models.

As shown in Figure \ref{fig:viewgeom}, the asteroid was observed over a range of viewing geometries during the fourteen years of observation. This is important for shape modelling, as the entire surface of the asteroid cannot be seen with a single viewing geometry. Viewing the asteroid's rotation from different aspect angles, and under different shadowing conditions, can greatly improve constraints on its shape and rotational state. For YORP detections it is also important to regularly view the asteroid at similar viewing geometries, where the lightcurve shape is similar, to constrain rotational phase offset measurements. The distribution of optical observations for PN9 is hence favourable as it includes both repeating and varied viewing geometries.

Rotational lightcurves were extracted using relative photometry, comparing the asteroid's brightness to a selection of stable background stars. In some cases, sidereal tracking was used if a desirable signal-to-noise ratio (SNR) could be achieved on the asteroid without its full width at half maximum (FWHM) profile exceeding atmospheric seeing. This ensured that a circular aperture with a radius of twice the FWHM profile could be used for photometry. Otherwise, the asteroid was differentially tracked and exposure times were set to avoid trailing of the background stars beyond atmospheric seeing. Consideration was also taken to ensure a sufficient temporal resolution was achieved, i.e. each exposure was not a significant fraction of the asteroid's rotation period. All images were processed using standard CCD reduction procedures with bias subtraction and flat field correction, along with dark field correction where necessary.

Our optical dataset includes lightcurves from ten different observatories. It should be noted that the choice of optical filter has a negligible impact on the shape and amplitude of relative lightcurves, hence we have included observations using a variety of broadband and clear filters. Information on the observations that were conducted at each observatory are given below.

\begin{table*}
        \centering  
        \begin{threeparttable}
        \caption{All optical lightcurves of 2000 PN9 considered in this study.}             
        \label{tab:obstable}      
        	\begin{tabular}{ccccccccccccc}
		\hline \hline \noalign{\smallskip}
ID & UT date   & {$R_\odot$} & {$\Delta_\oplus$} & {$\alpha$}  & $\lambda_O$ &  $\beta_O$  & Total   & Obs. & Filter & Included & Included & Reference\\ 
&  {[yyyy-mm-dd]}   & {[AU]}  &   {[AU]}   & {[$\degr$]} & {[$\degr$]} & {[$\degr$]} & [hour] & facility  & & (model) & (ph. off) & \\ \hline 
\noalign{\smallskip}
1 & 2006-03-10 & 1.098 & 0.072 & 75.55 & 101.9 & 58.89 & 0.3 &   ChO    &    B     &    & \ & 1 \\  
2 &  & "" & "" & "" & "" & "" & 0.5 &   ChO    &    V     &  &  &   1  \\  
3 &  & "" & "" & "" & "" & "" & 0.7 &   ChO    &    R     &  &  &   1  \\  
4 &  & "" & "" & "" & "" & "" & 0.3 &   ChO    &    I     &  & \ &   1  \\  
5 & 2006-03-20 & 1.099 & 0.249 & 59.37 & 125.2 & 57.85 & 4.2 &   ChO    &    R     & \textbullet & \textbullet &  1 \\  
6 & 2006-04-03 & 1.228 & 0.495 & 51.57 & 131.9 & 56.30 & 1.3 &   ChO    &    R     &  & &   1 \\  
7 & 2010-08-28 & 1.927 & 0.965 & 12.96 & 334.2 & 25.3 & 2.4 &   NTT    &    R     & & \textbullet   &    \\  
8 & 2010-08-29 & 1.920 & 0.956 & 12.87 & 333.5 & 25.0 & 3.3 &   NTT    &    R     & \textbullet & \textbullet &   \\  
9 & 2010-09-03 & 1.876 & 0.915 & 13.36 & 329.1 & 22.5 & 1.3 &   ESOD    &    R     &  &  &    \\  
10 & 2010-09-08 & 1.846 & 0.894 & 14.72 & 326.2 & 20.6 & 4.8 &   TMO    &    R     &  &  &  \\  
11 & 2010-09-09 & 1.839 & 0.890 & 15.18 & 325.5 & 20.0 & 5.8 &   TMO    &    R     &  &  &   \\  
12 & 2010-09-10 & 1.831 & 0.886 & 15.67& 324.7 & 19.5 & 5.7 &     TMO  &    R      &             &  &    \\ 
13 & 2010-10-14 & 1.558 & 0.944 & 37.80 & 307.3 & -1.2 & 3.9 &   NTT    &    R     & \textbullet  & \textbullet &  \\  
14 & 2011-03-10 & 0.956 & 0.118 & 105.00 & 56.3 & -17.8 & 1.5 &   ChO    &    R     &  & \textbullet &    \\  
15 & 2011-03-11 & 0.964 & 0.117 & 101.12 & 62.0 & -10.4 & 0.7 &   ChO    &    R     & \textbullet & \textbullet &    \\  
16 & 2011-03-13 & 0.981 & 0.122 & 92.43 & 72.5 & 4.2 & 1.8 &   ChO    &    R     & \textbullet & \textbullet &    \\  
17 & 2011-03-15 & 0.998 & 0.139 & 84.44 & 81.5 & 16.4 & 3.8 &   AbAO    &    clear     & \textbullet   & \textbullet &  \\  
18 & 2011-03-23 & 1.069 & 0.254 & 66.78 & 104.5 & 38.9 & 2.5 &   ChO    &    R     & \textbullet & \textbullet &    \\  
19 & & "" & "" & "" & "" & "" & 2.6 &   ChO    &    R     & \textbullet & \textbullet &    \\  
20 & 2011-03-27 & 1.105 & 0.321 & 62.38 & 111.1 & 42.8 & 4.2 &   AbAO    &    R     & \textbullet & \textbullet &    \\  
21 & 2015-06-18 & 2.398 & 2.070 & 24.92 & 349.6 & 30.4 & 3.5 &   PAL    &    R     &  & \textbullet &    \\  
22 & 2015-09-10 & 1.886 & 0.958 & 16.68 & 322.6 & 22.2 & 1.4 &   TSAO    &    R     &    &  & \\  
23 & 2015-09-11 & 1.879 & 0.855 & 17.15 & 321.9 & 21.7 & 2.5 &   CrAO    &    clear     & \textbullet & \textbullet &   \\  
24 & 2016-03-28 & 1.056 & 0.337 & 70.87 & 98.3 & 24.4 & 3.5 &   PDS    &    V     & \textbullet & \textbullet &   2 \\  
25 & & "" & "" & "" & "" & "" & 1.1 &   PDS    &    V     &  & \textbullet &   2  \\  
26 & 2016-03-29 & 1.065 & 0.351 & 69.54 & 100.1 & 26.0 & 2.4 &   PDS    &    V     & \textbullet & \textbullet &   2 \\  
27 &  & "" & "" & "" & "" & "" & 1.0 &   PDS    &    V     &  & & 2 \\  
28 & 2016-03-30 & 1.074 & 0.366 & 68.29 & 100.1 & 26.0 & 2.9 &   PDS    &    V     & \textbullet & \textbullet &   2 \\  
29 & 2016-03-31 & 1.083 & 0.381 & 67.12 & 103.3 & 28.9 & 3.3  &   PDS    &    V    & \textbullet & \textbullet &   2 \\  
30 &  & "" & "" & "" & "" & "" & 1.4 &   PDS    &    V     & \textbullet   & \textbullet & 2 \\  
31 & 2020-08-10 & 2.133 & 1.243 & 17.1 & 339.2 & 33.0 & 3.2 &   INT    &    V     &  & \textbullet &    \\
32 & 2020-08-11 & 2.127 & 1.231 & 16.9 & 338.7 & 32.9 & 4.1 &   INT    &    V     & \textbullet   & \textbullet &  \\ 
33 & 2020-11-01 & 1.513 & 1.217 & 40.8 & 305.8 & -3.3 & 2.6 &   NTT    &    V     &  & \textbullet  &    \\  
34 & 2020-11-02 & 1.504 & 1.223 & 41.0 & 305.9 & -3.7 & 3.0 &   NTT    &    V     & \textbullet & \textbullet &   \\  
35 & 2020-11-03 & 1.496 & 1.229 & 41.2 & 306.0 & -4.1 & 2.9 &   NTT    &    V     & \textbullet & \textbullet &  \\  
    \hline
	    \end{tabular}
	   \begin{tablenotes}\footnotesize
		\item \textbf{Notes.} Each lightcurve has a chronologically assigned ID, then the UT date at the beginning of the night, 
		the heliocentric ($R_\odot$) and geocentric ($\Delta_\oplus$) distances in AU, 
		the solar phase angle $(\alpha)$, 
		the observed ecliptic longitude $(\lambda_O)$, 
		the observed ecliptic latitude $(\beta_O)$,
		the total time over which the target was observed,
		the observing facility and the photometric filter.
		Points in the `Included (model)' column indicate which lightcurves were included in modelling, and points in the `Included (ph. off)' column indicate lightcurves that were included in the phase offset analysis.
		References to published lightcurves are listed.
		Observing facility key: 
		ChO -- Chuhuiv Observatory $0.7\,\rm m$ Telescope (121 - Kharkiv, Ukraine);
		ESOD -- European Southern Observatory Danish $1.54\,\rm m$ Telescope (809 - La Silla, Chile);
		NTT -- European Southern Observatory $3.6\,\rm m$ New Technology Telescope (809 - La Silla, Chile); 
		TMO -- Table Mountain Observatory $0.6\,\rm m$ Telescope (673 - California, USA);
		AbAO -- Abastumani Astrophysical Observatory $0.7\,\rm m$ Telescope (119 - Abastumani, Georgia);
		TSAO -- Tien-Shan Astronomical Observatory $1.0\,\rm m$ Telescope (N42 - Almaty, Kazakhstan);
		CrAO -- Crimean Astrophyscial Observatory $2.6\,\rm m$ Shain Telescope (095 - Nauchny, Ukraine);
		PAL -- Palomar Observatory $5.1\,\rm m$ Hale Telescope (675 - California, USA);
		PDS -- Palmer Divide Station $0.35\,\rm m$ (various) (U82 - California, USA);
		INT -- Isaac Newton Group $2.54\,\rm m$ Isaac Newton Telescope (950 - La Palma, Spain)
		\item \textbf{References.} (1) \cite{belskaya_polarimetry_2009}; (2) \cite{2016MPBu...43..240W}
	    \end{tablenotes}
	\end{threeparttable}
\end{table*}   

\subsubsection{Chuhuiv Observatory - 2006, 2011} \label{Chuhuiv}
The $0.7\, \rm m$ telescope at Chuhuiv Observatory (Kharkiv, Ukraine) was used to observe PN9 in March 2006 and March 2011. The asteroid was imaged in 2006 with a $375\times242$ pixel CCD with a field-of-view (FOV) of $10.5\arcmin\times8.0\arcmin$ using the Johnson-Cousins BVRI filters. In 2011, observations were conducted in the Johnson-Cousins R filter using a 1056 x 1027 pixel CCD with an FOV of $16.9\arcmin\times16.4\arcmin$.
The lightcurves resulting from the 2006 apparition were previously published in \cite{belskaya_polarimetry_2009}. 
\subsubsection{New Technology Telescope - 2010, 2020}
ESO's $3.6\, \rm m$ New Technology Telescope (NTT) at La Silla Observatory (Chile) was used to observe PN9 in 2010 and 2020 with the ESO Faint Spectrograph and Camera v.2 (EFOSC2). EFOSC2 has an FOV of $4.1\arcmin \times4.1\arcmin$ and a $2048\times2048$ pixel CCD. We used EFOSC2 in imaging mode with $2\times2$ binning, and images were co-added to increase the SNR on the asteroid. We observed PN9 using the Bessel R filter for two nights in August 2010 and one night in October 2010, and with the Bessel V filter for three nights in November 2020. 
\subsubsection{Danish $1.54\,\rm m$ Telescope - 2010}
We used the $1.54\, \rm m$ Danish Telescope at La Silla Observatory (Chile) to observe PN9 for one night in September 2010. We used the Danish Faint Object Spectrograph and Camera (DFOSC), which has an FOV of $13.3\arcmin\times13.3\arcmin$ and a usable CCD area of $2148\times2102$ pixels. DFOSC was used to image PN9 with $1\times1$ binning in the Bessel R filter. Images were co-added before lightcurve extraction.
\subsubsection{Table Mountain Observatory - 2010}
PN9 was observed over three nights with the Jet Propulsion Lab's $0.6\, \rm m$ telescope at Table Mountain Observatory (California, USA) in September 2010. We imaged PN9 with a $1024\times1024$ pixel CCD that has an FOV of $8.9\arcmin\times8.9\arcmin$ using the R filter and $1\times1$ binning and images were co-added for lightcurve extraction.
\subsubsection{Abastumani Observatory - 2011}
In March 2011, observations of PN9 were carried out with the $0.7\, \rm m$ Telescope at the Abastumani Astrophysical Observatory (Abastumani, Georgia). We imaged the asteroid without a filter using a $3072\times2048$ pixel CCD with an FOV of $44.4\arcmin\times29.6\arcmin$. 
\subsubsection{Hale Telescope - 2015}
We used the $5.1\, \rm m$ Hale telescope at Palomar Observatory (California, USA) to observe PN9 in June 2015. The telescope was equipped with the Large Format Camera (LFC), which has six $2048\times4096$ chips, each of which has an FOV of $6.1\arcmin\times12.3\arcmin$. We used the central CCD chip with $2\times2$ binning and the Bessel R filter to image PN9. Images were co-added for lightcurve extraction.
\subsubsection{Tien-Shan Observatory - 2015}
In September 2015, there were observations of PN9 with the $1.0\, \rm m$ telescope at the Tien-Shan Astronomical Observatory (Almaty, Kazakhstan). We used a $3072\times3072$ pixel CCD, which has an FOV of $18.9\arcmin\times18.9\arcmin$, with $2\times$2 binning using the Johnson R filter. 
\subsubsection{Shain Telescope - 2015}
The asteroid was observed in September 2015 with the $2.6\, \rm m$  Shain Telescope at the Crimean Astrophysical Observatory (Nauchny, Ukraine). We imaged PN9 with a $2048\times2048$ pixel CCD, which has an FOV of $9.5\arcmin\times9.5\arcmin$, using $2\times2$ binning without a filter. 
\subsubsection{Palmer Divide Station - 2016}
This analysis includes six published lightcurves from the Palmer Divide Station (California, USA). PN9 was observed with three $0.35\, \rm m$ Meade LX200GPS telescopes equipped with commercial CCDs using the Johnson V filter. These lightcurves were obtained through the Asteroid Lightcurve Data Exchange Format (ALCDEF) database \citep{2011MPBu...38..172W} and are discussed in \cite{2016MPBu...43..240W}. Note that observations taken with different telescopes during the same night are treated as separate lightcurves.
\subsubsection{Isaac Newton Telescope - 2020}
We observed PN9 over two nights in August 2020 with the $2.5\, \rm m$ Isaac Newton Telescope (La Palma, Spain). Imaging was conducted in the Harris V filter with $1\times1$ binning using the central chip of the Wide Field Camera. The CCD was windowed to give a $10'\times10'$ field with a resolution of $1820\times1820$. Images were co-added for lightcurve extraction.

\subsection{Planetary radar}

\begin{table*}
    \centering     
    \begin{threeparttable}
	\caption{Delay-Doppler observations of (23187) 2000 PN9}
	\label{tab:radtable}
	\begin{tabular}{ccccccccccc}
		\hline\hline
		\noalign{\smallskip}
		Obs. & {UT Date} & {RTT} & {Baud}         & Res. & {Start-Stop}           & {Runs} & Radar & Note \\
		& {[yyyy-mm-dd]} & {[s]}  & {[$\mu\rm s$]} & [m] & {[hh:mm:ss-hh:mm:ss]} & & model \\ \hline
		\noalign{\smallskip}
	        Arecibo & 2001-03-03 & 62 & CW   &    & 09:40:32-09:56:40 & 8  & &  \\
	                &            &    & CW   &    & 09:59:25-10:00:23 & 1  & & \\
	      	        &            &    & 4   & 600 & 10:02:38-10:16:19 & 3  & & Ranging\\
	      	        &            &    & 4.5 & 675 & 10:17:46-10:23:28 & 3  & & Ranging\\
	      	        &            &    & 0.5 & 75  & 10:27:14-10:36:39 & 5  & & Ranging\\ \\
	      	Goldstone & 2001-03-03  & 62  & 1.0  &  150   & 13:14:07-15:02:52 & 49 & \textbullet &\\ \\ 
	      	Arecibo & 2001-03-04 & 67 & CW &      & 09:04:03-09:16:47 & 6  & \textbullet &\\
	      	        &            &    & 0.2 & 30  & 09:18:59-09:24:35 & 3  & \textbullet &\\
	      	        &            &    & 0.1 & 15  & 09:27:01-09:38:27 & 3  & &\\
	      		    &            & 68 & 0.1 & 15  & 10:09:44-10:31:28 & 10 & \textbullet &\\ \\
	      	Arecibo & 2001-03-05 & 76 & CW &      & 09:05:55-09:12:43 & 3  & \textbullet &\\
	      	        &            & 77 & 0.2 & 30  & 09:15:42-10:52:44 & 38 & \textbullet &\\ \\
      		Goldstone & 2006-03-07 & 36  & 0.125  &  19  & 19:24:26-19:31:09 & 6 & \textbullet &\\
      		          &            &     & 0.125  &  19  & 19:31:49-20:30:26 & 48 & \textbullet &\\ \\
      		Goldstone & 2006-03-10 & 86 & CW &    & 12:02:53-14:42:14 & 58 & & Low SNR\\
      		\hline
	\end{tabular}
	\begin{tablenotes} \item \textbf{Notes.} ``Obs.'' is the facility with which the observations were made. ``UT Date'' is the start date of the observations in universal time. ``RTT'' is the signal's round trip time to the object and back. ``Baud'' is the baud length and ``Res'' is the delay resolution; continuous wave observations are marked as ``CW'' and do not have spatial resolution.  ``Start-Stop'' is the UT timespan in which the observations were made. ``Runs'' is the number of transmit-receive cycles that were completed.
	\end{tablenotes}
    \end{threeparttable}
\end{table*}

This analysis made use of delay-Doppler imaging and continuous wave echo power spectra obtained by planetary radar facilities. For delay-Doppler imaging a circularly polarised phase-modulated signal is transmitted, with the modulation pattern determined by a pseudo-random binary phase code \citep{1993RvMP...65.1235O, magri_radar_2007}. The modulation pattern allows for the determination of distance between the observer and the point on the asteroid that reflected the signal. The resolution of delay information is determined by the temporal resolution of the modulation by the pseudo-random code and is known as the baud length. A delay-Doppler image is constructed with delay in the vertical axis and Doppler shift in the horizontal axis. Continuous wave (CW) spectra do not contain delay information, and instead only describe the Doppler shift of the reflected signal in the two circular polarisation orientations. 
It is important to carefully select which radar datasets to include in the shape modelling process, as poor quality data can greatly increase computational cost without improving the model. For a summary of radar observations of PN9, see Table \ref{tab:radtable}.

\subsubsection{Arecibo Observatory - 2001} 
The William E. Gordon telescope at Arecibo (Puerto Rico, USA) was a 305 m fixed-dish radio telescope with a $2380\,\rm MHz$ radar transmitter. It was used to obtain radar observations of PN9 on 3, 4 and 5 March 2001. The delay-Doppler imaging on 3 March were for ranging and ephemeris correction, hence they were excluded from the analysis. On 4 March, imaging was mostly conducted with a baud length of $0.1\,\rm\mu s$ giving a \textasciitilde $15\,\rm m$ resolution, with some further imaging at $0.2\,\rm\mu s$ (\textasciitilde $30\,\rm m$). On 5 March, a baud length of $0.2\,\rm\mu s$ (\textasciitilde $30\,\rm m$) was used for imaging. The observers submitted astrometric measurements which were used to refine the orbital solution. The CW spectra from 4 and 5 March were included in the analysis. Observations from 3 March were not obtained until late in the modelling process. Due to limited computational resources and the presence of data from subsequent days, these data were not added to the model. All but ten minutes of the delay-Doppler imaging is at extremely low resolution and would not offer a significant contribution to the model, however we recommend that the continuous wave spectra are included in any future work.

\subsubsection{Goldstone Solar System Radar - 2001, 2006} 
The Goldstone Solar System Radar (GSSR) facility consists of the fully steerable 70 m DSS-14 "Mars" antenna equipped with an $8560\,\rm MHz$ transmitter. DSS-14 is located in the Mojave Desert (California, USA) and is a part of the Deep Space Network. Delay-Doppler imaging of PN9 was conducted by GSSR on 3 March 2001 and 7 March 2006. The 2001 imaging used a baud length of $1.0\,\rm\mu s$ (\textasciitilde $150\,\rm m$) while the 2006 imaging used a baud length of $0.125\,\rm\mu s$ (\textasciitilde $19\,\rm m$). In 2006, GSSR also obtained CW spectra of PN9. These were not included in the analysis, as there are already higher quality radar data for this epoch and viewing geometry. Astrometric measurements from both 2001 and 2006 were used for ephemeris correction.
\section{Physical modelling and spin-state analysis} \label{sec:modelling}
\subsection{Lightcurve-only modelling } \label{sec:LCmodel}
The majority of observational data for PN9 are in the form of optical lightcurves. As modelling with radar data is an iterative and computationally intensive process, and fitting procedures are highly sensitive to input parameters, it was more efficient to first perform a lightcurve-only analysis of PN9. This can allow the use of predetermined constraints on rotation period, pole orientation and shape, which greatly improves efficiency when later modelling the object with a combination of lightcurve and radar data. 

The lightcurve-only modelling includes observations marked in Table \ref{tab:obstable}. Lightcurves that were not included were unsuitable due to poor temporal resolution, gaps in rotational coverage or low SNR.

\begin{figure}
	\resizebox{\hsize}{!}{\includegraphics[width=12cm, clip=true]{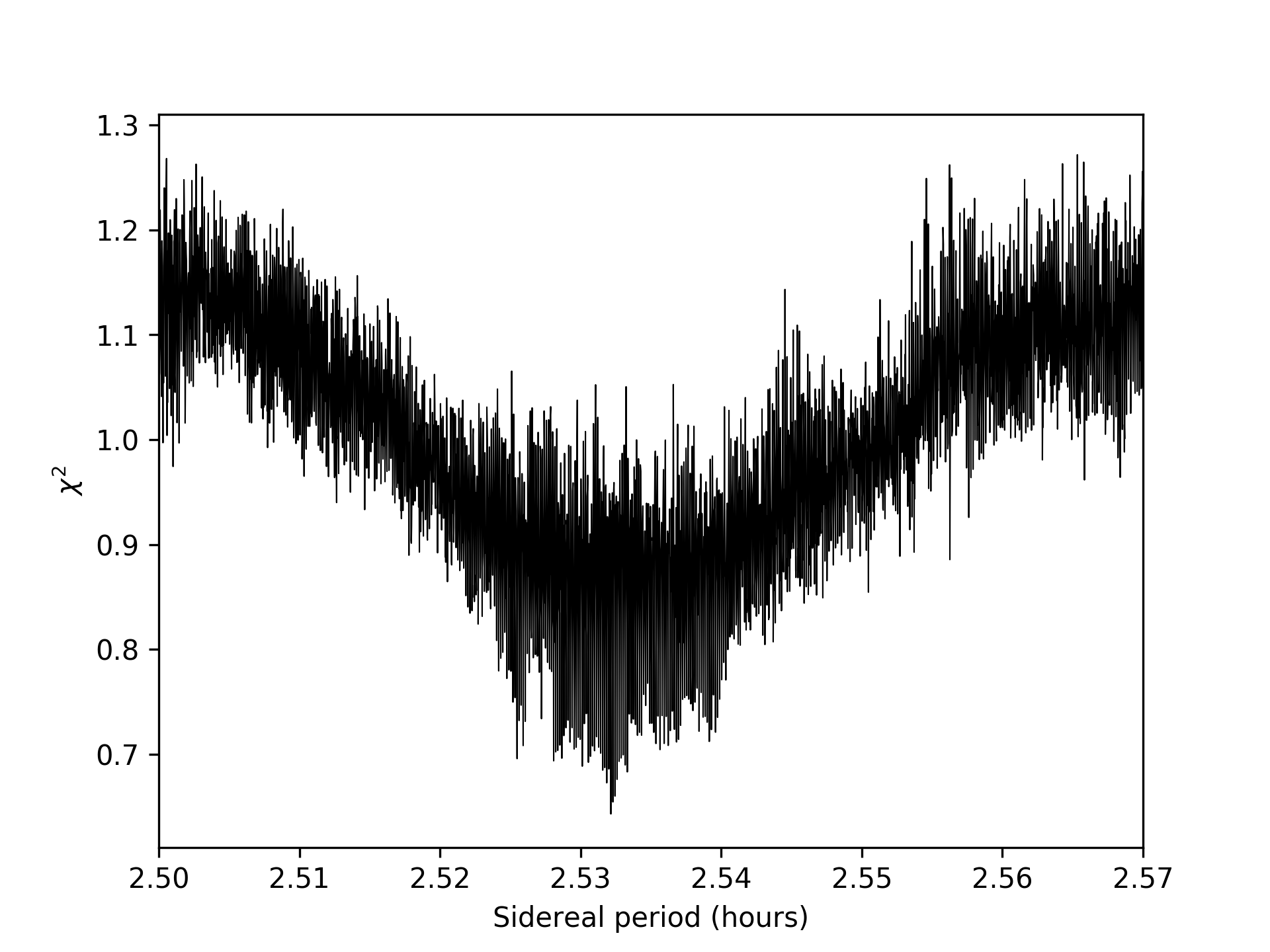}}
	\caption{The results of a period search for asteroid (23187) 2000 PN9. For each period in the range shown, lightcurve data were used to optimise a model for six different rotational poles in the celestial sphere. The lowest $\chi^2$ across the six models was recorded for the period being used. The best-fit sidereal rotation period for 2000 PN9 from this scan is $2.532\pm0.008$ h.}
	\label{fig:periodogram}
\end{figure}

A search for PN9's sidereal rotation period was conducted between 2.500 and 2.570 h, a range based on the previously reported synodic periods \citep{2007OAP....20...43G, belskaya_polarimetry_2009,2016MPBu...43..240W}, using the convex inversion routines described by \cite{Kaasalainen_optimization_1_2001} and \cite{kaasalainen_optimization_2_2001}. For each iteration over the period scan range, a shape model was generated for six random and unique rotational poles. Each shape model was then optimised to best fit the lightcurve data across the period range. The results of this scan, shown in Figure \ref{fig:periodogram}, identify a best-fit rotation period of $2.532\pm0.008$ h.

A further period scan was conducted over a wider but more coarse range of periods, searching for solutions between 1 h and 10 h. Solutions were found close to integer multiples of the 2.532 h period, which were discounted due to their corresponding shape models being both physically extreme and inconsistent with radar imaging data.

Using an input period of $2.532$ h from the period scan, we conducted a search for the asteroid's rotational pole. For each pole on a $5\degr\times5\degr$ grid covering the celestial sphere, the model's period and convex shape were optimised. This scan assumed principal axis rotation. The scan was then repeated with the addition of YORP acceleration, for a range of YORP factors from $-10^{-8}\,\rm rad/day^{2}$ to $\rm10^{-8}\,\rm rad/day^{2}$ in steps of $2\times10^{-9}\,\rm rad/day^{2}$. The goodness-of-fit for the global best solution of each YORP step does not converge towards any YORP value. This indicates that no YORP solution is found, hence only solutions with constant period rotation are considered in this section. A search for YORP with a radar-derived model of the asteroid is discussed in section \ref{sec:yorp}. 

\begin{figure}
\centerline{\includegraphics[width=\columnwidth]{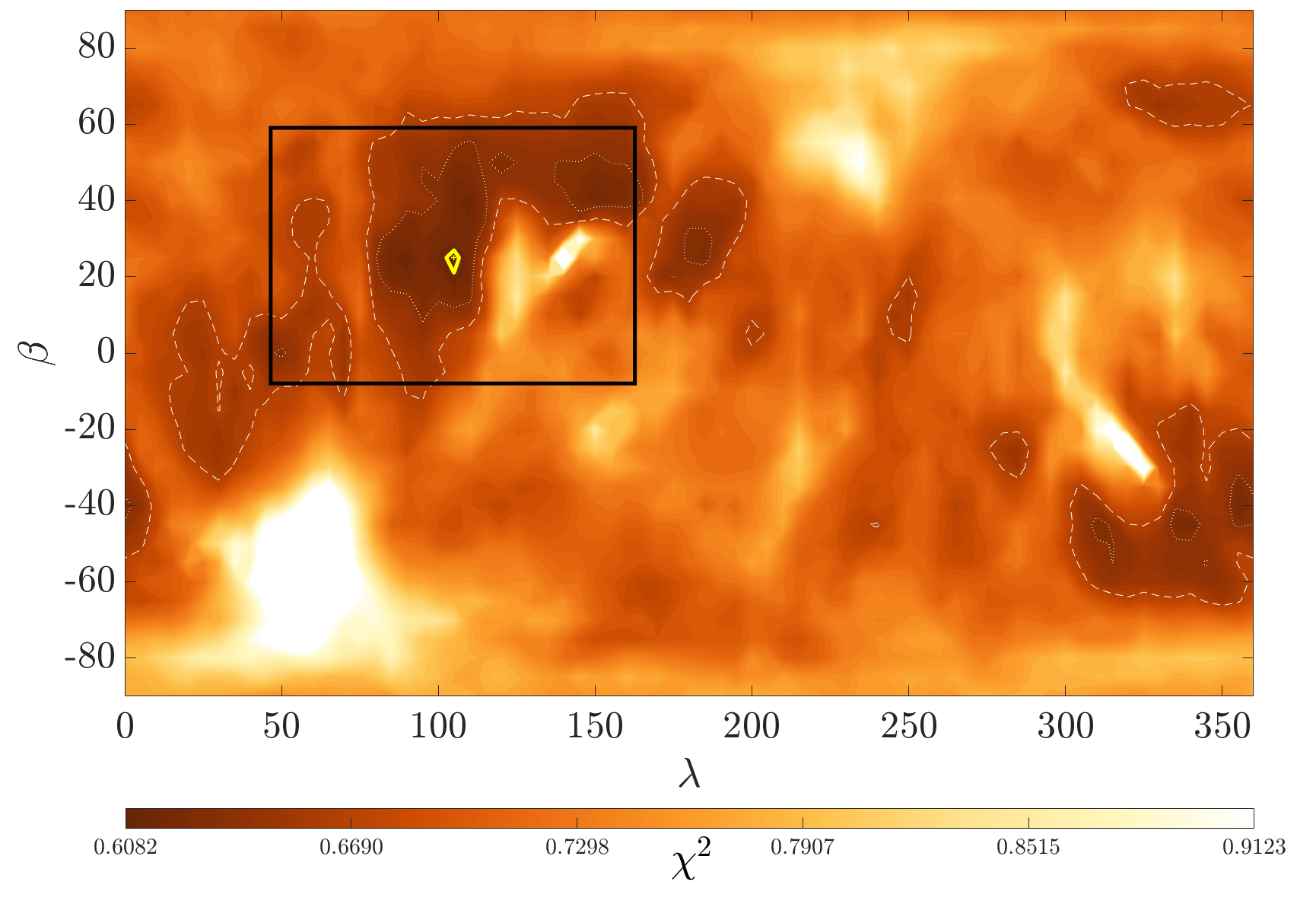}}
\centerline{\includegraphics[width=\columnwidth]{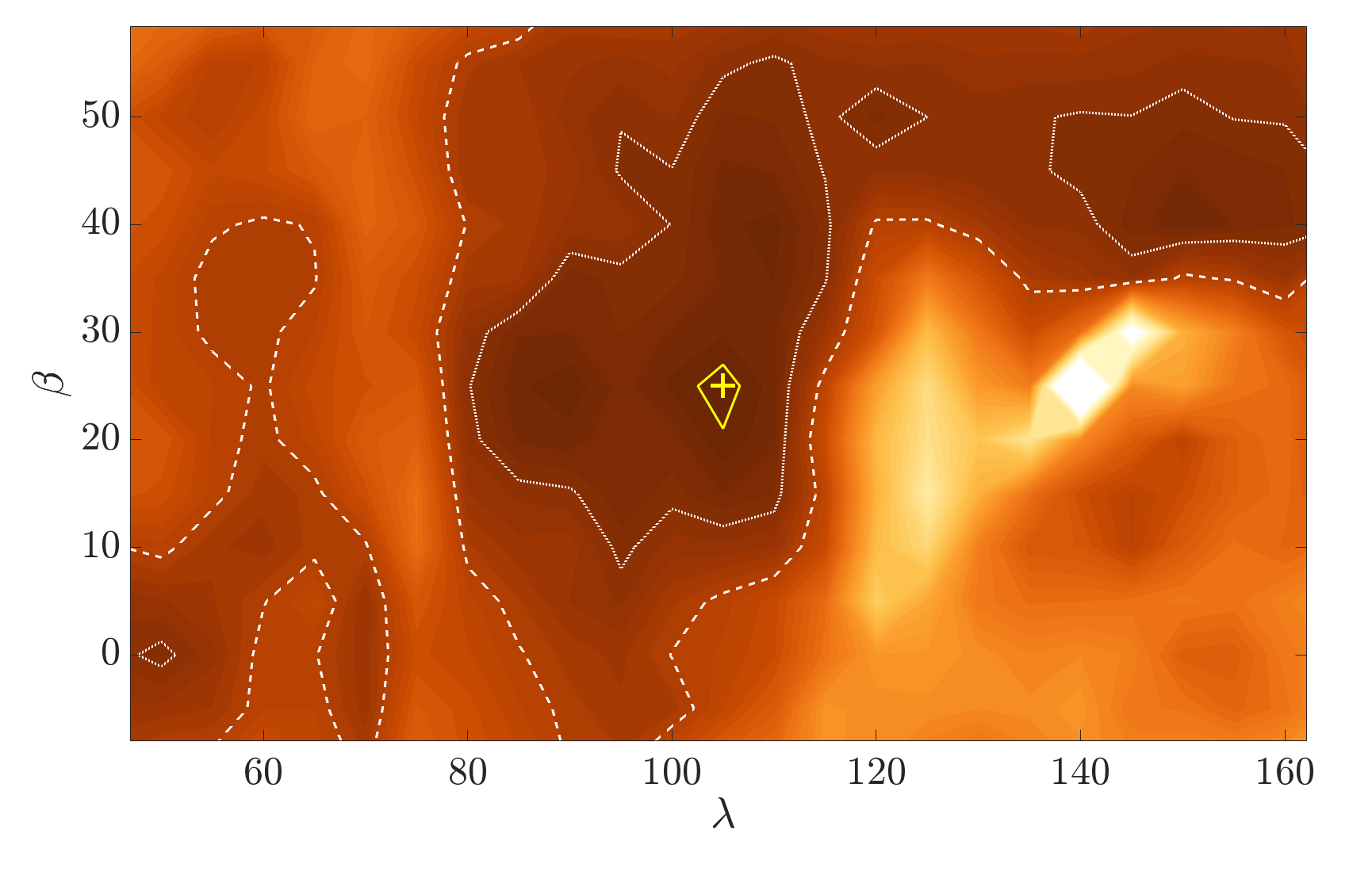}}
\caption{The results of a search for the rotational pole of (23187) 2000 PN9 using convex inversion of lightcurve data. For each pole solution in ecliptic coordinates $\lambda$ and $\beta$, the $\chi^2$ fit of the solution is plotted for a colour range where the global minimum $\chi^2$ is black and solutions 50\% greater than the minimum are white. The best solution is marked with a yellow `+'. The yellow line and the white dotted and dashed lines enclose regions where $\chi^2$ is within $1\%$, $5\%$ and $10\%$ of the best solution respectively. The top panel shows the full celestial sphere, with the region enclosed by the black rectangle shown in the bottom panel.}
\label{fig:convpoles}
\end{figure}

\begin{figure*}
    \includegraphics[width=14cm,trim=0 0 0 0,clip=true]{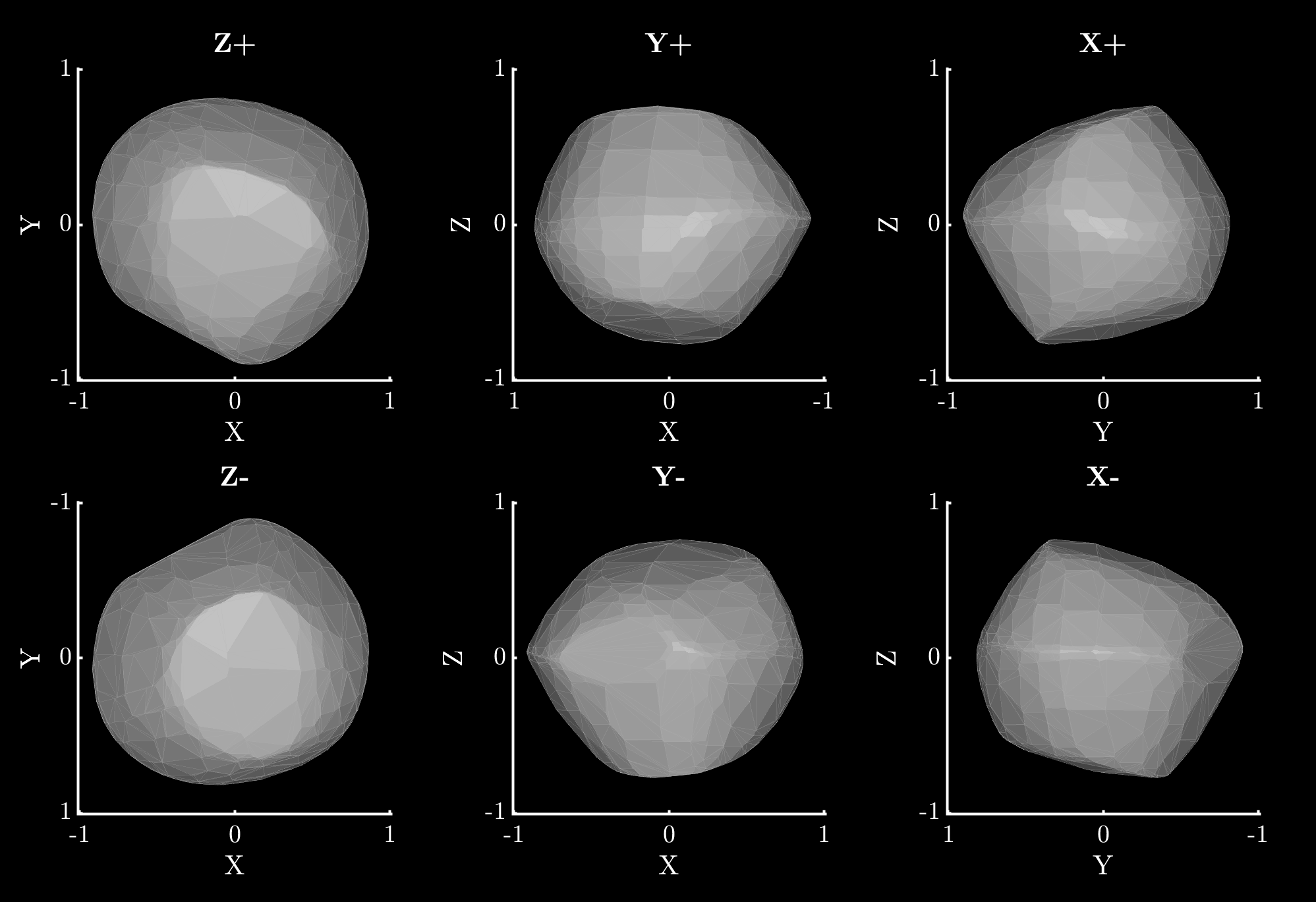}
    \caption{The best-fit convex hull model of (23187) 2000 PN9 with the rotational pole $\lambda=105\degr$ $\beta=+25\degr$. This model assumes principal-axis rotation and a constant period of 2.532 h. The top row shows the model from the positive end of the Z, Y and X axes in the body-centric co-ordinate system. The bottom row shows the model from the negative end of the Z, Y and X axes. The Z-axis is aligned with the rotational pole, which is the shortest axis of inertia. The X-axis is arbitrarily set such that it is viewed from the positive end for the plane-of-sky at $T_{0}$. Axis lengths are in arbitrary units, as lightcurve inversion does not produce scaled models.}
    \label{fig:convhull}
\end{figure*}

Figure \ref{fig:convpoles} shows the results of the constant-period pole scan, which does not converge to a single region in the celestial sphere. The best model's rotational pole lies at ecliptic longitude $\lambda=105\degr$ and ecliptic latitude $\beta=+25\degr$ with a rotation period of $2.532$ h. Models within 5\% of the best solution have poles corresponding to opposite regions of the celestial sphere, and are consistent in shape and period. This result shows that the orientation of the rotational axis is constrained, although the data are insufficient for distinguishing between prograde and retrograde rotation. This is a common issue when modelling with low-amplitude lightcurves produced by highly symmetrical objects. The convex hull model of the global best solution, shown in Figure \ref{fig:convhull}, indicates that the asteroid is an oblate spheroid with signs of an equatorial ridge. There is a flattened section on the equator which could be interpreted as a crater, but could also be caused by a prominence or large boulder. The polar regions are also flattened, although this could be an artefact caused by uncertainty in the Z-axis. As shown in Figure A1, the model is able to reproduce the shape of most lightcurves, although in some cases there is a small phase offset and a mis-match in amplitude. Since there is no coherent progression in phase offset, the phase offsets are the result of period uncertainty which can be reduced with radar modelling. Convex hull models struggle to reproduce low-amplitude lightcurves due to the heightened dependence on surface features, which are not modelled, while general uncertainties in shape and pole can suppress or amplify brightness variation cased by the asteroid's overall shape.

\subsection{Combined radar \& lightcurve model} \label{sec:shape}
Further modelling of PN9 was conducted using a combination of lightcurve and radar data using the \textsc{shape} software package \citep{Hudson:1993uo, magri_radar_2007}. As discussed in Section \ref{sec:LCmodel}, it is efficient to begin with well-constrained input parameters describing the asteroid's shape and spin-state. In this case, those estimates are taken from the convex inversion analysis.

An input model was constructed for \textsc{shape} comprising a triaxial ellipsoid with principal axis rotation. Both the convex hull model and inspection of delay-Doppler images indicate that PN9 has a spheroid shape, so a sphere with a $1.9\,\rm km$ diameter was used as the input model. This diameter was chosen as it was close to the $2\,\rm km$ estimate from an earlier unpublished analysis of the radar data \citep{2006SASS...25..169B}, but also in agreement with a reported $1.6\pm0.3\,\rm km$ diameter based on an estimated optical albedo of $0.24\pm0.06$ \citep{belskaya_polarimetry_2009}. The rotation period was set to 2.532 h, as previously determined in section (\ref{sec:LCmodel}).

A $10\degr\times10\degr$ grid of poles was set up, covering the celestial sphere. For each fixed pole, the model's global shape and period were optimised to fit the radar data marked in Table \ref{tab:radtable}. The ellipsoid model for each pole was then converted to a vertex model with 1000 vertices and 1996 facets, to allow for the fitting of surface features through the adjustment of individual facets. The continuous wave (CW) spectra were removed at this stage, as the model was sufficiently well-constrained and there was a risk of over-fitting to noise. 

\begin{figure}
	\resizebox{\hsize}{!}{\includegraphics[width=12cm, clip=true]{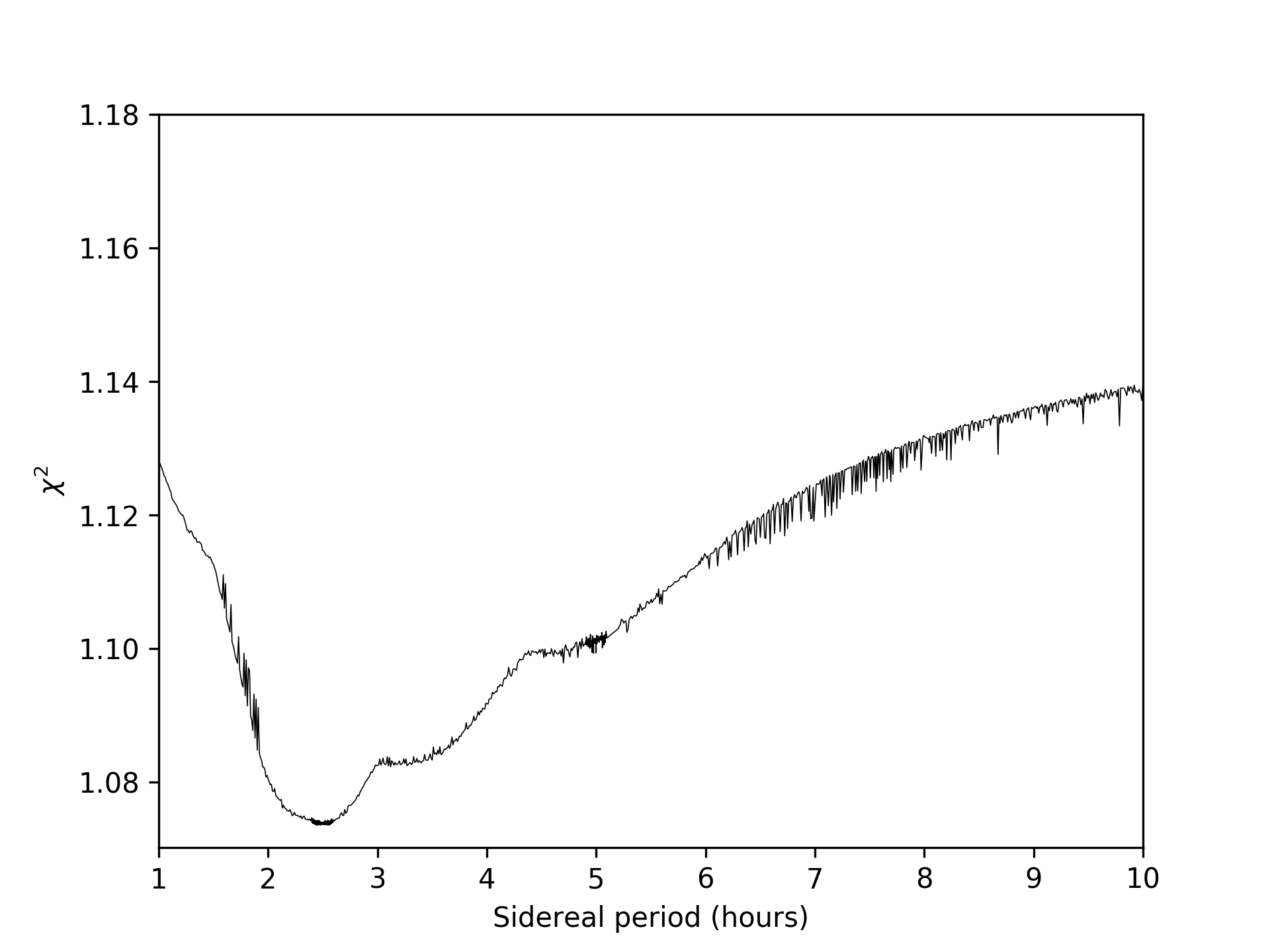}}
	\caption{The results of a period search for asteroid (23187) 2000 PN9 with radar data. \textsc{shape} was used to optimise an ellipsoid model for six different rotational poles in the celestial sphere to fit both continuous-wave and delay-Doppler data. The lowest $\chi^2$ across the six models was recorded for the period being used, with a higher temporal resolution close to 2.5 h, 5 h and 7.5 h. This is a coarse scan intended to demonstrate a global minimum close to 2.5 h independently of lightcurve data.}
	\label{fig:radpscan}
\end{figure}

Early iterations of PN9's vertex model remained in good agreement with the convex hull model's shape, pole and period. As the radar model's initial parameters were derived from lightcurve data, a coarse radar period scan was conducted to confirm that the radar data independently favour a 2.5 h solution. This period scan utilised the same six-pole strategy described in Section \ref{sec:LCmodel}. Computational limitations restricted this to a coarse resolution that can only show global minima, and not identify local minima required for a precise period measurement. The resolution was increased around multiples of 2.5 h, as these are the most likely alternate solutions. As shown in Figure \ref{fig:radpscan}, a coarse period scan with radar data indicates a clear global minimum close to 2.5 h.

As a visual inspection of radar data shows a shape that is  consistent with the convex hull model, it can be said that the lightcurve and radar datasets independently favour the same shape and period for PN9.

Having confirmed this, lightcurve data were then progressively introduced in subsequent fitting runs to produce a combined radar and lightcurve model. The full subset of lightcurves, marked in Table \ref{tab:obstable}, were not all included until the final iterations of modelling. Each additional lightcurve causes a significant increase in computation time whilst yielding diminishing returns. It is therefore most efficient to gradually introduce the lightcurve dataset as the model improves.

During the modelling process, various penalties were applied with \textsc{shape} to discourage certain features. The first penalty prevents excessive deviation of the centre of mass from the origin of the body-centric coordinate system. The second penalty prevents large divergence between the model's Z-axis and the axis of maximum inertia. A third penalty disallows non-principal axis rotation. A fourth penalty is used to suppress unphysical spikes that can occur when fitting a vertex model. Finally, a fifth penalty was applied to discourage deep concavities.  

These penalties increase the $\chi^2$ fit value when penalised features are encountered, meaning that \textsc{shape} is less likely to produce those features as it optimises models to produce a smaller $\chi^2$ value. Each of these penalties is given a strength, with larger penalties more strongly discouraging features. The first three penalties were given a relatively high strength, and the latter two penalties were low in strength to ensure they only discouraged unphysical features without restricting the construction of craters, ridges and boulders.

The results of the \textsc{shape} pole scan with both radar and lightcurve data are shown in Figure \ref{fig:radpoles}. The  pole is again constrained to two opposite regions, with the best solution at ecliptic longitude $\lambda=96\degr$ and ecliptic latitude $\beta=+30\degr$. Solutions within 1\% of the global best fit in both the northern and southern hemispheres have consistent shapes and periods, again indicating an uncertainty as to whether PN9's rotation is prograde or retrograde. While a retrograde solution cannot be completely ruled out, the prograde solution is more clearly favoured in a lightcurve-only analysis (Fig. \ref{fig:convpoles}). As the two solutions have identical rotation periods and very similar shapes, any qualitative analysis of the two solutions will yield the same conclusions. As such, the subsequent sections of this paper will only consider the prograde solution. 

\begin{figure}
    	\resizebox{\hsize}{!}{
        \includegraphics[width=12cm, trim=0.5cm 2.5cm 1.5cm 2.5cm, clip=true]{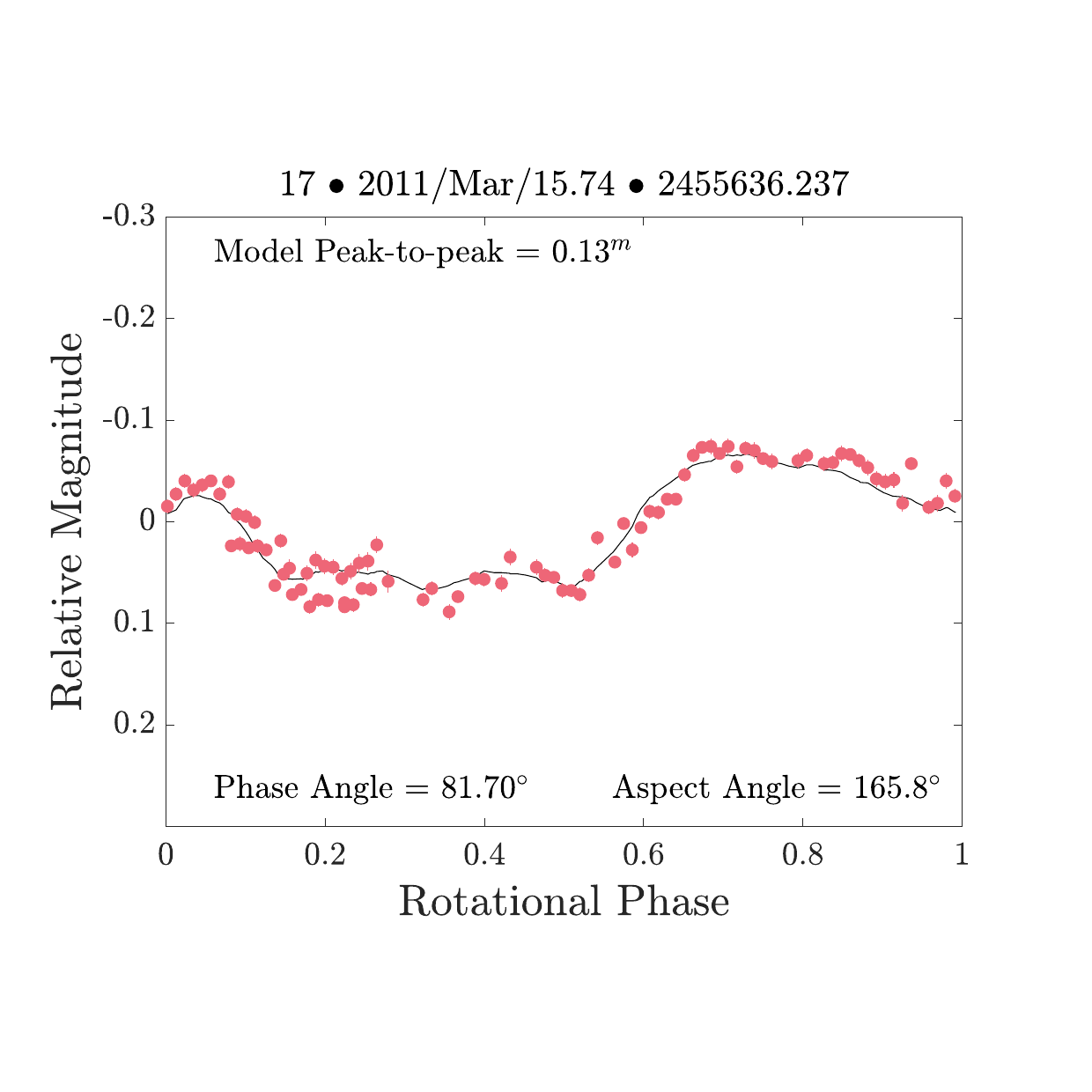}
        \includegraphics[width=12cm, trim=0.5cm 2.5cm 1.5cm 2.5cm, clip=true]{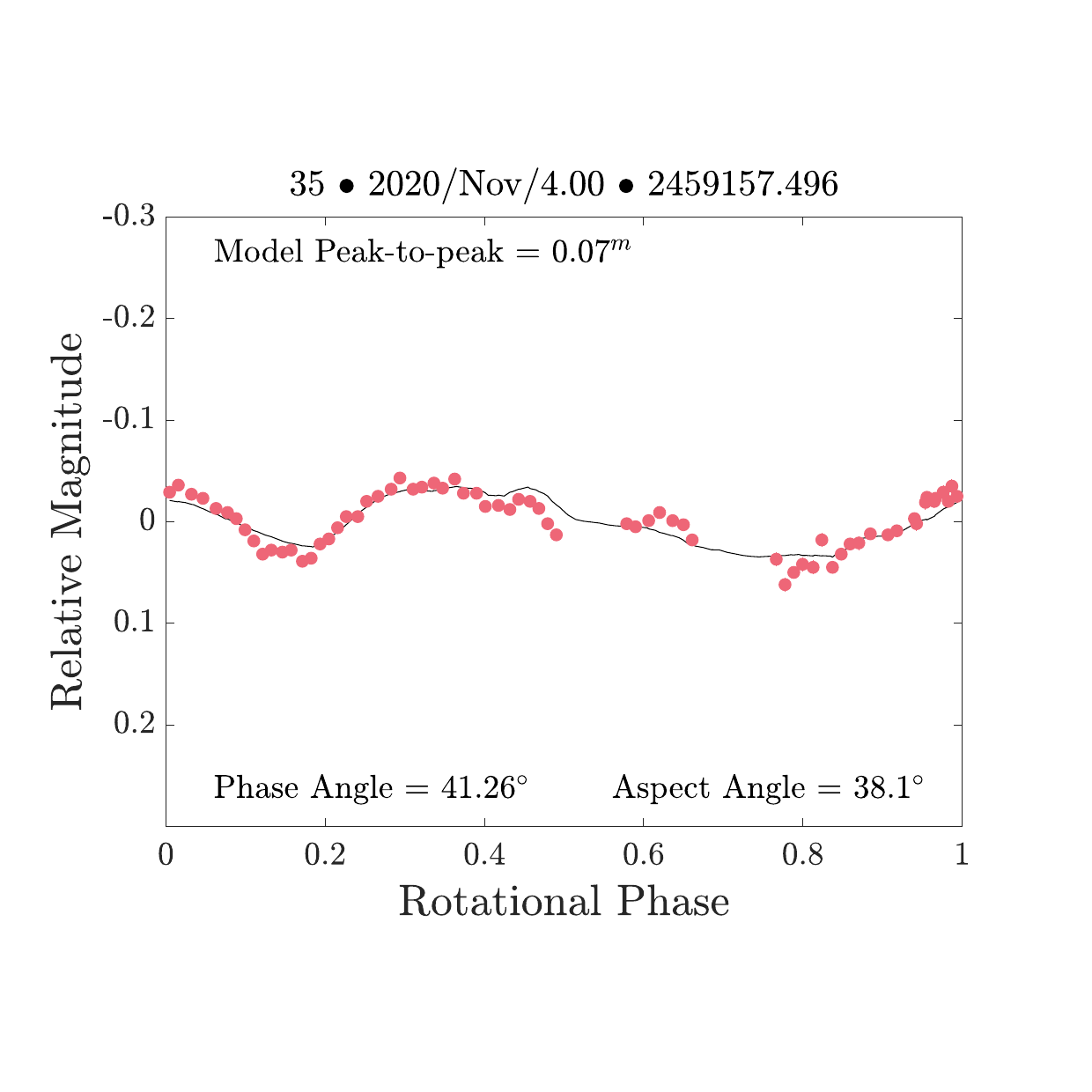}
    }
	\caption{A comparison of observational data (red points) and the corresponding synthetic lightcurve (solid black line) for lightcurves 17 and 35 (Table \ref{tab:obstable}). The synthetic lightcurves were generated using the combined radar and lightcurve model for (23187) 2000 PN9, with a combination of the Lambertian and Lommel-Seelinger scattering models. For plots of all 35 lightcurves, see Figure A8.}
	\label{fig:radLC17}
\end{figure}

The best-fitting model was re-modelled with 2000 vertices to allow for closer fitting of surface features to the 15 m resolution radar imaging, although this yielded a negligible improvement in the $\chi^2$ fit. Both the optical and radar observations cover the entire surface of the asteroid, leaving no `unseen' surface area in either of the two wavelength regimes. The geometric parameters of this model are presented in Table \ref{tab:modgeom} and the shape model is shown in Figure \ref{fig:radmod}. The lightcurve fits, shown in Figure A8, are an improvement upon those produced by the convex hull model. The majority of lightcurves are fitted well in terms of shape, phase and amplitude, with the few poor fits generally corresponding to low-quality lightcurves that were not used in the modelling process.

\begin{figure}
    	\resizebox{\hsize}{!}{
        \includegraphics[width=12cm, trim=0.5cm 2.5cm 1.5cm 2.5cm, clip=true]{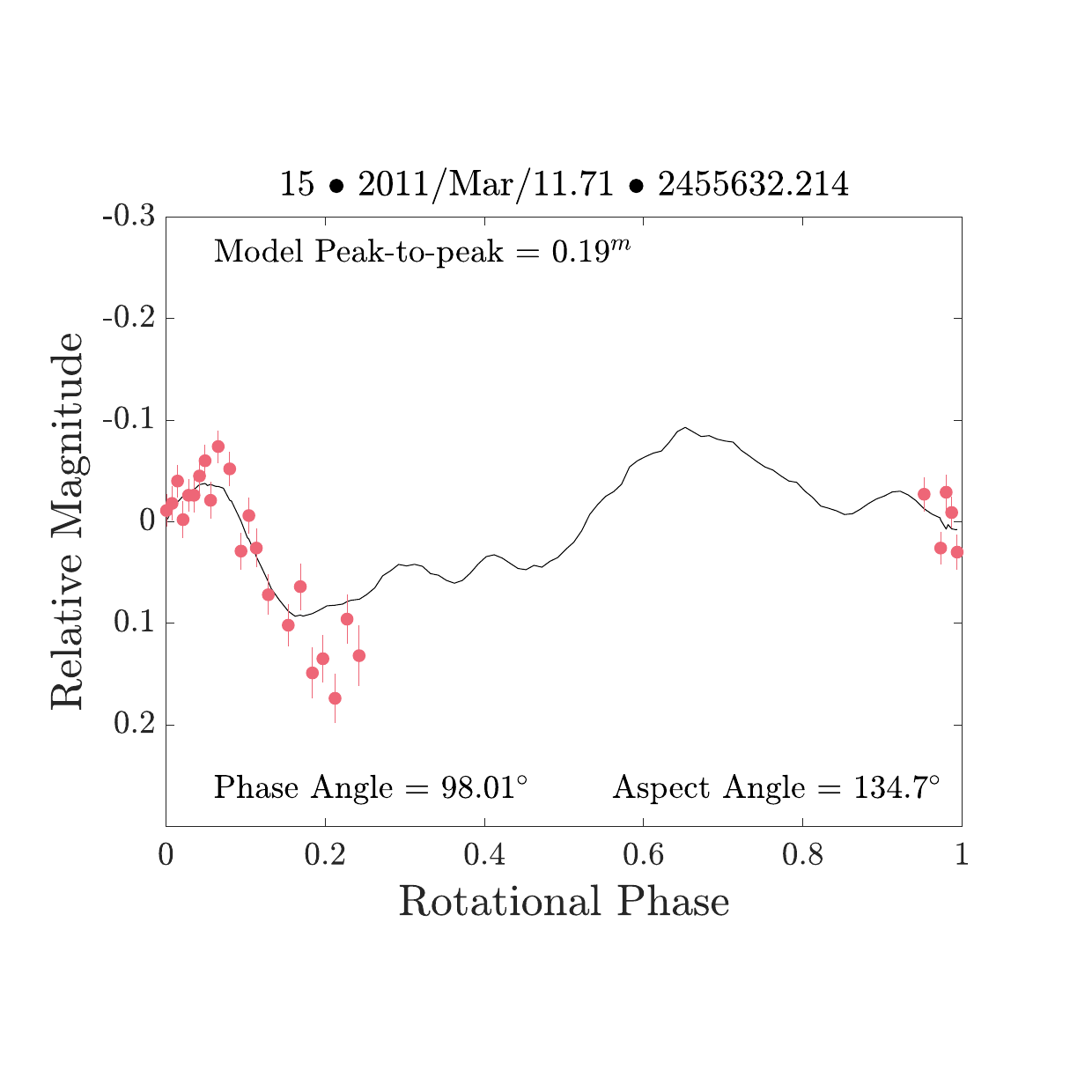}
        \includegraphics[width=12cm, trim=0.5cm 2.5cm 1.5cm 2.5cm, clip=true]{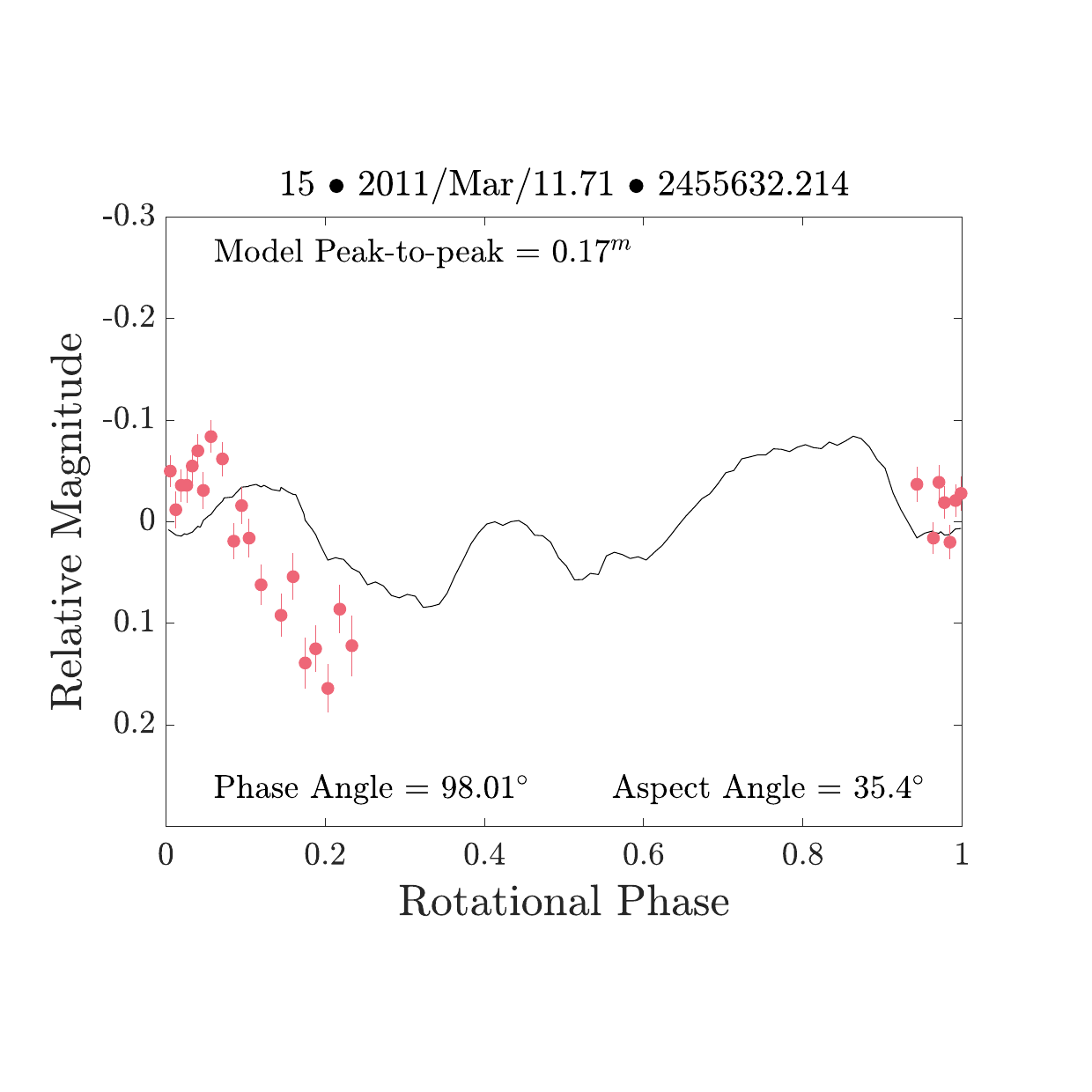}
    }
	\caption{A comparison of lightcurve fits produced by the prograde (left panel) and retrograde (right panel) models of (23187) 2000 PN9 for lightcurve 15. The prograde model produces a better fit than the retrograde model, which has a smaller amplitude and a minor phase offset compared to the data.}
	\label{fig:radLCcomp}
\end{figure}

\begin{figure}
\centerline{\includegraphics[width=9.5cm, trim=0.5cm 2.5cm 1.5cm 2.5cm,clip=true]{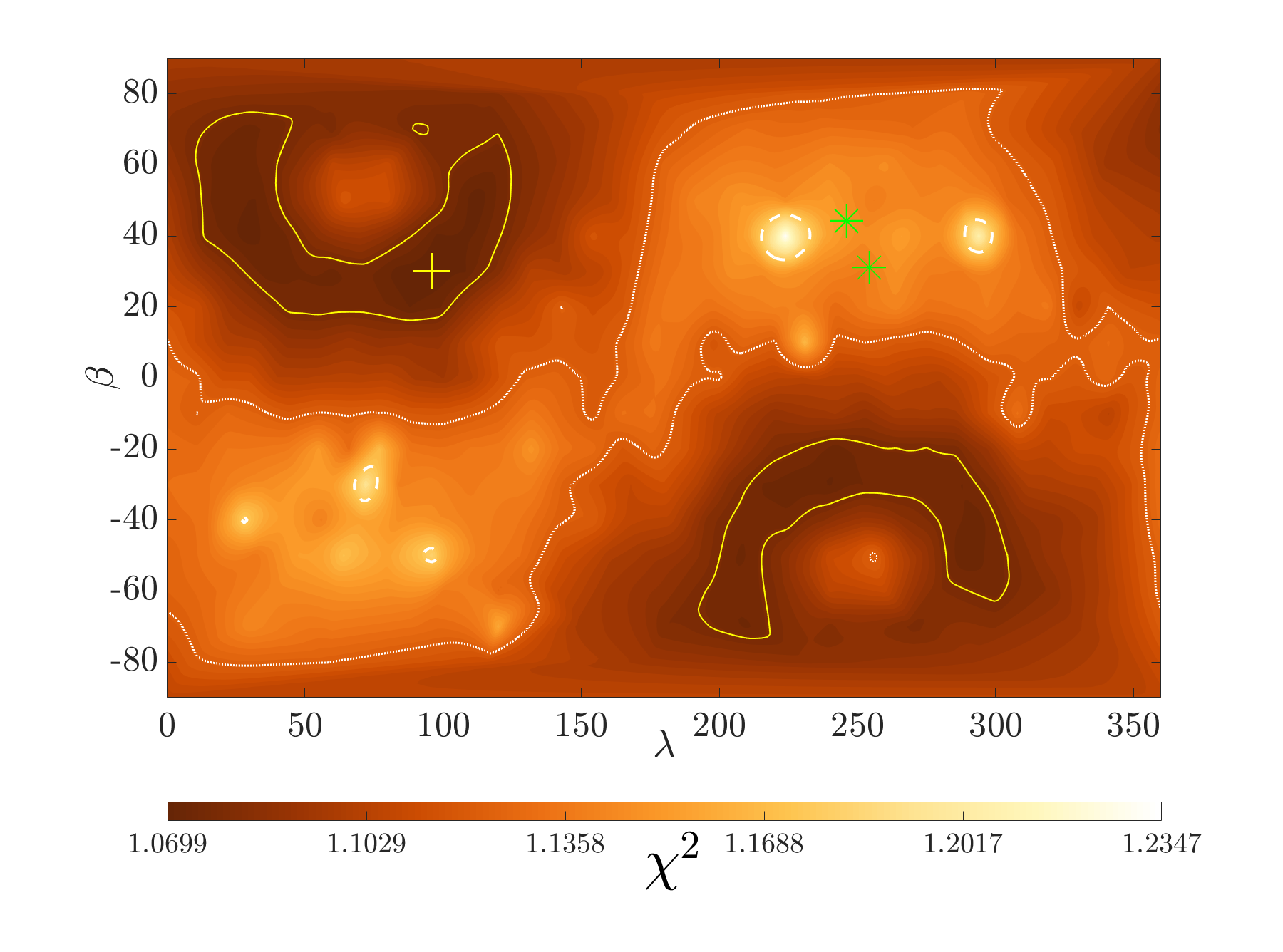}}
\caption{The results of a search for the rotational pole of (23187) 2000 PN9 using \textsc{shape} with radar and lightcurve data. For each pole solution in ecliptic coordinates $\lambda$ and $\beta$, the $\chi^2$ fit of the solution is plotted for a colour range where the global minimum $\chi^2$ is black and the maximum is white. The best solution is marked with a yellow `+'. The yellow lines enclose regions where $\chi^2$ is within $1\%$ of the best solution, and dotted and dashed white lines enclose regions where $\chi^2$ is within $5\%$ and $10\%$ of the best solution respectively. Green stars indicate the ecliptic coordinates of the observer's line-of-sight for each date where delay-Doppler imaging was taken.}
\label{fig:radpoles}
\end{figure}

\begin{table}
    \begin{threeparttable}
	\caption{Summary of parameters for the prograde radar and lightcurve model of (23187) 2000 PN9.}
	\label{tab:modgeom}
	\begin{tabular}{cc}
		\hline \hline \noalign{\smallskip}
		Parameter & Value \\
		\hline \noalign{\smallskip}
        $\lambda$ & $96\pm36^{\circ}$\\ \\
        $\beta$ & $+30\pm17^{\circ}$ \\ \\
        P & $2.53216\pm0.00015$ h\\ \\
		Max. extent along (x, y, z) & $1.82 \times 1.82 \times 1.77\rm ~km$ \\
        ($\pm$) & ($0.08 \times 0.07 \times 0.11\rm ~km$) \\ \\
		Surface area & $9.61\pm0.80\rm ~km^2$ \\ \\
		Volume & $2.62\pm0.34\rm ~km^3$ \\ \\
  		DEEVE dimensions (2a, 2b, 2c) & $1.73 \times 1.73 \times 1.68\rm ~km$\\
        ($\pm$) & ($0.10 \times 0.09\times 0.06\rm ~km$) \\ \\
		$\rm D_{eq}$ & $1.71\pm0.07\rm ~km$ \\ \\
		\noalign{\smallskip} \hline
	\end{tabular}%
    \begin{tablenotes} \item \textbf{Notes.}  The maximum extents are measured along the three axes of a body-centric coordinate system. ``DEEVE'' stands for dynamically equivalent equal-volume ellipsoid.  $D_{eq}$ is the diameter of a sphere that has a equal volume to the model.
    \end{tablenotes}
    \end{threeparttable}
\end{table}

\begin{figure*}
    \includegraphics[width=14cm,trim=0 0 0 0,clip=true]{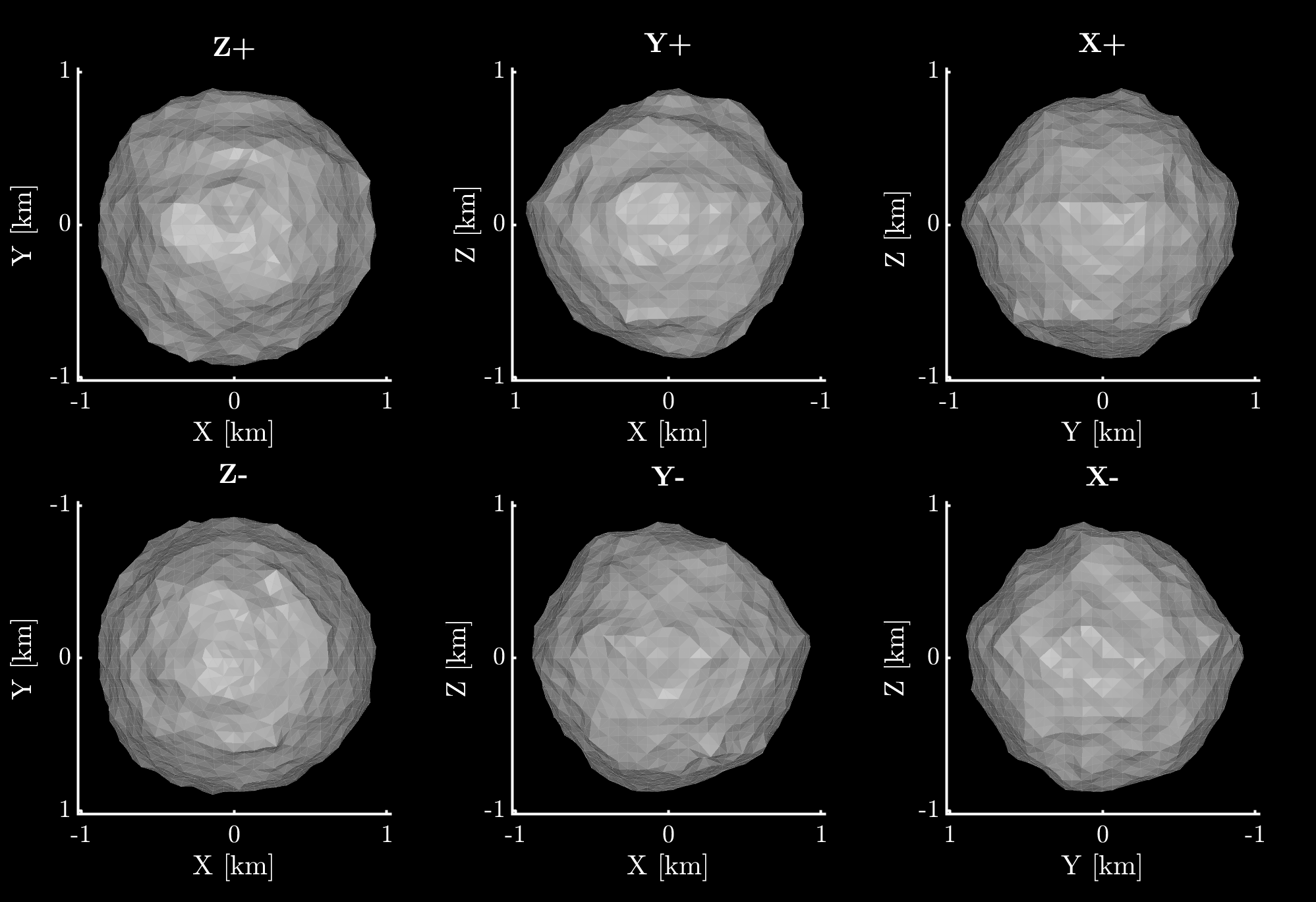}
    \caption{The best-fit shape model of (23187) 2000 PN9 constructed with radar and lightcurve data. This model has its rotational pole at ecliptic longitude $\lambda=96\degr$ and ecliptic latitude $\beta=+30\degr$, and a sidereal rotation period of $2.53216\pm0.00015\rm~h$. The top row shows the model from the positive end of the Z, Y and X axes in the body-centric co-ordinate system. The bottom row shows the model from the negative end of the Z, Y and X axes. The Z-axis is aligned with the rotational pole, which is the shortest axis of inertia. The X-axis is arbitrarily set such that it is viewed from the positive end for the plane-of-sky at $T_{0}$. Axis lengths are given in kilometres. It should be noted that the model for the antipode solution has a very similar shape, such that any discussion of this model's features will also apply to the antipode model.}
    \label{fig:radmod}
\end{figure*}

\begin{figure*}
\includegraphics[width=13cm,trim=0 0 0 0,clip=true]{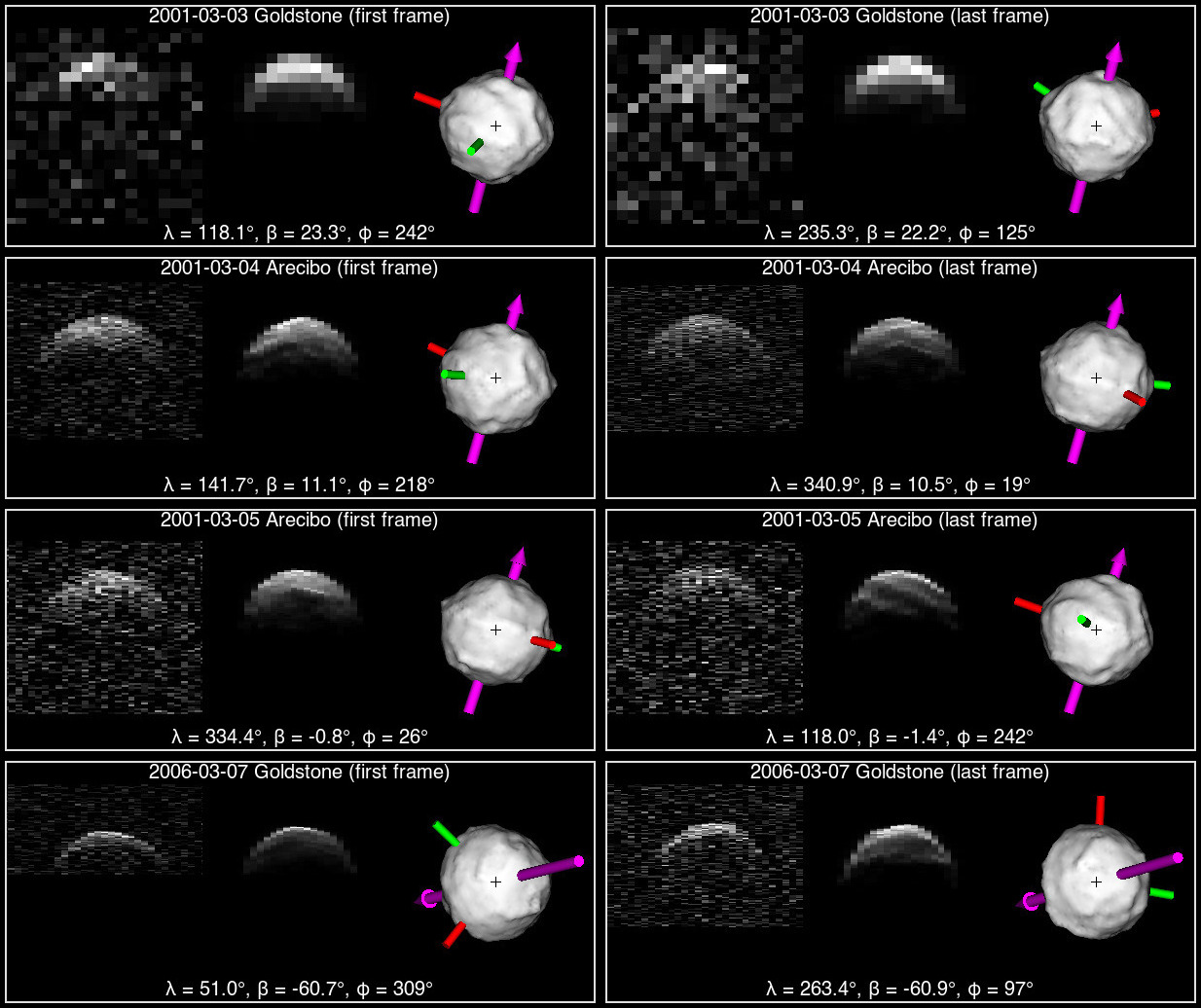}
\caption{A comparison between delay-Doppler observations and the combined radar and lightcurve model for (23187) 2000 PN9, showing the first and last frame of each included dataset. Each three-panel image comprises the observational data (first panel), a synthetic echo (second panel) and a plane-of-sky projection of the model (third panel). On the first two panels, delay increases towards the bottom of the vertical axis and Doppler frequency increases along the horizontal axis. The plane-of-sky projections (third panel) are displayed with the celestial north at the top and east to the left, in an equatorial coordinate system. The rotation axis, which is closely aligned with the z-axis in the body-centric coordinate system, is marked with a purple arrow. The axes of minimum and intermediate inertia are indicated by red and green rods respectively. The body-fixed longitude $\lambda$ and latitude $\beta$ for the radar line-of-sight, and the rotational phase $\phi$, are labelled for each image. These values were determined using the radar shape model's spin-state. The projected centre of mass is marked with a cross. The full sets of radar imaging data are shown in Figures A2 through A5.}
\label{fig:ofs_preview}
\end{figure*}

\subsection{Differences and limitations of the models} \label{sec:modelcomp}
A comparison of the convex inversion and \textsc{shape} pole scans (Figs. \ref{fig:convpoles} and \ref{fig:radpoles}) shows that the latter method produces clearer convergence towards the global pole solution for PN9.

For highly symmetrical asteroids such as PN9, optical lightcurves will be dominated by surface features and observations will be more affected by instrumental performance and atmospheric conditions. Combining lightcurves taken with different filters can amplify these issues, especially when considering scattering effects on the asteroid. In the case of PN9, lightcurves included in the analysis were taken in the V, R and clear filters. Separating these into different subsets to produce independent models is not a viable option, as each of the subsets alone does not provide an adequate number of rotations and viewing geometries to produce a good model.

For asteroids where global shape dominates both optical and radar features (e.g. \cite{2021MNRAS.507.4914Z}), combining the data will better constrain the model. For asteroids like PN9, where surface properties are dominant, it can be difficult to reconcile the radar and optical data.

Radar can penetrate several wavelengths into the surface of an asteroid, and is thus sensitive to features within the top layer of material. Optical observations, however, only represent the surface of the asteroid. If there are any surface features that do not correlate with sub-surface features, such as buried rocks, radar echoes will be produced from features that are not visible on the surface \citep{2016Icar..269...38V}, hence there can be a disparity between optical and radar observations. The heightened importance of scattering laws and albedo introduces further complexity, resulting in a model that is a compromise between fitting both the optical and radar data.

Searches using only the radar data were also conducted, although the pole was poorly constrained without the wide range of viewing geometries afforded by the lightcurves. The shape model also benefits from the inclusion of lightcurve data, as the wide range of viewing geometries results in shadowing effects that can be used to better constrain the surface.

Observations that only cover a range of low sub-observer latitudes (i.e. equatorial views of the asteroid) can cause inaccuracy in shape models. When modelling with \textsc{shape}, this can cause models to assume a more spherical shape caused by uncertainty in the rotation axis. While the combined radar and lightcurve model for PN9 is highly spherical, the lightcurve and radar data span a sufficient range of viewing geometries to eliminate concerns as to whether PN9 could be more oblate than the model suggests. Goldstone radar imaging data from 2006 (Fig. A5) are particularly useful in this regard, as they correspond to a sub-observer latitude of -61$\degr$ over 148$\degr$ of rotation. 

\subsection{Disk-integrated properties}
Continuous wave (CW) spectra can be used to determine the asteroid's circular polarisation ratio (SC/OC), whereby the echo power is recorded in both the same circular (SC) and opposite circular (OC) polarisations.

Arecibo observations of PN9 from 4 and 5 March 2001 give SC/OC ratios of $0.234\pm0.003$ and $0.235\pm0.006$ respectively, which is consistent with the mean SC/OC ratio of $0.270\pm0.079$ for S and Q class NEAs \citep{benner_surfrough_2008}. OC radar cross sections were also measured on these dates, returning $0.20\pm0.05\rm ~km^2$ and $0.18\pm0.05\rm ~km^2$ on 4 and 5 March respectively. The radar albedo, which is determined by dividing the OC cross-section by the model's projected area, was determined to be an average of $0.08\pm0.08$ on both days. This is consistent with the mean radar albedo of $0.19\pm0.06$ for S and Q type NEAs reported in \citet{2022PSJ.....3..222V}.

The SC/OC ratio is often taken as an analogue for structural complexity near the surface. The SC component is determined by surface roughness at scales comparable to the sampled wavelength. For mirrorlike backscattering, the SC component would be zero. A surface that is very rough on scales comparable to the transmitted signal's wavelength will have a stronger SC component \citep{ostro2002}. For the Arecibo CW observations of PN9, this scale is 13 cm. Results from the OSIRIS-REx mission, however, have cast doubt upon this interpretation of SC/OC ratios. Circular polarisation ratios suggest that Bennu is relatively smooth above the cm-scale \citep{nolan_bennu_2013}. However, the spacecraft data show that Bennu's surface is much rougher than this with larger scale boulders than expected \citep{2019NatAs...3..341D}. Eros and Itokawa, which are both S-complex asteroids that have been visited by spacecraft, have SC/OC ratios of $0.22\pm0.06$ and $0.26\pm0.04$ respectively \citep{2001M&PS...36.1697M,2004M&PS...39..407O}. Due to dissimilar formation processes, these asteroids exhibit differences in surface roughness \citep{2019Icar..325..141S}. PN9 is not thought to share a formation process with either of these asteroids, hence taxonomic and polarimetric similarities do not guarantee a similar surface to Eros or Itokawa. Didymos, which is a recently-visited S-complex asteroid with an SC/OC ratio of $0.20\pm0.02$ \citep{benner_surfrough_2008}, is thought to be YORP evolved \citep{2022PSJ.....3..160M}. Despite these similarities, a direct comparison is not advised. Didymos likely experienced spin-breakup to form Dimorphos, while it is not clear if PN9 has previously broken up and reformed. While radar polarimetry can be used to reliably infer the surface roughness of small bodies (e.g. \cite{2021PSJ.....2...30H}), caution should be taken in assuming the surface roughness of PN9 from its SC/OC ratio. 
 

To determine the optical albedo of PN9, the HG photometric system \citep{1989aste.book...524} was fit to Minor Planet Center (MPC) astrophotometric data reported in the Johnson V-band using a Monte Carlo resampling method, obtaining $\text{H}=15.947 \pm 0.036$ and $\text{G}=0.108 \pm 0.016$. MPC astrophotometry has previously been shown to provide valuable data to constrain the phase curves of asteroids \citep[e.g.][]{2013PhDT.......276W}. The MPC astrophotometry does not report individual photometric uncertainties, and so each data point was resampled with a standard deviation equal to the maximum observed light curve amplitude of 0.181 to account for rotational variability in the data. Using the absolute magnitude and the radar-derived diameter $\rm D_{eq} = 1.71 \pm 0.07$$\rm ~km$, the optical albedo is calculated as $0.25 \pm 0.02$, consistent with the polarimetrically-derived albedo from \citet{belskaya_polarimetry_2009}. Caution must be used when inferring physical properties from phase curve-derived parameters of NEAs, however, due to the potential for changing aspect to introduce additional brightness modulations in the phase curve that are unrelated to the scattering behaviour of the surface material \citep{2022MNRAS.513.3076J}.

\subsection{Rotational phase analysis and the search for YORP}\label{sec:yorp}
Minute changes to an asteroid's rotation period can be detected through rotational phase analysis. A constant-period model of the asteroid can be used to generate synthetic lightcurves, which can then be compared with optical lightcurves. Any difference in rotational phase between the observed and synthetic lightcurves indicates a change in spin state.

Asteroids undergoing constant rotational acceleration due to the YORP effect will show a quadratic increase in rotational phase offset against time. Step changes in rotation period caused by mass-lofting, impacts, or repeated planetary perturbations, would cause sporadic changes in phase offset. 

The best-fit model of PN9 presented in Section \ref{sec:shape}, which was constructed from a combination of radar and lightcurve data, was used to produce synthetic lightcurves corresponding to the 18 optical lightcurves used in this analysis.

To generate  each synthetic lightcurve, a ray-tracing algorithm was used to determine the illumination of each of the model's facets through a full rotation at the appropriate viewing geometry. The scattering model that was used is a combination of the Lambertian and Lommel-Seelinger models \citep{kaasalainen_optimization_2_2001}. The sum of facet fluxes was then used to calculate the expected relative brightness of the asteroid, accounting for self-shadowing effects. This was converted to a relative magnitude, then both the synthetic and observed lightcurve magnitudes were offset to oscillate about a common zero point. The synthetic lightcurves were then shifted in phase in steps of 0.5$^{\circ}$ and the $\chi^2$ fit of the shifted synthetic lightcurves to the observed lightcurves was measured. For each lightcurve, the shift that produced the best overall fit was taken to be the rotational phase offset between the constant period model and the actual rotational phase of the asteroid.

As PN9 is a highly symmetrical asteroid, brightness variations due to rotation are extremely small with lightcurve amplitudes often being as small as $\sim$0.05 magnitudes. Without observing clear turning points that can be reliably linked between lightcurves, it is difficult to detect a coherent progression in phase offsets caused by YORP. As PN9's lightcurves are extremely sensitive to surface detail and scattering parameters, it is not always possible to identify turning points that repeat across both observed and synthetic lightcurves. There are, however, a small number of clear and repeated turning points within our dataset that can be used for a phase offset analysis.

\begin{figure}
\centerline{\includegraphics[width=\columnwidth]{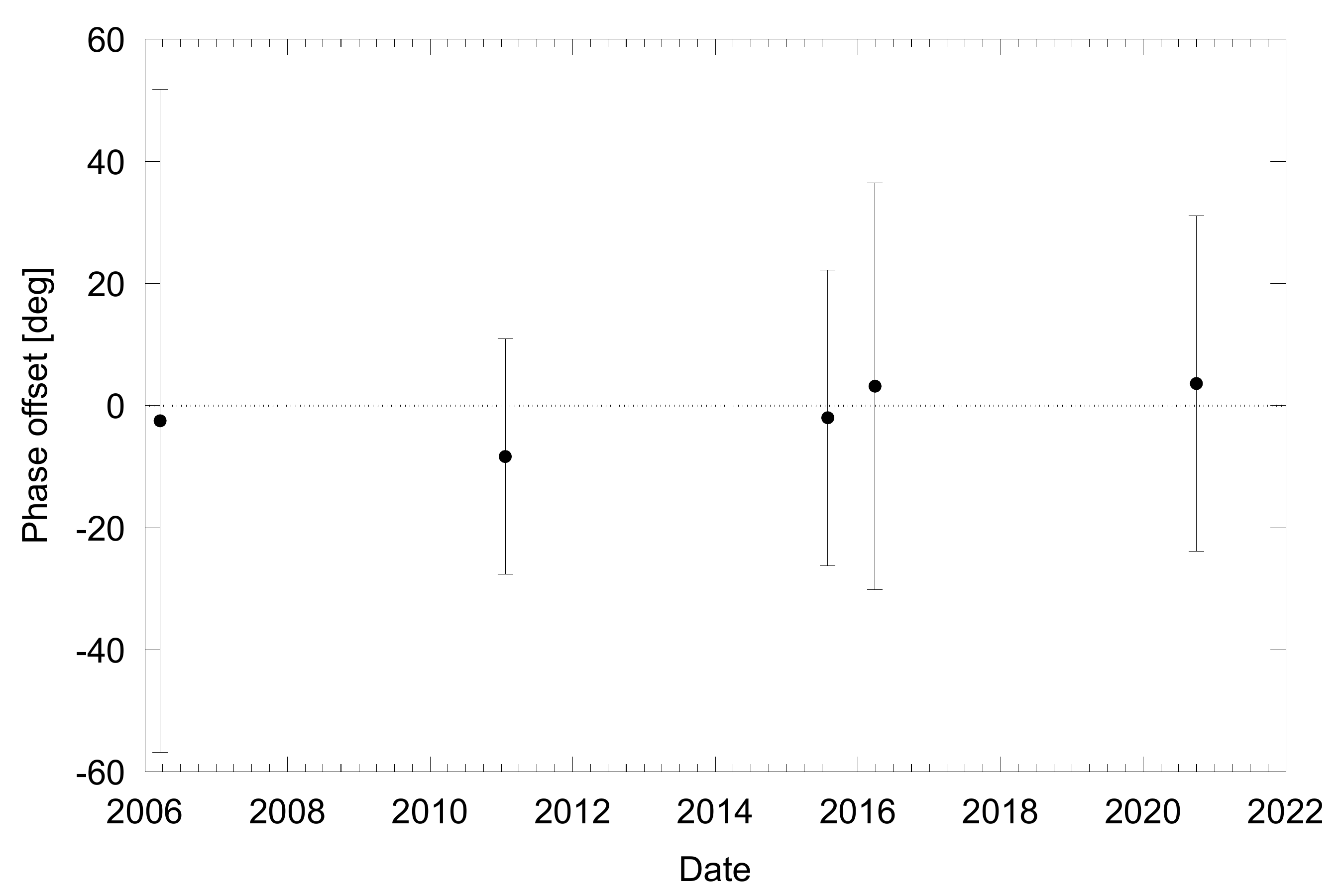}}
\caption{Phase offset measurements for the best-fit combined radar and lightcurve model of (23187) 2000 PN9 where $T_0$=2453815.29199 (March 2006). Phase offsets were measured against the `ph. off' subset of lightcurves marked in Table \ref{tab:obstable}. Phase offset measurements were averaged from groups of lightcurves, with groups being arranged such that there are a maximum of 180 days between consecutive lightcurves within a group. The straight dashed line represents a constant period model for reference.}
\label{fig:phoff}
\end{figure}

Figure \ref{fig:phoff} shows the measured phase offsets for each epoch, where temporally clustered measurements are averaged. A total of 24 lightcurves were included in the YORP fitting process, which are indicated in \ref{tab:obstable}. The excluded lightcurves produced unacceptably large phase uncertainties. The best-fit YORP strength for PN9 is $0.2\pm1.6\times10^{-8}\,\rm rad/day^{2}$, which is comparable in magnitude to the smallest confirmed YORP detections, and in line with expectations for a $\sim$2 km asteroid. Although this measurement is poorly constrained, and it is not possible to rule out constant-period rotation, it does place an upper limit on YORP acceleration. 

We also considered a case where YORP acceleration between each apparition induces close to 360$^{\circ}$ of additional rotation. This would produce an apparent phase offset of 0$^{\circ}$ at each apparition by bringing the asteroid's rotation back into phase with a constant-period model. For this to be the case, YORP acceleration would have to be close to an integer multiple of $4.8\times10^{-6}\,\rm rad/day^{2}$. This value is greater than the current strongest published YORP detection of $3.49\times10^{-6}\,\rm rad/day^{2}$ with (54509) YORP \citep{lowry_direct_2007,taylor_spin_2007}. Considering that the diameter PN9 is $\sim$15$\times$ greater than that of (54509) YORP, and that PN9 has high global symmetry, a YORP torque of this magnitude is considered to be unlikely. Our analysis therefore finds no compelling evidence for rotational acceleration of PN9, within the limits of the data. We discuss the potential significance of this below.

\subsection{Geophysical properties}\label{sec:geophysical}

\begin{figure}
\centerline{\includegraphics[width=\columnwidth]{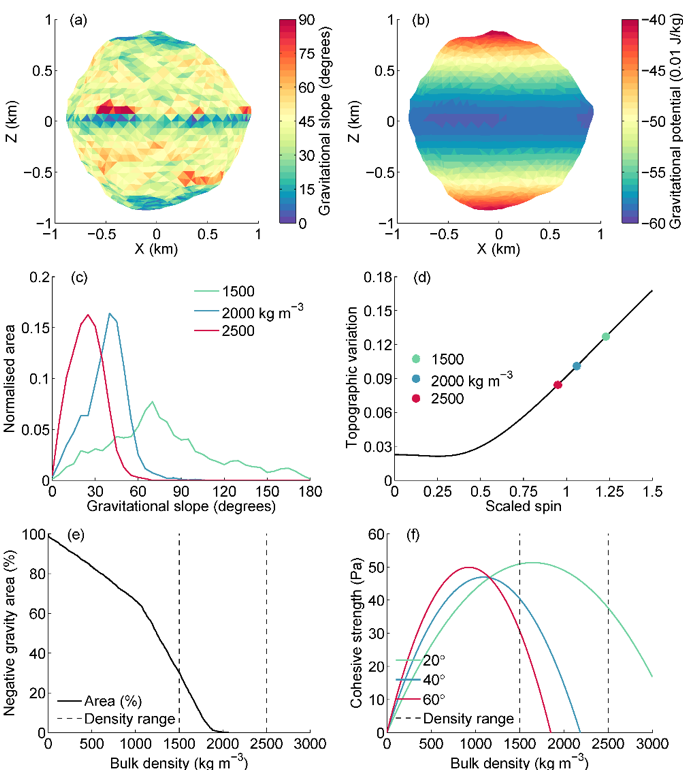}}
\caption{Geophysical analysis of asteroid (23187) 2000 PN9. (a) Gravitational slopes and (b) gravitational potential computed assuming a bulk density of 2000 kg m$^{-3}$. (c) Areal distribution from (a) for three different values of bulk density. (d) Topographic variation from (b) for three different values of bulk density. (e) Negative effective gravity area as a function of bulk density. (f) Cohesive strength as a function of bulk density and angle of friction. The vertical dashed lines in (e) and (f) show the bulk density range for a typical S-type rubble-pile asteroid.}
\label{fig:geophys}
\end{figure}

The rapid spin-rate of PN9 implies that it could undergo frequent landslide and mass shedding events, and/or undergo structural failure. To investigate the spin-stability of PN9, we applied several geophysical analyses to the radar-derived shape model following the methods previously applied to asteroids (68346) 2001 KZ66 and (2102) Tantalus in \cite{2021MNRAS.507.4914Z} and \cite{2022MNRAS.515.4551R}, respectively. In particular, gravitational slopes, gravitational potential, and topographic variation were determined by applying a polyhedron gravity field model modified for rotational centrifugal forces \citep{1997CeMDA..65..313W, 2014Natur.512..174R, 2014Icar..234...53R, 2019Icar..329..207R}, and body-average cohesive forces were evaluated using the Druger-Prager failure criterion \citep{2007Icar..187..500H}. These calculations were performed over a bulk density range of 1500 to 2500 kg m$^{-3}$ to cover the expected values for an S-type rubble-pile asteroid \citep{2012P&SS...73...98C}.

Figure \ref{fig:geophys} summarises the results of these analyses and indicates that PN9 is qualitatively very similar to asteroid Tantalus (i.e. Figure 12 of \cite{2022MNRAS.515.4551R}). For instance, a minimum bulk density of $\sim$2070 kg m$^{-3}$ is required to prevent rotational mass shedding (Fig. \ref{fig:geophys}e) and a cohesive strength of up to $\sim$50 Pa (Fig. \ref{fig:geophys}f) is required to prevent rotational structural failure (versus ~2200 kg m$^{-3}$ and ~45 Pa for Tantalus, respectively). As shown in Figure \ref{fig:geophys}c, for a nominal bulk density of 2000 kg m$^{-3}$ the gravitational slopes peak at $\sim$40$\degr$ and there is prominent latitudinal banding in the gravitational potential (Fig. \ref{fig:geophys}b). This facilitates mass movement from PN9’s poles to its equator \citep{2015Icar..247....1S} and the Sq/Q-type classification of PN9 may indicate recent re-surfacing caused by YORP spin-up \citep{2018Icar..304..162G}. If PN9 happens to be spinning-up by YORP, then the conditions for landslides, mass shedding, and structural failure become more easily met, which could eventually lead to the formation of a small moon (e.g. \cite{walsh_rotational_2008}).

Our data contain no evidence to suggest that PN9 already has a natural satellite. While the presence of a secondary can be difficult to detect with optical imaging, radar observations are particularly effective at identifying multiple systems (e.g. \cite{2011Icar..216..241B, 2019LPI....50.2945T}). A preliminary analysis finds there are no consistent peaks in the continuous wave spectra, nor any visible satellites in delay-Doppler radar images. Any sufficiently bright secondary with a diameter above 19 m would be seen in individual images. The maximum photometric contribution from an undetected satellite would thus be on the order of $10^{-4}$ magnitudes. We note that as a secondary would most likely have formed from a previous breakup of the primary, its composition - and hence brightness - would be comparable to that of the primary. Moons of top-shaped asteroids typically have $\sim$1\% of the mass of their primary \citep{2022ApJ...937L..36H}. Assuming equal densities, in the case of PN9 this would correspond to a $\sim$370 m moon. A secondary of this size would be detectable in any of the radar imaging data.


\section{Discussion}\label{sec:analysis}

\begin{table*}
    \centering
    \begin{threeparttable}
	\caption{Comparison of top-shaped NEAs with published physical models that are based on radar or spacecraft data. \label{tab:YORPoids}} 	
	\begin{tabular}{lllccccccc}
		\hline\hline
		\noalign{\smallskip}
		\multicolumn{2}{c}{Asteroid} & Period & Diameter & Volume & Rotational & Type & SC/OC & Multiplicity & Ref. \\
		\multicolumn{2}{c}{}& [$h$] & [$km$] & {[$km^3$]}& Pole {($\lambda$, $\beta$) [$^\circ$]} & & & &\\ \hline
		\noalign{\smallskip}
		    (2102) & Tantalus$^{\dagger}$ & 2.391 & 1.3 & 1.05 & (180, +24) & Sr & 0.19 &    & 1 \\
            (3200) & Phaethon$^{\ddagger}$ & 3.604 & 6.4 & 75 & (316, -50) & B & 0.19 & & 2,3,4 \\
	        (23187) & 2000 PN9 & 2.532 & 1.82 & 2.627 & (096, +30) & S/Sq/Q & 0.23 &   & this work, 5,6  \\
	        (65803) & Didymos & 2.260 & 0.84 & 0.249 &  (310, -84) & Sq & 0.2 & Binary & 7,8,9 \\
	        (66391) & Moshup$^{\S}$ & 2.765 & 1.53 & 1.195 & (326, -65) & S & 0.45 & Binary & 10,11\\
	        (101955) & Bennu & 4.296 & 0.57 & 0.062 & (086, -60)  & B & 0.18 &  & 12,13   \\
	        (136617) & 1994 CC & 2.389 & 0.69 & 0.125 & (336, +22) & Sq & 0.40, 0.50 & Triple & 14  \\
            (153591) & 2001 SN263 & 3.426 & 2.9 & 8.2 & (309, -80) & B & 0.17 & Triple & 15 \\
	        (162173) & Ryugu & 7.633 & 0.88 & 0.377 & (179, -87) & Cg & N/A &  & 16,17  \\
	        (185851) & 2000 DP107 & 2.775 & 0.99 & 0.337 & (294, +78) & C & 0.25 & Binary & 7,18,19,  \\
            (276049) & 2002 CE26 & 3.293 & 3.65 & 21.7 & (317, -20) & C & 0.21 & Binary & 20 \\
	        (341842) & 2008 EV5 & 3.725 & 0.42 & 0.035 & (189, -84) & C/X & 0.38 & & 21,22,23 \\
	\end{tabular}
	\begin{tablenotes} \item \textbf{Notes.} ``Period'' is the sidereal rotation period of the asteroid. ``Diameter'' gives the maximum equatorial diameter. ``Volume'' is derived from the physical model of the asteroid. ``Rotational Pole'' denotes the spin-axis orientation of the asteroid in the ecliptic coordinate system. ``Type'' denotes the taxonomic classification(s) each asteroid has been given. ``SC/OC'', also known as the circular polarisation ratio, is the ratio between same circular and opposite circular polarised radar echo. ``Multiplicity'' denotes the number of known bodies in the asteroid system. Inclusion in this list is determined by the shape of the primary or `Alpha' body, and physical parameters refer to the primary. \item $\dagger$ Retrograde model; $\ddagger$ Values for the shape and spin-state are preliminary as of December 2022; $\S$ Also known as 1999 KW4. \item \textbf{References.} (1) \cite{2022MNRAS.515.4551R}; (2) Marshall (priv. comm.); (3) \cite{2019P&SS..167....1T}, (4) \cite{1985MNRAS.214P..29G}; (5) \cite{2014Icar..228..217T}; (6) \cite{2019Icar..324...41B}; (7) \cite{benner_surfrough_2008}; (8) \cite{2020Icar..34813777N}; (9) \cite{2018P&SS..157..104C}, (10) \cite{ostro_radar_2006}; (11) \cite{2004Icar..170..259B}; (12) \cite{2019Natur.568...55L}; (13) \cite{nolan_bennu_2013};  (14) \cite{2011Icar..216..241B}; (15) \cite{2015Icar..248..499B}; (16) \cite{2019Sci...364..268W}; (17) \cite{2019Sci...364..252S}; (18) \cite{2015AJ....150...54N}; (19) \cite{2003Icar..163..363D}; (20) \cite{2006Icar..184..198S}; (21) \cite{busch_radar_2011}; (22) \cite{2008DPS....40.2821S}; (23) \cite{2011Icar..216..184R}
	\end{tablenotes}
	\end{threeparttable}
\end{table*}

Our analyses with optical and radar data show that PN9 has a spinning-top shape, which is characteristic of a rapidly rotating rubble pile \citep{2018ARA&A..56..593W}. 

Top-shaped or `YORPoid' asteroids are believed to be ubiquitous within the inner Solar System, due to the rate they are being discovered through radar and spacecraft imaging. The number of well-modelled examples, however, is relatively low.

In addition to this work, we have identified eleven top-shaped asteroids that have published models with full geometric parameters. These are summarised in Table \ref{tab:YORPoids}. Some objects, such as (2867) \v{S}teins \citep{2010Sci...327..190K} and (29075) 1950 DA \citep{2007Icar..190..608B, TJZ_thesis} were excluded as their top-like shapes exhibit global asymmetries that differentiate them from more definitive examples such as Bennu \citep{2019Natur.568...55L} or Moshup \citep{ostro_radar_2006}. It should be noted that a larger number of candidates were identified, but do not have publicly available models and/or geometric parameters. We have compiled an informal list of these objects which can be accessed online\footnote{\url{http://astro.kent.ac.uk/~YORP/spintop.html}}. 

In comparison to other top-shaped asteroids, several features of PN9 stand out. The majority of top-shaped asteroids listed in Table \ref{tab:YORPoids} are multiple systems. As discussed in Section \ref{sec:geophysical}, there is no evidence to suggest that PN9 has any satellites. To date, PN9 is the second largest top-shaped solitary asteroid with a fully developed shape model. In comparison to other top-shaped asteroids, PN9 has higher levels of global symmetry and a less pronounced equatorial ridge or bulge. This could be a result of PN9's greater mass, or the presence of internal cohesive forces.

Top-shaped asteroids are poor candidates for YORP detection. Their highly symmetrical shapes produce low-amplitude lightcurves that do not vary significantly between different viewing geometries. This makes it difficult to constrain the rotational pole and period, and increases the importance of accurate surface fitting and the performance of scattering models. While radar observations can somewhat mitigate the limitation, the only confirmed YORP detection on a top-shaped asteroid to date is derived from both radar and Hubble Space Telescope observations of (101955) Bennu \citep{Nolan:2019eib}. Nevertheless, they may be crucial in distinguishing between components of `normal YORP' (NYORP), which is dominated by global shape, and `tangential YORP' (TYORP), which is driven by irregularity across an asteroid's surface \citep{golubov_tangential_2012}. Strong YORP detections on globally symmetric asteroids, which should have very small NYORP components, would imply a strong TYORP component. Separating the components of observational YORP detections can only be possible if both extremes are studied, as opposed to the current bias towards YORP analyses of highly asymmetric asteroids which have significant NYORP components.

Our analysis of PN9 includes 14.5 years of lightcurve coverage and shows that it is not currently experiencing significant rotational acceleration. Small YORP torques or sporadic changes to rotation period, however, cannot be ruled out with the current data. 

The YORP effect is thought to be a key mechanism in the production of spinning-top rubble piles and binary systems. In the `spin-up' configuration YORP torque can steadily increase an asteroid's spin rate until it experiences physical deformation to become a top-shaped `YORPoid'. 

An analysis has been performed of the spin-driven evolution of (101955) Bennu and (162173) Ryugu by \cite{2020Icar..35213946H}, who find that reshaping at longer periods is driven by changes to surface structure, while reshaping at shorter periods is driven by the failure of internal structures. Ryugu and Bennu, which are both C-complex asteroids, have measured bulk densities of $1190\rm ~kg~m^{-3}$ \citep{2019NatAs...3..352S, 2019Sci...364..268W}. As an S-complex asteroid it is likely that PN9 has a higher density than this \citep{2012P&SS...73...98C}, which would suggest that a higher spin rate is required to induce rotational deformation. PN9's 2.53 h rotation period, which is close to the 2.2 h spin barrier for cohesionless asteroids \citep{2000Icar..148...12P}, favours the failure of internal structure being primarily responsible for any recent deformation PN9 has experienced. 

YORP-driven deformation of near-Earth asteroids is likely to be self-limited by various mechanisms. As an asteroid approaches or crosses the spin-limit barrier, surface regolith may migrate from the poles towards the equator \citep{2019Icar..317..354H}. In order to conserve angular momentum, the asteroid's period must increase, countering the YORP spin-up. Due to the YORP effect's strong dependence on shape, spin-driven reshaping into a more symmetrical top shape will decrease the strength of YORP torques. This self-governing must not occur in all cases, however, as it has been demonstrated that the YORP effect can form binaries through rotational breakup \citep{walsh_rotational_2008}. 

Rotational breakup does not always produce a multiple system. Material can re-accrete towards the equator, producing an equatorial ridge \citep{2022ApJ...937L..36H}, while the orbital evolution of satellites can lead to them migrating outward until they are lost. As PN9 is near-spherical and does not have a prominent equatorial ridge, there is no indication that it has previously experienced spin-breakup or lost a satellite. 

It is also possible that PN9 is an example of an asteroid that is trapped in a state of rotational equilibrium, where normal and tangential YORP components enforce a constant rotation period over long time periods \citep{2019AJ....157..105G}. If a significant fraction of asteroids are found to have near-zero YORP acceleration, it would confirm the existence of `sinks' that halt the YORP cycle. This would have a significant impact on theories of asteroid evolution. YORP equilibrium states are, however, expected to be seen in systems that are more physically complex than PN9 \citep{2015MNRAS.449.2489B, 2016MNRAS.458.3977G}.

In the next century, PN9 will not come within 100 lunar distances of Earth. This is beyond the range of current and near-future radar facilities, limiting any future observations to optical and infrared telescopes. The best opportunity to observe PN9 until at least 2030 will be from the northern hemisphere in mid-2025, when medium-sized telescopes will be able to image the asteroid over several rotations. Larger northern telescopes should be able to image PN9 in early 2024 and early 2029, while facilities in the southern hemisphere are limited to the aforementioned mid-2025 apparition until after 2030. These observations could be used to better constrain PN9's pole and extend the baseline in the search for YORP, while any improvements to the physical model may improve upon the current phase offset measurements. It is unlikely that further ground-based observations will result in a YORP detection for PN9, however the current constraints could be significantly improved.
      
The non-detection of rotational acceleration of PN9, combined with its highly symmetrical shape and short rotation period, suggest that if it is indeed YORP-evolved then it is an example of self-limitation. In order to better understand the physical evolution of near-Earth asteroids, it is essential to understand the factors that determine if YORP spin-up of a rubble pile will self limit or continue past the spin-breakup barrier and form a binary. As YORPoids are unfavourable targets for YORP detection, analyses of objects that are in the late stages of YORP evolution are under-represented. Further study of these asteroids with future ground-based optical and radar facilities, as well as spacecraft observations, are essential to better understanding the influence of YORP on evolutionary pathways for small bodies.

\section*{Acknowledgements}
We thank all staff at the various observatories that contributed towards this work. LD, SCL, AR, BR, SLJ, TZ, CS, AF and SFG all acknowledge support from the UK Science and Technology Facilities Council. YNK was partially funded by ALLEA through Funding Line 1 of the European Fund for Displaced Scientists (EFDS). IB thanks the PAUSE program for scientists in danger for its support. Part of this research was carried out at the Jet Propulsion Laboratory, California Institute of Technology, under contract with the National Aeronautics and Space Administration (80NM0018D0004). This work was based in part on observations collected at the European Organisation for Astronomical Research in the Southern Hemisphere under ESO programmes 185.C-1033(D, C) and 106.C-0794(A). It was also based in part on service observations made with the Isaac Newton Telescope operated on the island of La Palma by the Isaac Newton Group of Telescopes in the Spanish Observatorio del Roque de los Muchachos of the Instituto de Astrofísica de Canarias under programme I/2020B/05. Observations at the Danish 1.54m telescope at the ESO La Silla observatory were performed as part of the MiNSTEp project, and supported by the Danish Natural Science Research Council (FNU). This research has been funded in part by the Aerospace Committee of the Ministry of Digital Development, Innovations and Aerospace Industry of the Republic of Kazakhstan (Grant No. BR 11265408). This work uses data obtained from the Asteroid Lightcurve Data Exchange Format (ALCDEF) database, which is supported by funding from NASA grant 80NSSC18K0851. This research made use of Astropy, \citep{astropy:2013, astropy:2018}, Matplotlib \citep{Hunter:2007}, IRAF Community Distribution\footnote{\url{https://iraf-community.github.io}} and the NASA/JPL HORIZONS ephemeris tool\footnote{\url{https://ssd.jpl.nasa.gov/horizons/}}. We thank Chris Magri for providing the \textsc{shape} software package and Sean Marshall for providing preliminary results for (3200) Phaethon. 

\section*{Data Availability}
The PDS lightcurves of 2000 PN9 first published in \cite{2016MPBu...43..240W} are available on the Asteroid Lightcurve Data Exchange Format (ALCDEF) database \citep{2011MPBu...38..172W} at \url{https://alcdef.org/}. Lightcurves that have not been previously published will be made available at \url{https://vizier.u-strasbg.fr}. The shape models presented in this work will be submitted to the Database of Asteroid Models from Inversion Techniques (DAMIT) at \url{https://astro.troja.mf f.cuni.cz/projects/damit/}. The radar data are available from the authors upon request.
 



\bibliographystyle{mnras}
\bibliography{ms} 





\bsp	
\label{lastpage}
\end{document}


\label{firstpage}
\pagerange{\pageref{firstpage}--\pageref{lastpage}}
\maketitle

\appendix
\section{Additional Figures}

\begin{figure*}
	
	\resizebox{\hsize}{!}{	
		\includegraphics[width=.25\textwidth, trim=0.5cm 2.5cm 1.5cm 2.5cm, clip=true]{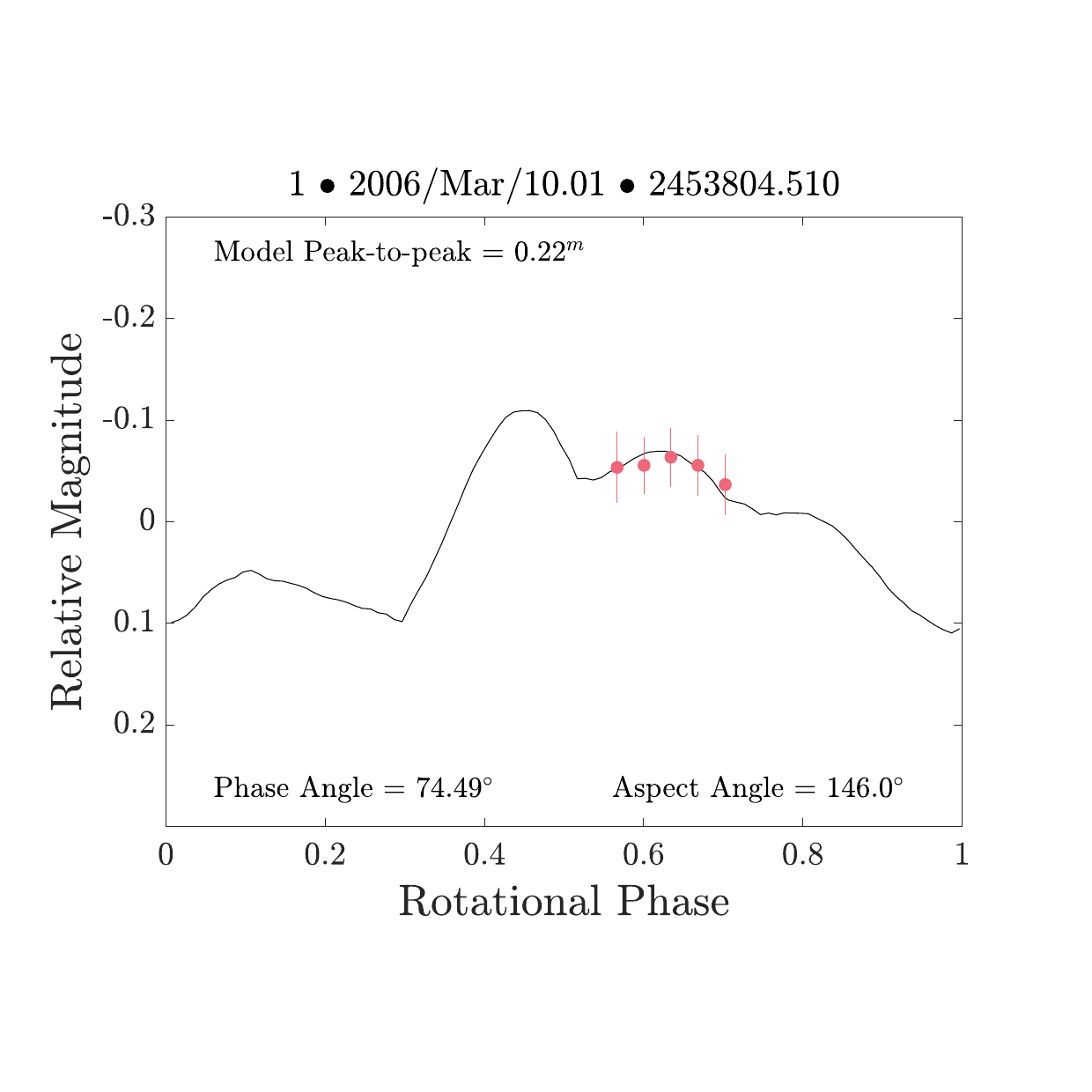}
        \includegraphics[width=.25\textwidth, trim=0.5cm 2.5cm 1.5cm 2.5cm, clip=true]{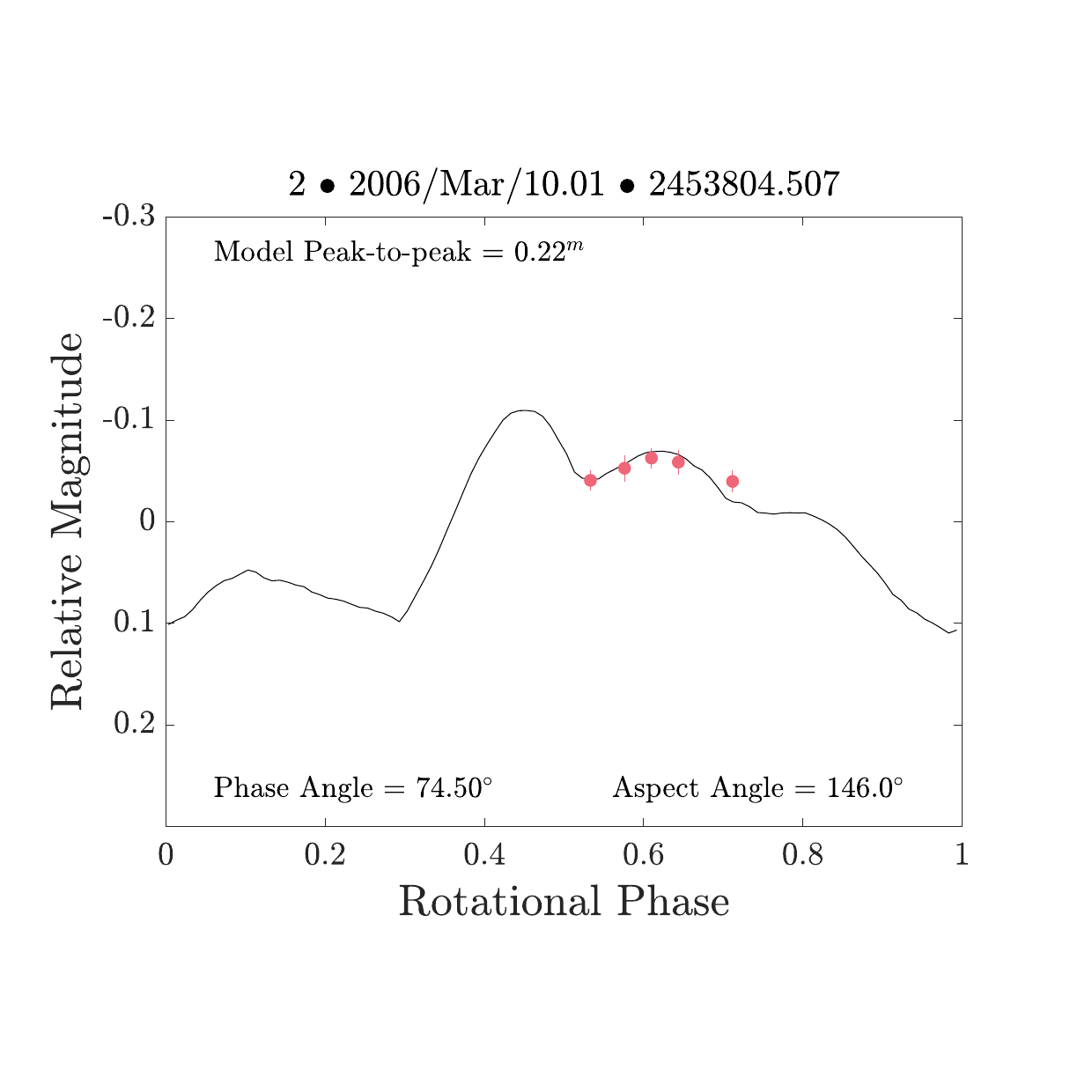}
        \includegraphics[width=.25\textwidth, trim=0.5cm 2.5cm 1.5cm 2.5cm, clip=true]{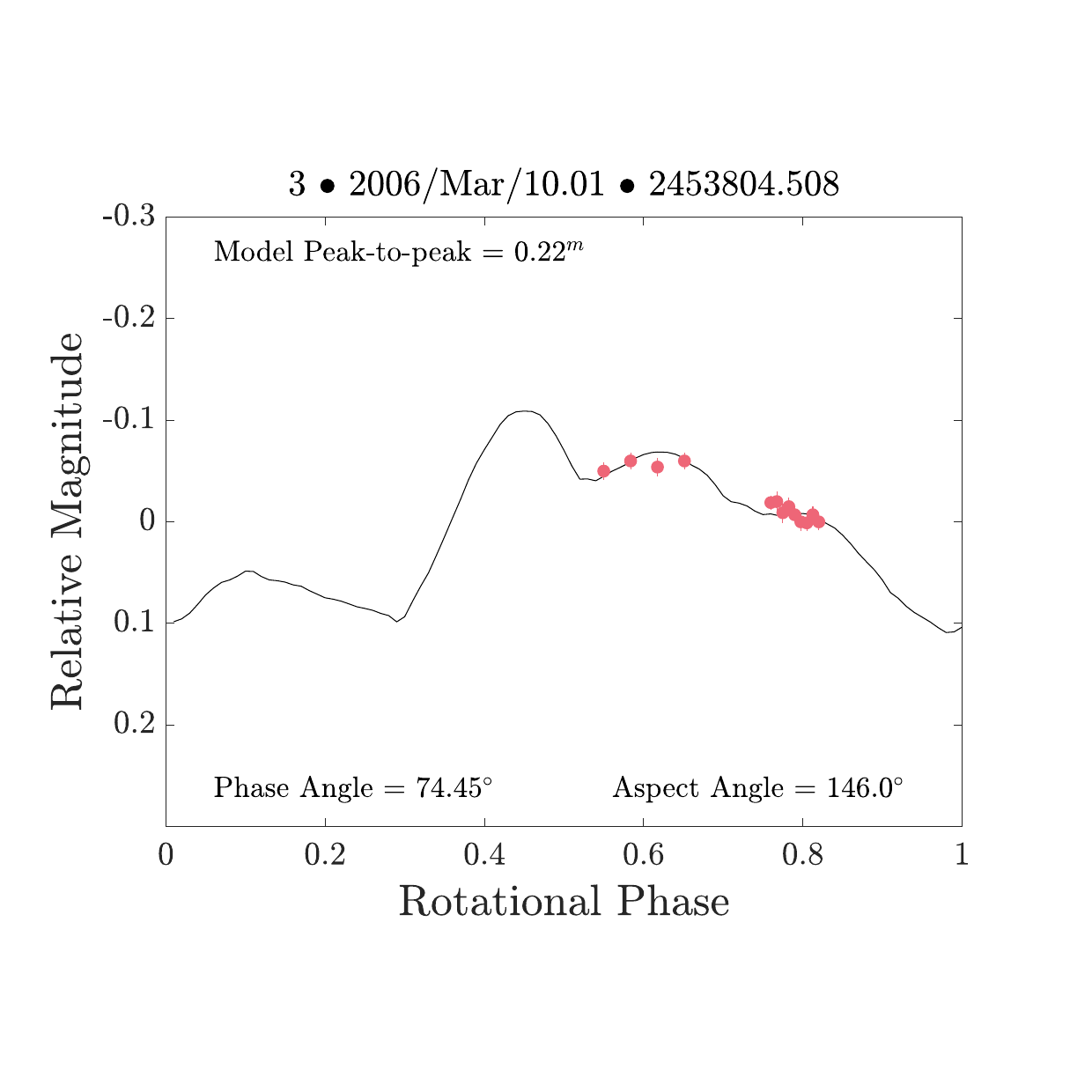}
        \includegraphics[width=.25\textwidth, trim=0.5cm 2.5cm 1.5cm 2.5cm, clip=true]{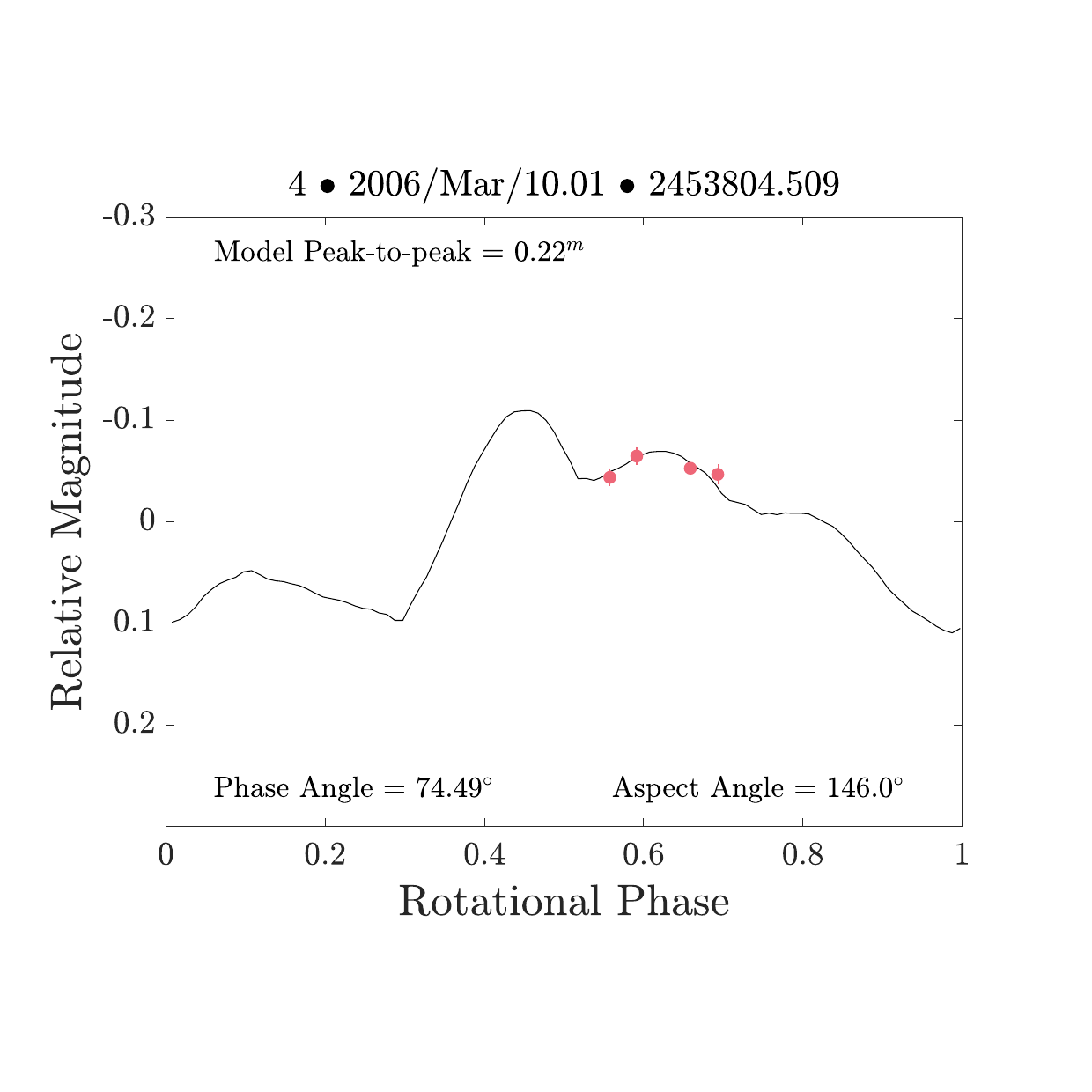}
	}

	\resizebox{\hsize}{!}{	

        \includegraphics[width=.25\textwidth, trim=0.5cm 2.5cm 1.5cm 2.5cm, clip=true]{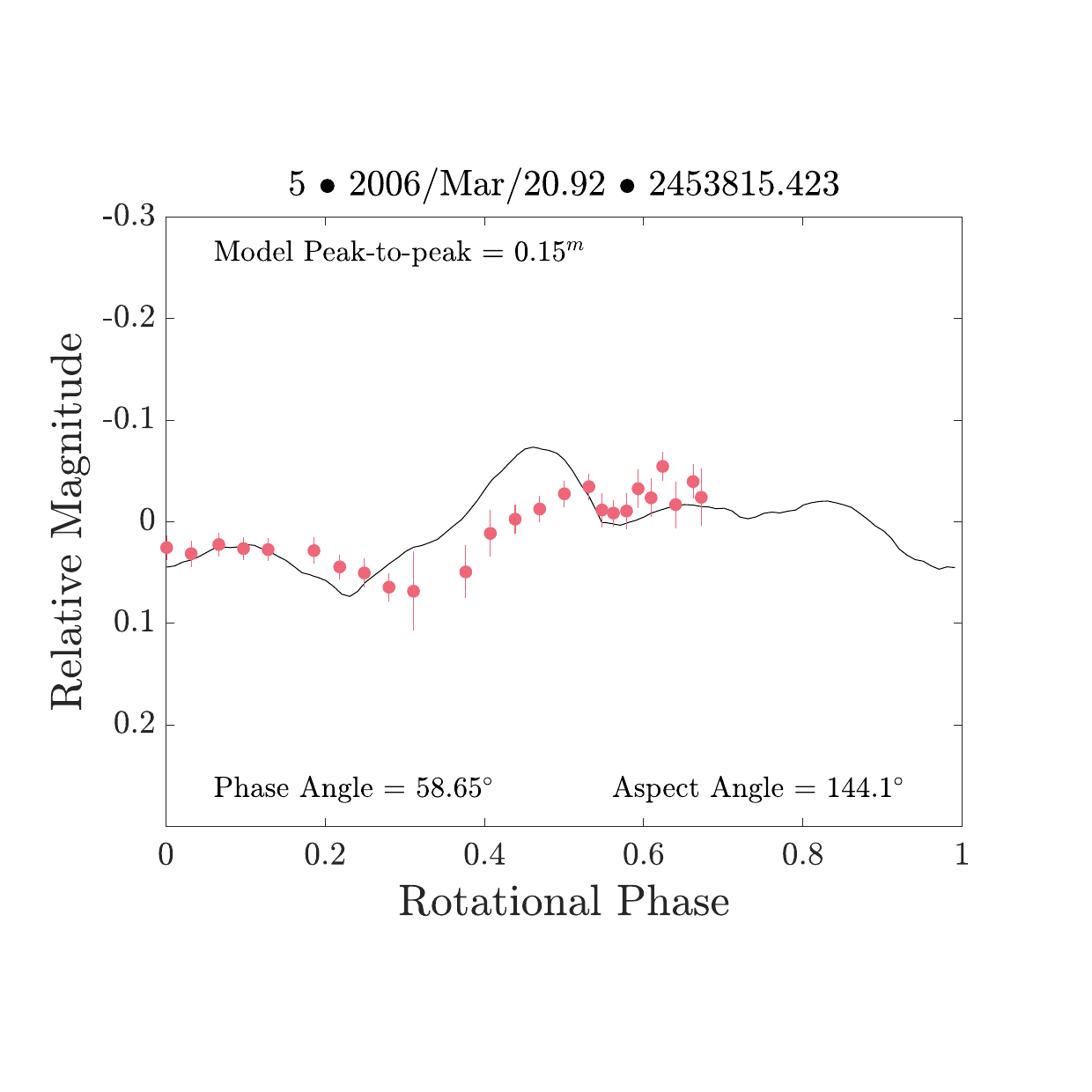}
        \includegraphics[width=.25\textwidth, trim=0.5cm 2.5cm 1.5cm 2.5cm, clip=true]{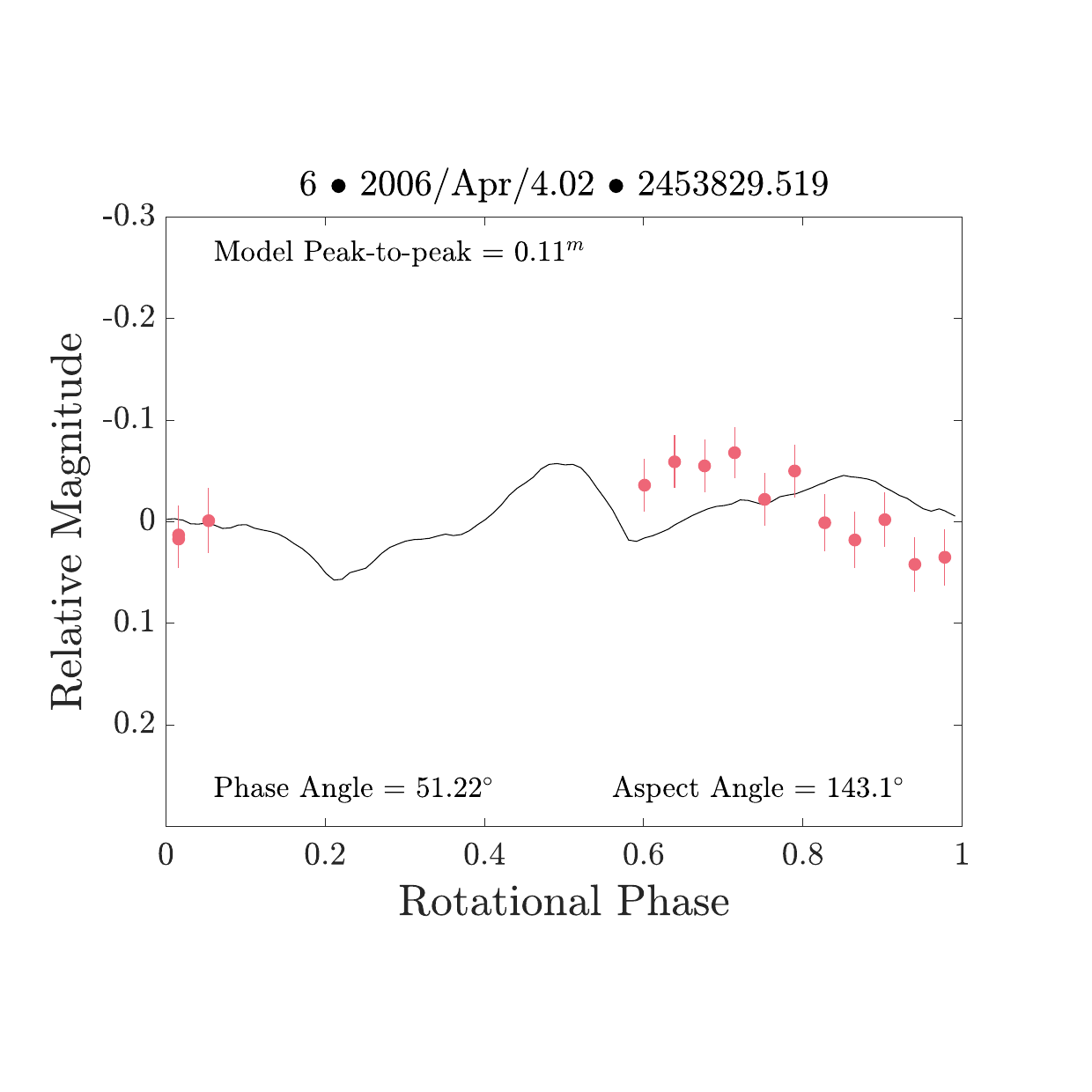}
        \includegraphics[width=.25\textwidth, trim=0.5cm 2.5cm 1.5cm 2.5cm, clip=true]{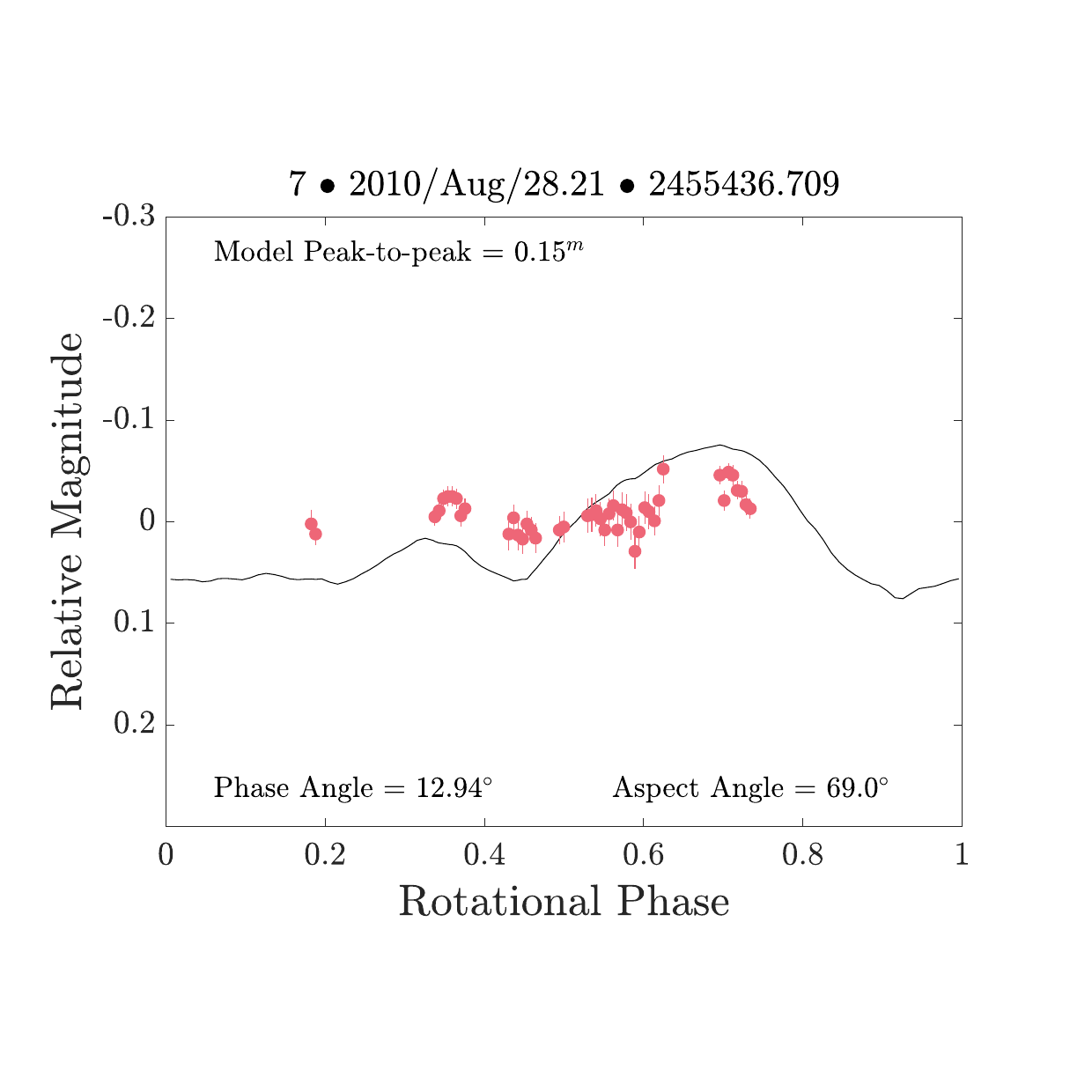}
        \includegraphics[width=.25\textwidth, trim=0.5cm 2.5cm 1.5cm 2.5cm, clip=true]{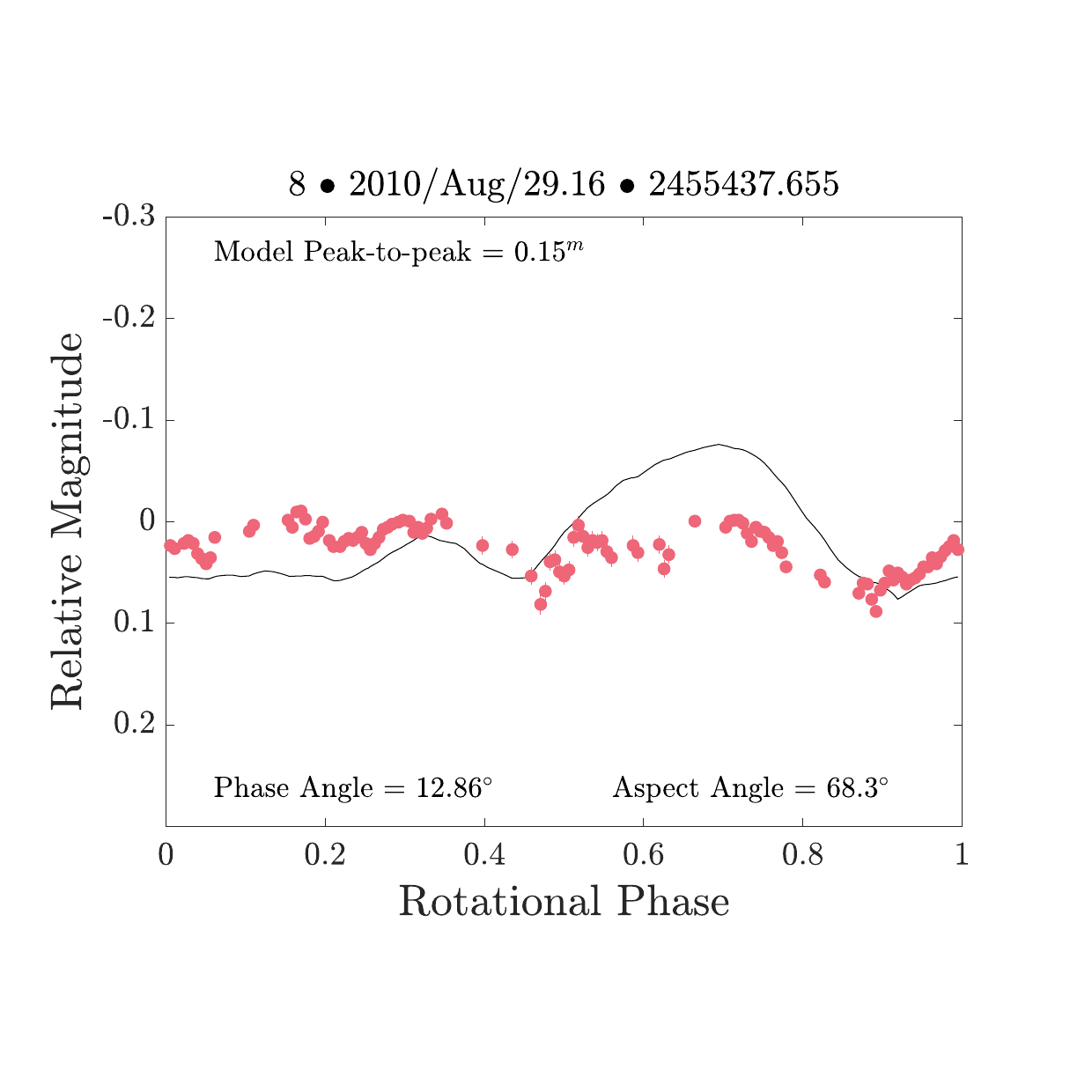}
	}
	
	\resizebox{\hsize}{!}{	

        \includegraphics[width=.25\textwidth, trim=0.5cm 2.5cm 1.5cm 2.5cm, clip=true]{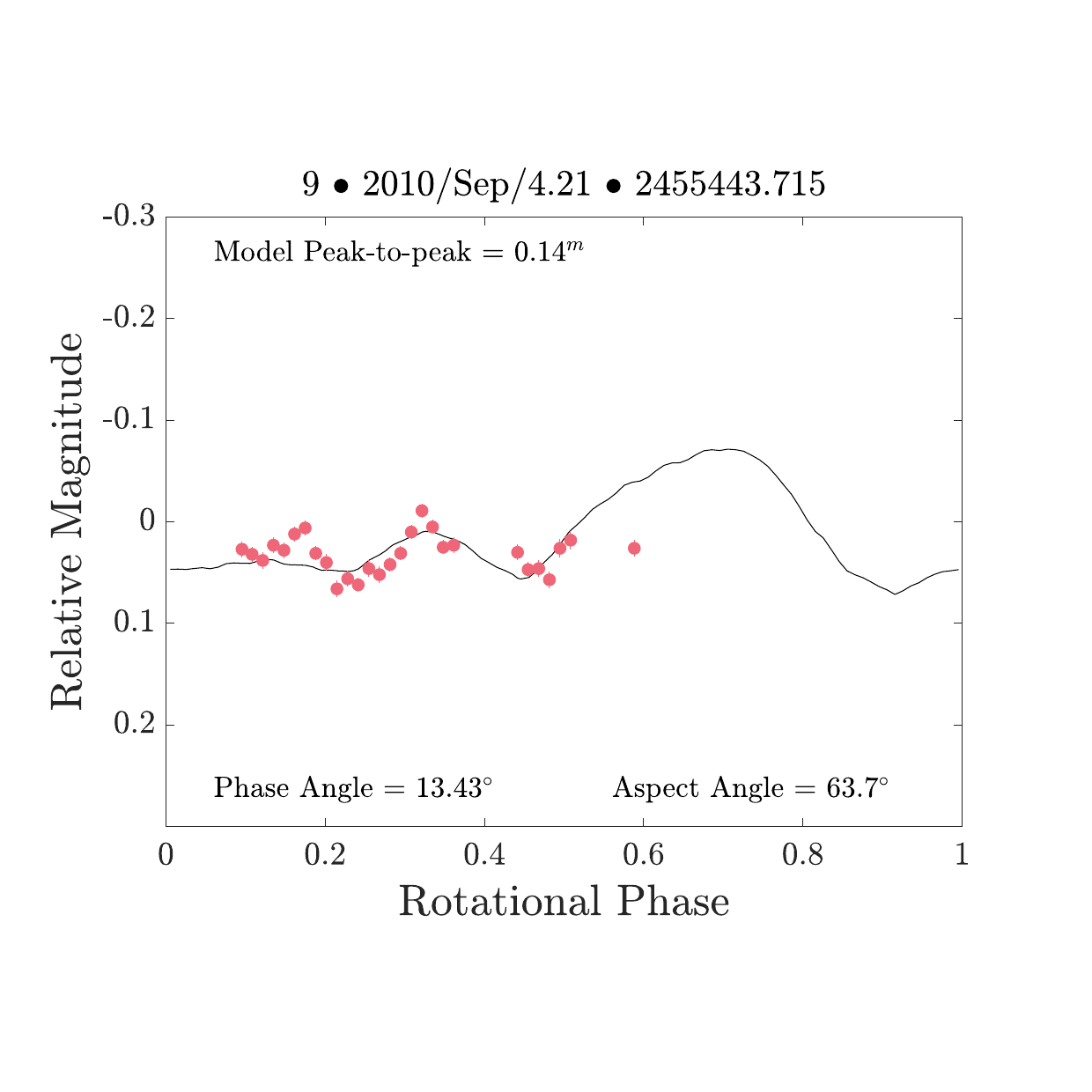}
        \includegraphics[width=.25\textwidth, trim=0.5cm 2.5cm 1.5cm 2.5cm, clip=true]{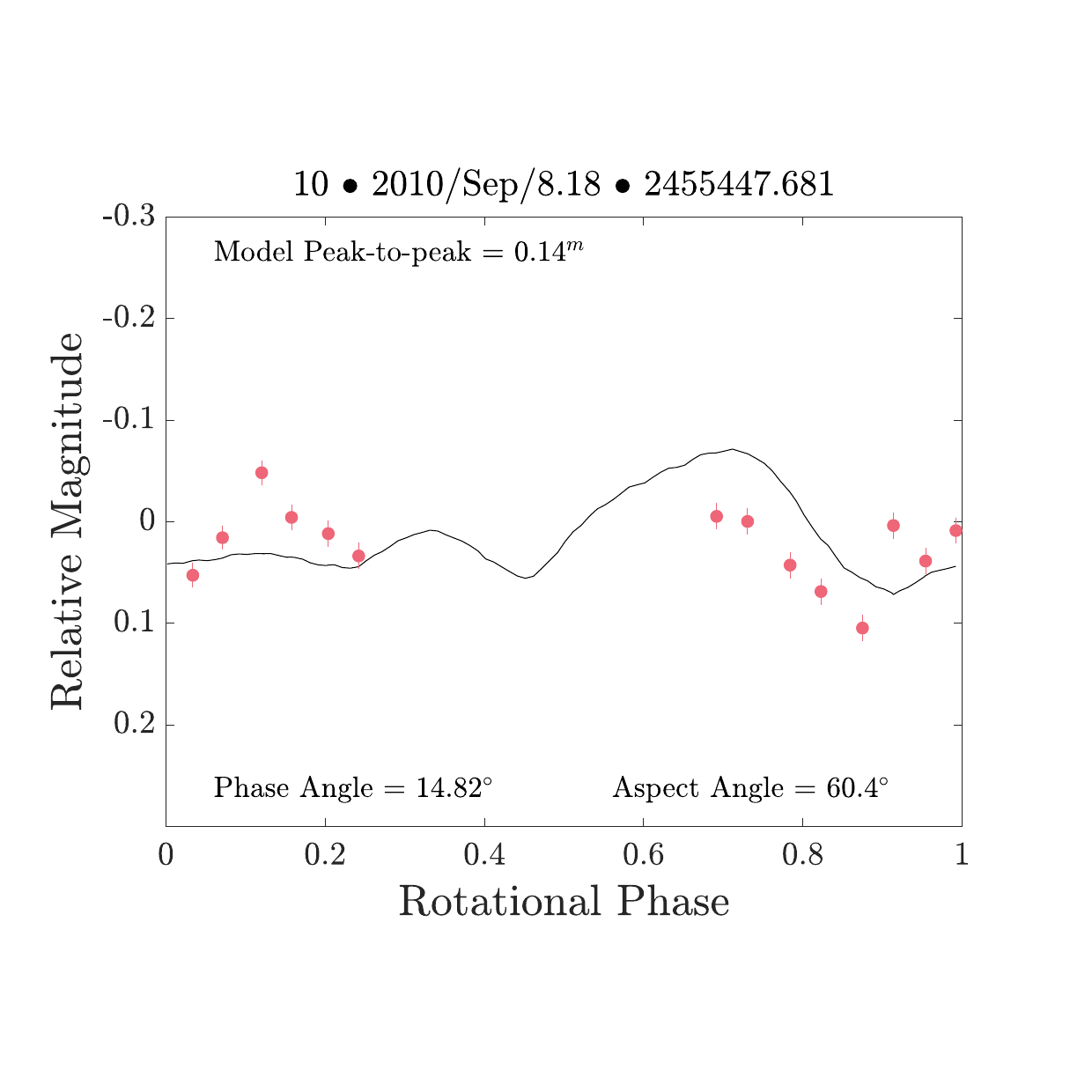}
        \includegraphics[width=.25\textwidth, trim=0.5cm 2.5cm 1.5cm 2.5cm, clip=true]{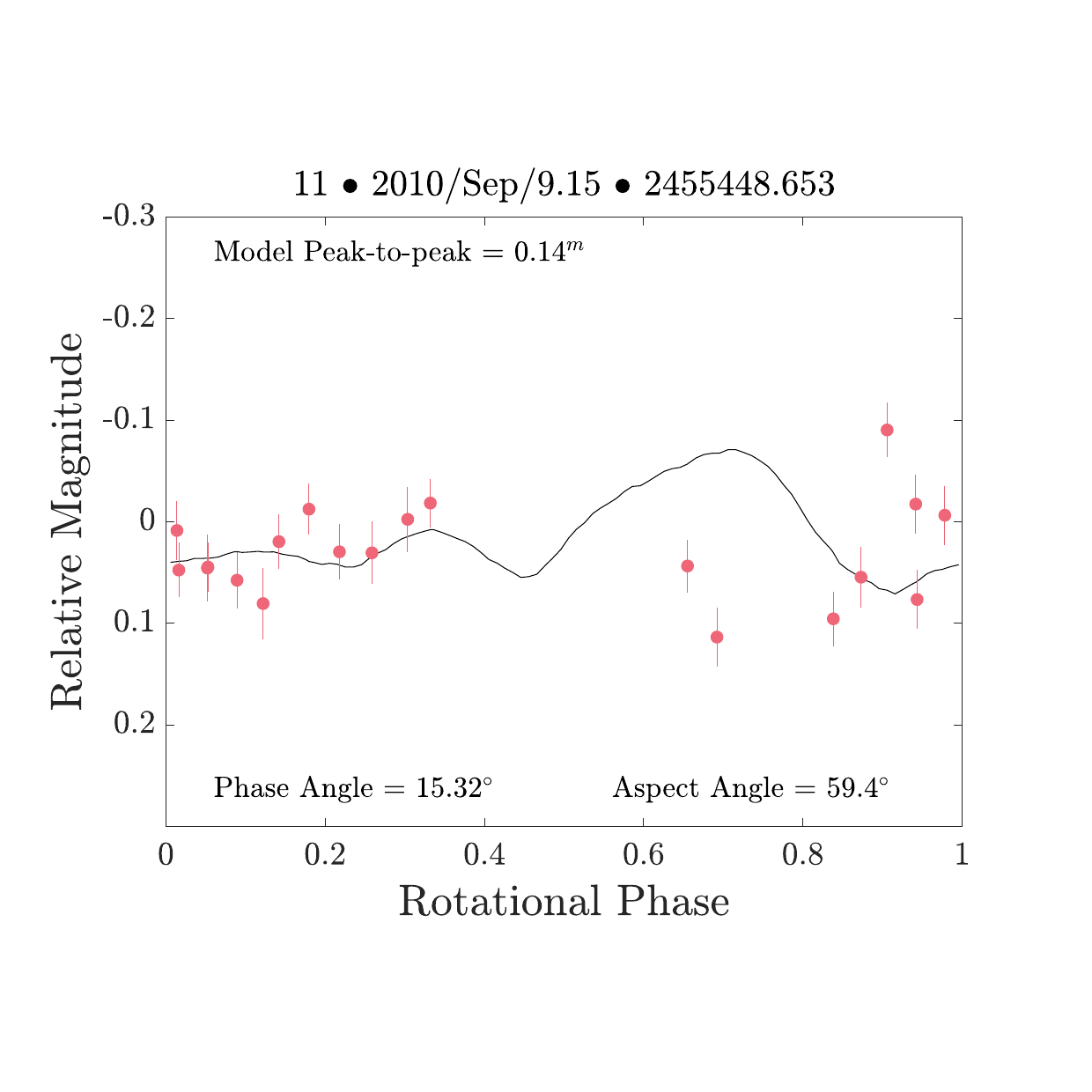}
        \includegraphics[width=.25\textwidth, trim=0.5cm 2.5cm 1.5cm 2.5cm, clip=true]{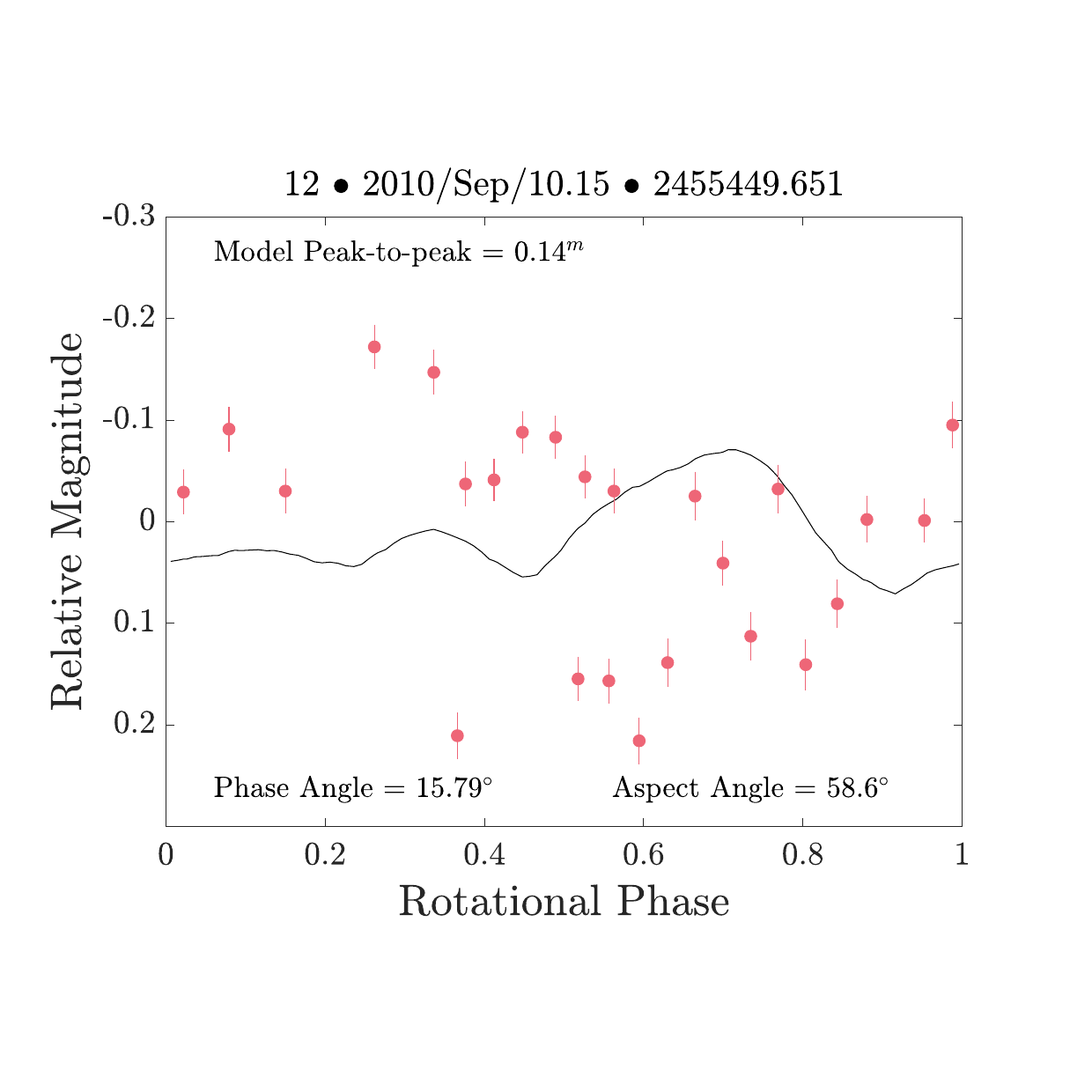}
	}

	\resizebox{\hsize}{!}{	
		\includegraphics[width=.25\textwidth, trim=0.5cm 2.5cm 1.5cm 2.5cm, clip=true]{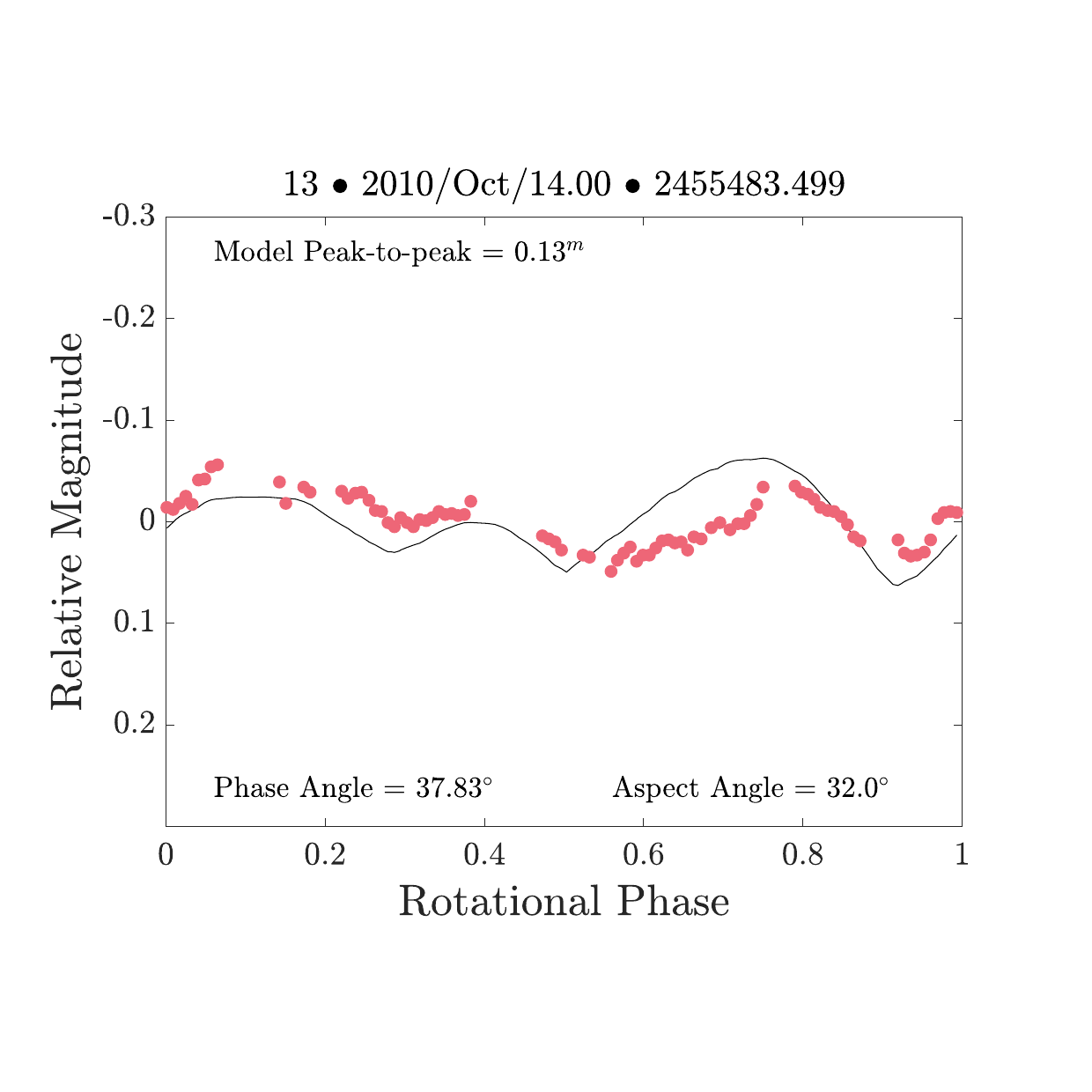}
        \includegraphics[width=.25\textwidth, trim=0.5cm 2.5cm 1.5cm 2.5cm, clip=true]{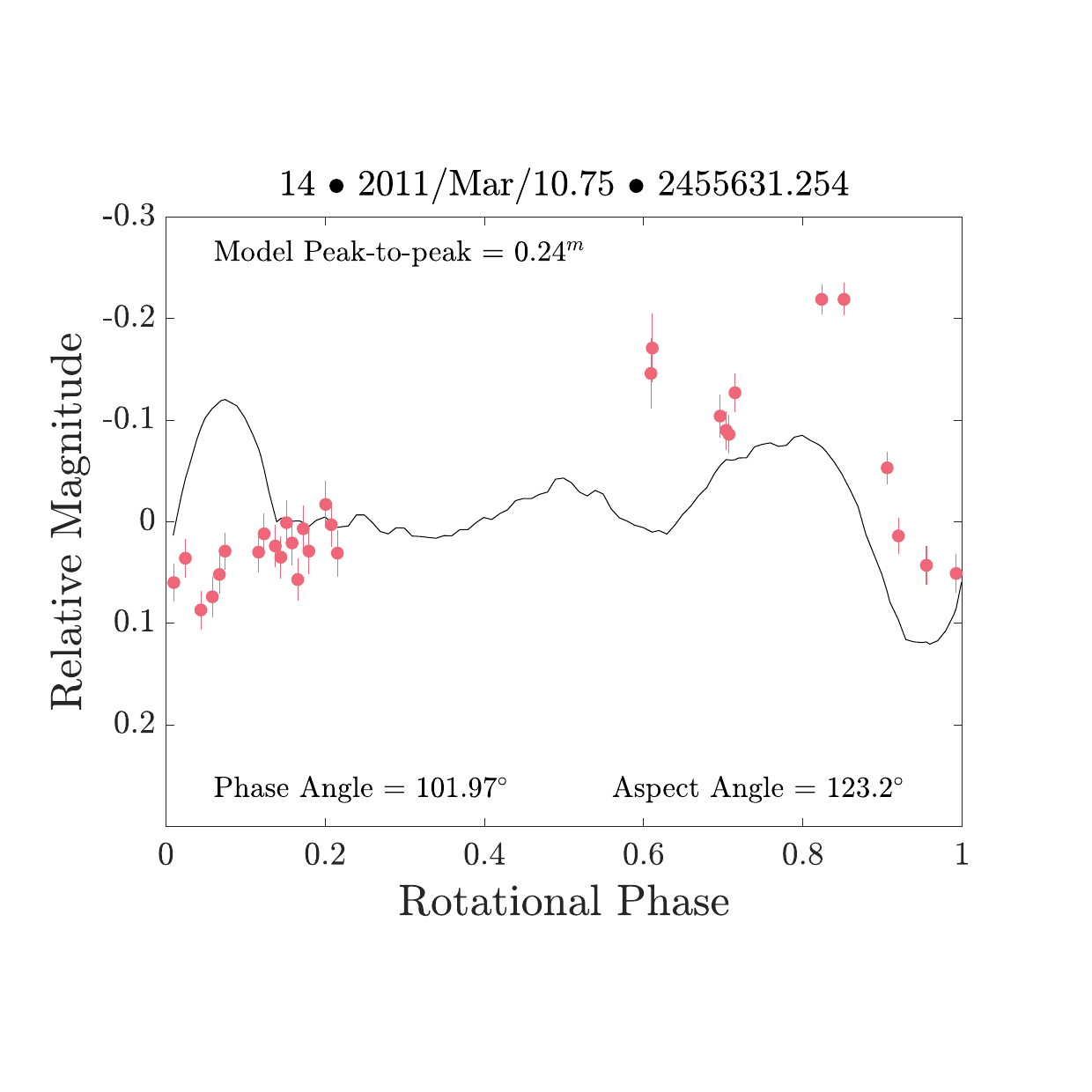}
        \includegraphics[width=.25\textwidth, trim=0.5cm 2.5cm 1.5cm 2.5cm, clip=true]{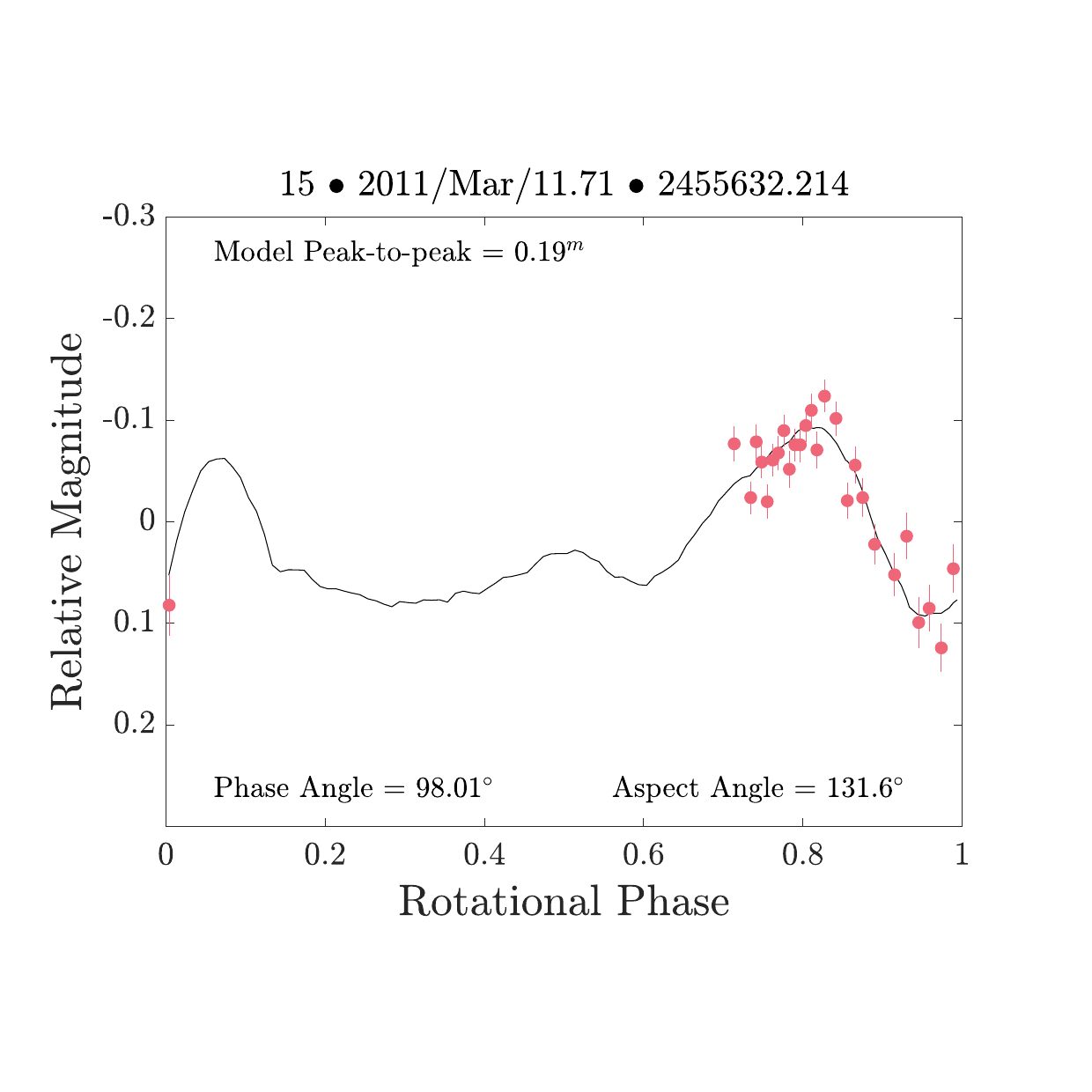}
        \includegraphics[width=.25\textwidth, trim=0.5cm 2.5cm 1.5cm 2.5cm, clip=true]{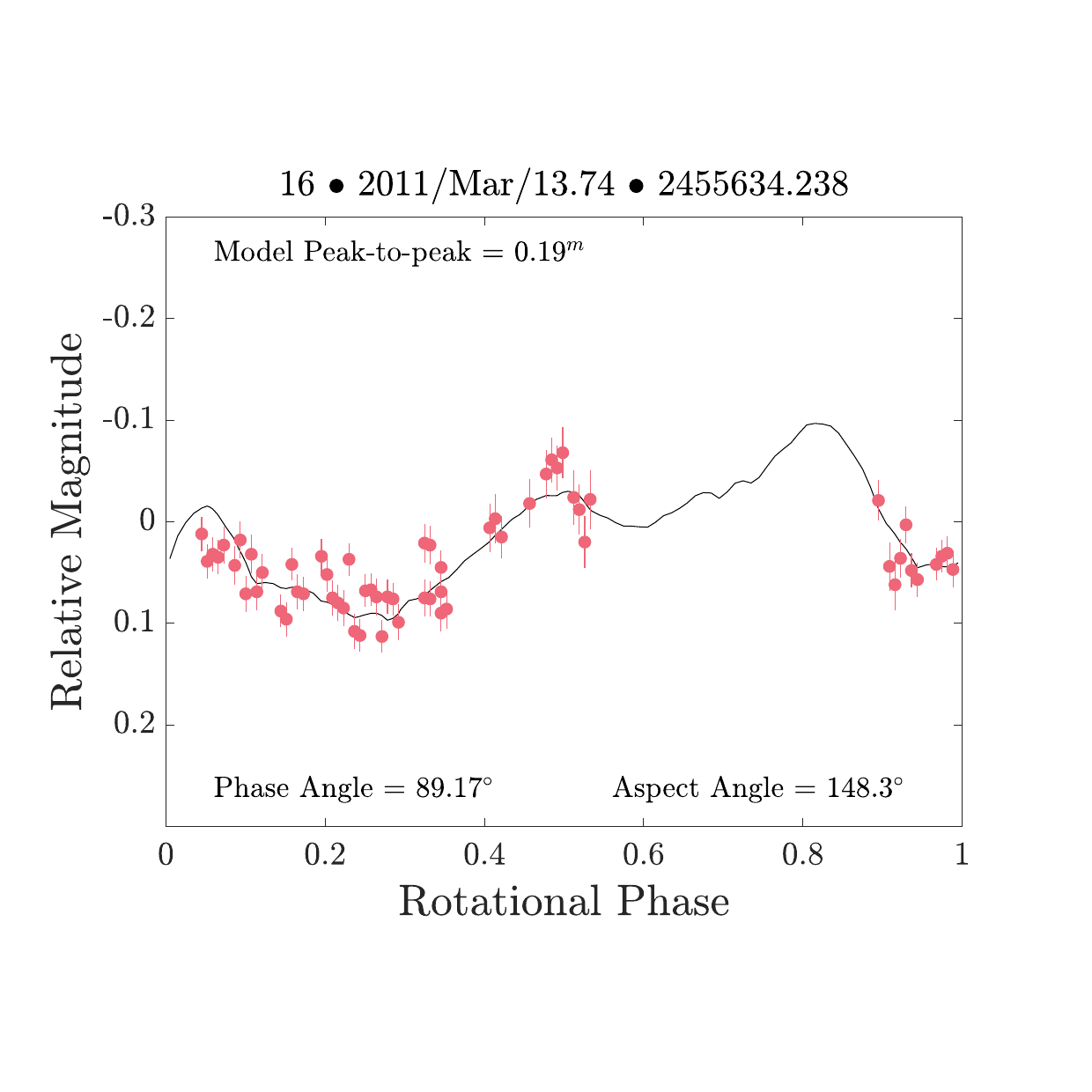}
	}

	\resizebox{\hsize}{!}{	

        \includegraphics[width=.25\textwidth, trim=0.5cm 2.5cm 1.5cm 2.5cm, clip=true]{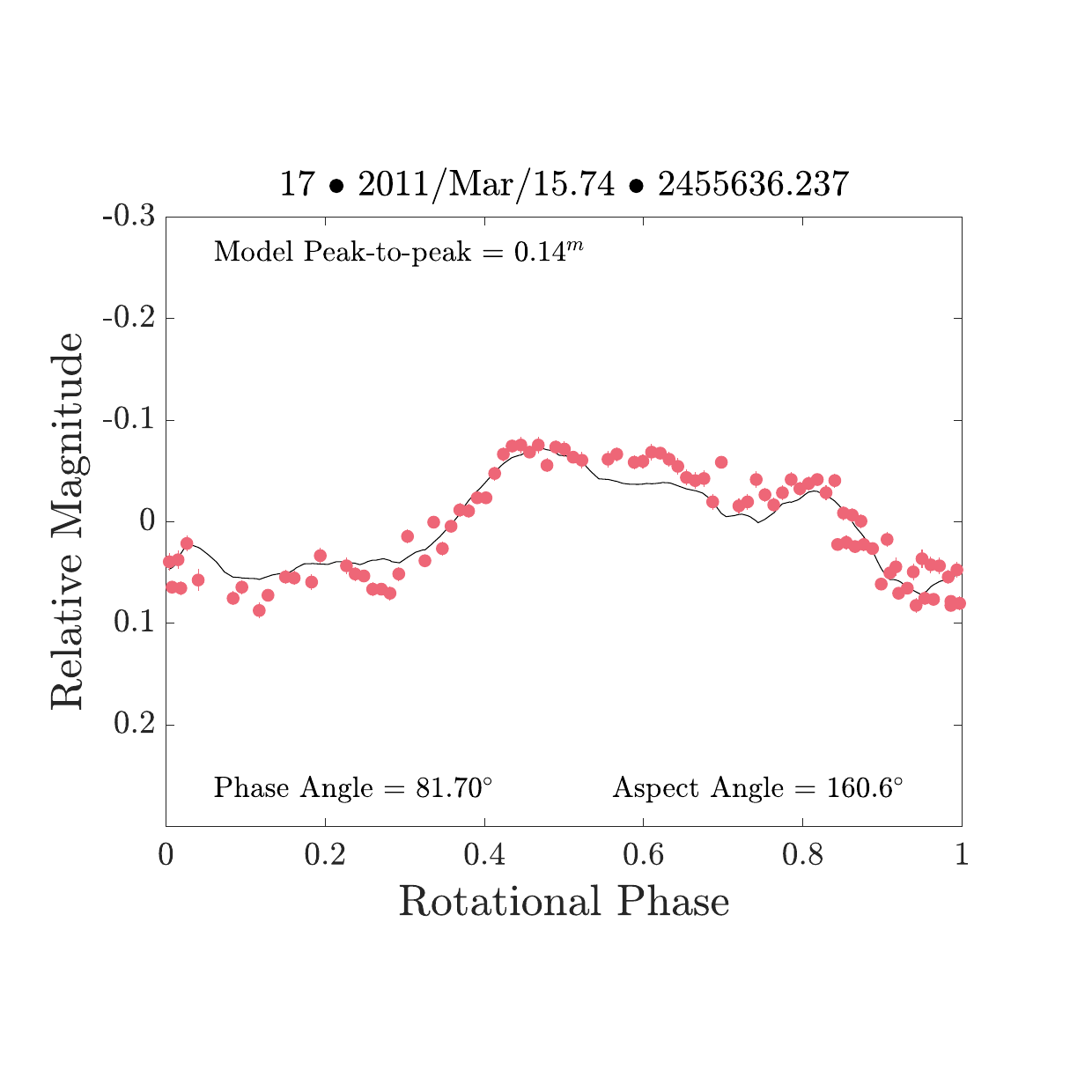}
        \includegraphics[width=.25\textwidth, trim=0.5cm 2.5cm 1.5cm 2.5cm, clip=true]{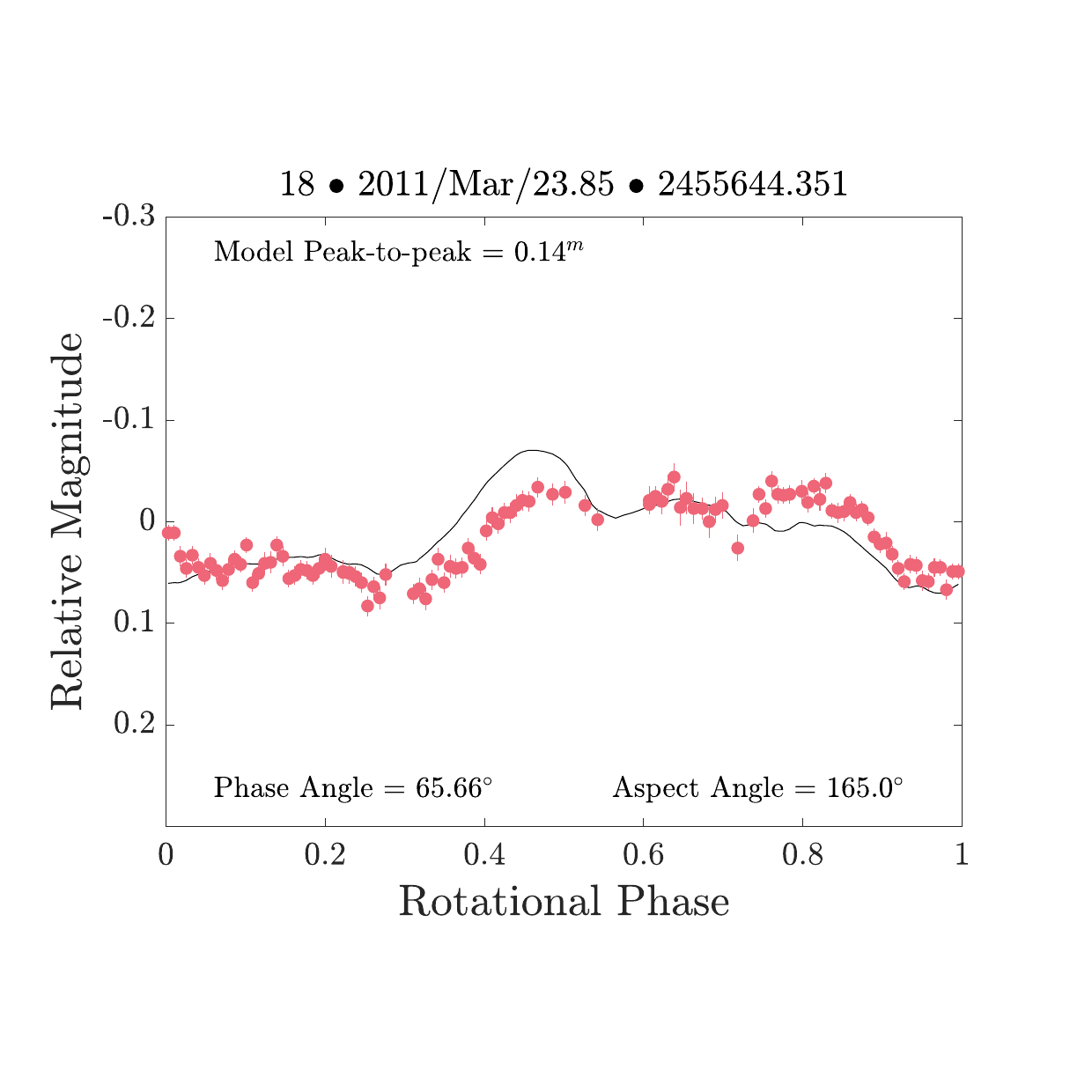}
        \includegraphics[width=.25\textwidth, trim=0.5cm 2.5cm 1.5cm 2.5cm, clip=true]{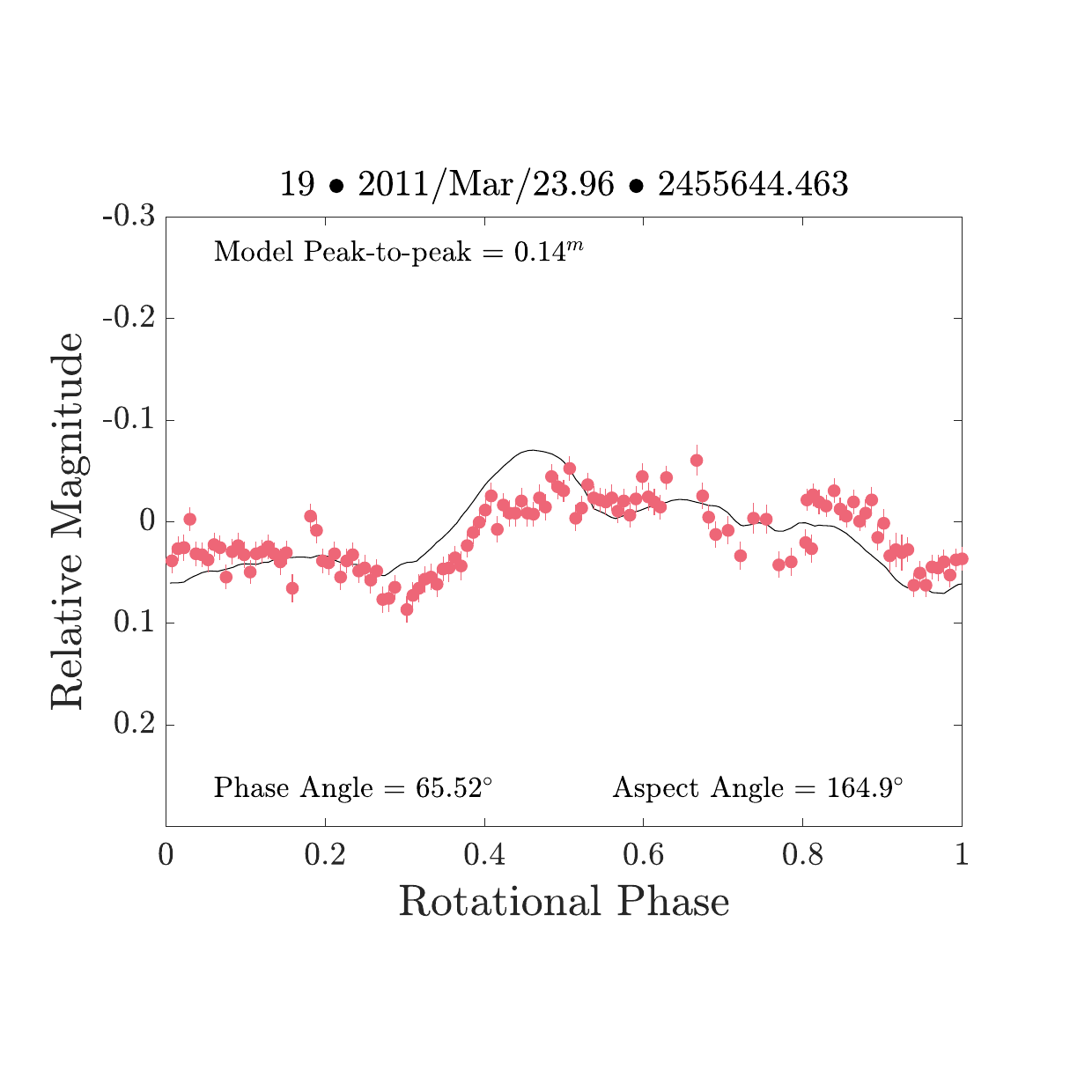}
        \includegraphics[width=.25\textwidth, trim=0.5cm 2.5cm 1.5cm 2.5cm, clip=true]{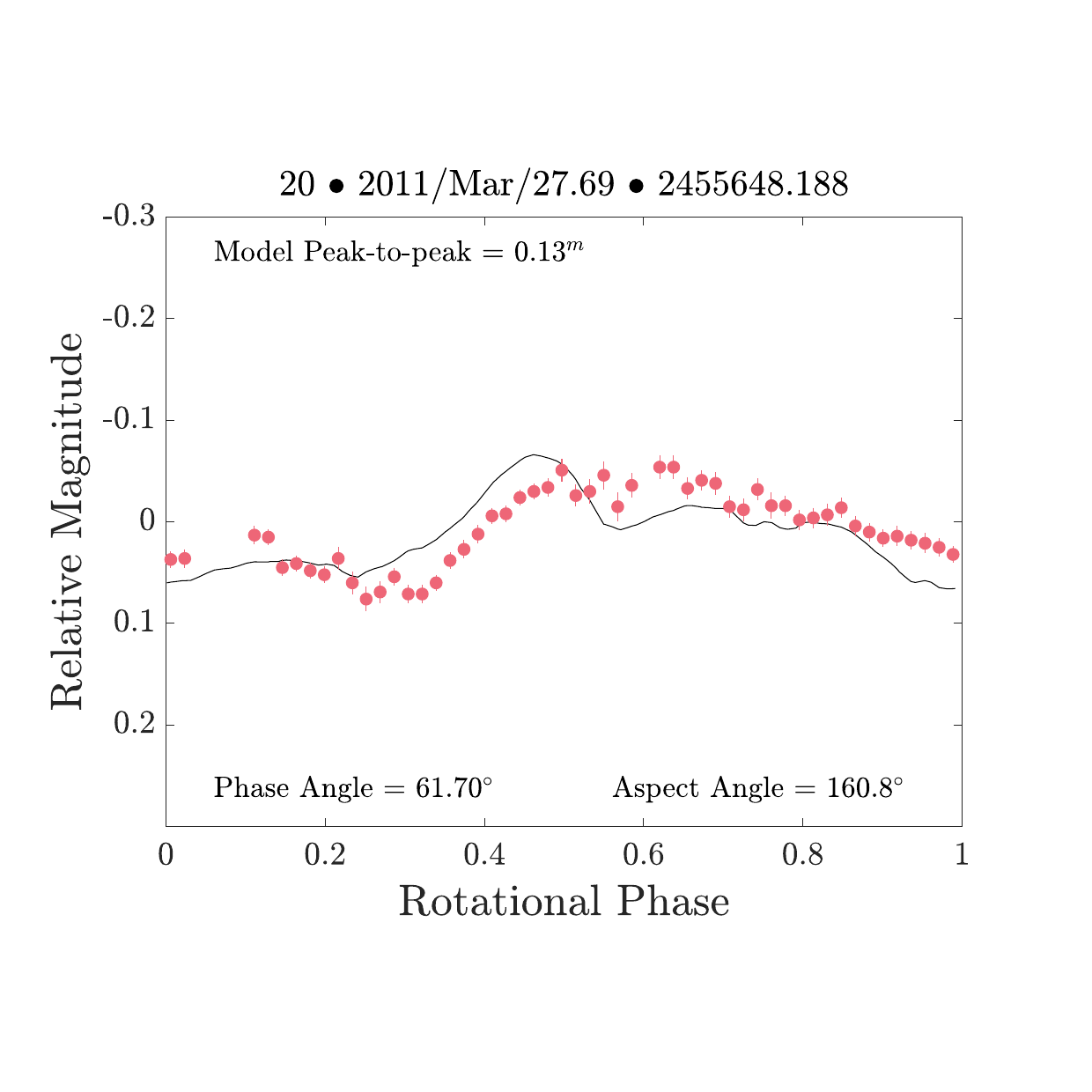}
	}
 	
	\caption{A comparison of the observed and synthetic lightcurves of (23187) 2000 PN9 for each lightcurve listed in Table 1. The synthetic lightcurves were generated using the convex hull model presented in Section 3.1 using the Lommel-Seelinger scattering model. Synthetic lightcurves are plotted as solid black lines, and observational data as red points. Lightcurve 12, which was not used in the analysis, has possible issues with the CCD that may be responsible for the poor fit.}
	\label{fig:synthLC_radar1}
\end{figure*}

\begin{figure*}	
	\resizebox{\hsize}{!}{	
        \includegraphics[width=.25\textwidth, trim=0.5cm 2.5cm 1.5cm 2.5cm, clip=true]{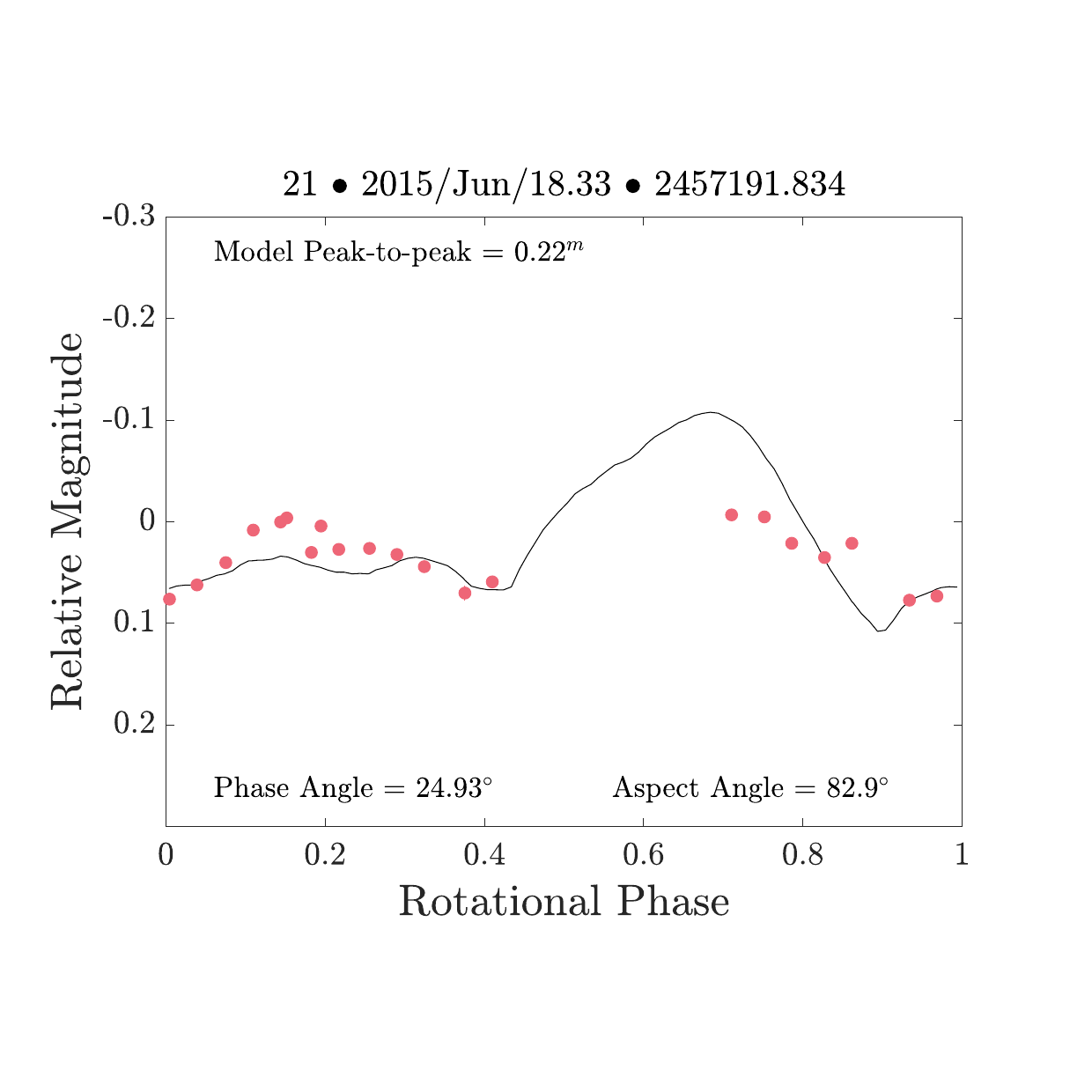}
        \includegraphics[width=.25\textwidth, trim=0.5cm 2.5cm 1.5cm 2.5cm, clip=true]{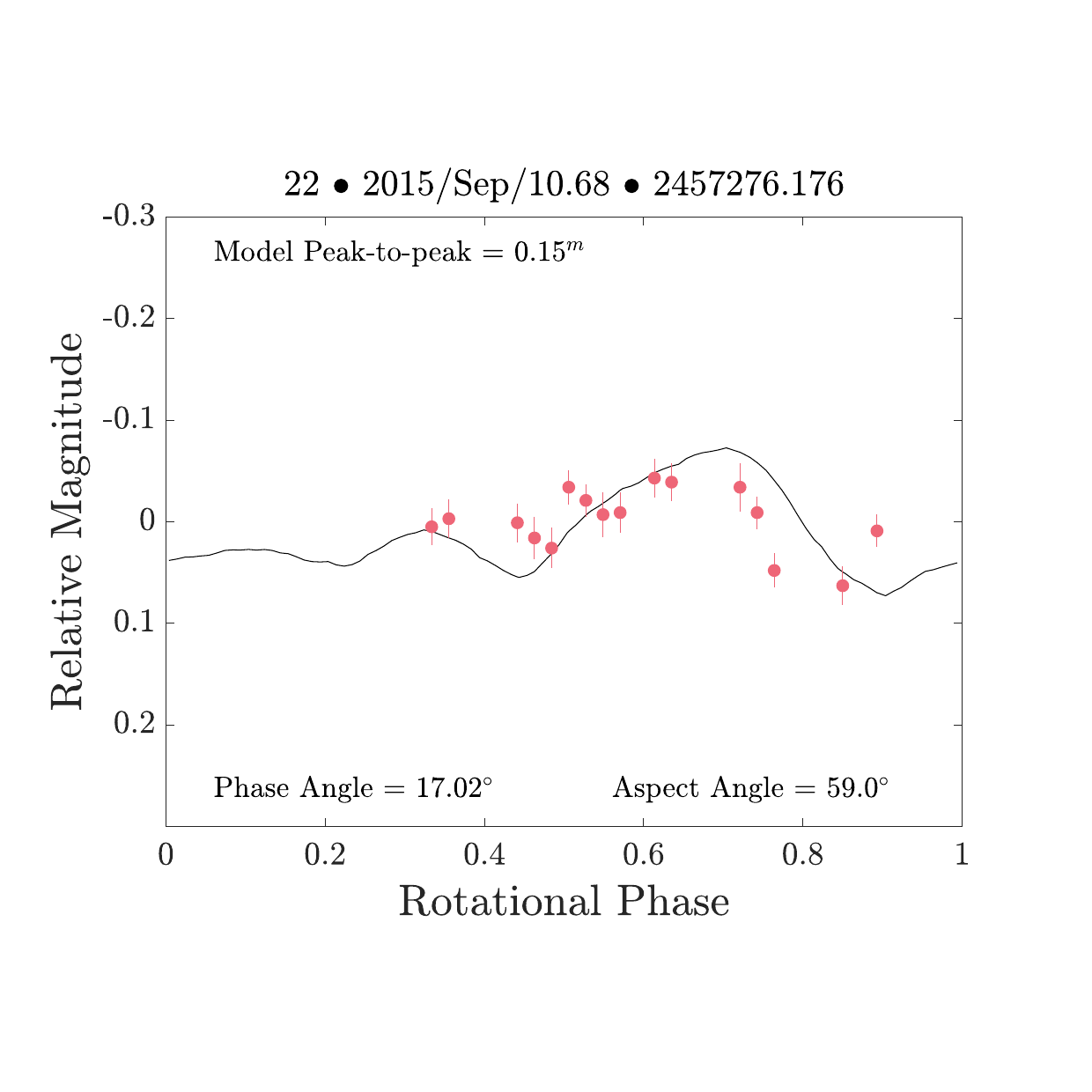}
        \includegraphics[width=.25\textwidth, trim=0.5cm 2.5cm 1.5cm 2.5cm, clip=true]{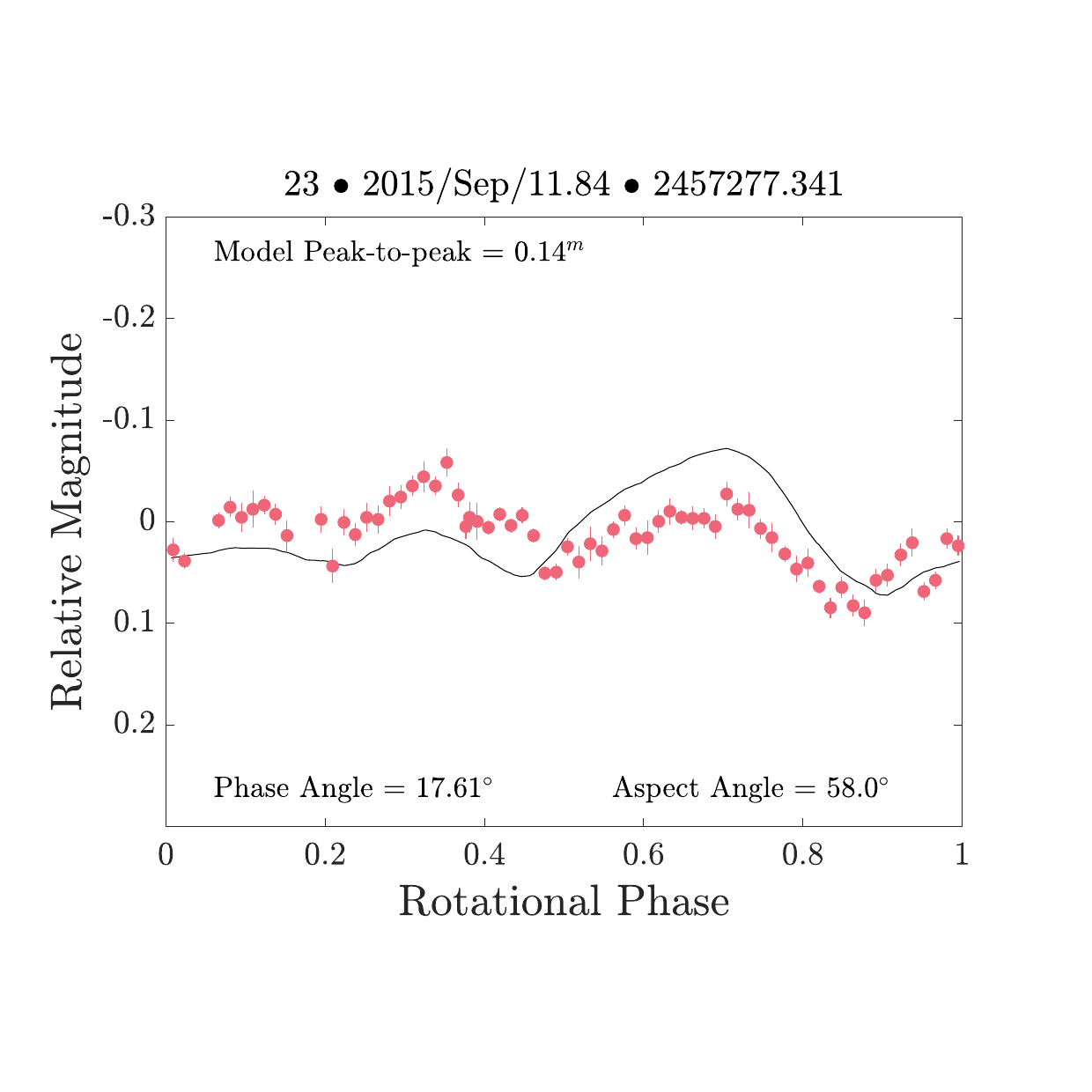}
        \includegraphics[width=.25\textwidth, trim=0.5cm 2.5cm 1.5cm 2.5cm, clip=true]{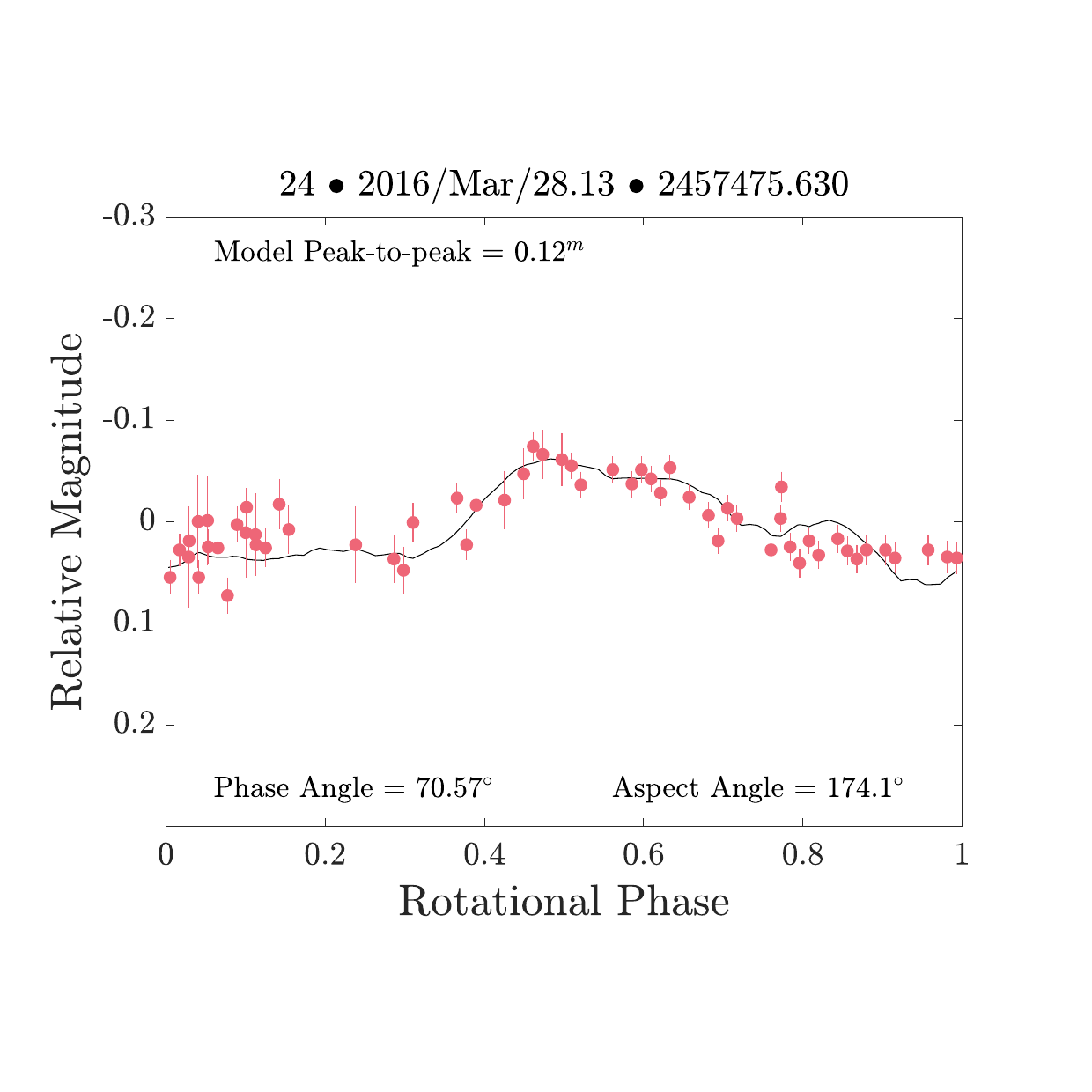}
	}
 
	\resizebox{\hsize}{!}{	
		\includegraphics[width=.25\textwidth, trim=0.5cm 2.5cm 1.5cm 2.5cm, clip=true]{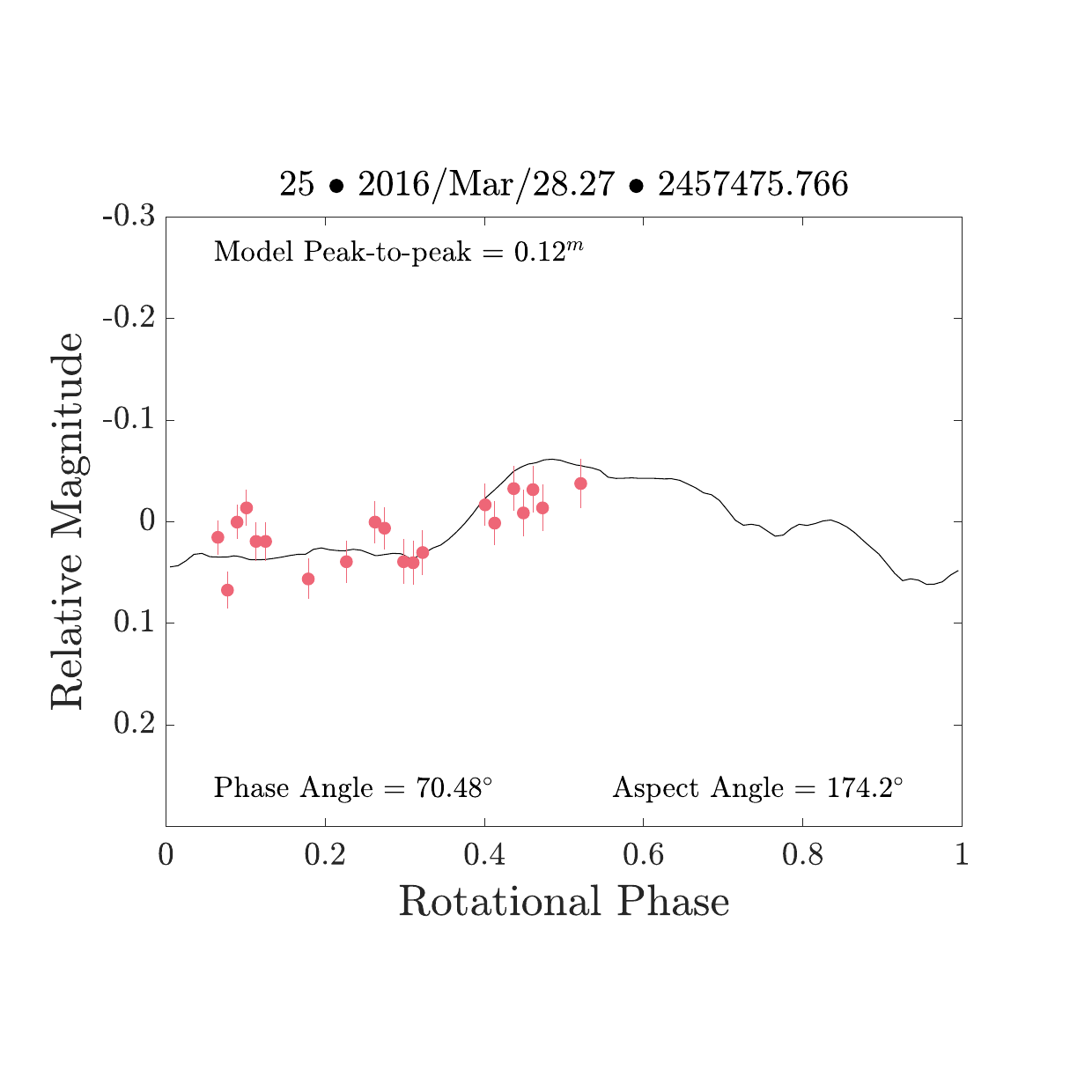}
        \includegraphics[width=.25\textwidth, trim=0.5cm 2.5cm 1.5cm 2.5cm, clip=true]{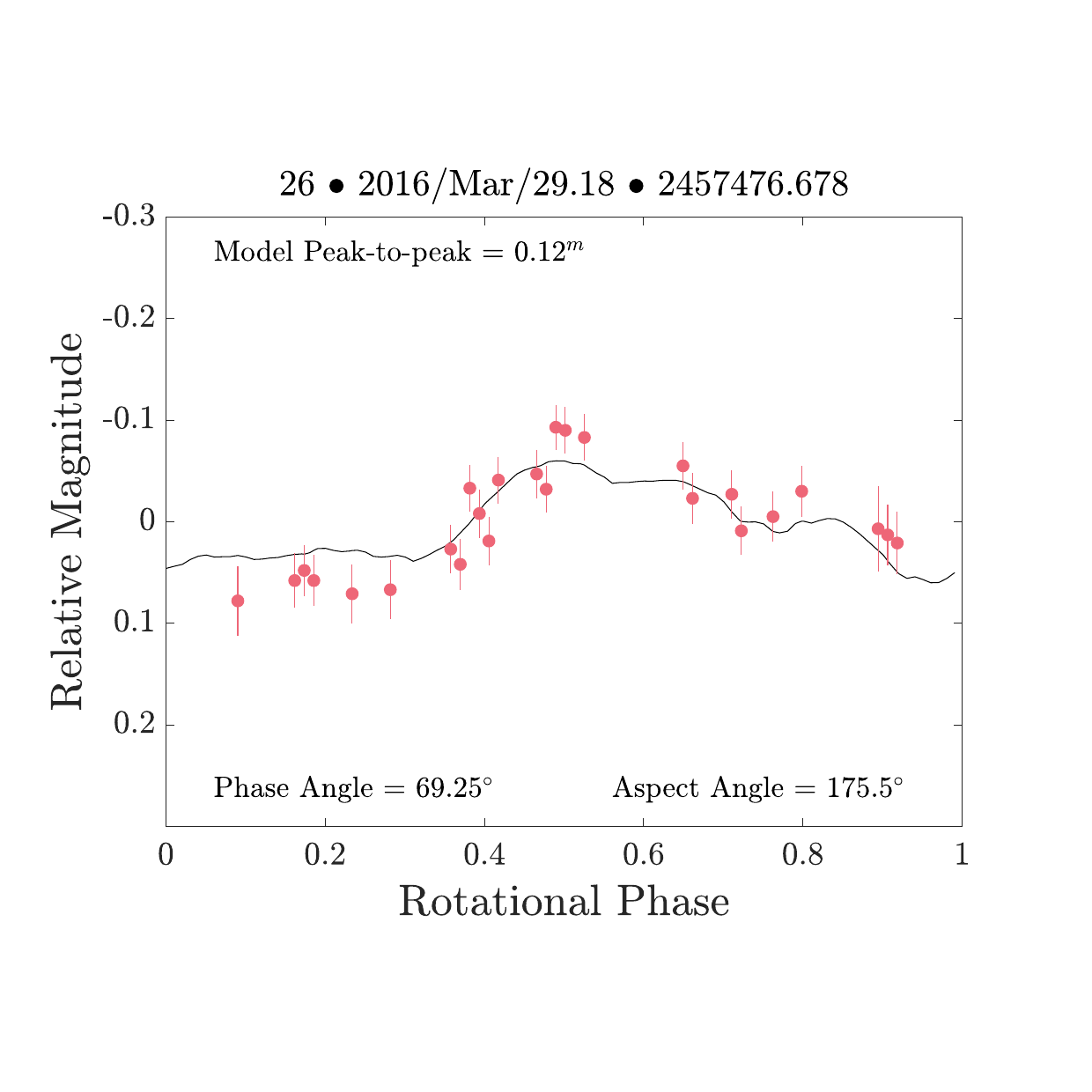}
        \includegraphics[width=.25\textwidth, trim=0.5cm 2.5cm 1.5cm 2.5cm, clip=true]{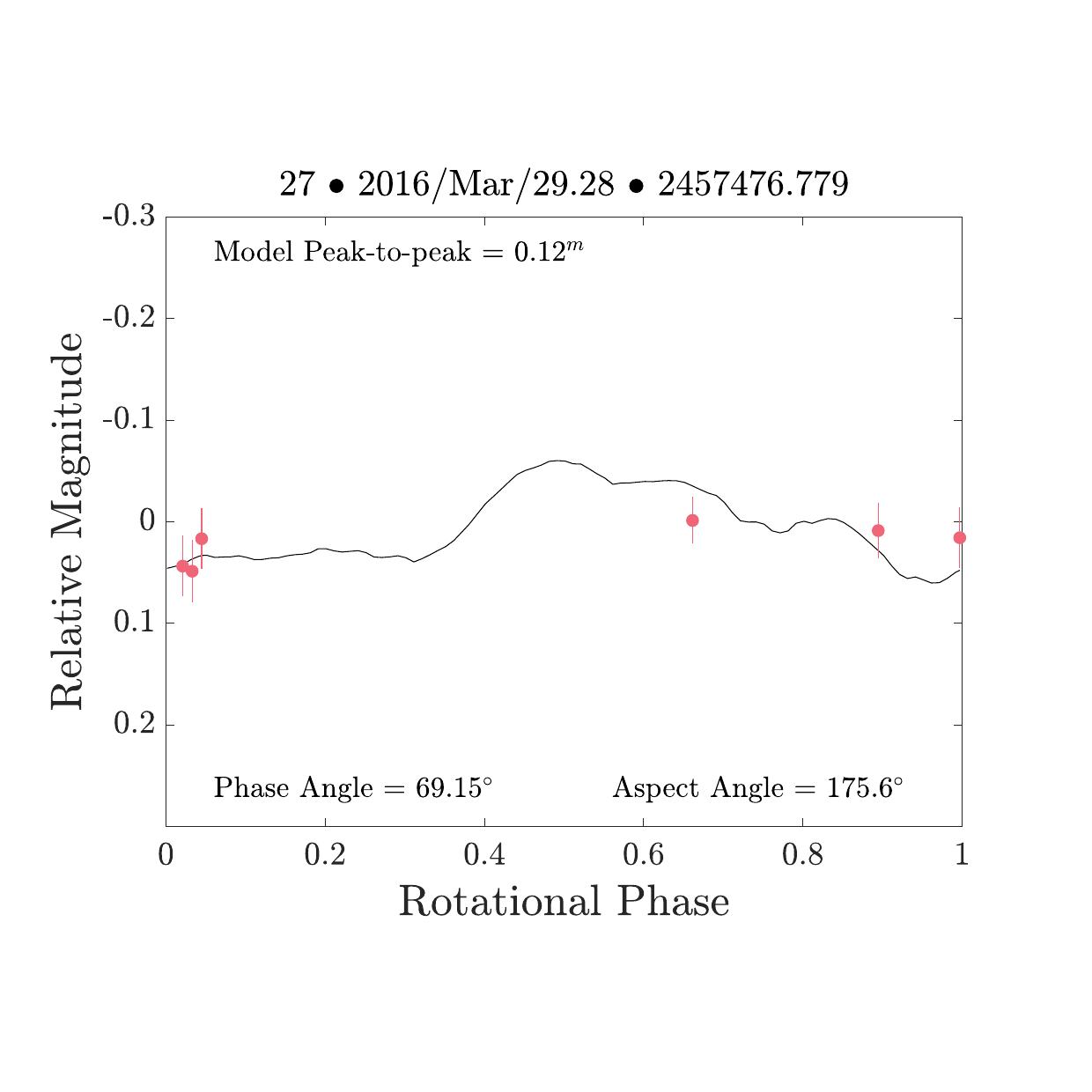}
        \includegraphics[width=.25\textwidth, trim=0.5cm 2.5cm 1.5cm 2.5cm, clip=true]{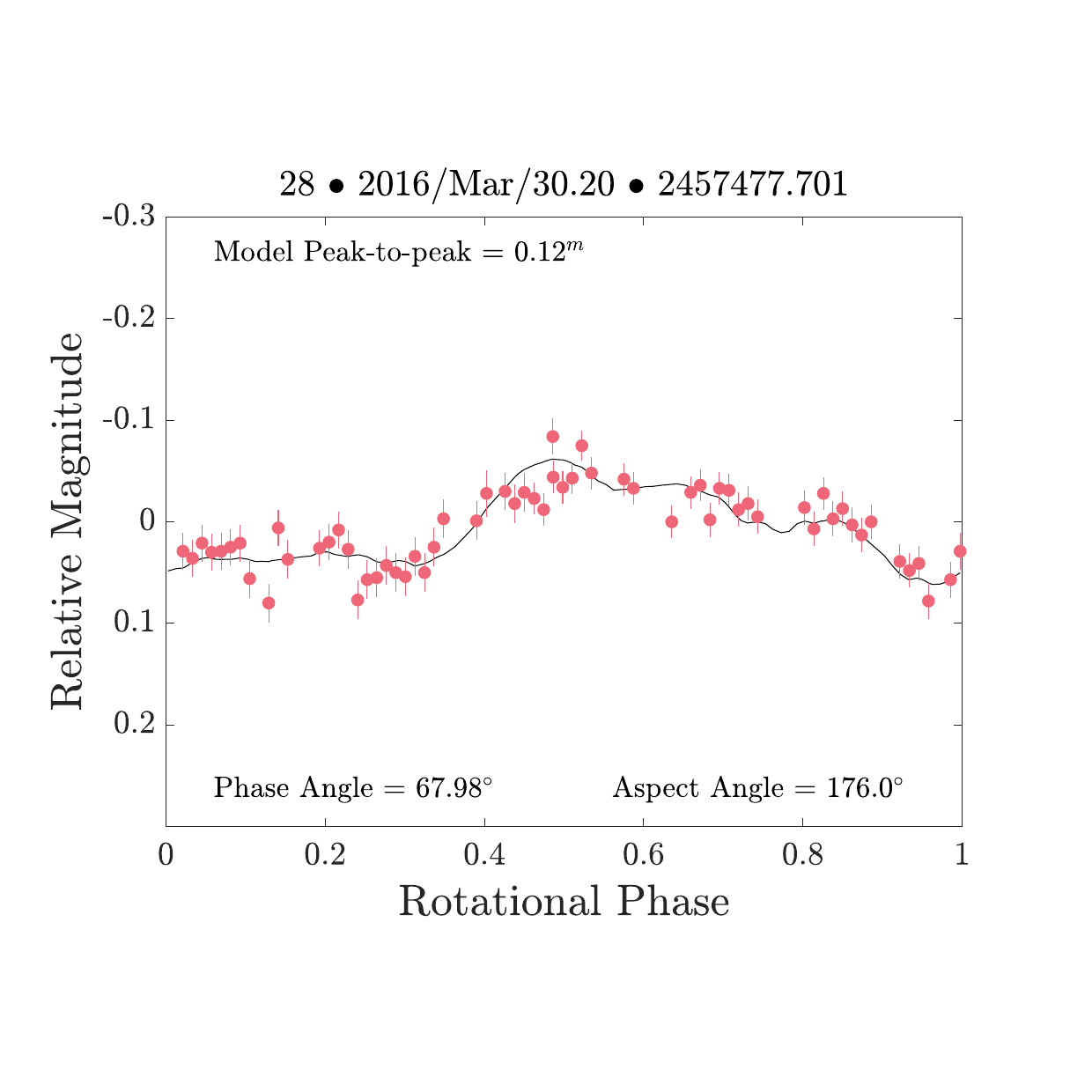}
	}
 
	\resizebox{\hsize}{!}{	
        \includegraphics[width=.25\textwidth, trim=0.5cm 2.5cm 1.5cm 2.5cm, clip=true]{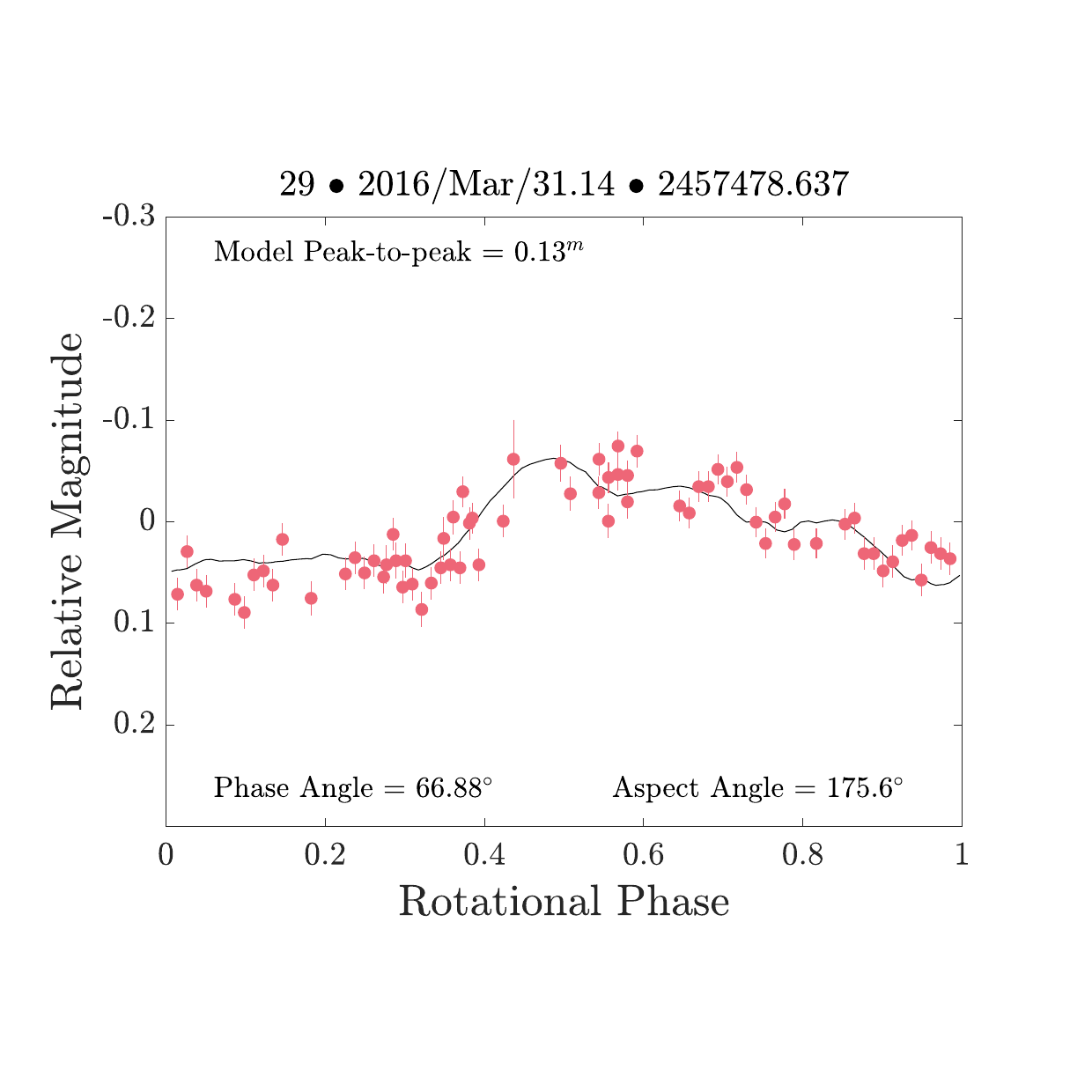}
        \includegraphics[width=.25\textwidth, trim=0.5cm 2.5cm 1.5cm 2.5cm, clip=true]{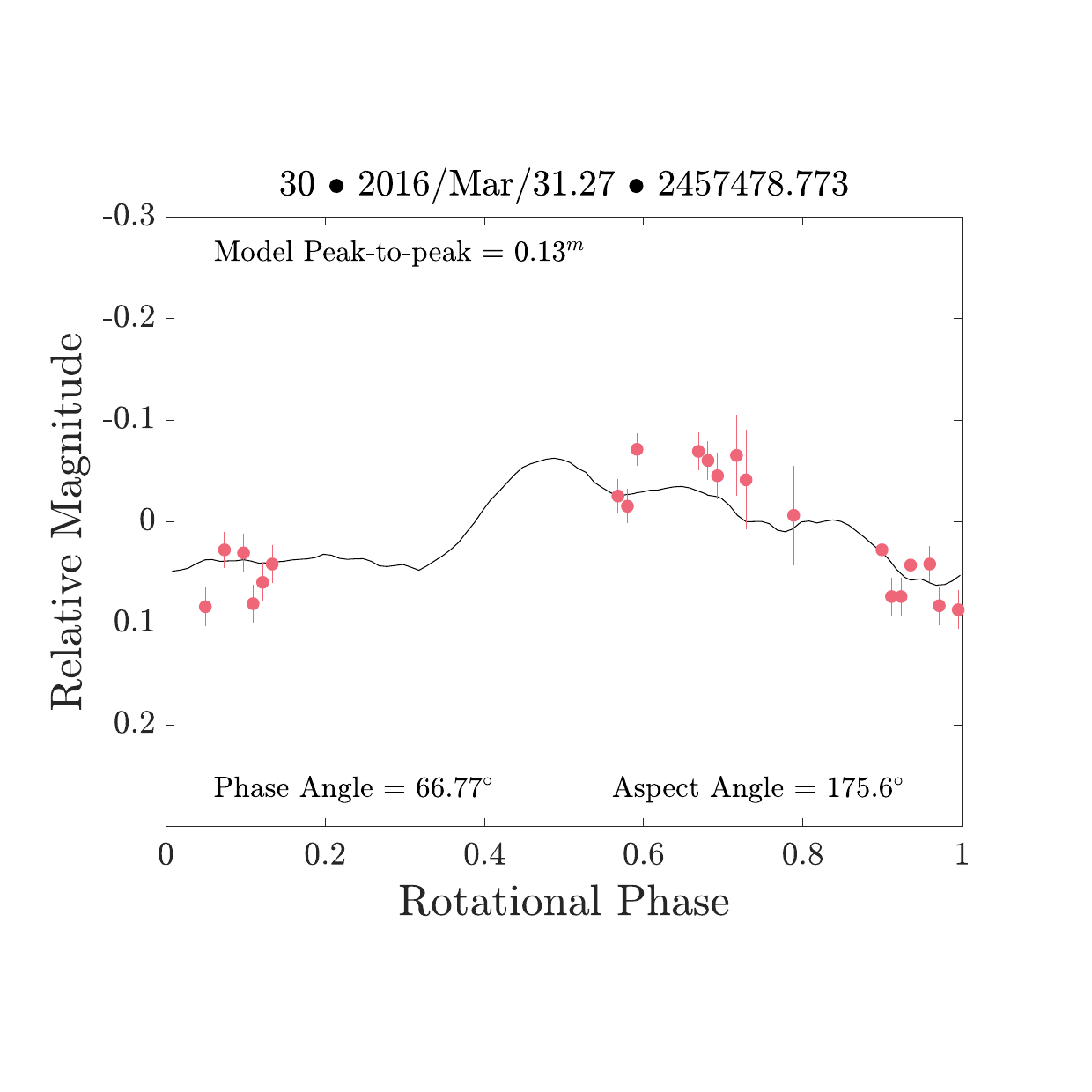}
        \includegraphics[width=.25\textwidth, trim=0.5cm 2.5cm 1.5cm 2.5cm, clip=true]{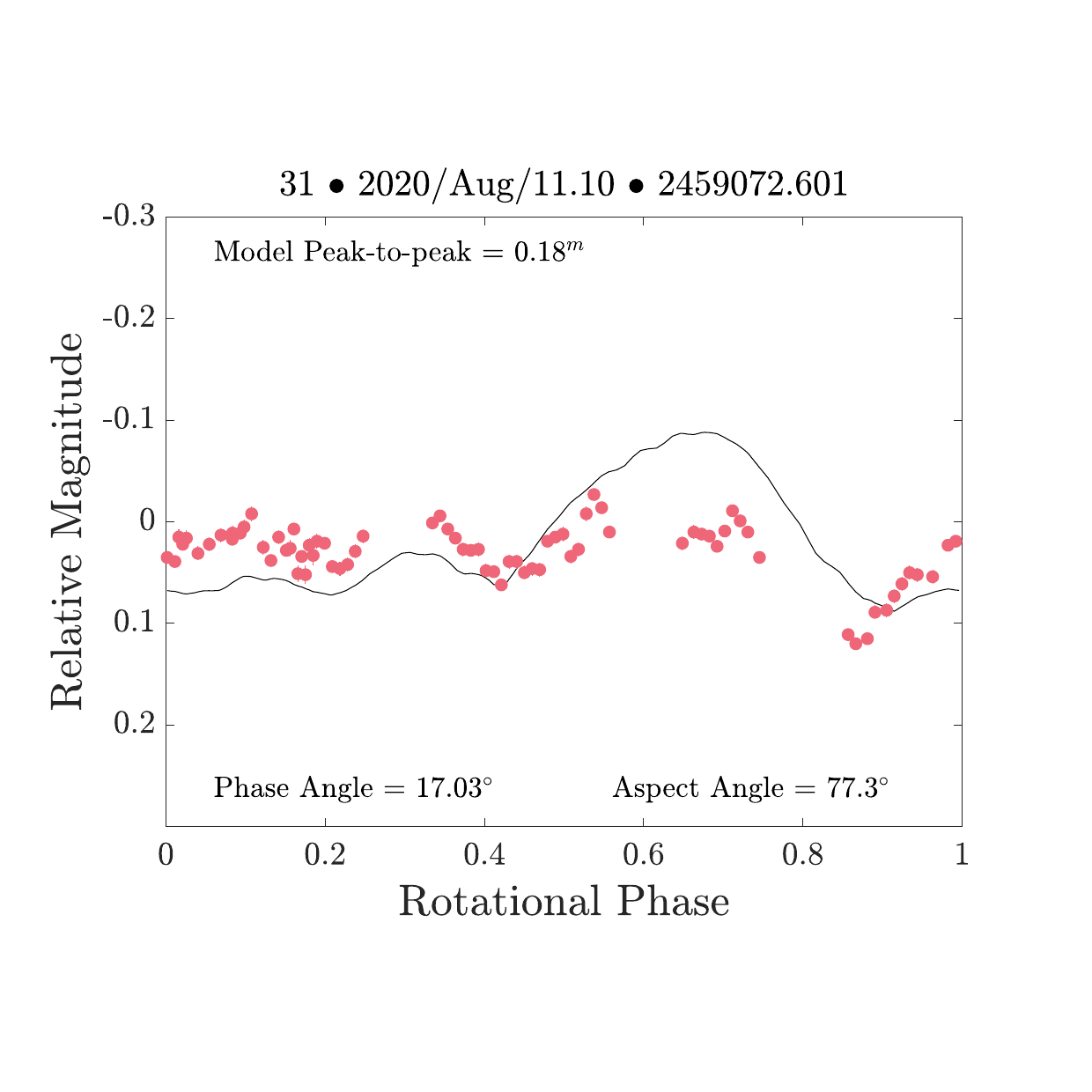}
        \includegraphics[width=.25\textwidth, trim=0.5cm 2.5cm 1.5cm 2.5cm, clip=true]{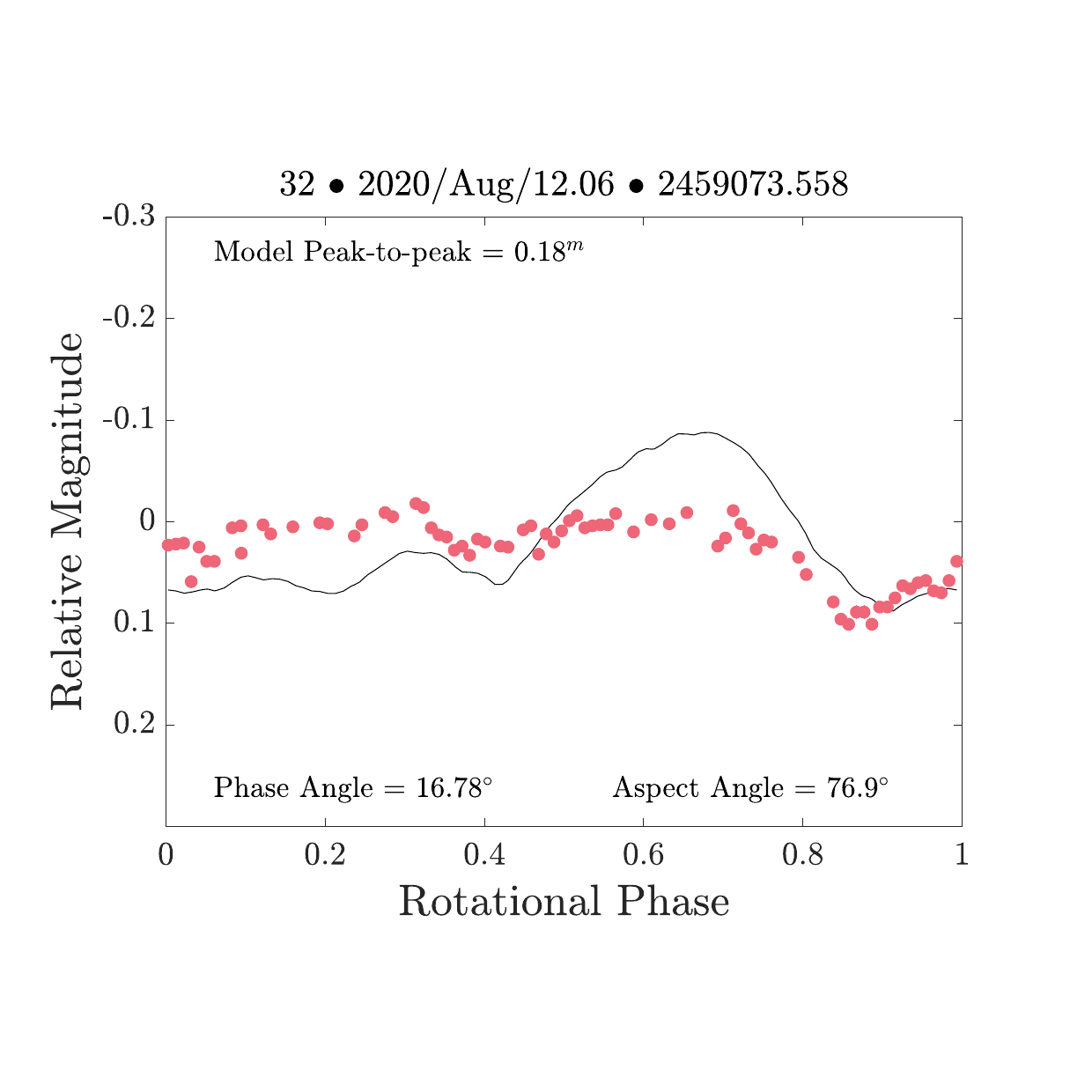}
	}

	\resizebox{0.75\textwidth}{!}{	
        \includegraphics[width=.25\textwidth, trim=0.5cm 2.5cm 1.5cm 2.5cm, clip=true]{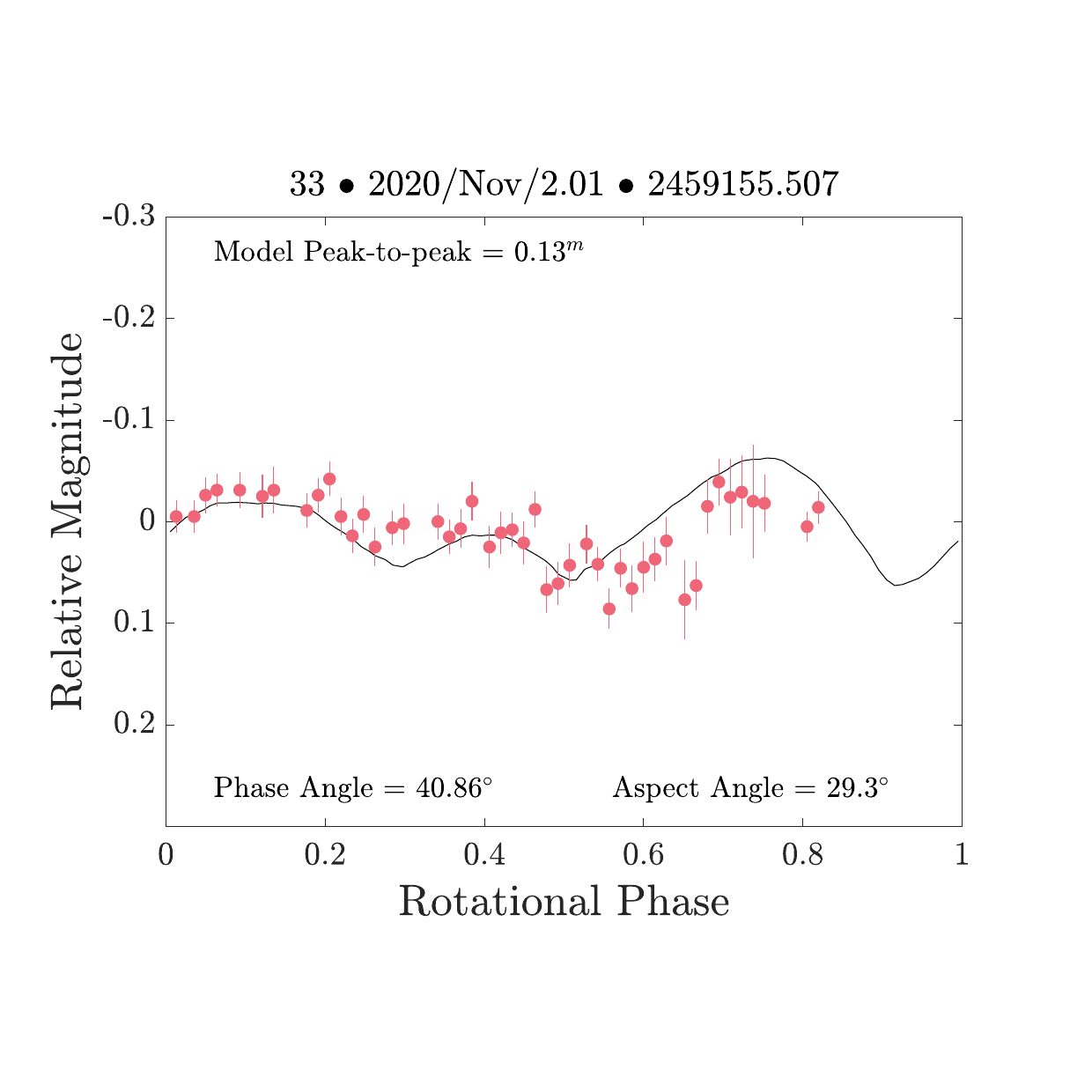}
        \includegraphics[width=.25\textwidth, trim=0.5cm 2.5cm 1.5cm 2.5cm, clip=true]{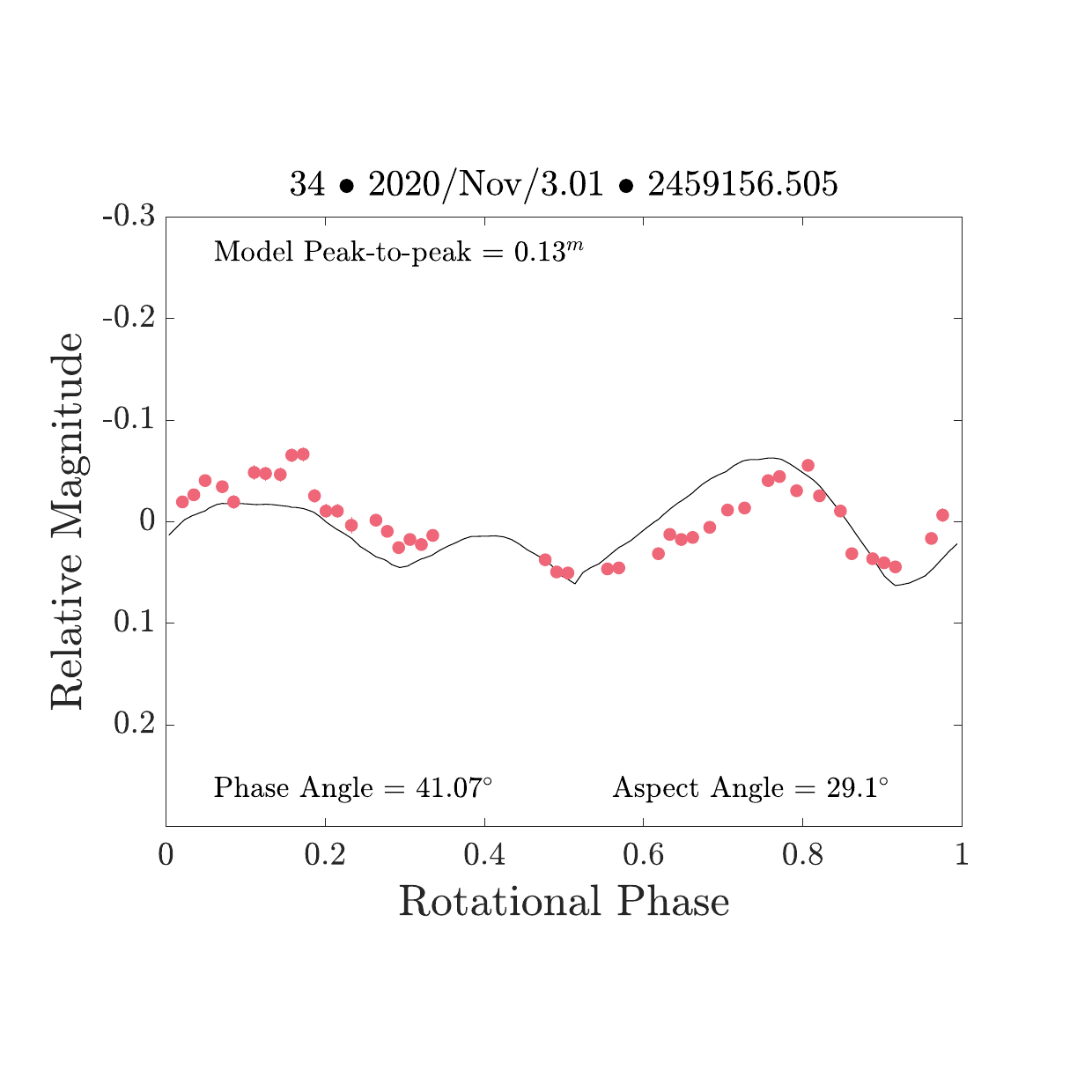}
        \includegraphics[width=.25\textwidth, trim=0.5cm 2.5cm 1.5cm 2.5cm, clip=true]{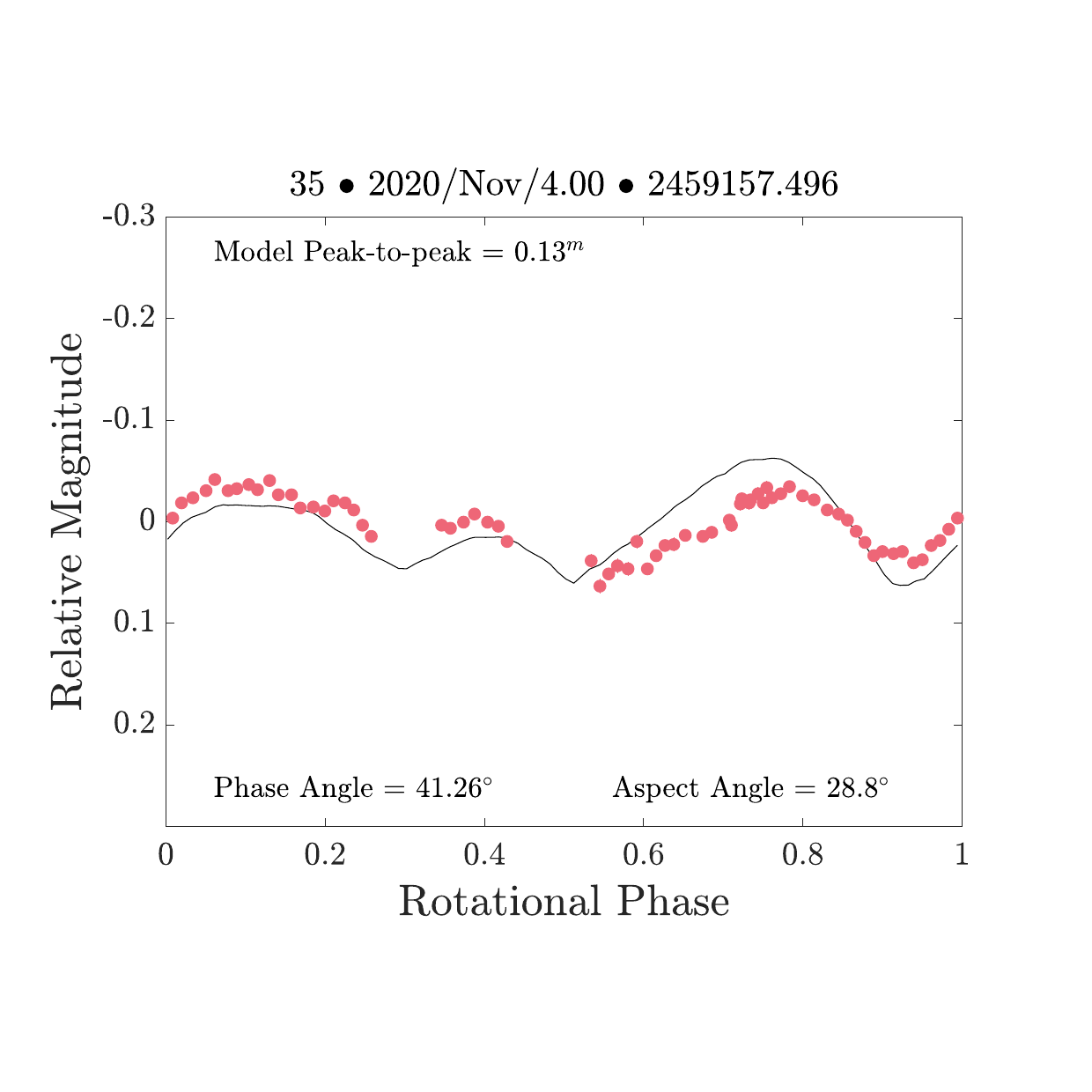}
	}
	\caption*{(continued)}
\end{figure*}

\begin{figure*}
    \centering
 	\includegraphics[width=0.6\textwidth]{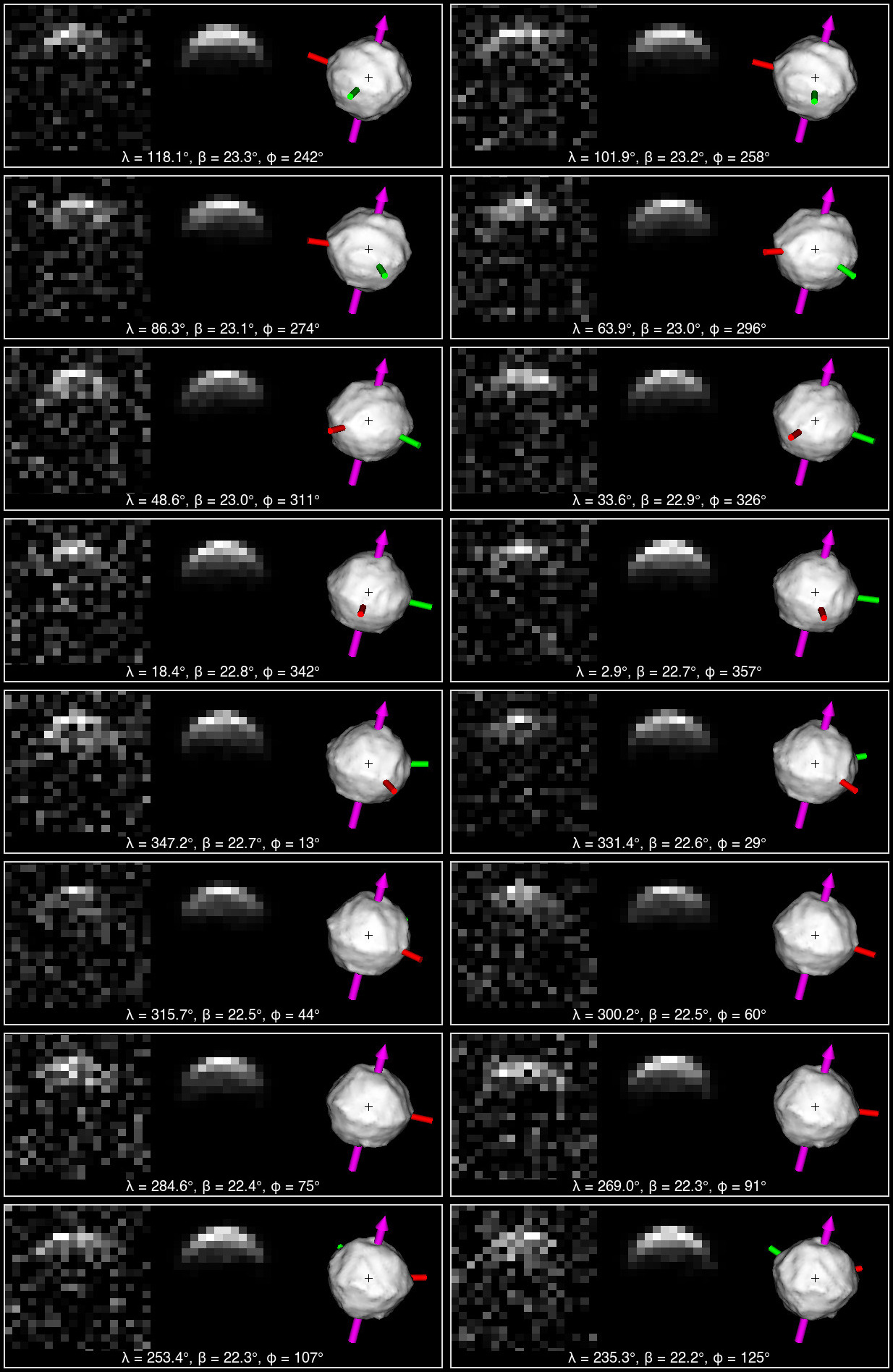}
	\caption{A comparison of the combined radar and lightcurve model of asteroid (23187) 2000 PN9 to delay-Doppler radar data. Each three-panel image comprises the observational data (left panel), a synthetic echo (centre panel) and a plane-of-sky projection of the model (right panel). On the delay-Doppler images (left and centre panels), delay increases towards the bottom of the vertical axis and Doppler frequency increases along the horizontal axis. The plane-of-sky projections (right panel) are displayed with the celestial north at the top and east to the left, in an equatorial coordinate system. The rotational axis, which is closely aligned with the z-axis in the body-centric coordinate system, is marked with a purple arrow. The axes of minimum and intermediate inertia are indicated by red and green rods respectively. The body-fixed longitude $\lambda$ and latitude $\beta$ for the radar line-of-sight, and the rotational phase $\phi$, are labelled for each image. These values were determined using the radar shape model's spin-state. The projected centre of mass is marked with a cross. The data in this figure correspond to radar observations with Goldstone on 2001-03-03, with a resolution of 1.0 $\mu$s $\times$ 6.15 Hz.}
	\label{fig:mont_g01}
\end{figure*}

\begin{figure*}
    \centering
	\includegraphics[width=0.6\textwidth]{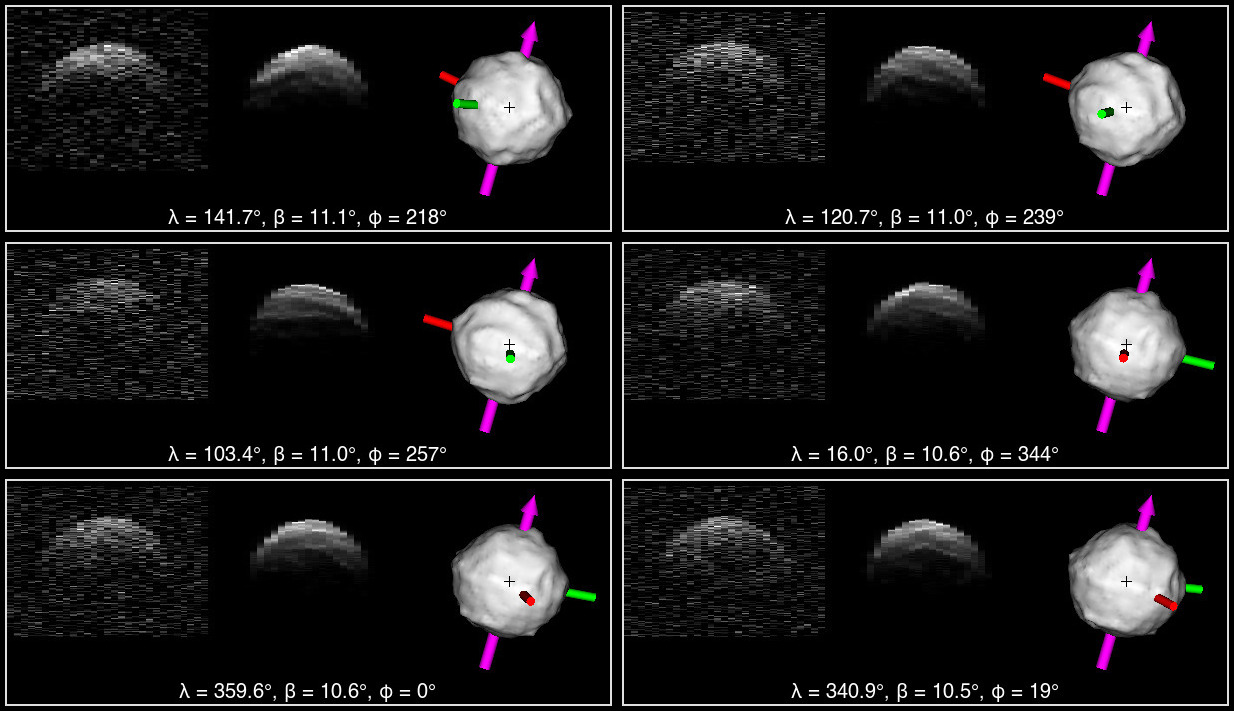}
	\caption{This figure is the same as Figure \ref{fig:mont_g01}, but for observations with Arecibo on 2001-03-04. The first image has a resolution of 0.2 $\mu$s $\times$ 1.00 Hz, and the remaining five are at a resolution of 0.1 $\mu$s $\times$ 1.00 Hz.}	
	\label{fig:mont_a01a}
\end{figure*}

\begin{figure*}
    \centering
	\includegraphics[width=0.6\textwidth]{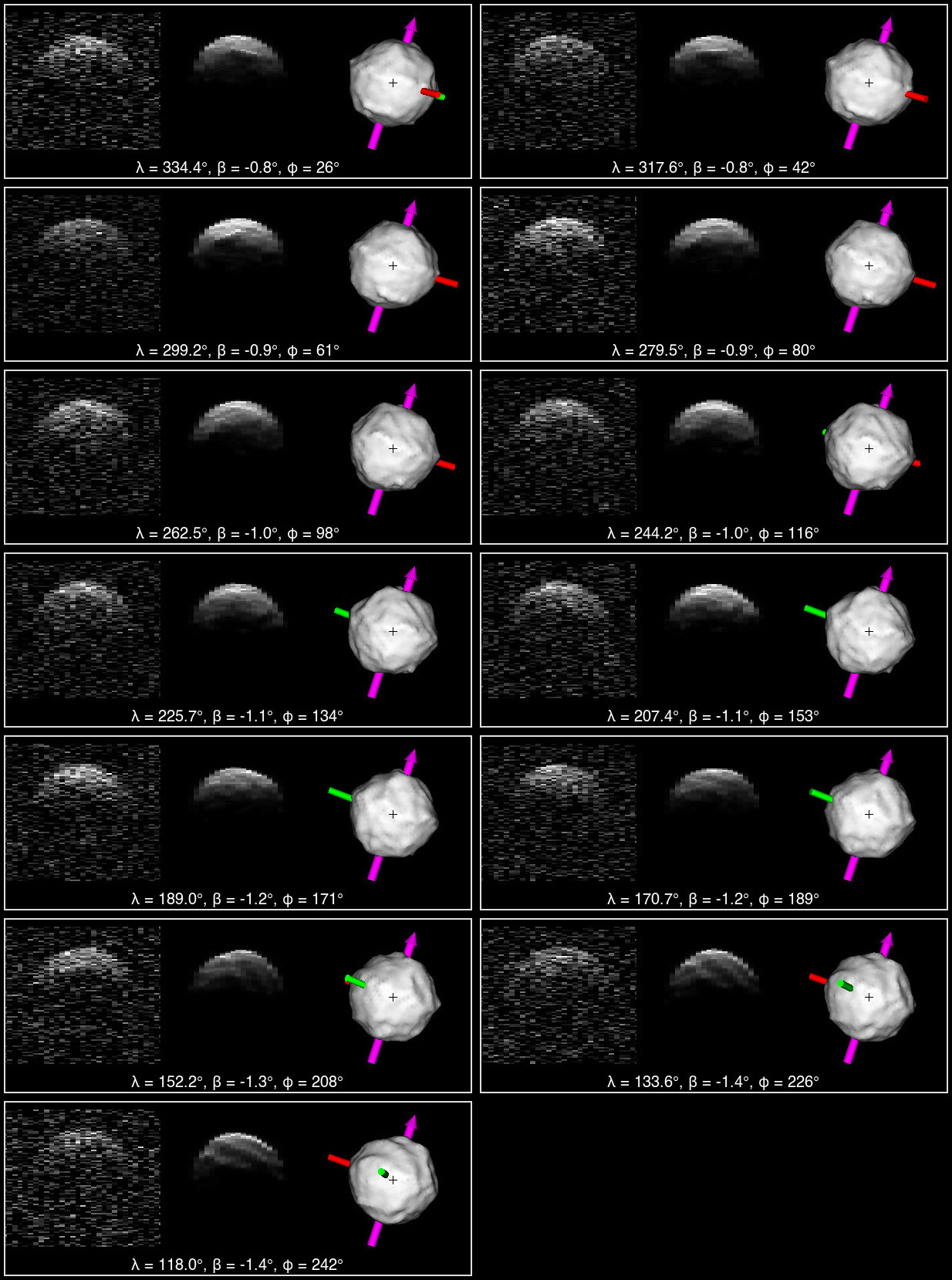}
	\caption{This figure is the same as Figure \ref{fig:mont_g01}, but for observations with Arecibo on 2001-03-05 with a resolution of 0.2 $\mu$s $\times$ 1.00 Hz.}
	\label{fig:mont_a01b}
\end{figure*}

\begin{figure*}
    \centering
	\includegraphics[width=0.6\textwidth]{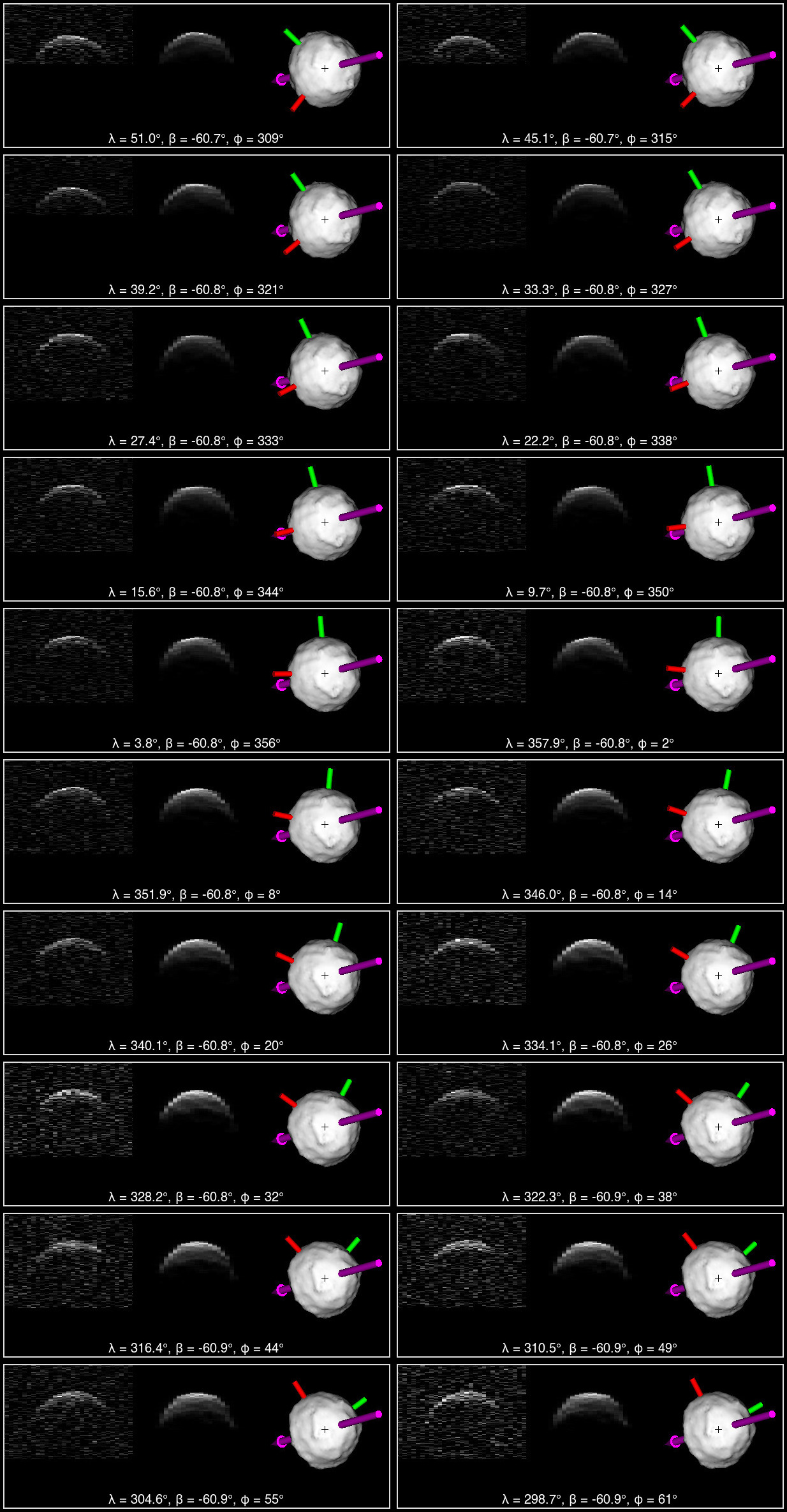}
	\caption{This figure is the same as Figure \ref{fig:mont_g01}, but for observations with Goldstone on 2006-03-07 with a resolution of 0.125 $\mu$s $\times$ 2.00 Hz.}
	\label{fig:mont_g06a}
\end{figure*}

\begin{figure*}
    \centering
	\includegraphics[width=0.6\textwidth]{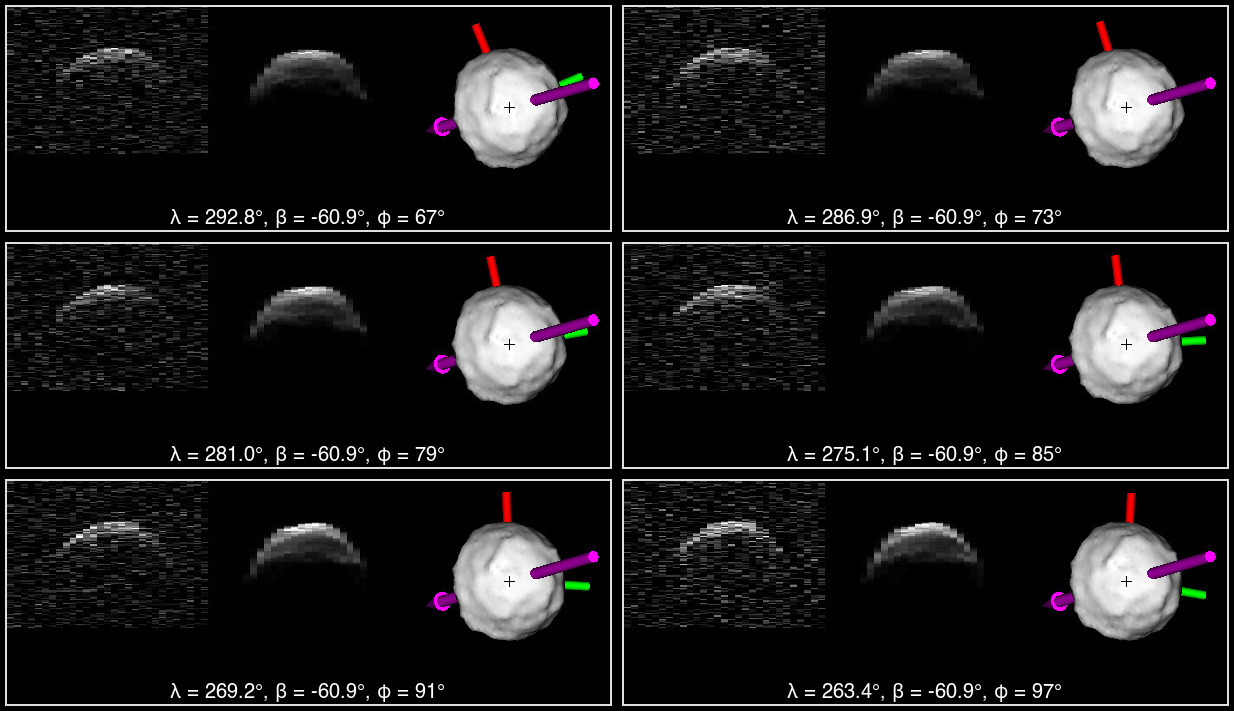}
	\caption*{(continued)}	
	\label{fig:mont_g06b}
\end{figure*}

\begin{figure*}
\centering
\begin{subfigure}{0.33\textwidth}
  \centering
  \includegraphics[width=0.99\linewidth]{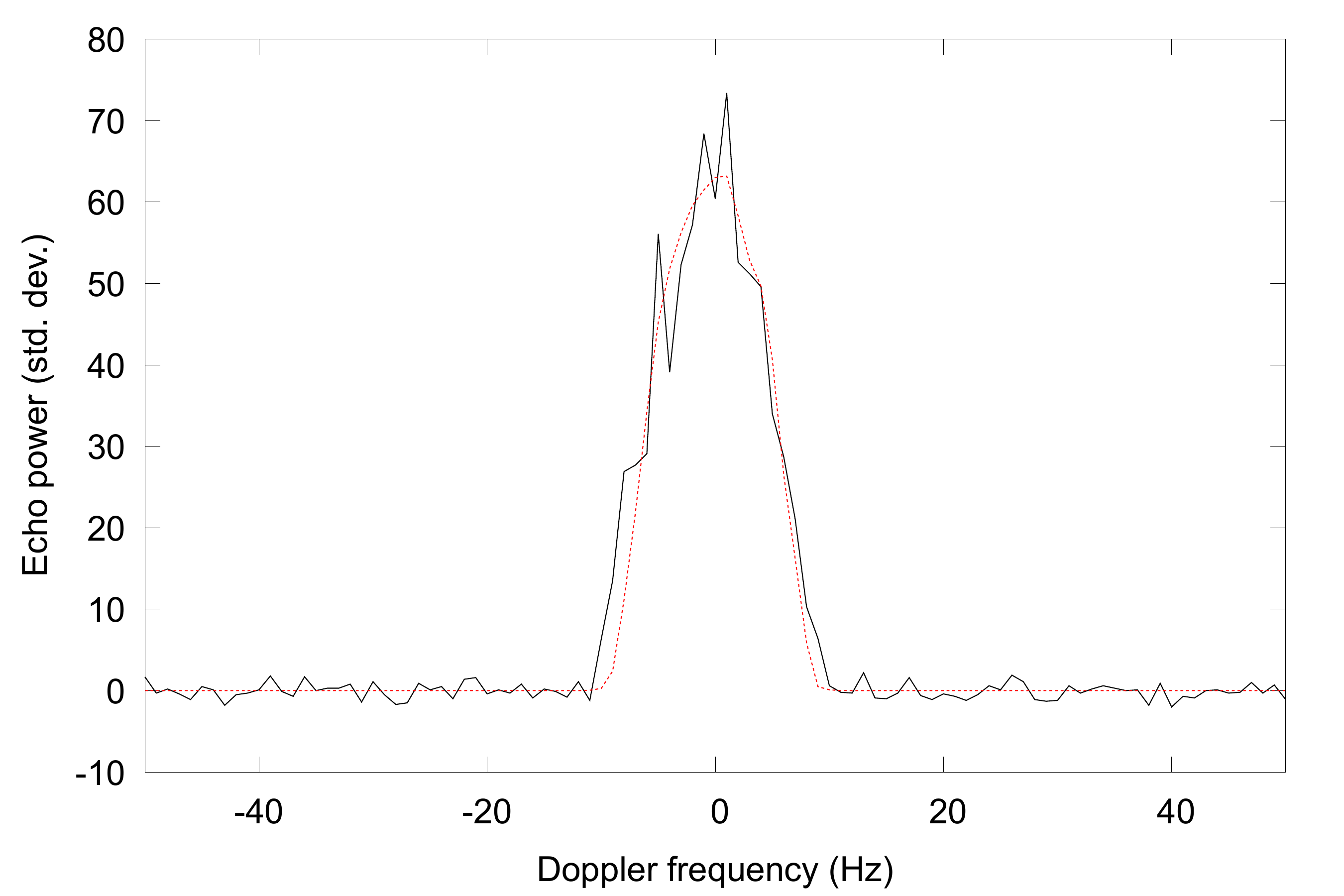}
\end{subfigure}%
\begin{subfigure}{0.33\textwidth}
  \centering
  \includegraphics[width=0.99\linewidth]{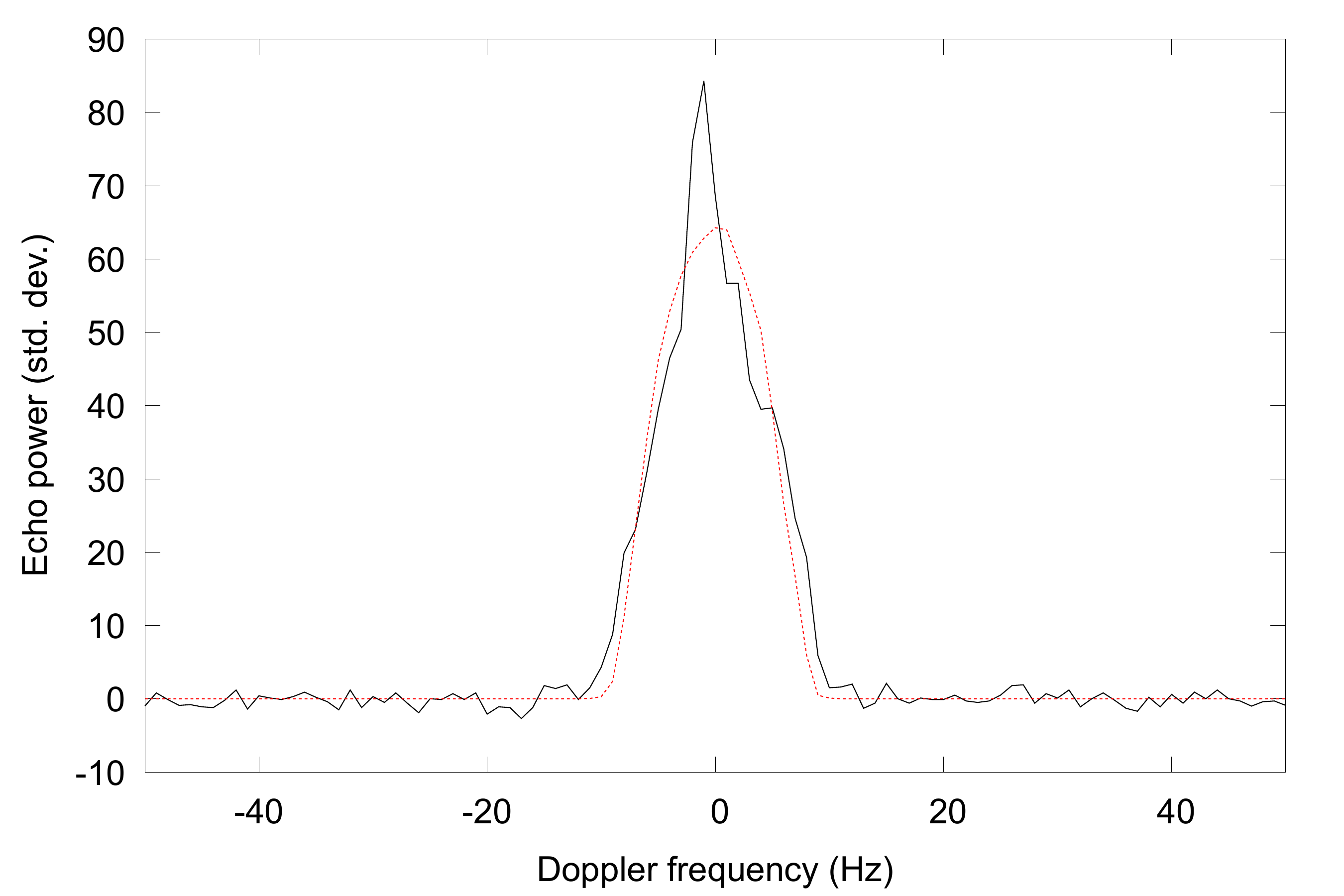}
\end{subfigure}
\begin{subfigure}{0.33\textwidth}
  \centering
  \includegraphics[width=0.99\linewidth]{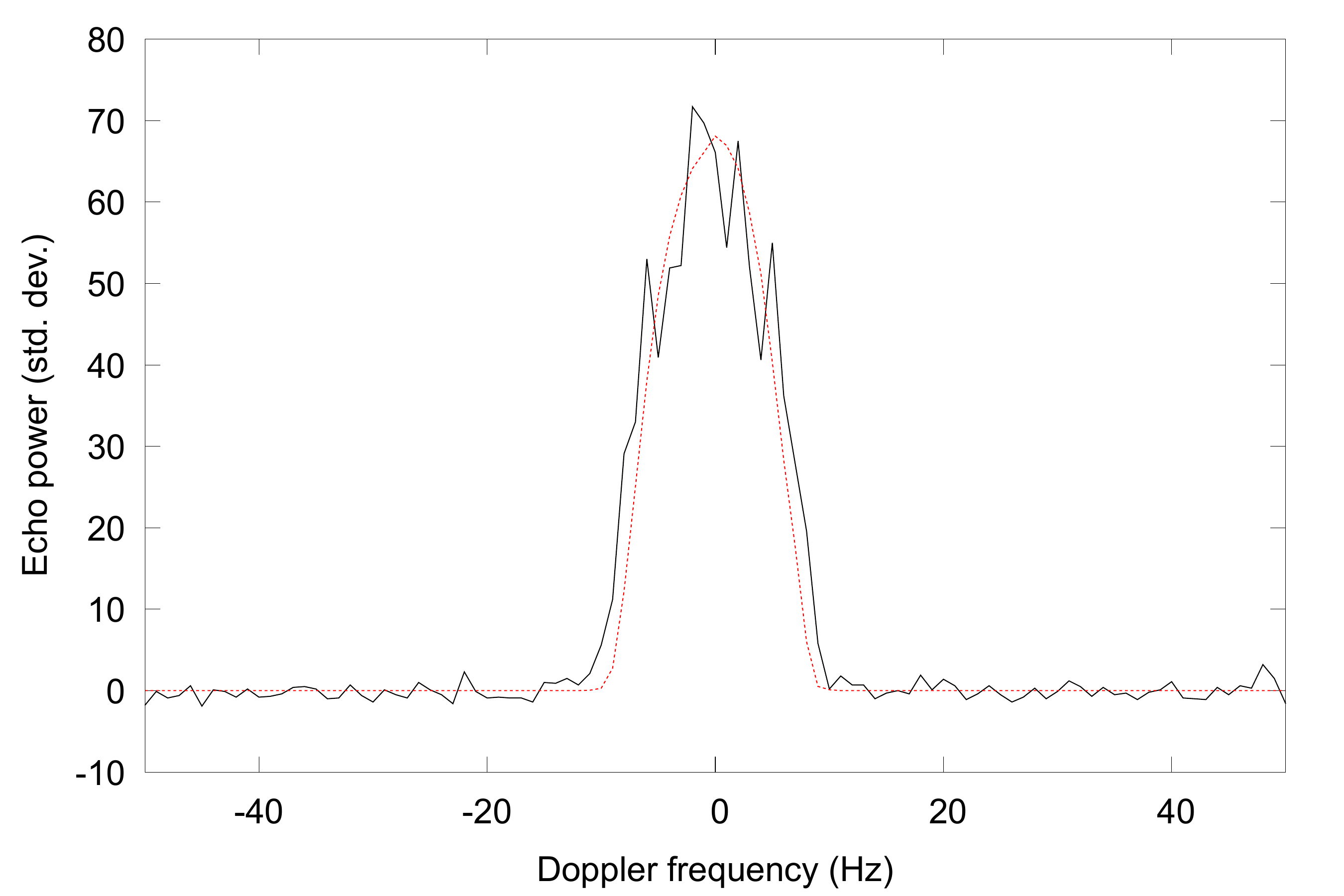}
\end{subfigure}
\begin{subfigure}{0.33\textwidth}
  \centering
  \includegraphics[width=0.99\linewidth]{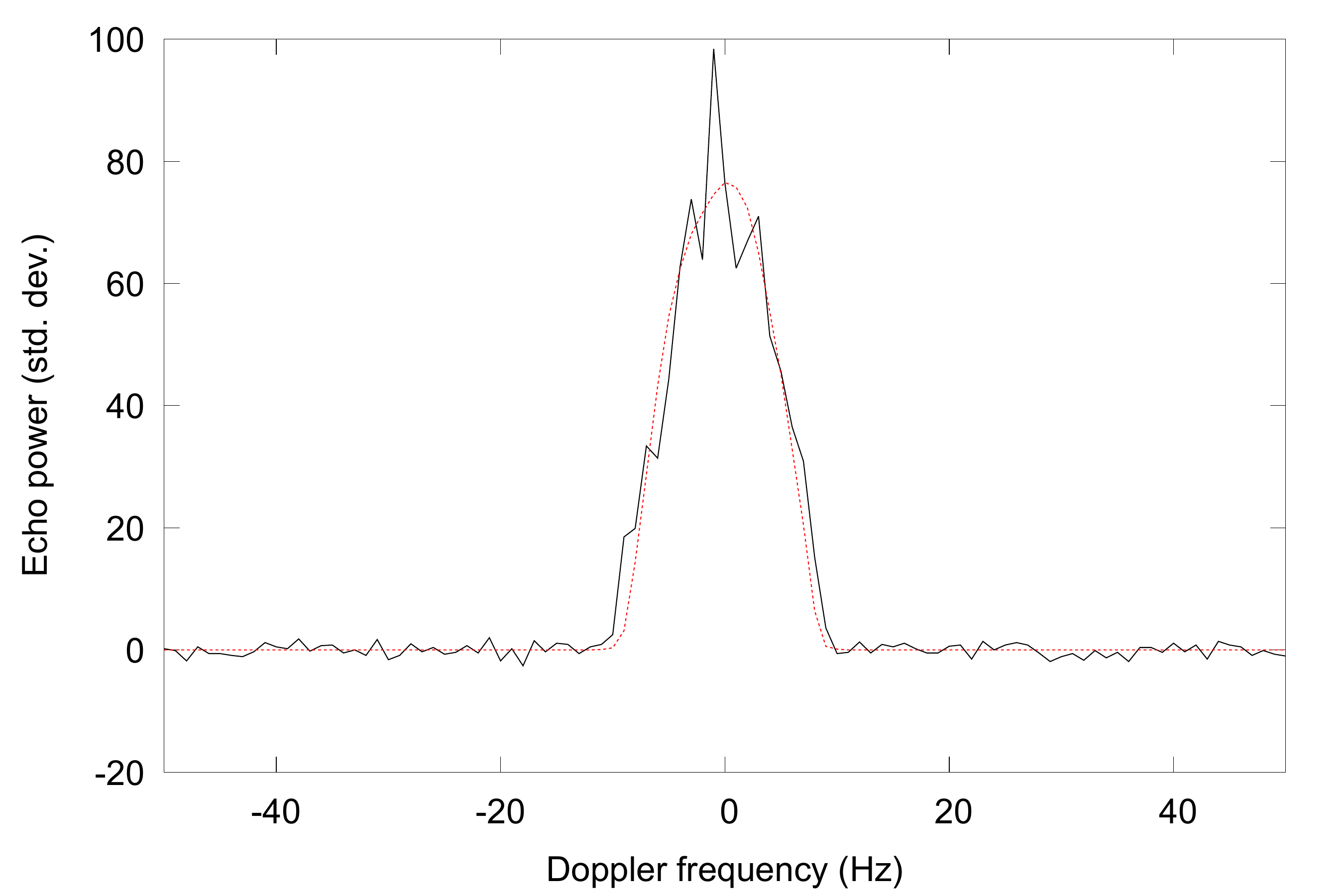}
\end{subfigure}
\begin{subfigure}{0.33\textwidth}
  \centering
  \includegraphics[width=0.99\linewidth]{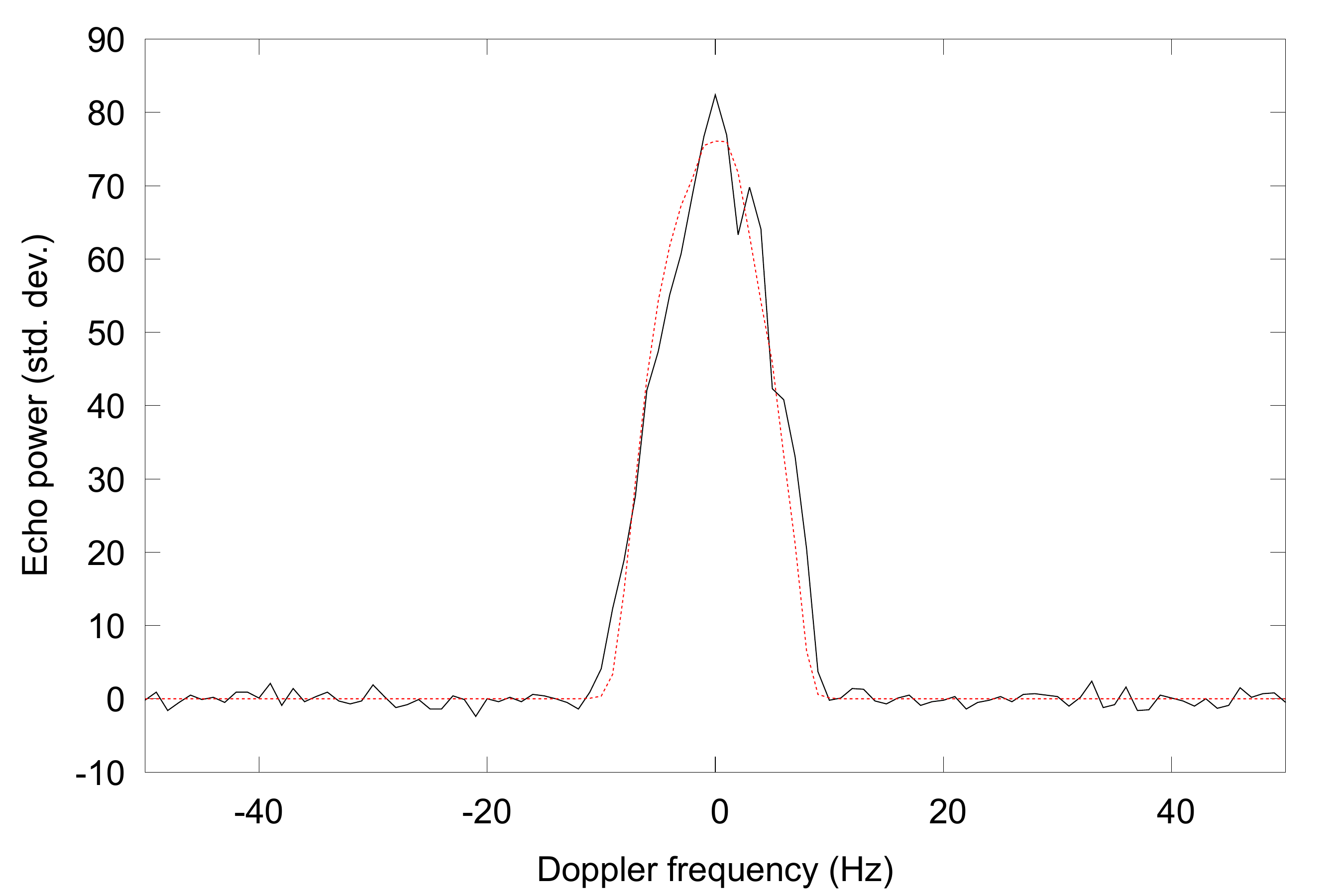}
\end{subfigure}
\begin{subfigure}{0.33\textwidth}
  \centering
  \includegraphics[width=0.99\linewidth]{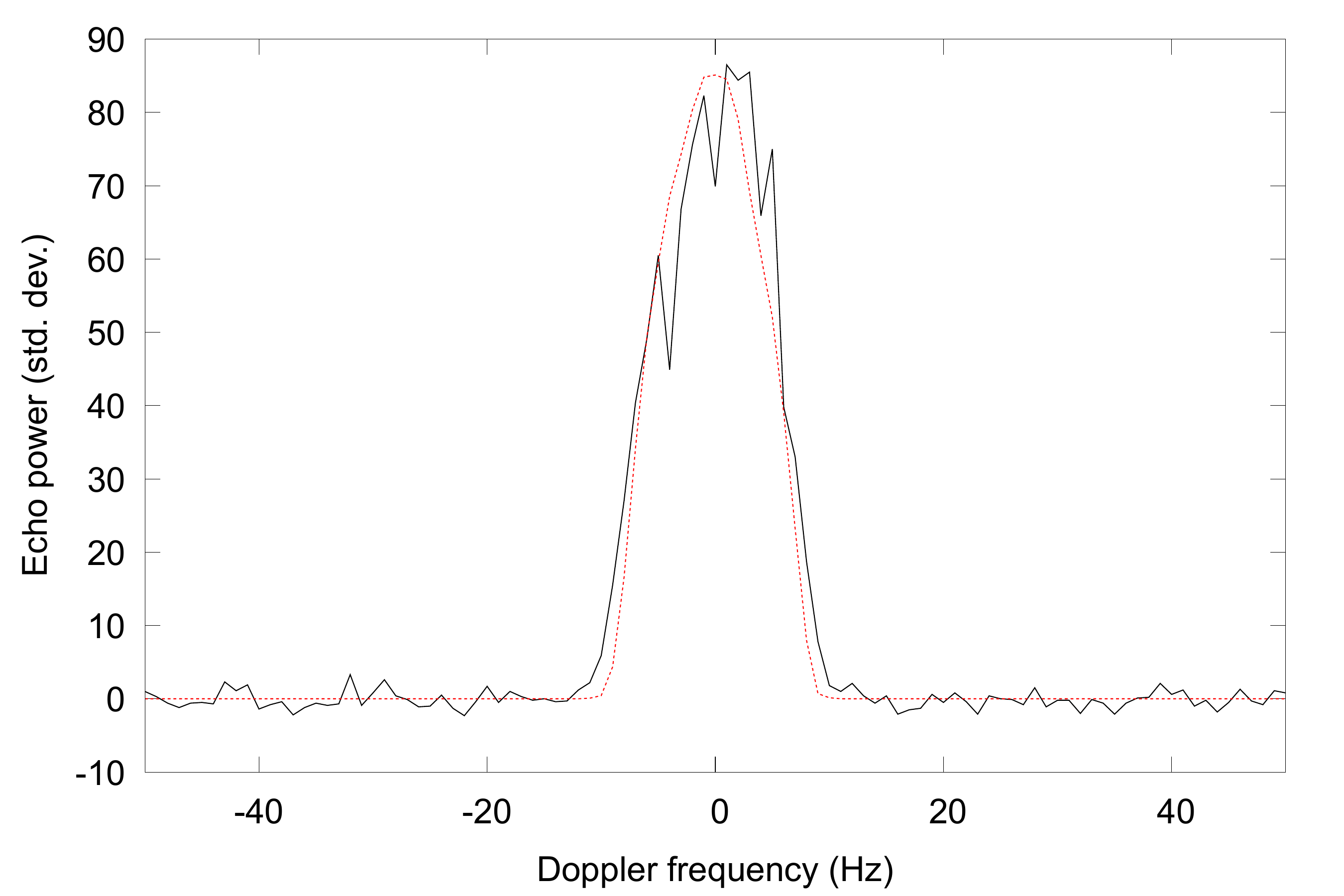}
\end{subfigure}
\caption{A comparison of observed and simulated continuous-wave power spectra for observations with Arecibo on 2001-03-04. The best-fit radar model (Figure 9)
was used to generate a synthetic echo for each corresponding observation. The solid black lines represents the observation (received OC spectra) while the dashed red lines represent the simulated echo. Echo power is measured in units of noise standard deviations and Doppler frequency offset, measured relative to the centre of mass, is measured in Hz. The frequency resolution is 1.0 Hz.}
\label{fig:arecibocw}
\end{figure*}

\begin{figure*}
\centering
\begin{subfigure}{0.33\textwidth}
  \centering
  \includegraphics[width=0.99\linewidth]{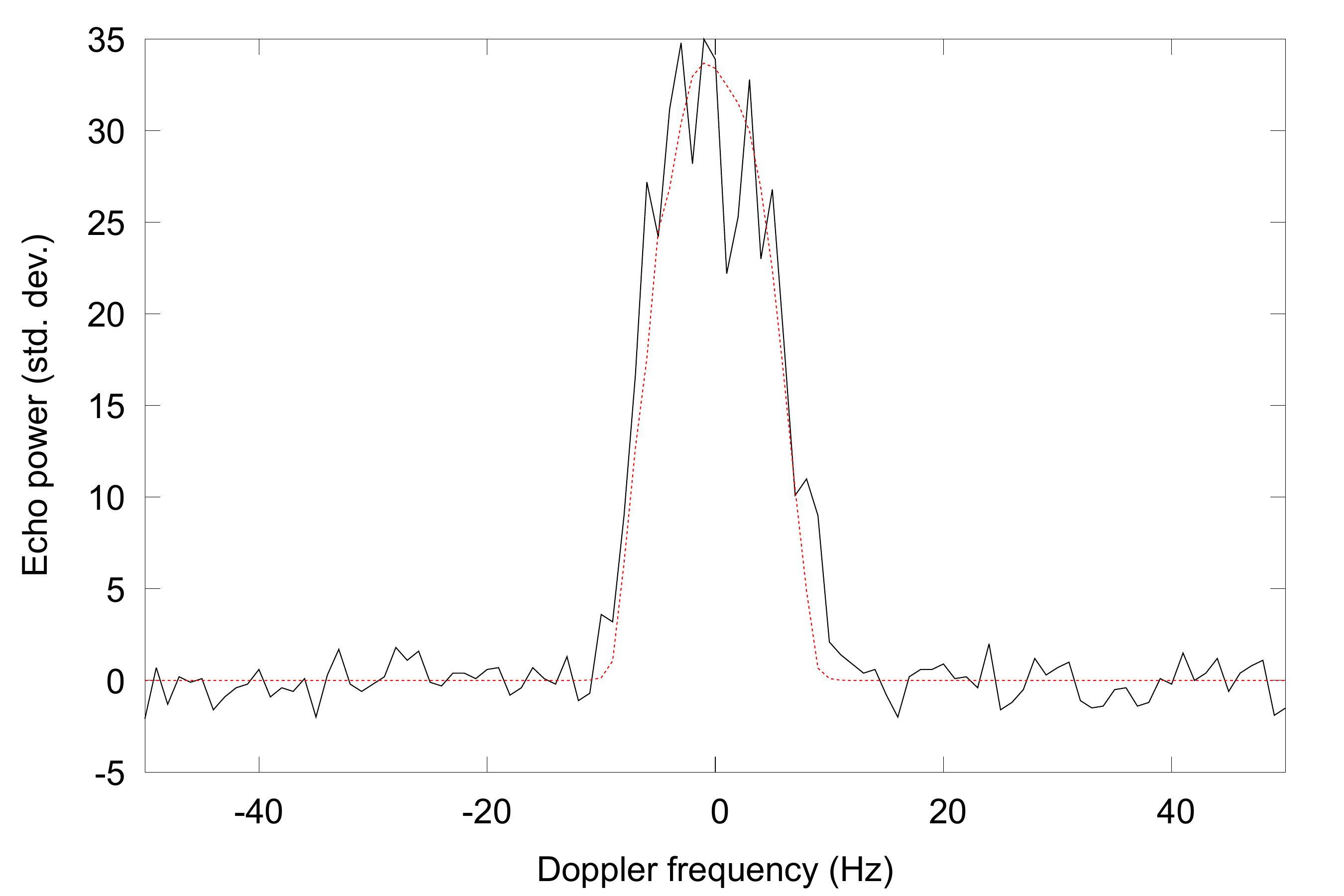}
\end{subfigure}%
\begin{subfigure}{0.33\textwidth}
  \centering
  \includegraphics[width=0.99\linewidth]{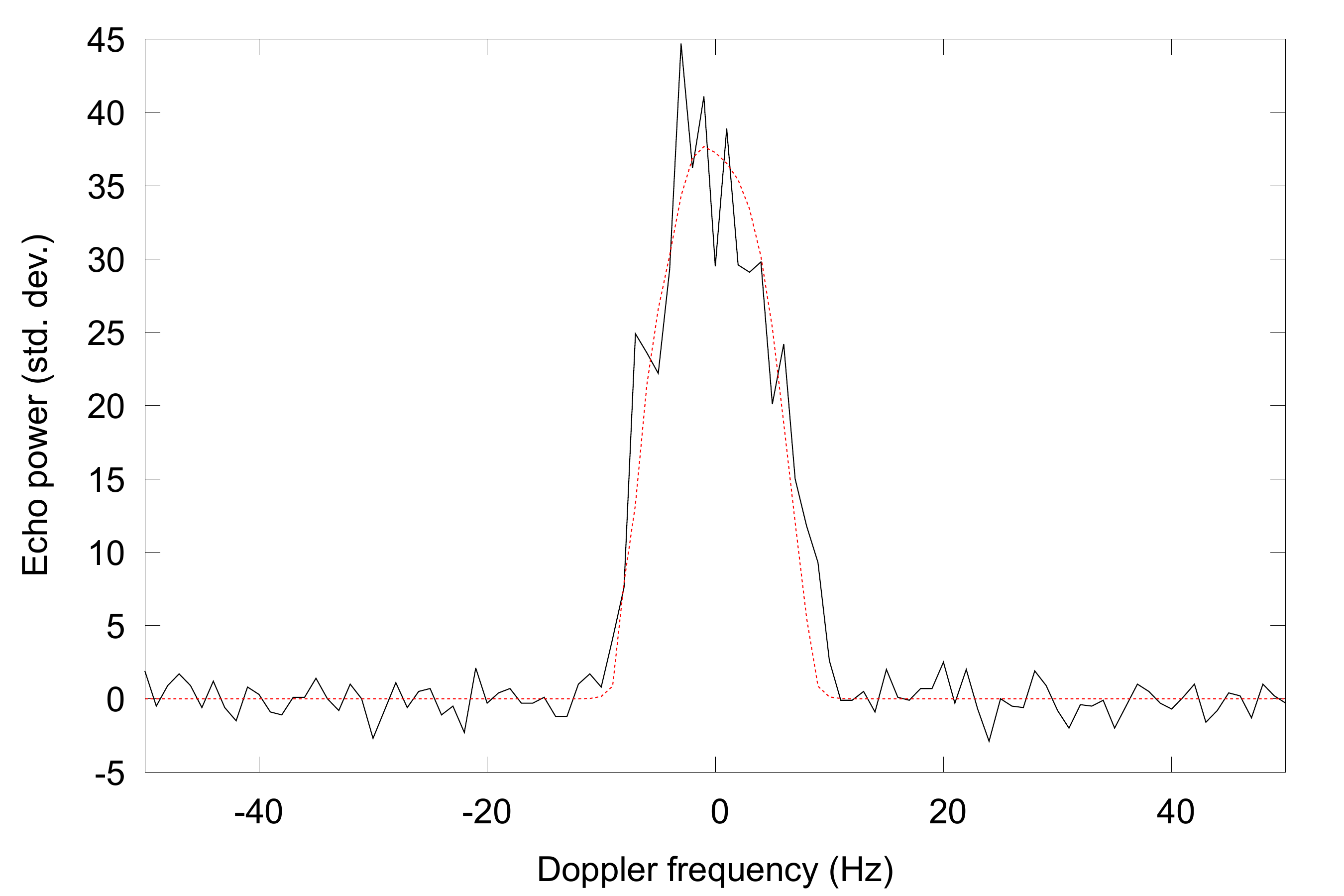}
\end{subfigure}
\begin{subfigure}{0.33\textwidth}
  \centering
  \includegraphics[width=0.99\linewidth]{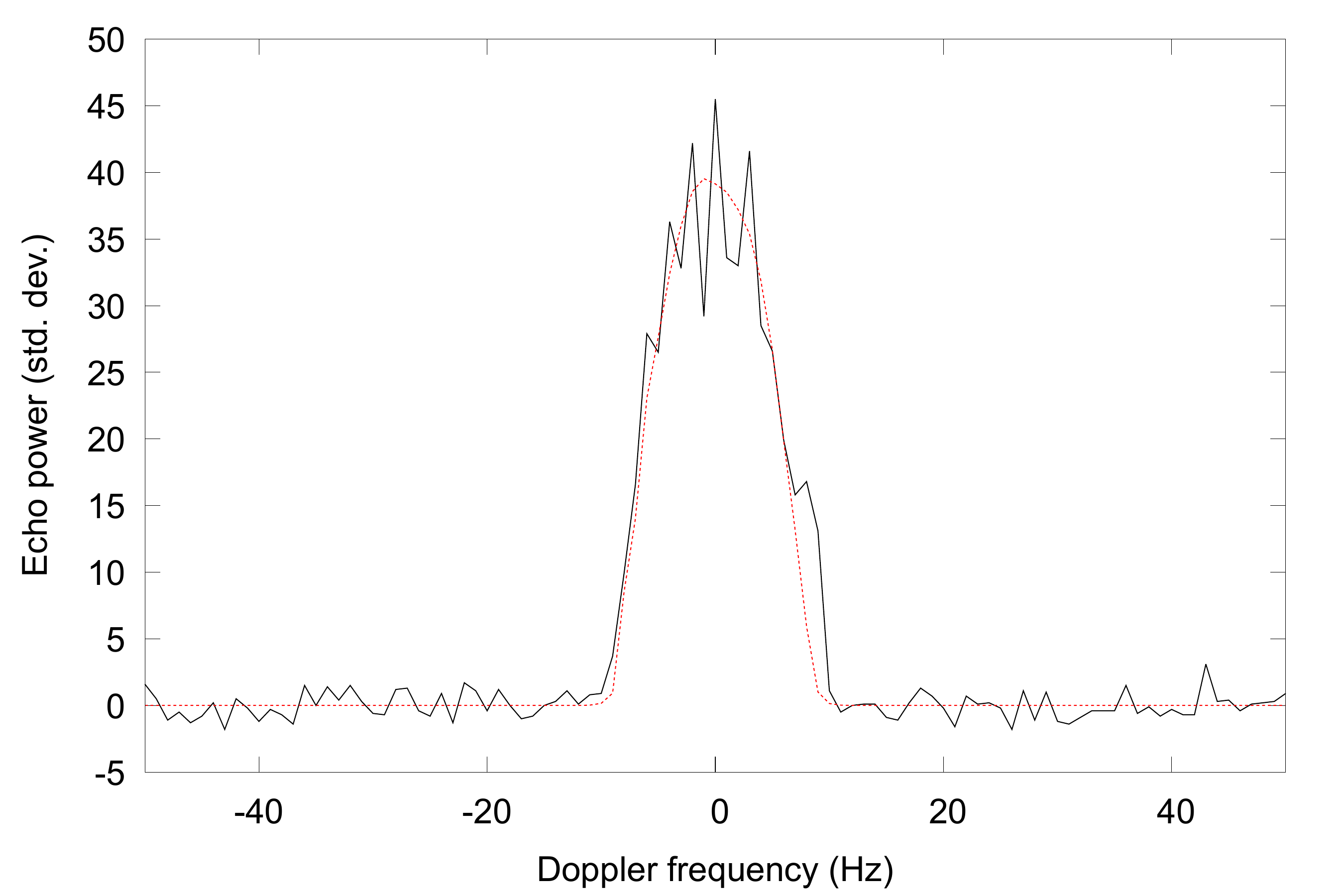}
\end{subfigure}
\caption{The same as Figure \ref{fig:arecibocw}, but for observations with Arecibo on 2001-03-05. The frequency resolution is 1.0 Hz.}
\label{fig:goldstoneocw}
\end{figure*}

\begin{figure*}
	
	\resizebox{\hsize}{!}{	
		\includegraphics[width=.25\textwidth, trim=0.5cm 2.5cm 1.5cm 2.5cm, clip=true]{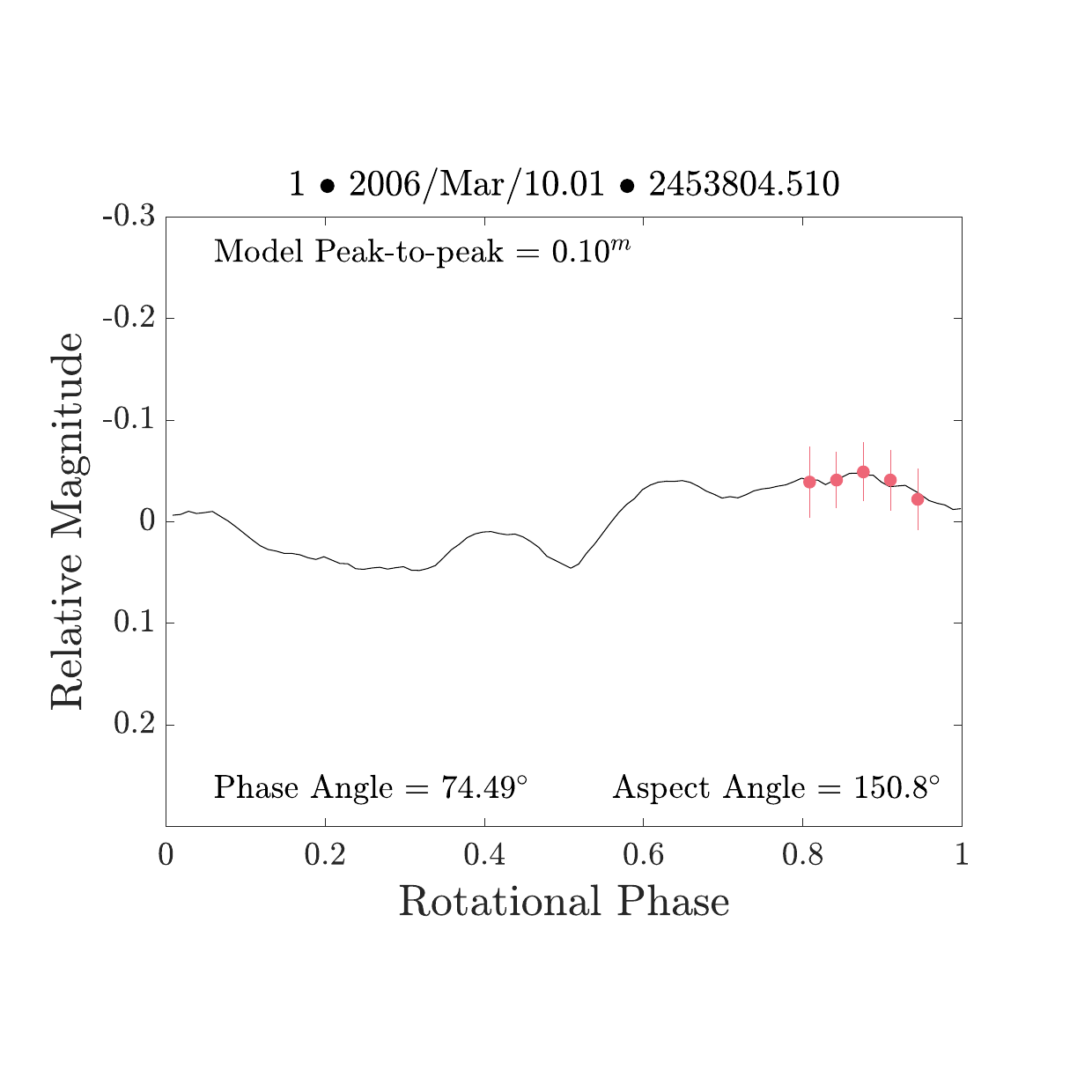}
        \includegraphics[width=.25\textwidth, trim=0.5cm 2.5cm 1.5cm 2.5cm, clip=true]{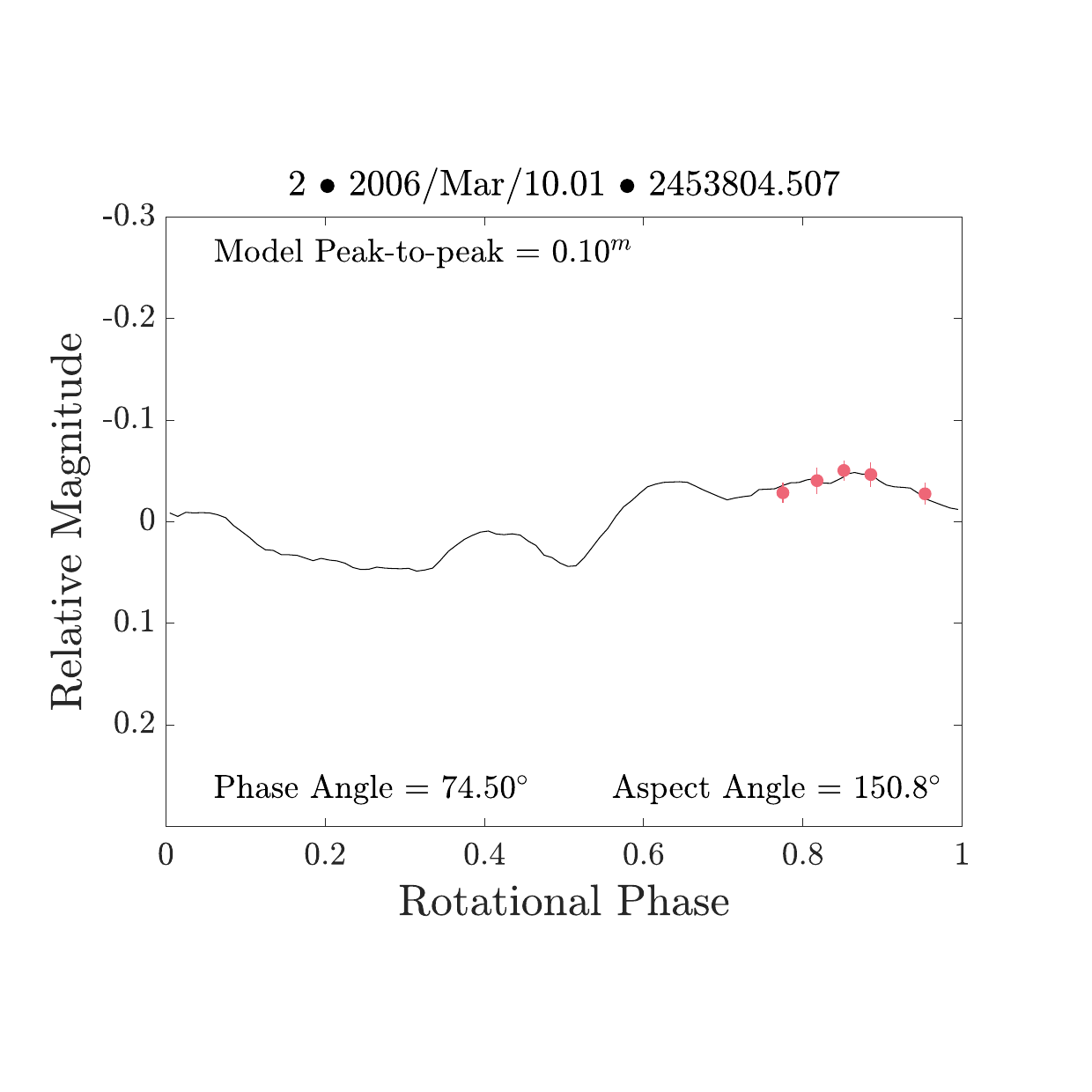}
        \includegraphics[width=.25\textwidth, trim=0.5cm 2.5cm 1.5cm 2.5cm, clip=true]{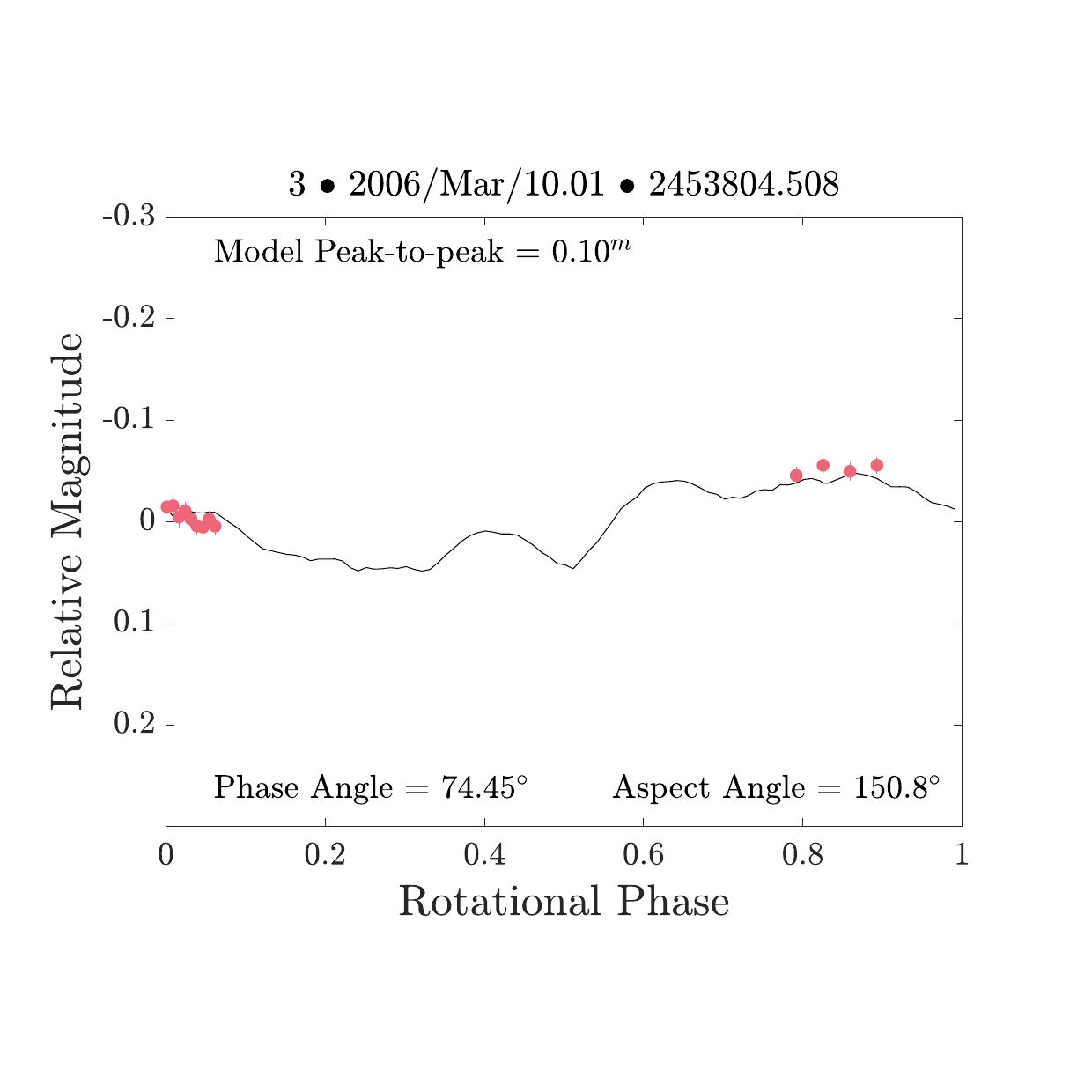}
        \includegraphics[width=.25\textwidth, trim=0.5cm 2.5cm 1.5cm 2.5cm, clip=true]{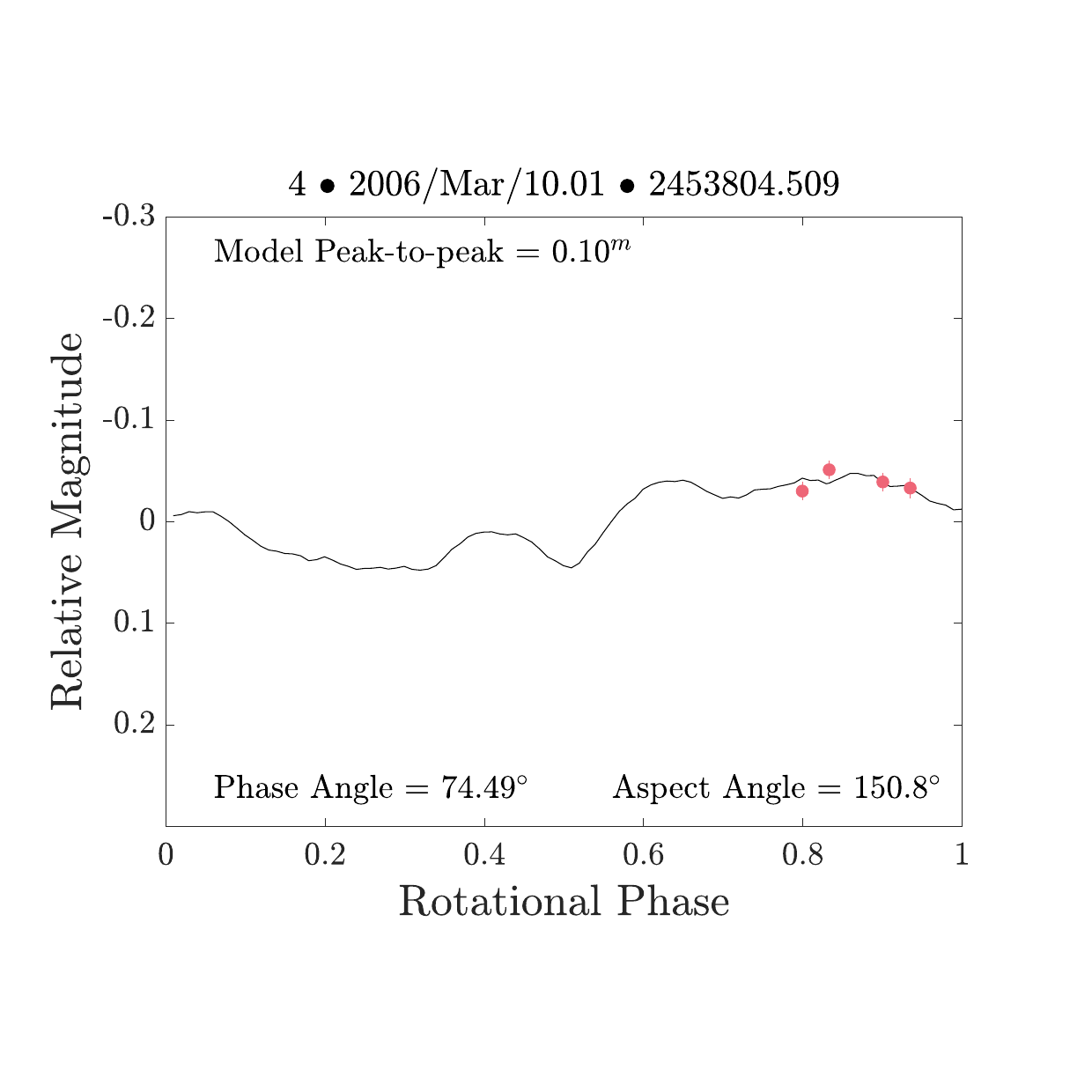}
	}
	
	\resizebox{\hsize}{!}{	

        \includegraphics[width=.25\textwidth, trim=0.5cm 2.5cm 1.5cm 2.5cm, clip=true]{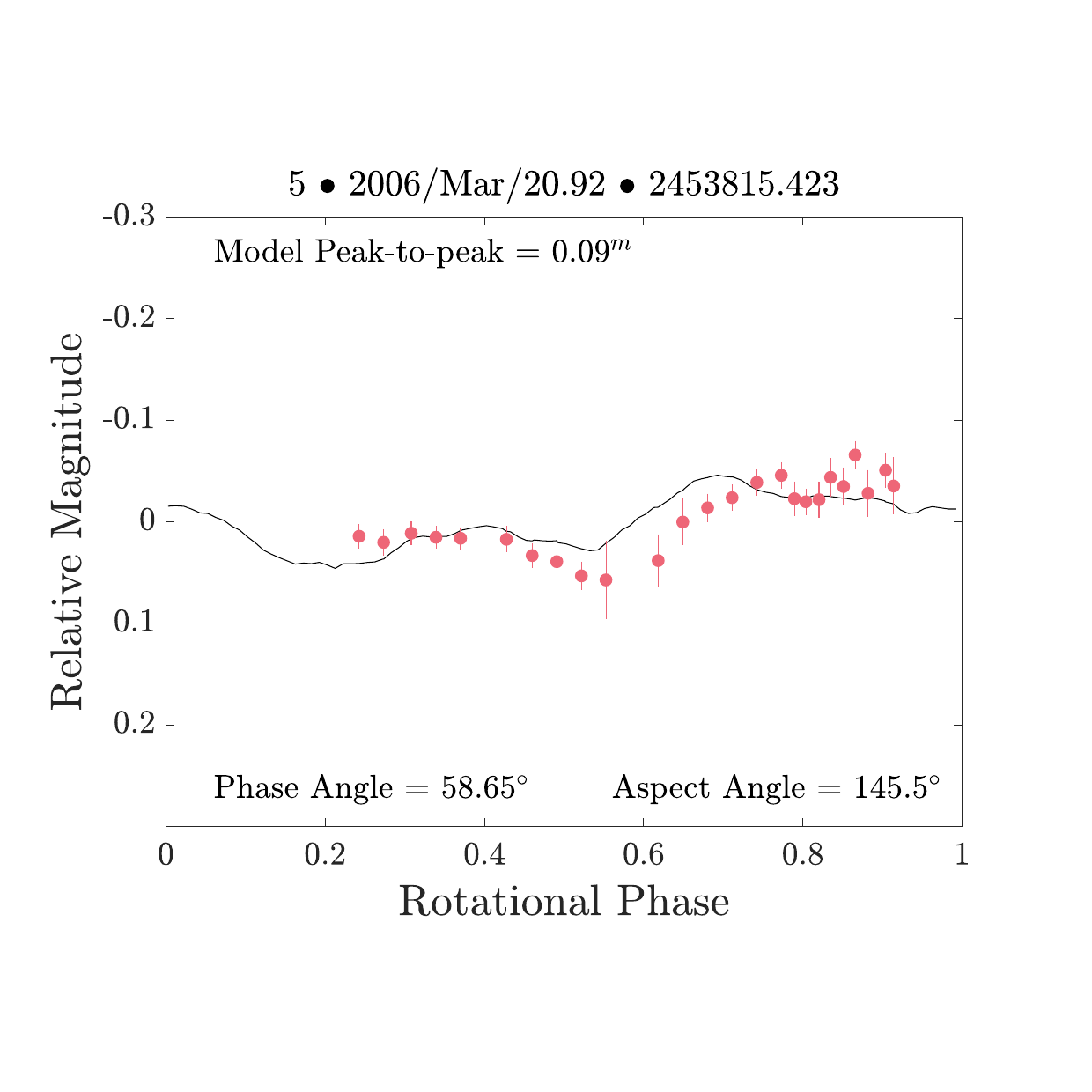}
        \includegraphics[width=.25\textwidth, trim=0.5cm 2.5cm 1.5cm 2.5cm, clip=true]{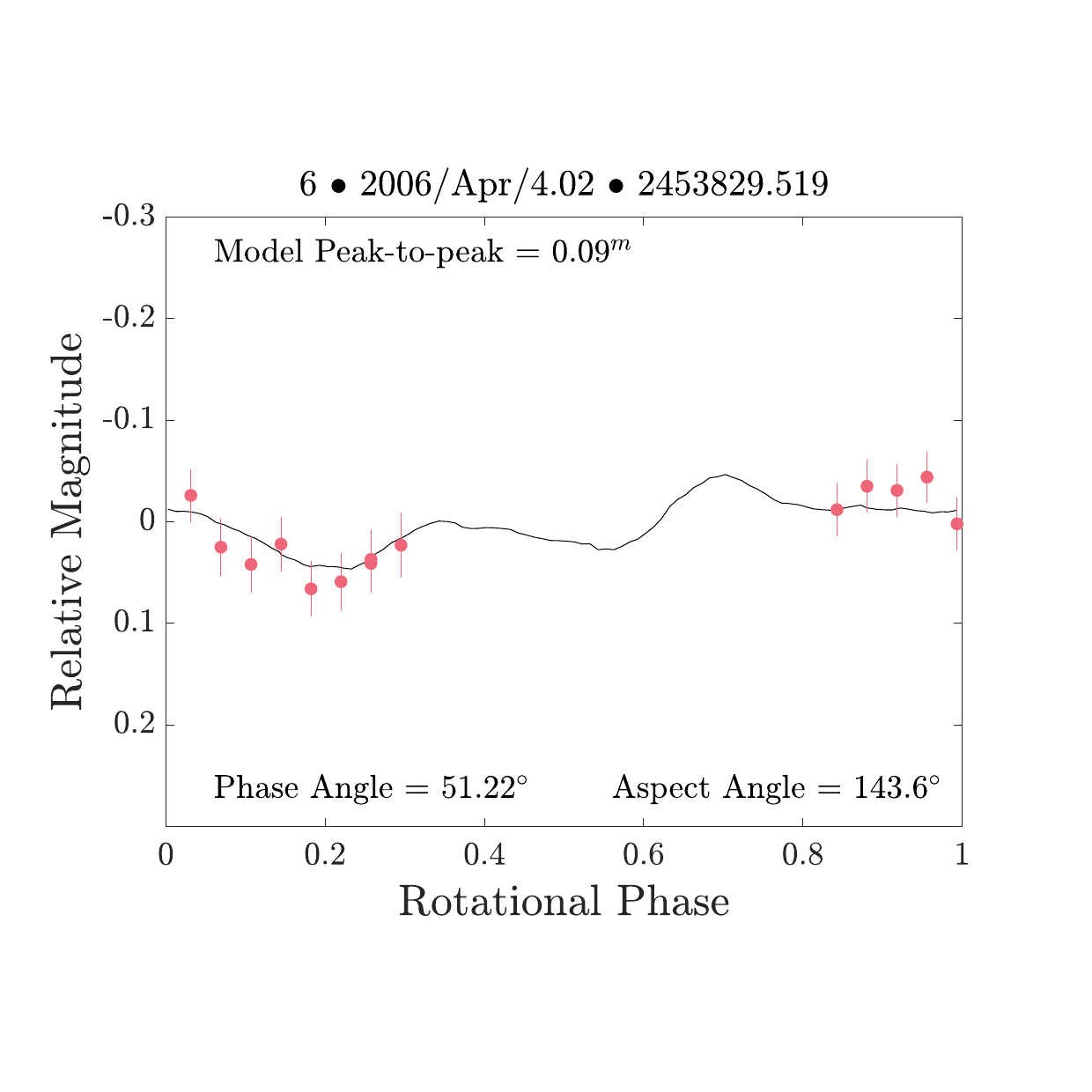}  
        \includegraphics[width=.25\textwidth, trim=0.5cm 2.5cm 1.5cm 2.5cm, clip=true]{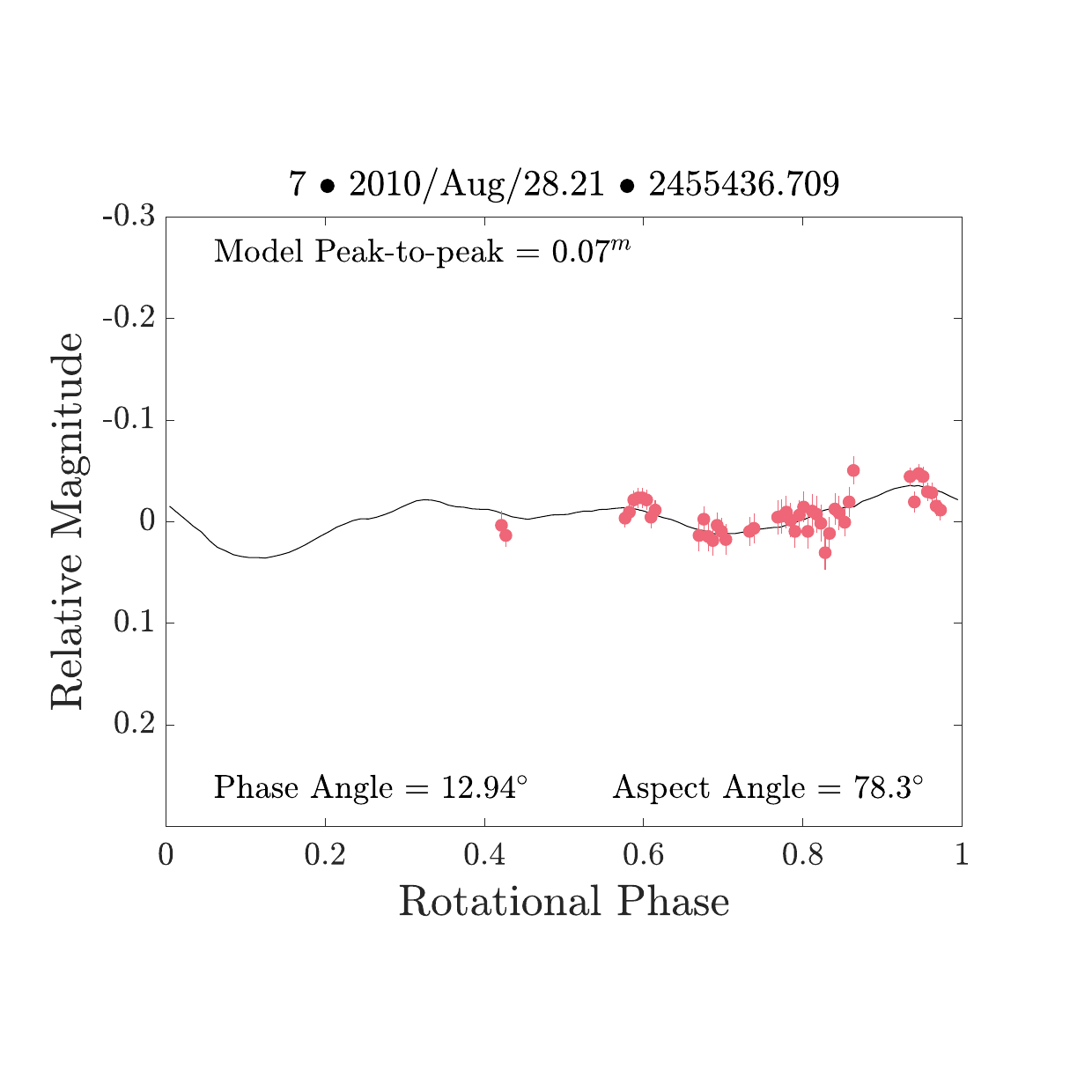}
        \includegraphics[width=.25\textwidth, trim=0.5cm 2.5cm 1.5cm 2.5cm, clip=true]{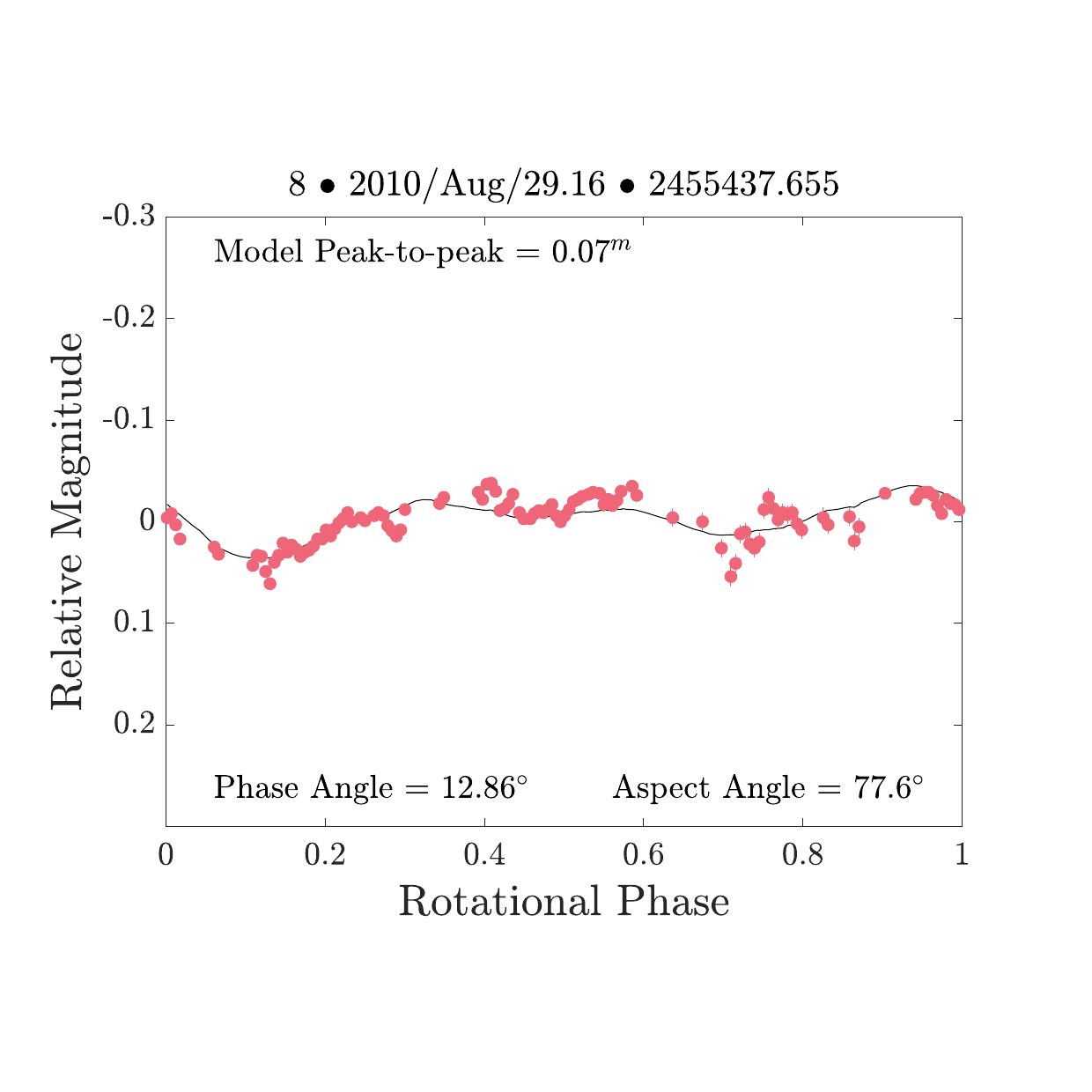}
	}
	
	\resizebox{\hsize}{!}{	
        \includegraphics[width=.25\textwidth, trim=0.5cm 2.5cm 1.5cm 2.5cm, clip=true]{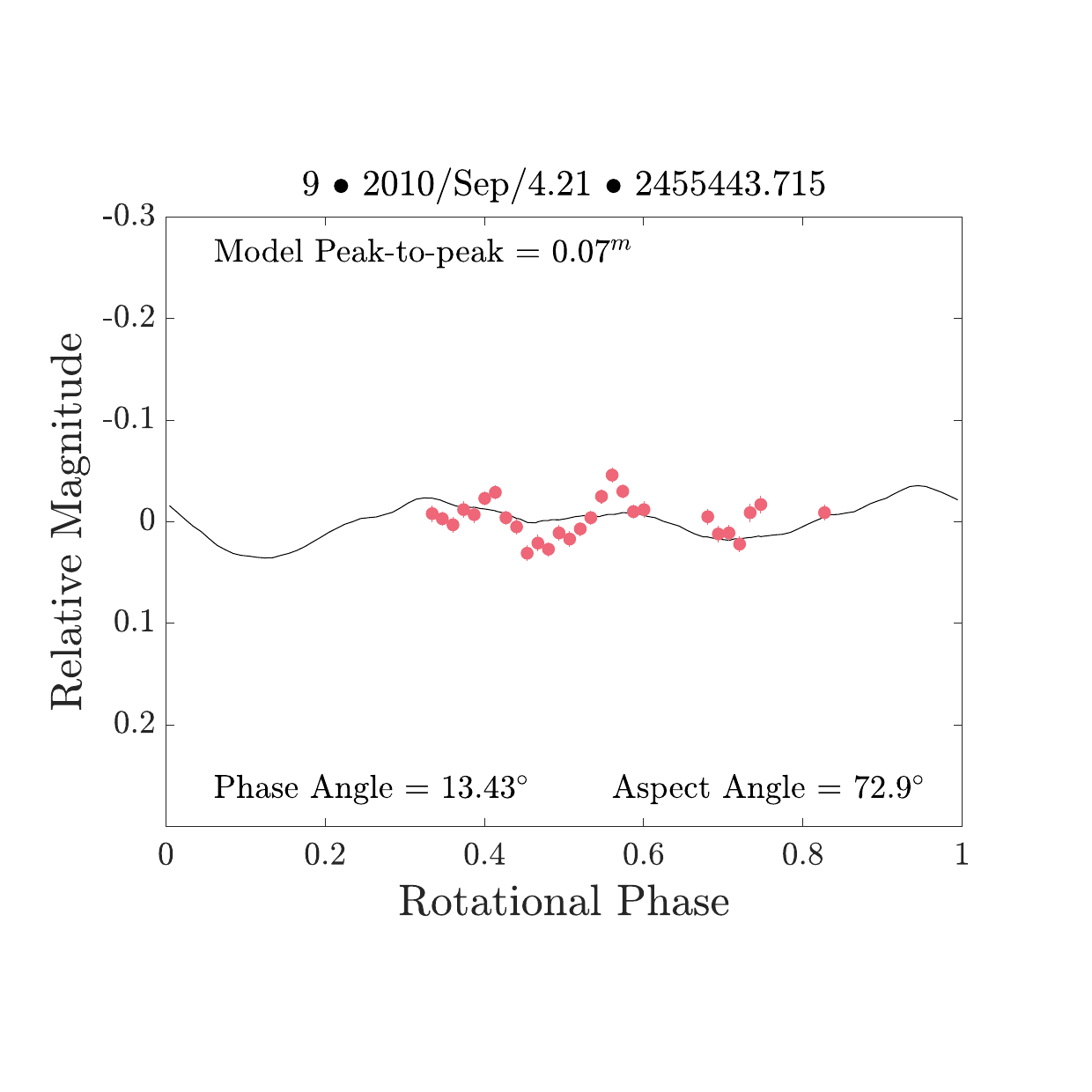}
        \includegraphics[width=.25\textwidth, trim=0.5cm 2.5cm 1.5cm 2.5cm, clip=true]{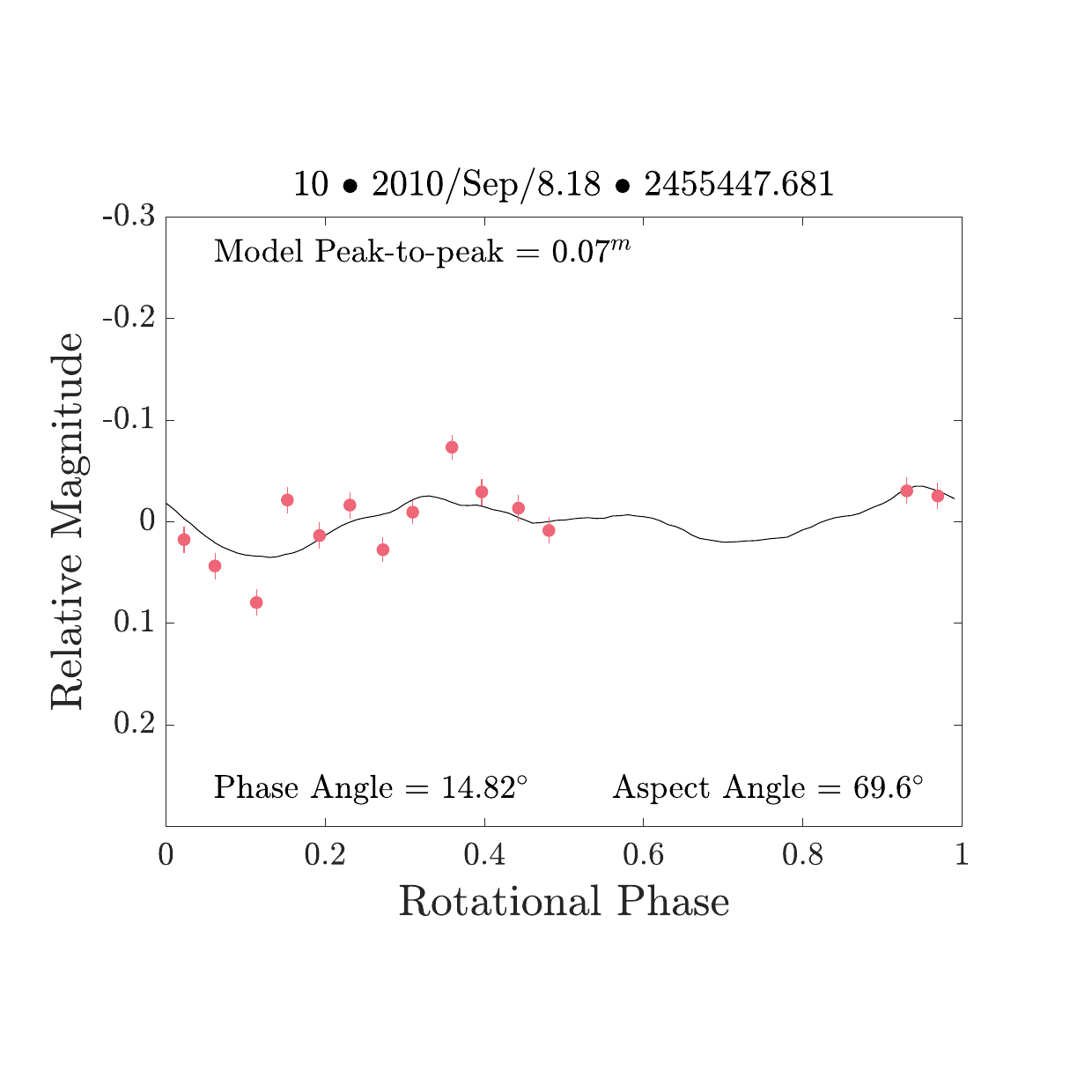}        
        \includegraphics[width=.25\textwidth, trim=0.5cm 2.5cm 1.5cm 2.5cm, clip=true]{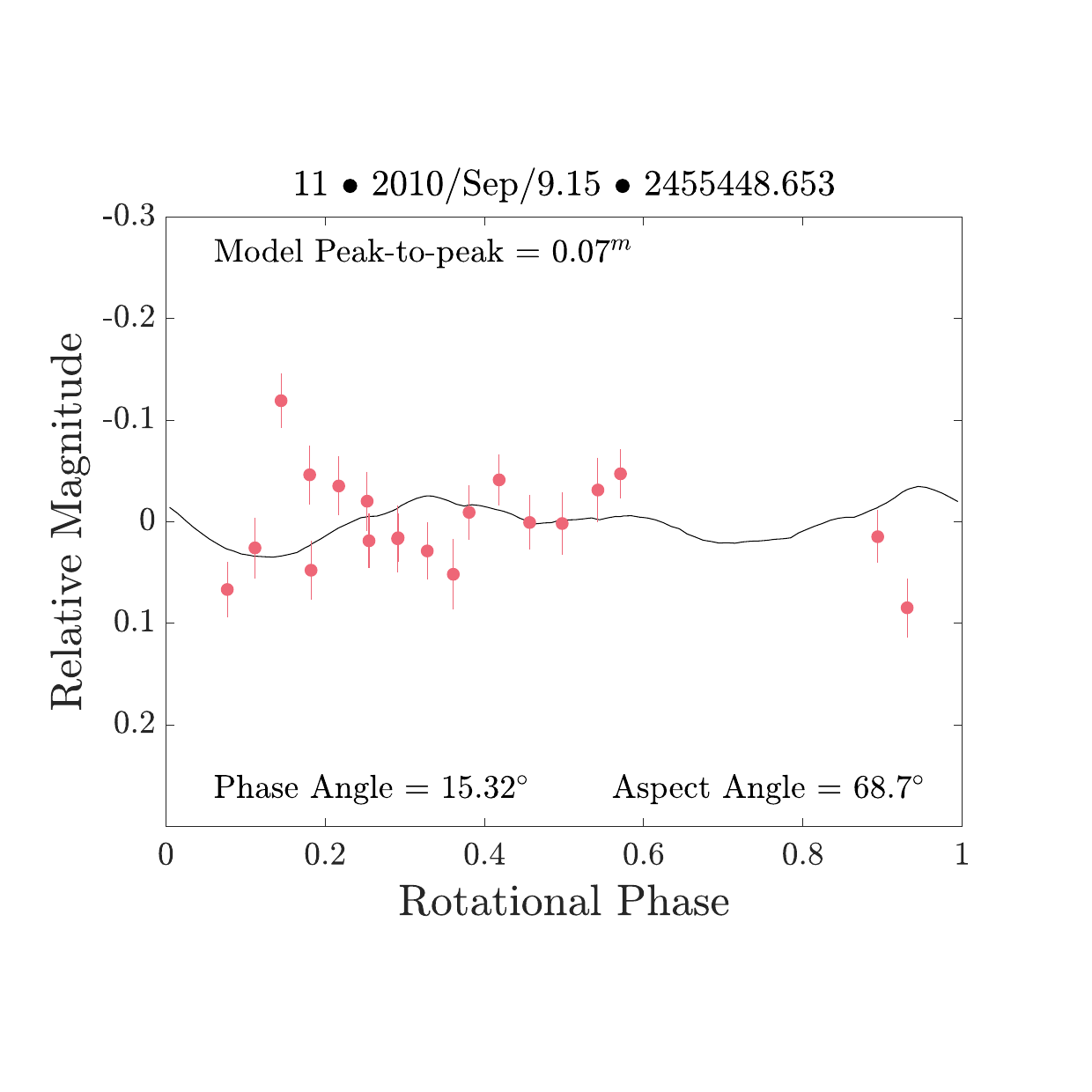}
        \includegraphics[width=.25\textwidth, trim=0.5cm 2.5cm 1.5cm 2.5cm, clip=true]{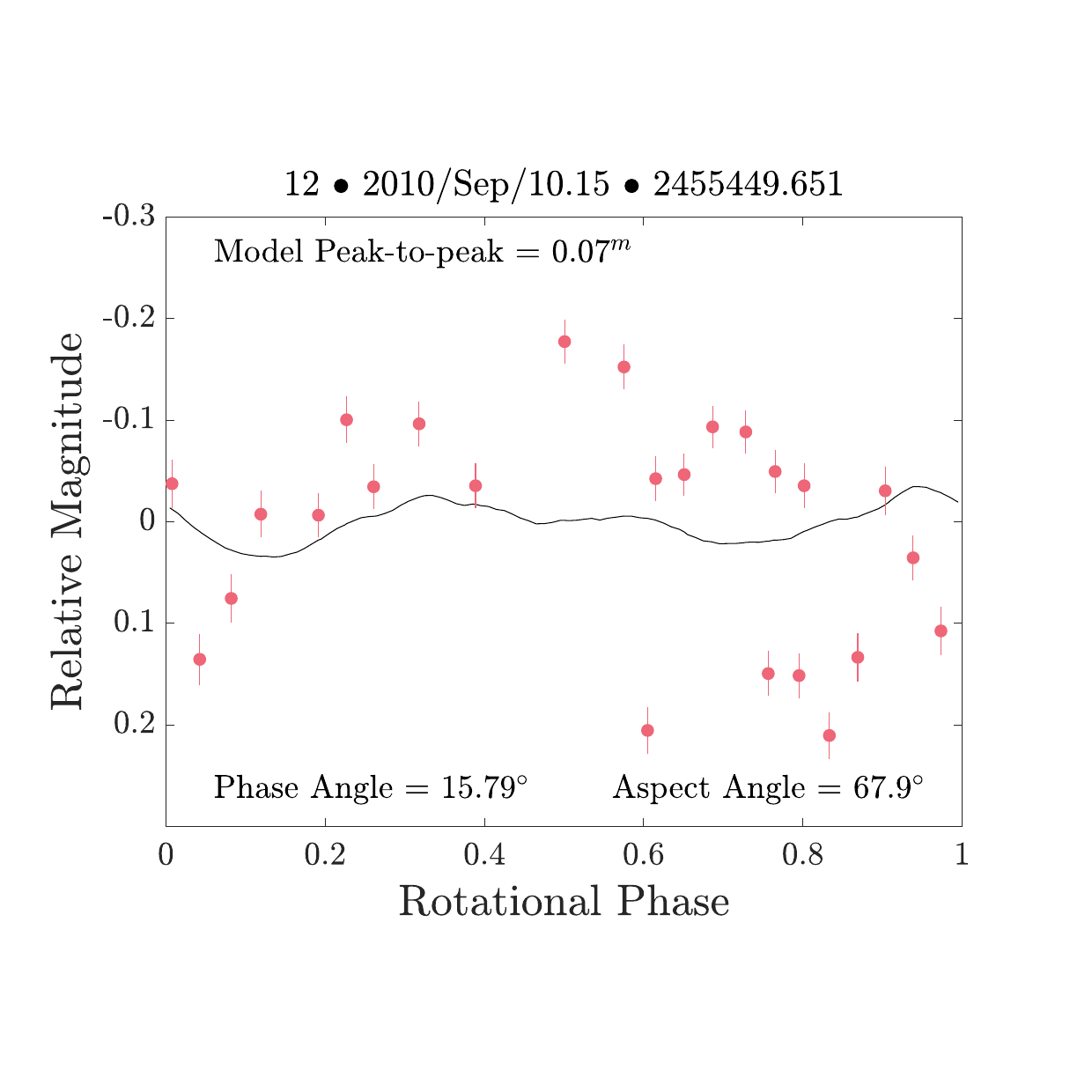}
	}
	
	\resizebox{\hsize}{!}{	
        \includegraphics[width=.25\textwidth, trim=0.5cm 2.5cm 1.5cm 2.5cm, clip=true]{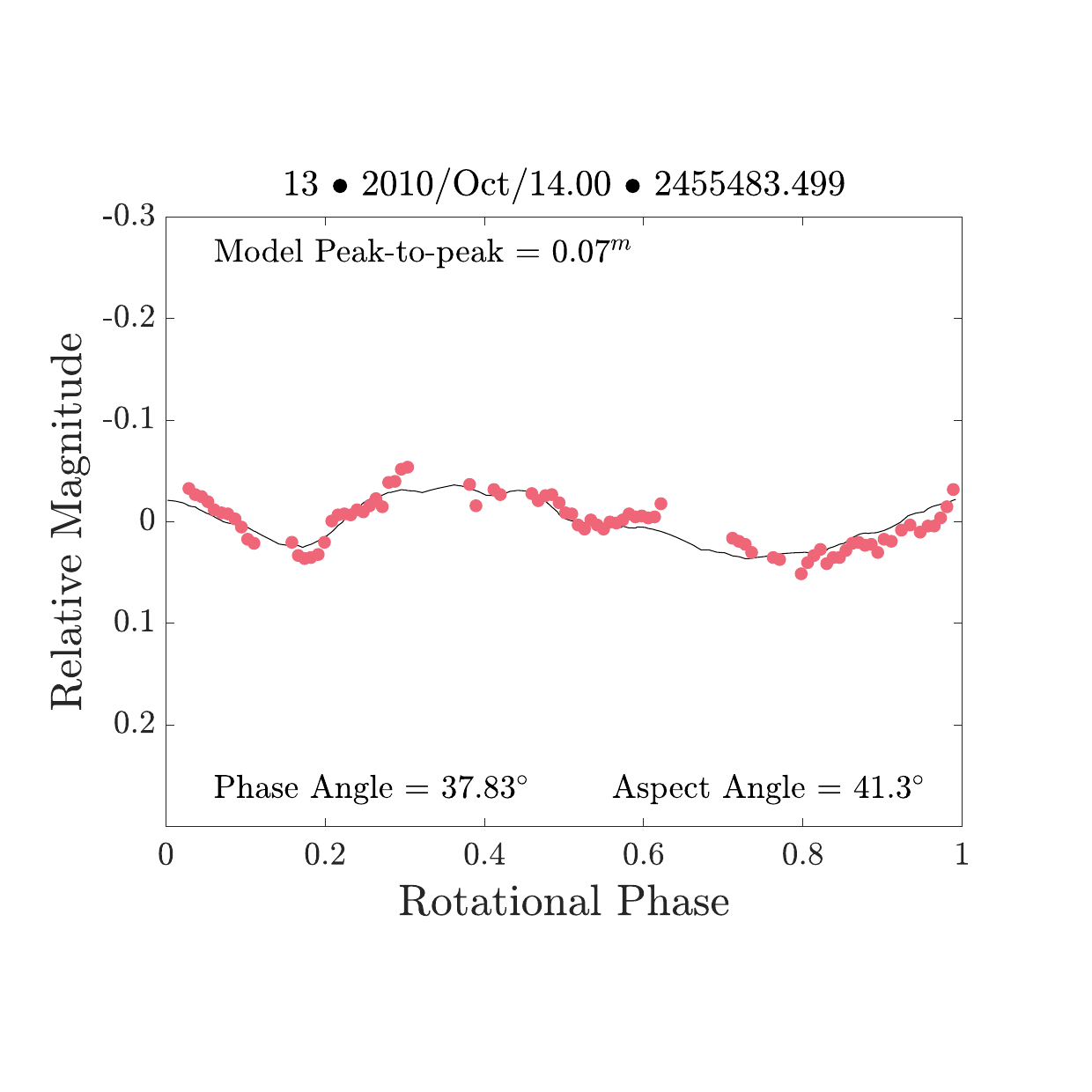}
        \includegraphics[width=.25\textwidth, trim=0.5cm 2.5cm 1.5cm 2.5cm, clip=true]{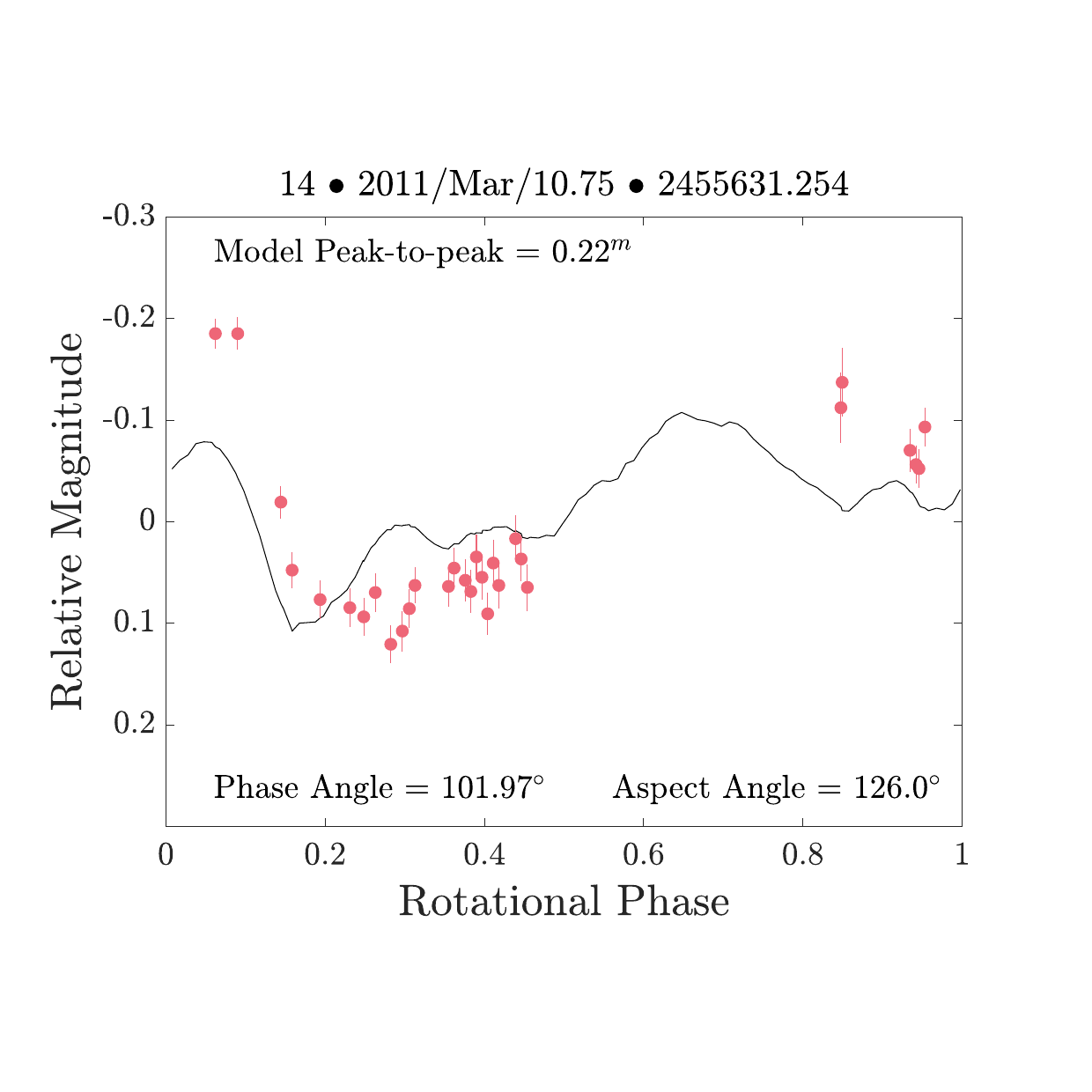}
        \includegraphics[width=.25\textwidth, trim=0.5cm 2.5cm 1.5cm 2.5cm, clip=true]{images/23187_radarLCfit_15.pdf}
        \includegraphics[width=.25\textwidth, trim=0.5cm 2.5cm 1.5cm 2.5cm, clip=true]{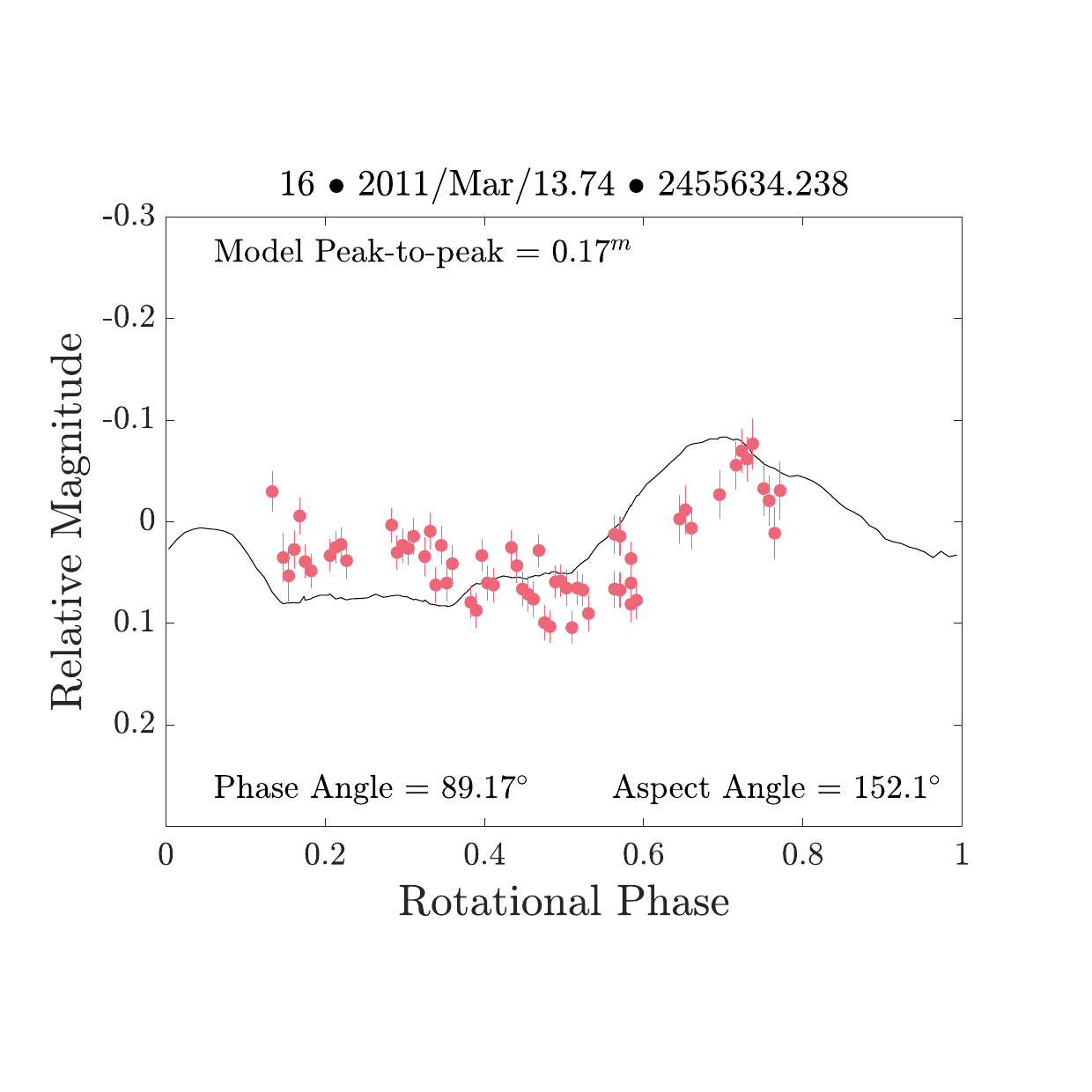}
    }
        
	\resizebox{\hsize}{!}{	
        \includegraphics[width=.25\textwidth, trim=0.5cm 2.5cm 1.5cm 2.5cm, clip=true]{images/23187_radarLCfit_17.pdf}
        \includegraphics[width=.25\textwidth, trim=0.5cm 2.5cm 1.5cm 2.5cm, clip=true]{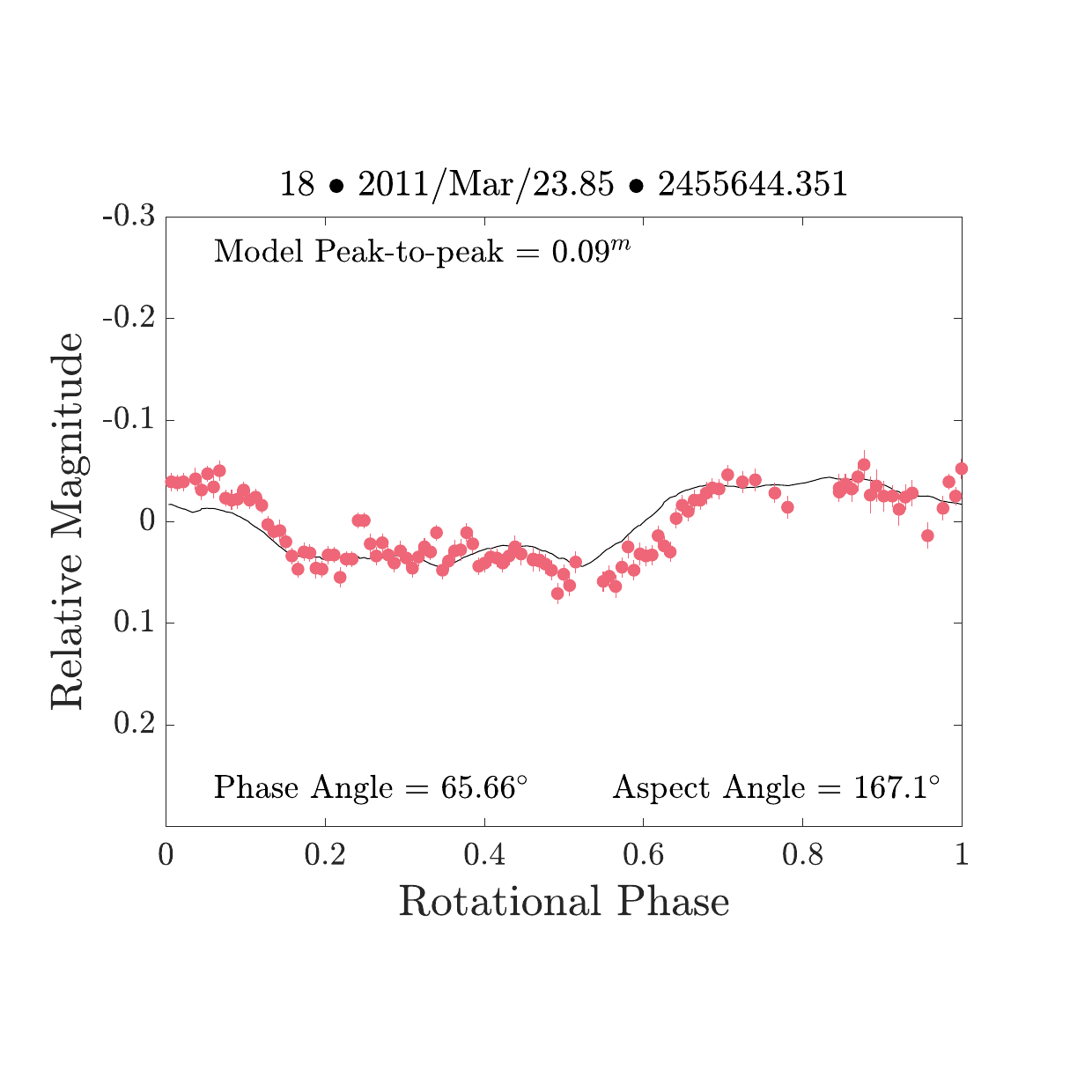}
        \includegraphics[width=.25\textwidth, trim=0.5cm 2.5cm 1.5cm 2.5cm, clip=true]{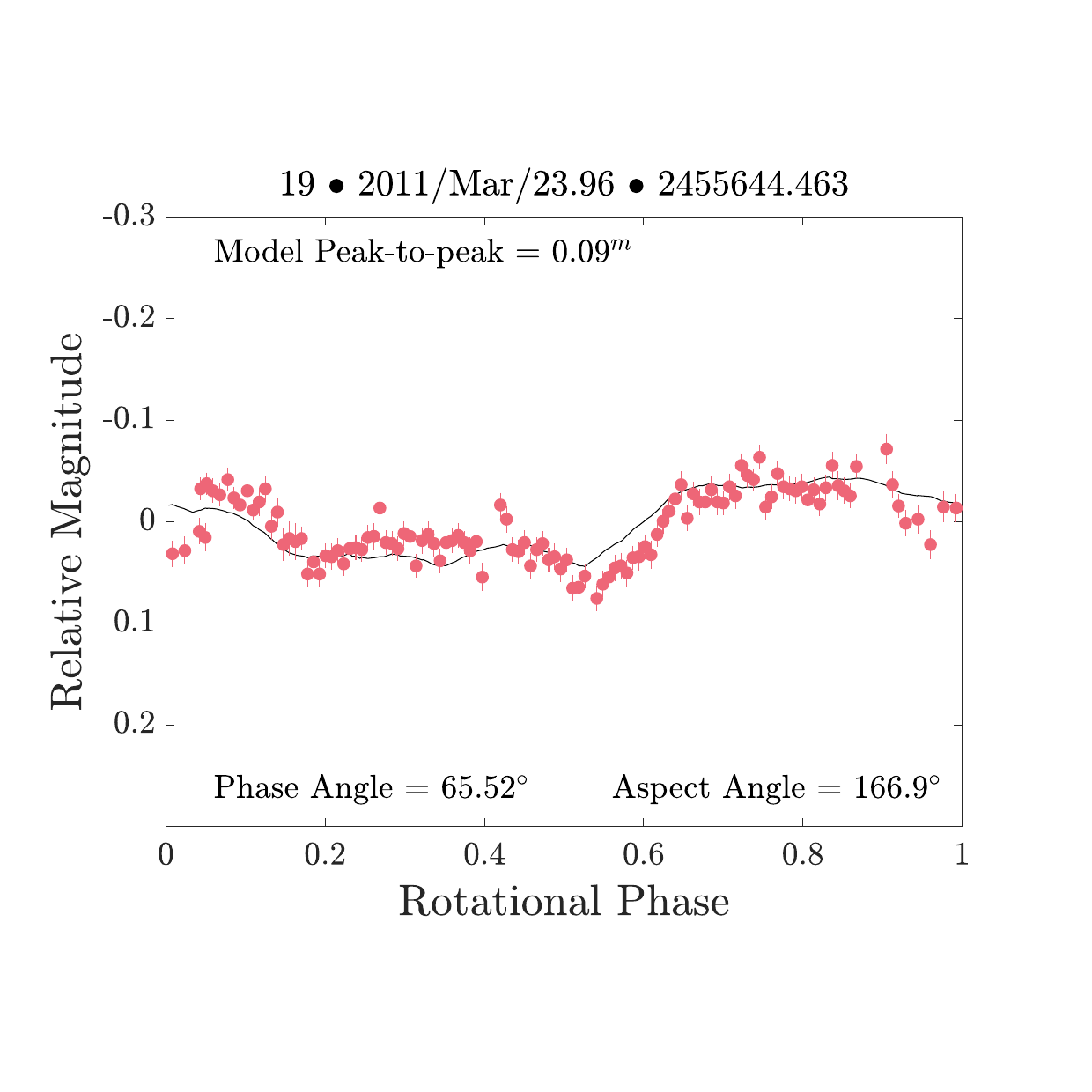}        
        \includegraphics[width=.25\textwidth, trim=0.5cm 2.5cm 1.5cm 2.5cm, clip=true]{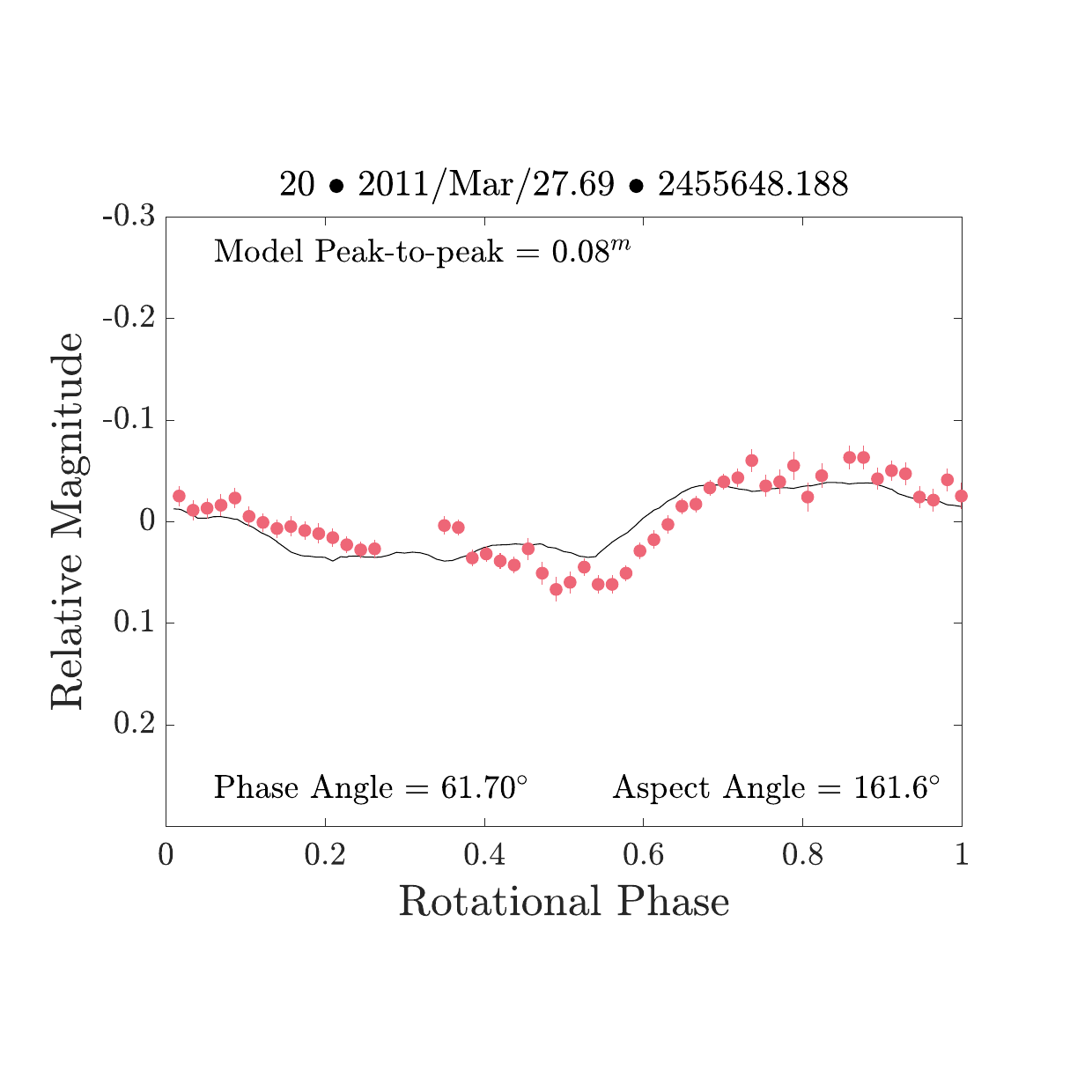}
	}
	\caption{A comparison of the observed and synthetic lightcurves of (23187) 2000 PN9 for each lightcurve listed in Table 2. The synthetic lightcurves were generated using the combined radar and lightcurve model presented in Section 3.2 using a combination of the Lambertian and Lommel-Seelinger scattering models. Synthetic lightcurves are plotted as solid black lines, and observational data as red points. Lightcurve 12, which was not used in the analysis, has possible issues with the CCD that may be responsible for the poor fit.}
	\label{fig:synthLC_radar2}
\end{figure*}

\begin{figure*}
	
	\resizebox{\hsize}{!}{	
        \includegraphics[width=.25\textwidth, trim=0.5cm 2.5cm 1.5cm 2.5cm, clip=true]{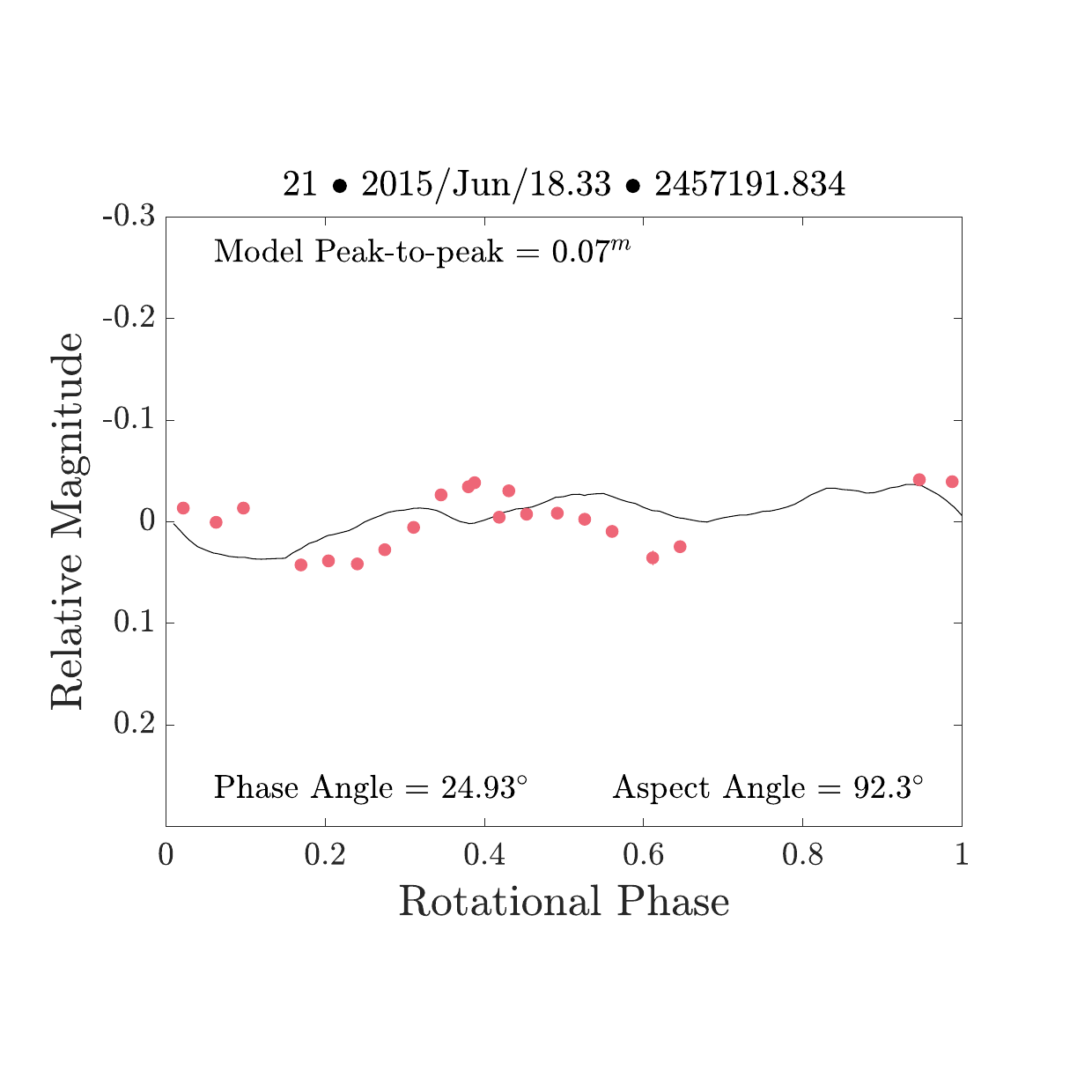}
        \includegraphics[width=.25\textwidth, trim=0.5cm 2.5cm 1.5cm 2.5cm, clip=true]{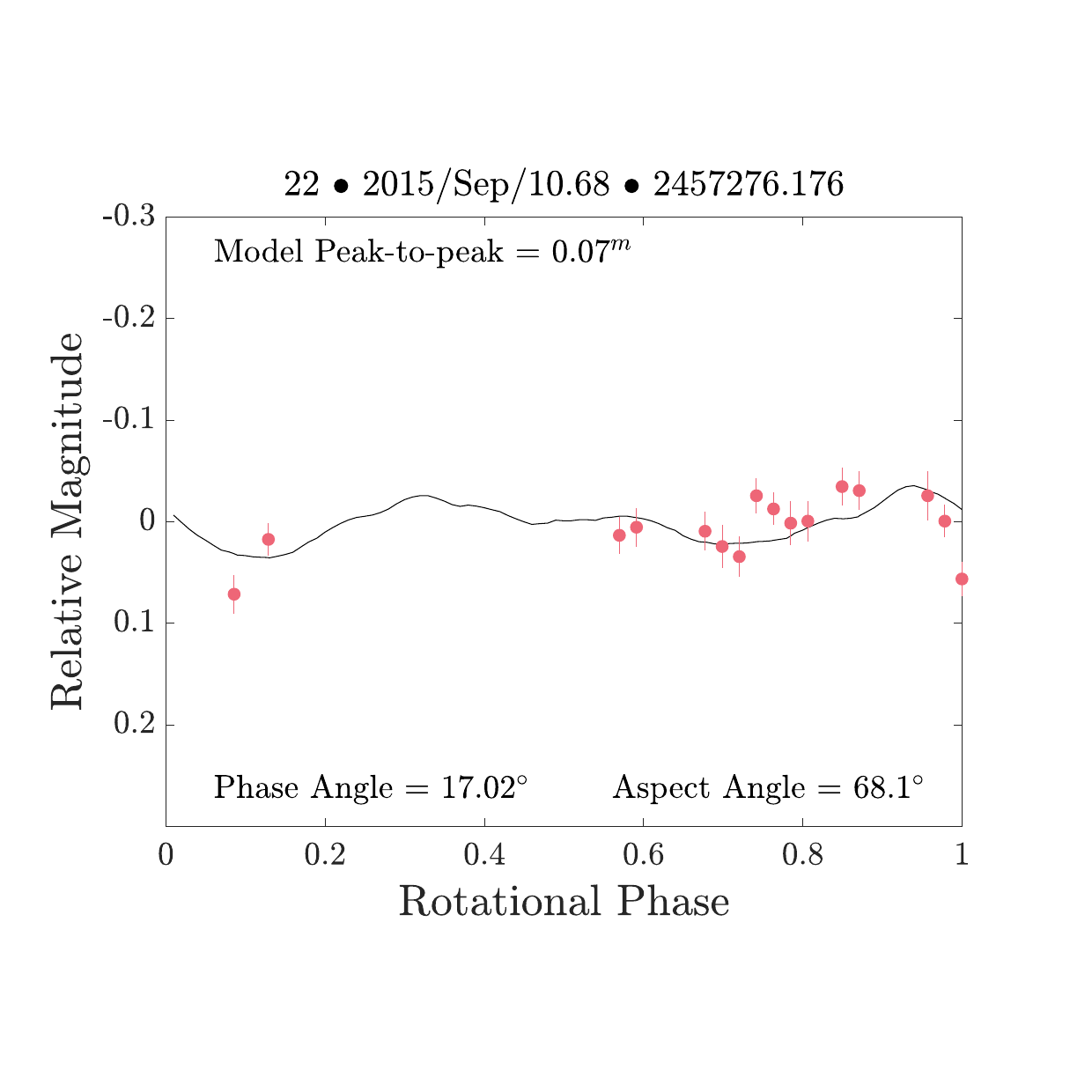}
        \includegraphics[width=.25\textwidth, trim=0.5cm 2.5cm 1.5cm 2.5cm, clip=true]{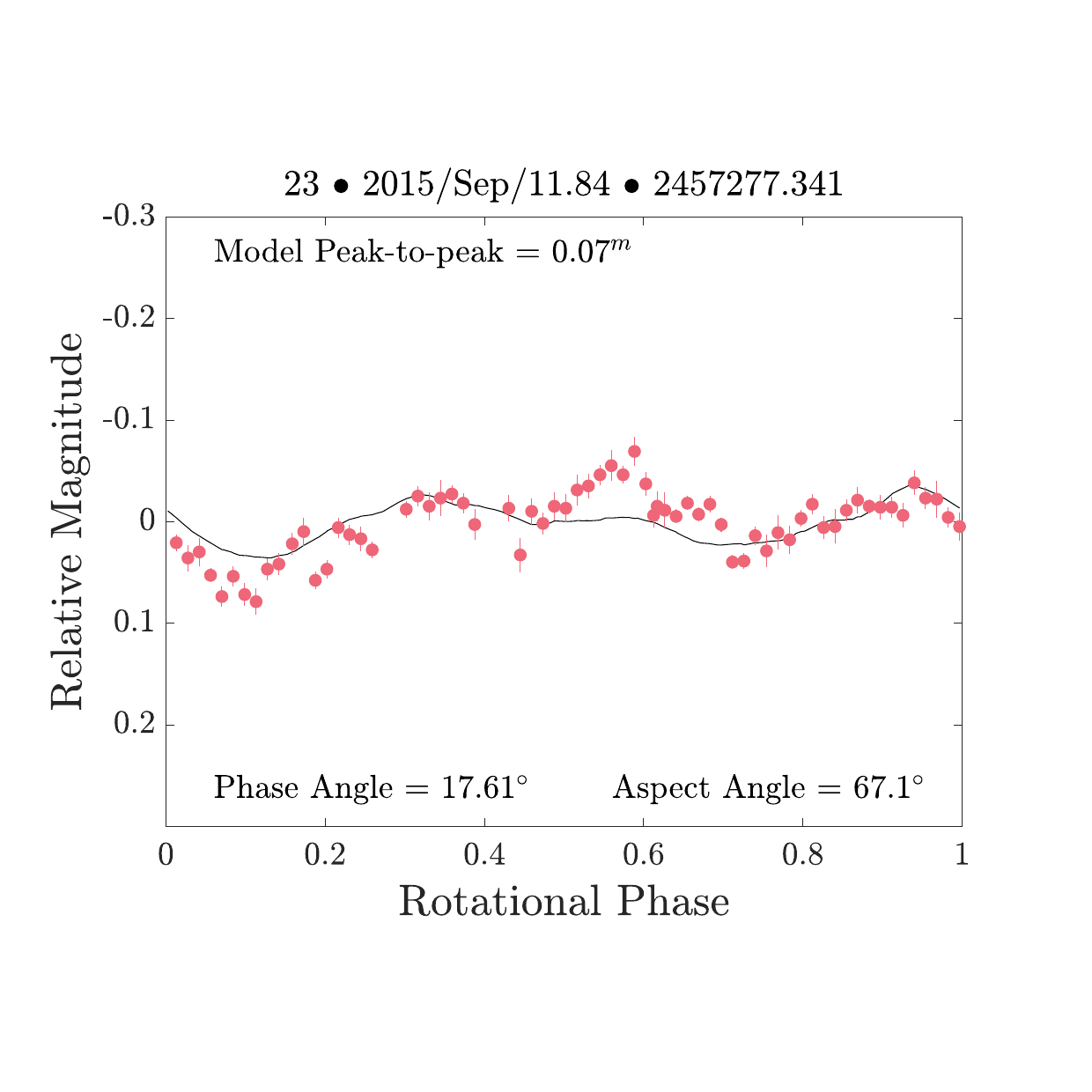}        
        \includegraphics[width=.25\textwidth, trim=0.5cm 2.5cm 1.5cm 2.5cm, clip=true]{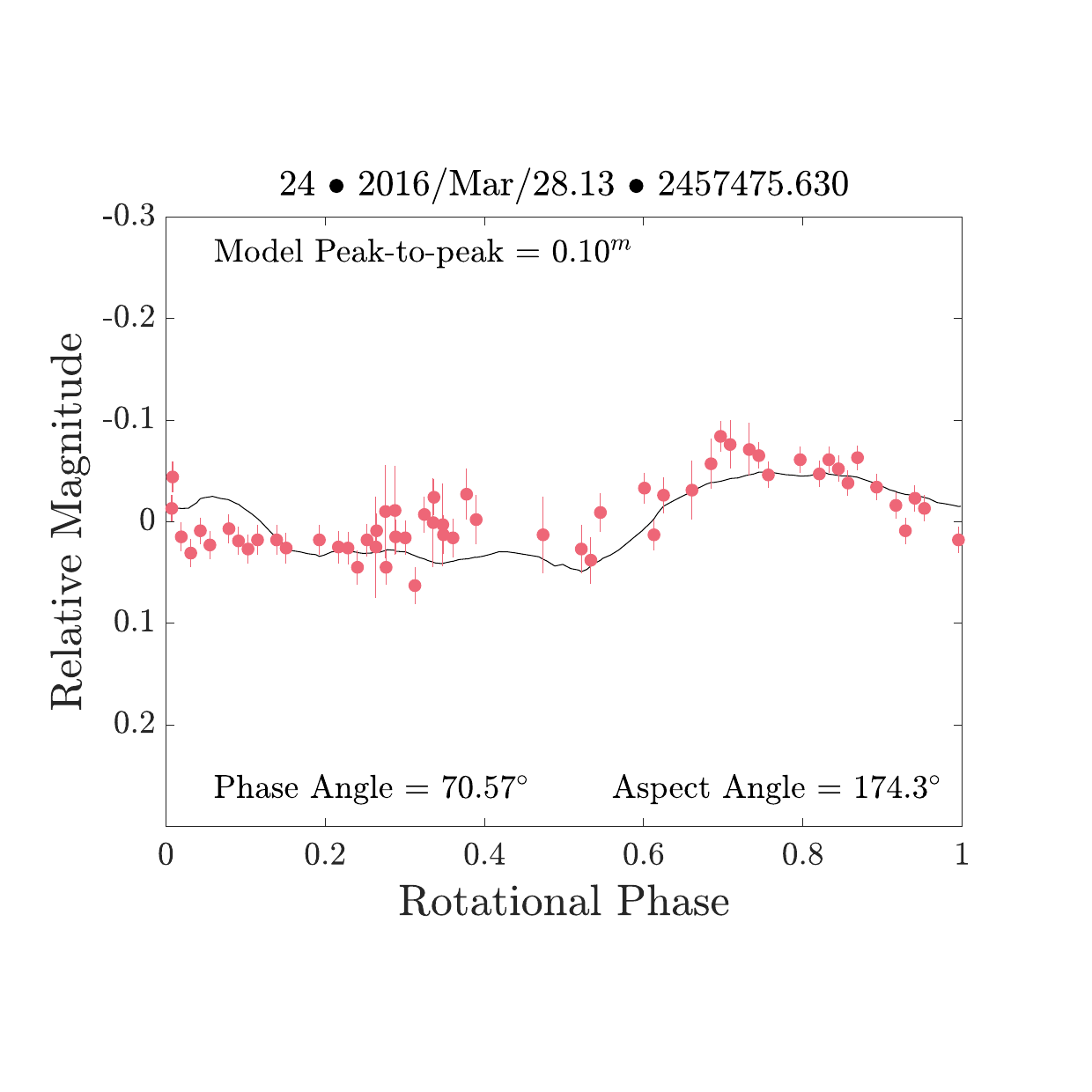}
	}
	
	\resizebox{\hsize}{!}{	
        \includegraphics[width=.25\textwidth, trim=0.5cm 2.5cm 1.5cm 2.5cm, clip=true]{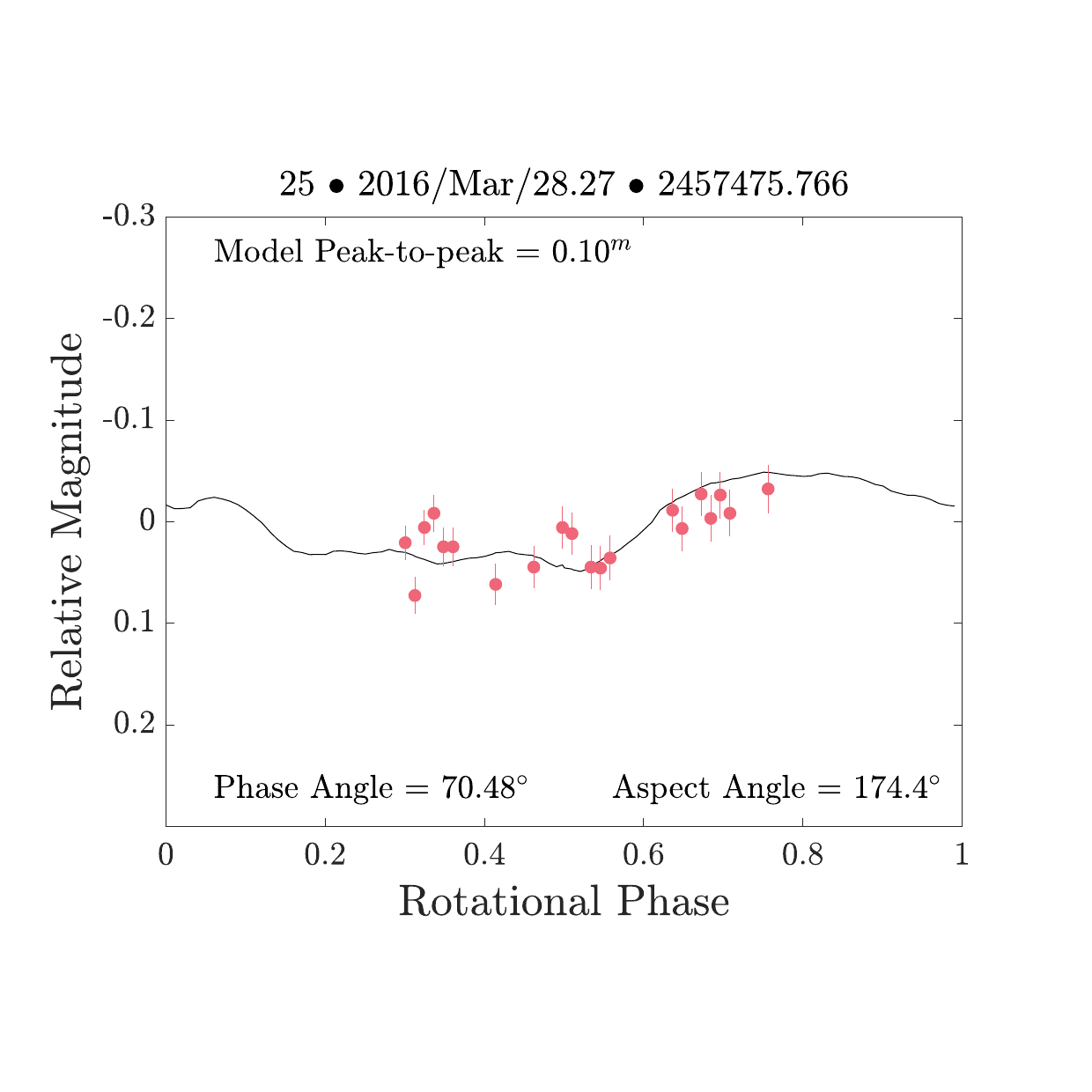}
        \includegraphics[width=.25\textwidth, trim=0.5cm 2.5cm 1.5cm 2.5cm, clip=true]{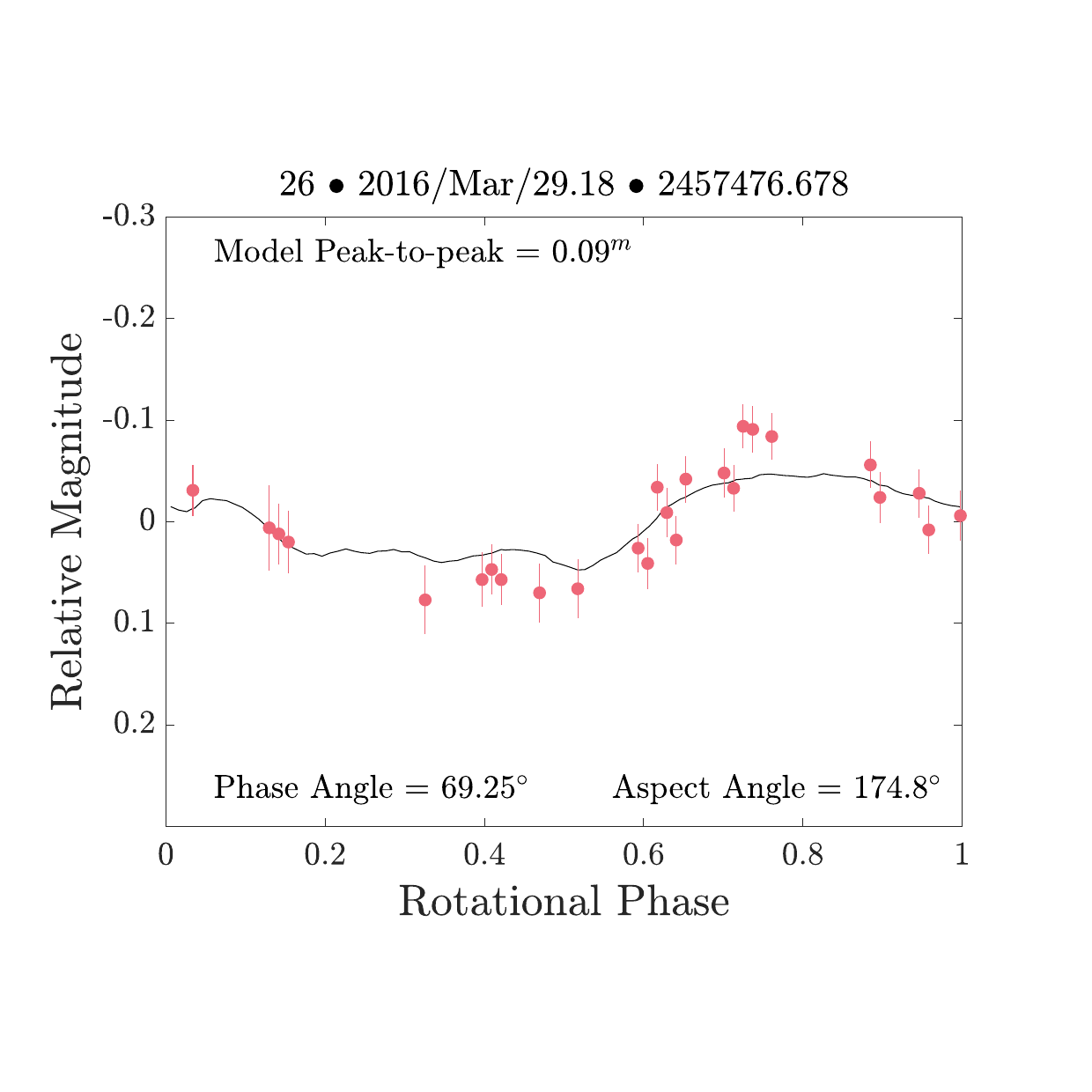}		
        \includegraphics[width=.25\textwidth, trim=0.5cm 2.5cm 1.5cm 2.5cm, clip=true]{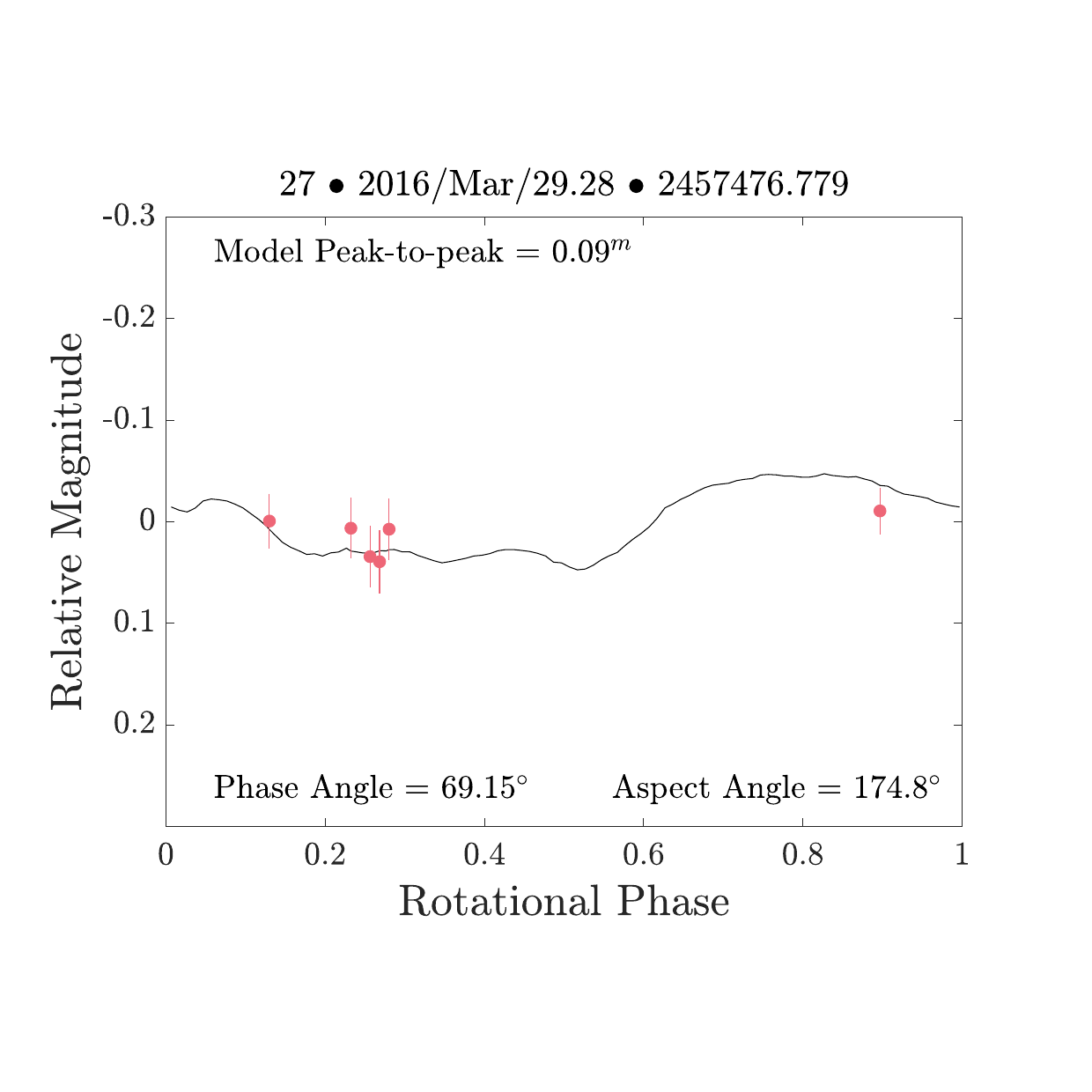}
        \includegraphics[width=.25\textwidth, trim=0.5cm 2.5cm 1.5cm 2.5cm, clip=true]{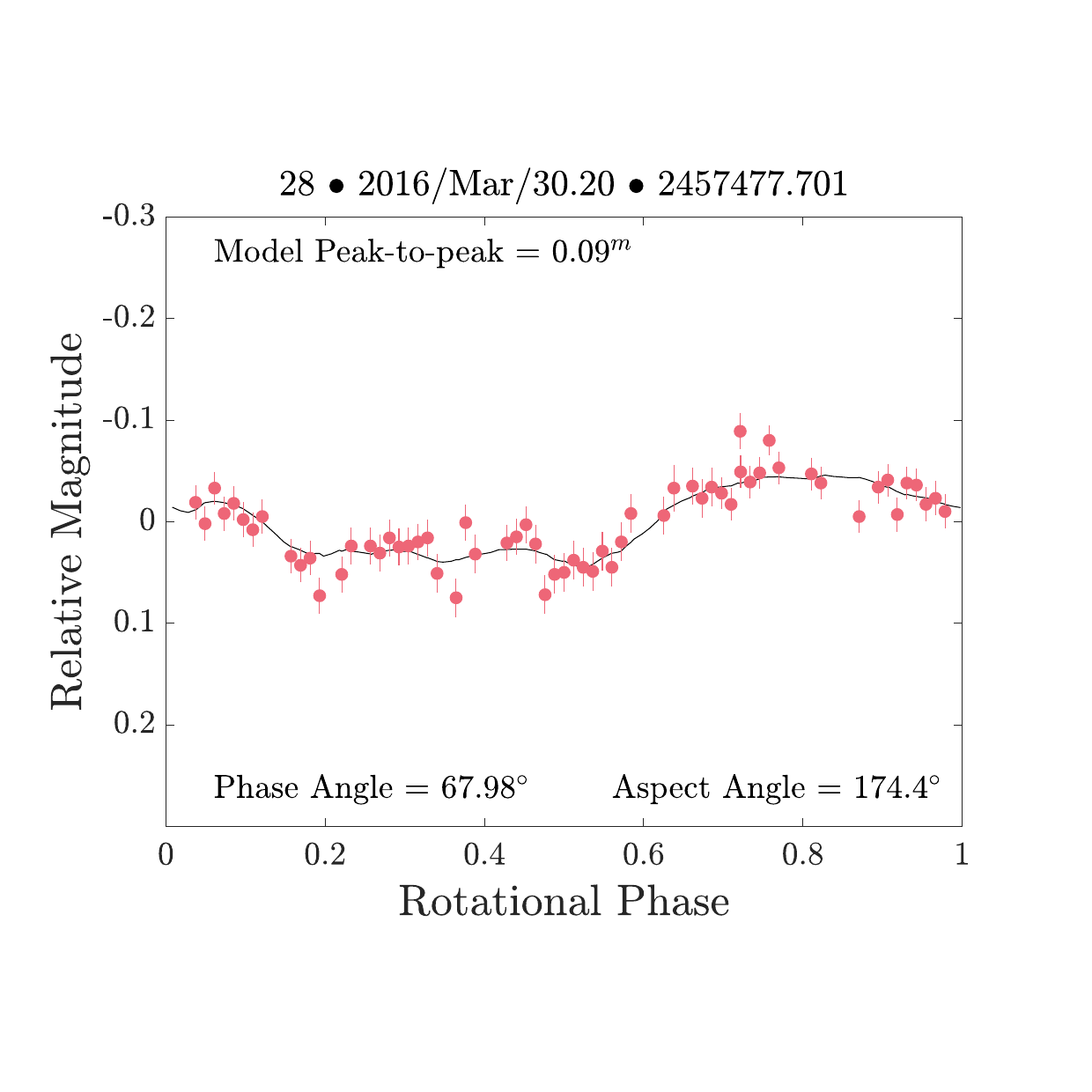}
	}	
 
    \resizebox{\hsize}{!}{	
        \includegraphics[width=.25\textwidth, trim=0.5cm 2.5cm 1.5cm 2.5cm, clip=true]{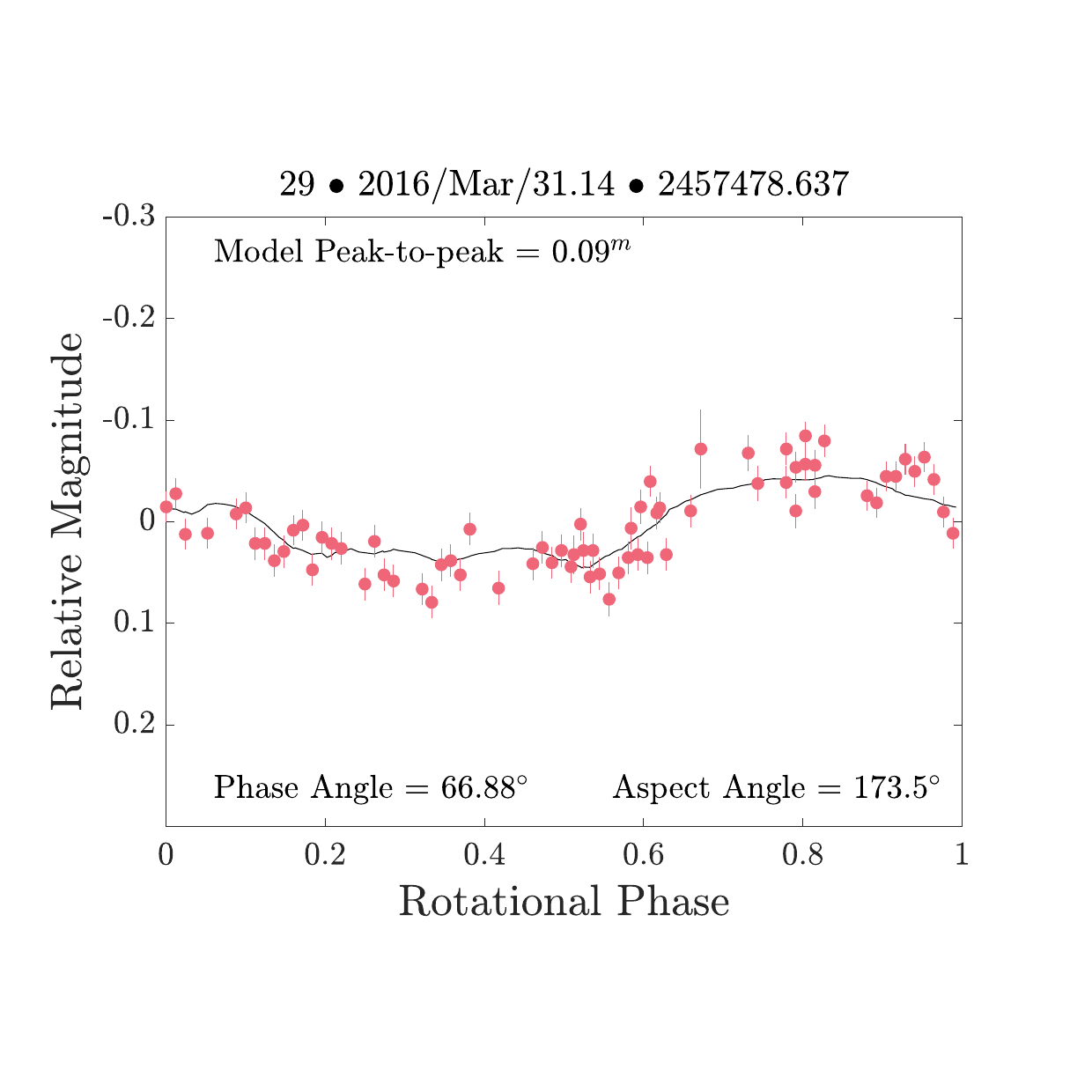}        
        \includegraphics[width=.25\textwidth, trim=0.5cm 2.5cm 1.5cm 2.5cm, clip=true]{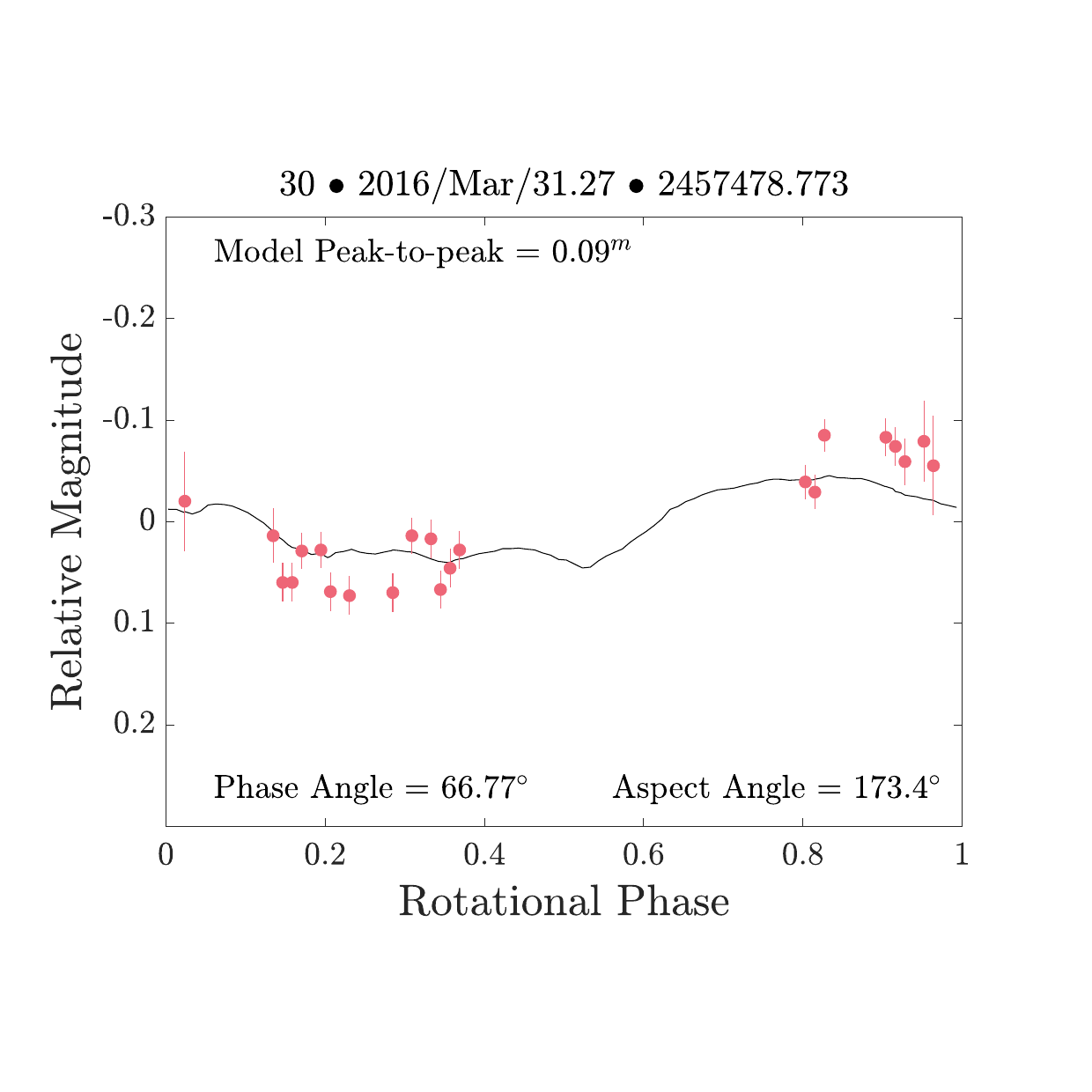}
        \includegraphics[width=.25\textwidth, trim=0.5cm 2.5cm 1.5cm 2.5cm, clip=true]{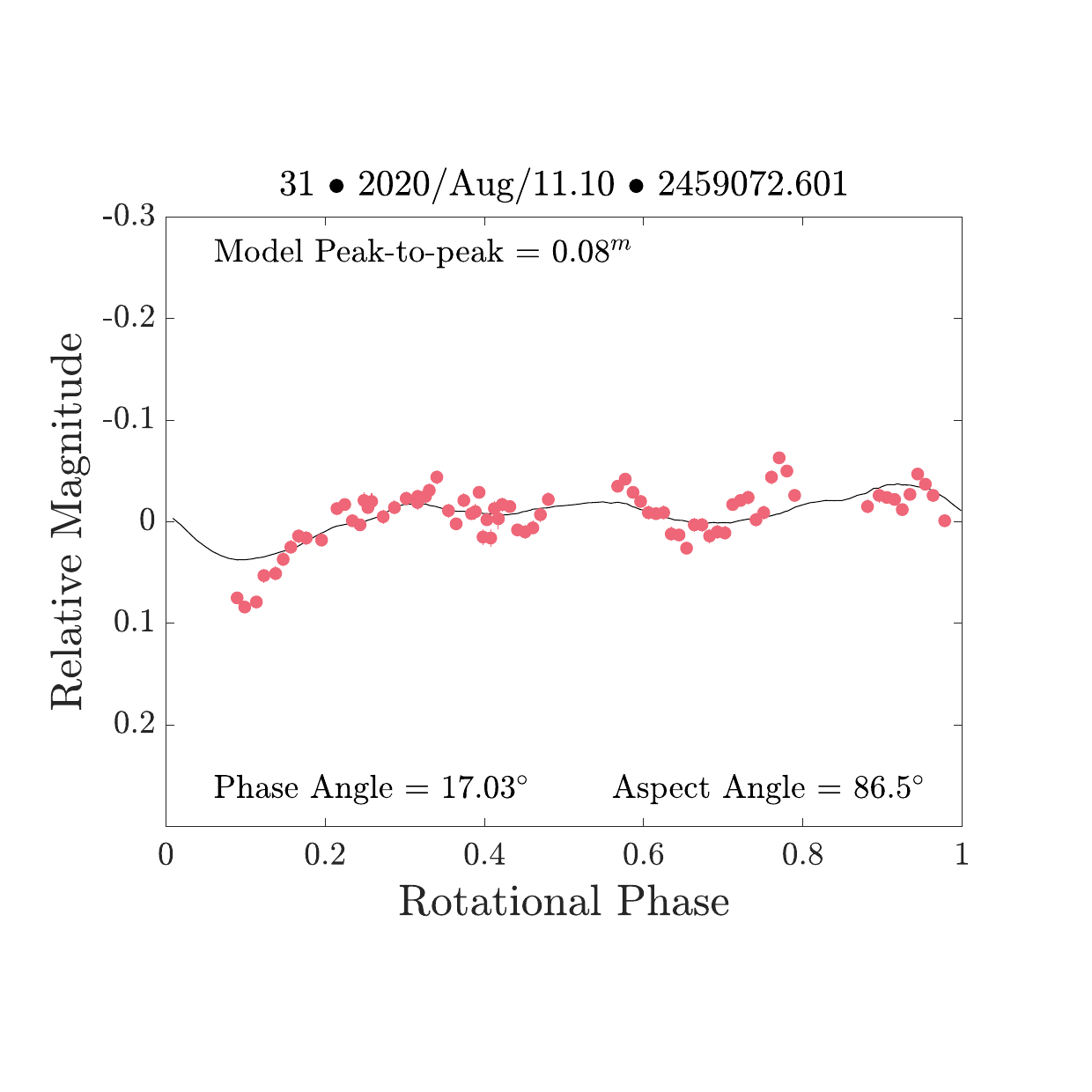}
        \includegraphics[width=.25\textwidth, trim=0.5cm 2.5cm 1.5cm 2.5cm, clip=true]{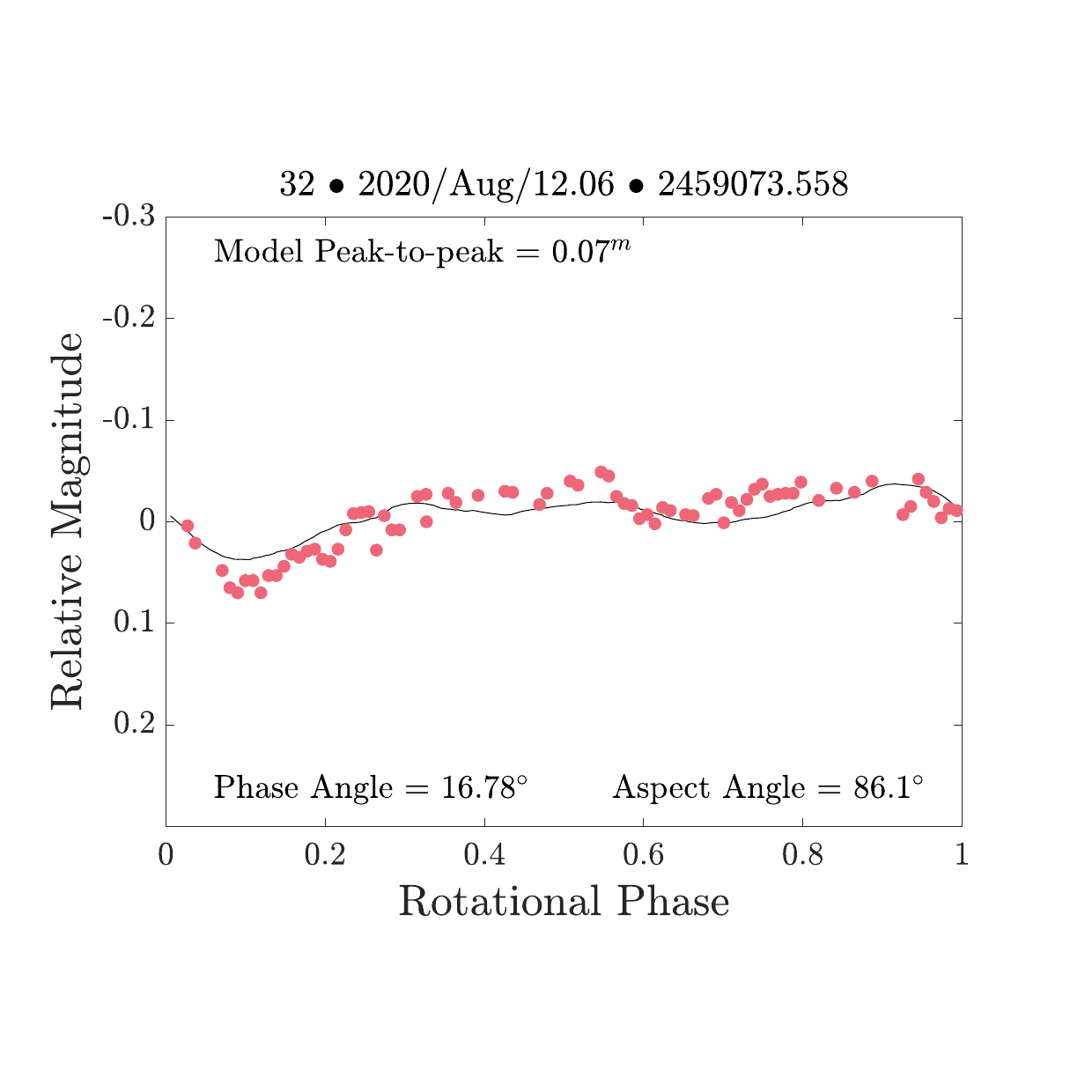}
	}

	\resizebox{0.75\textwidth}{!}{	
        \includegraphics[width=.25\textwidth, trim=0.5cm 2.5cm 1.5cm 2.5cm, clip=true]{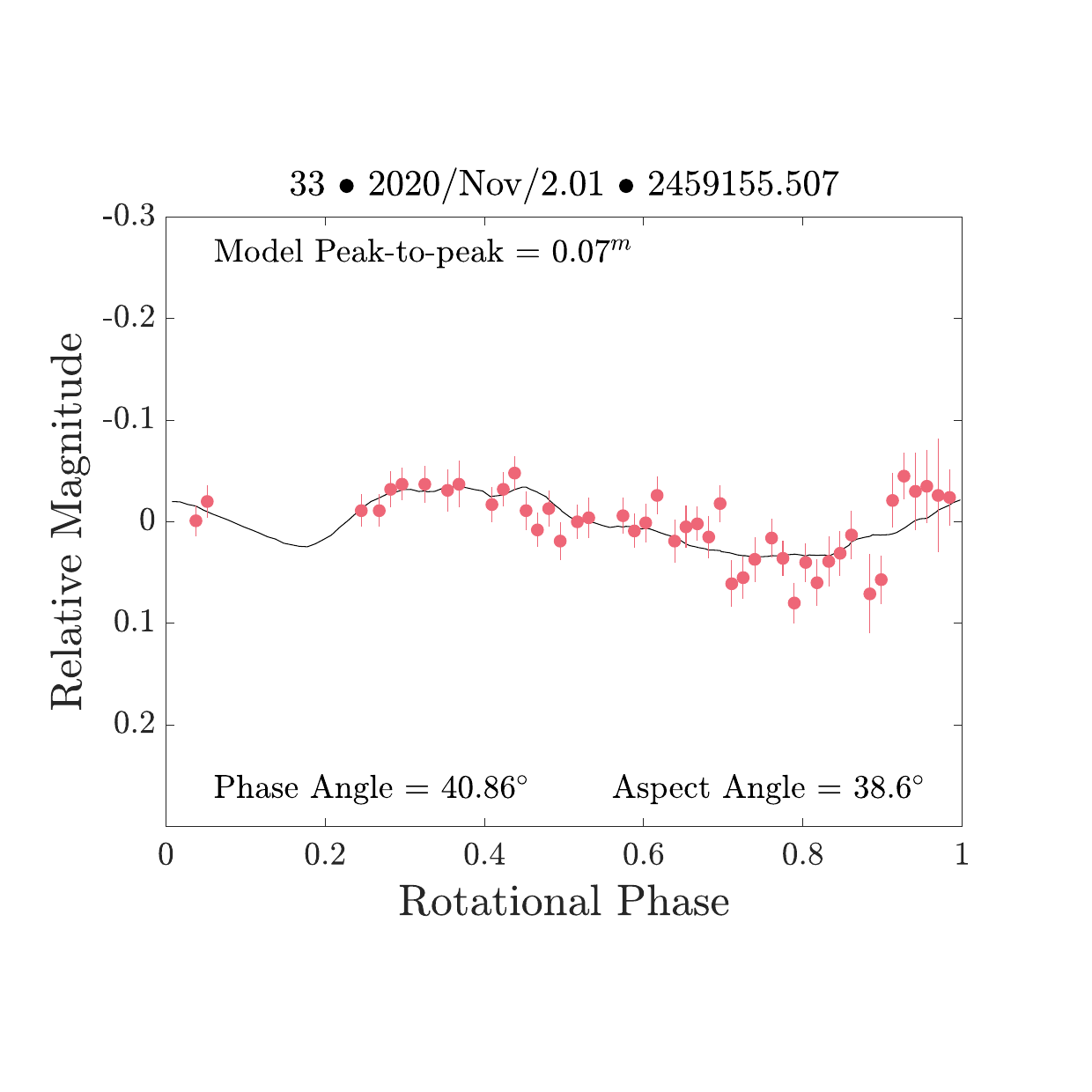}
        \includegraphics[width=.25\textwidth, trim=0.5cm 2.5cm 1.5cm 2.5cm, clip=true]{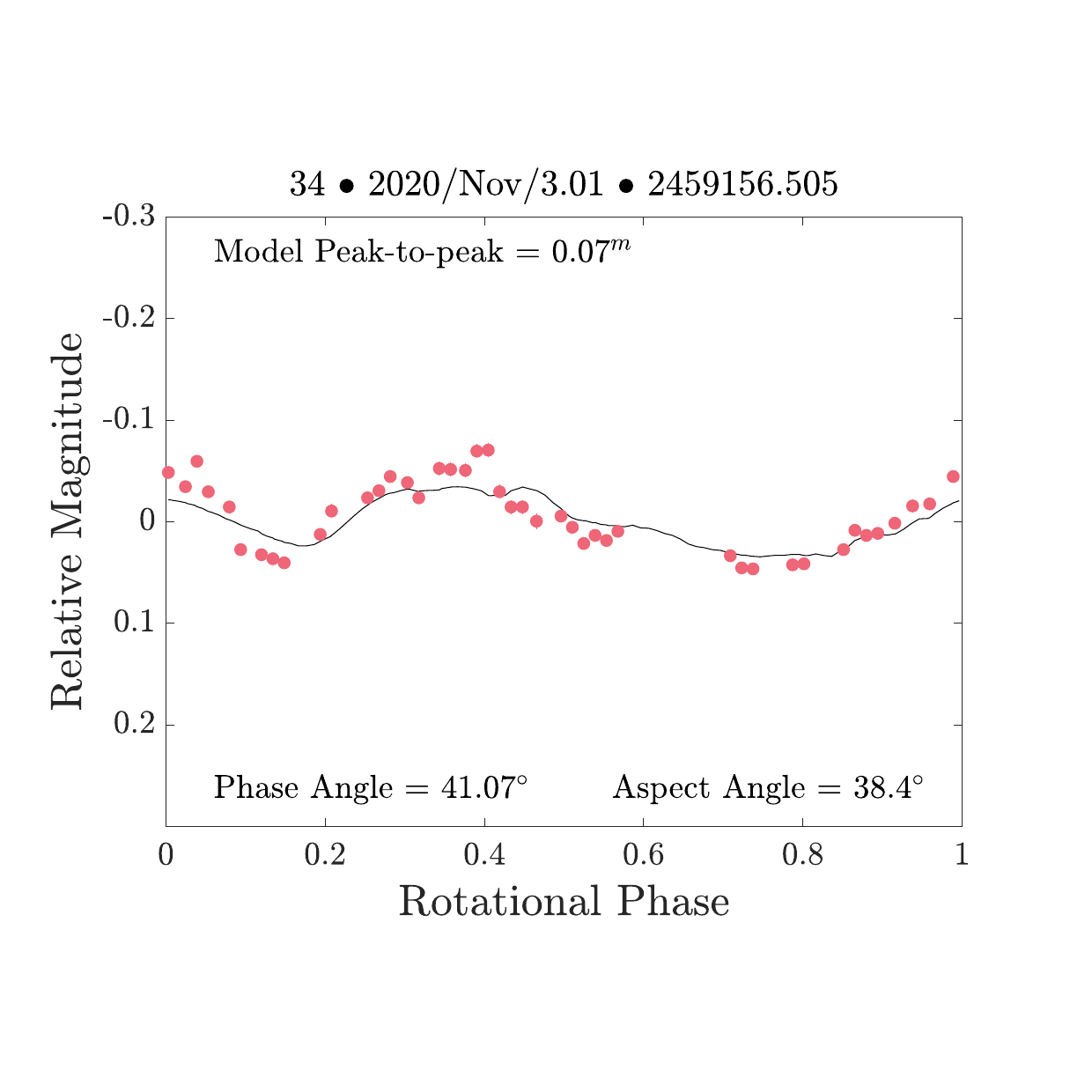}
        \includegraphics[width=.25\textwidth, trim=0.5cm 2.5cm 1.5cm 2.5cm, clip=true]{images/23187_radarLCfit_35.pdf}        
	}	
	\caption*{(continued)}
\end{figure*}


\label{lastpage}